%% file: main.tex
\documentclass[a4paper,12pt,twoside,openright]{book}

\usepackage[T1]{fontenc} 

\usepackage[left=3.5cm,right=3.5cm,top=3.5cm,bottom=3.5cm]{geometry} 
\usepackage{xcolor} 
\definecolor{chapterblue}{RGB}{35,35,110}
\definecolor{linkblue}{rgb}{0,0,0.8}
\definecolor{citeblue}{rgb}{0,0,0.9}

\usepackage[hidelinks,colorlinks=true,linkcolor=linkblue, citecolor=citeblue, urlcolor={blue!80!black}]{hyperref} 

\usepackage{newtxtext,newtxmath} 
\usepackage{titlesec} 
\usepackage{microtype} 

\usepackage{pdfpages}

\titlespacing{\chapter}{0pt}{0pt}{40pt}
\titleformat{\chapter}[display]
{\normalfont\bfseries\color{chapterblue}}
{\filleft\hspace*{-60pt}
	\rotatebox[origin=c]{90}{
		\normalfont\color{black}\Large
		\textls[180]{\textsc{\chaptertitlename}}
	}\hspace{10pt}
	{\setlength\fboxsep{0pt}
		\colorbox{chapterblue}{\parbox[c][3cm][c]{2.5cm}{
				\centering\color{white}\fontsize{80}{90}\selectfont\thechapter}}}
}
{10pt}
{\titlerule[2.5pt]\vskip3pt\titlerule\vskip4pt\LARGE\sffamily}

\usepackage[backend=biber, sorting=none, style=numeric-comp]{biblatex} 
\AtEveryBibitem{
    \clearfield{urlyear}
    \clearfield{urlmonth}
    \clearfield{extra}
    \clearlist{language}
    \clearfield{note}
}
\usepackage{amsmath,amsfonts} 
\usepackage{graphicx} 
\usepackage{wrapfig} 
\usepackage{subfig} 
\usepackage{multirow} 
\usepackage{makecell} 
\usepackage{rotating} 
\usepackage{url} 
\usepackage{algorithm} 
\usepackage{algorithmic} 
\usepackage{minitoc} 
\usepackage{nicefrac} 
\usepackage[acronym,nonumberlist,nopostdot,toc]{glossaries} 
\glsdisablehyper 
\makeglossaries 
\input{glossary_acronyms} 
\usepackage{qrcode} 
\usepackage{upgreek} 
\usepackage{listings} 
\AtEndDocument{
  \clearpage
  \includepdf[pages=1,fitpaper]{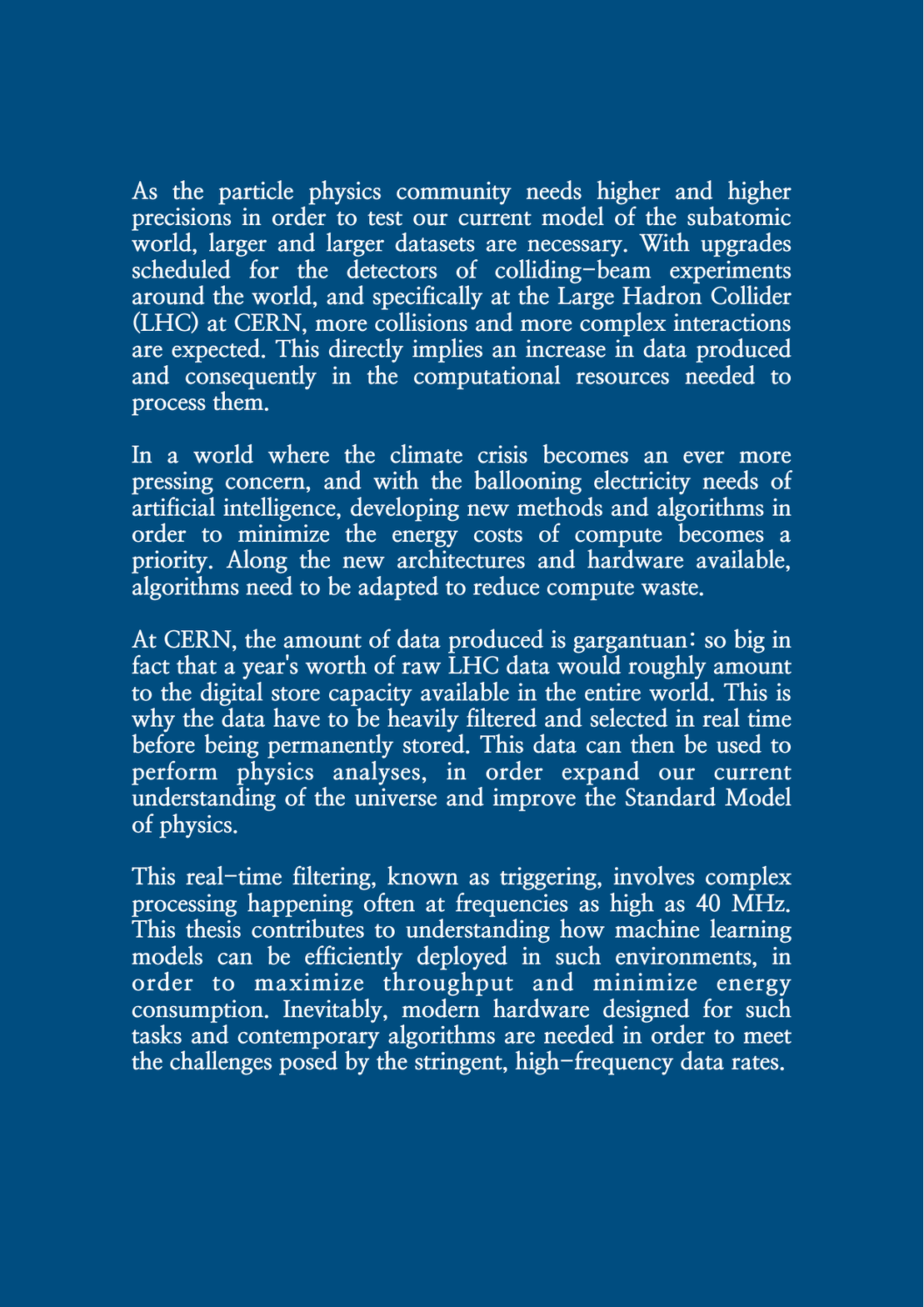}
}

\begin{document}

\frontmatter

\includepdf[pages=1,fitpaper]{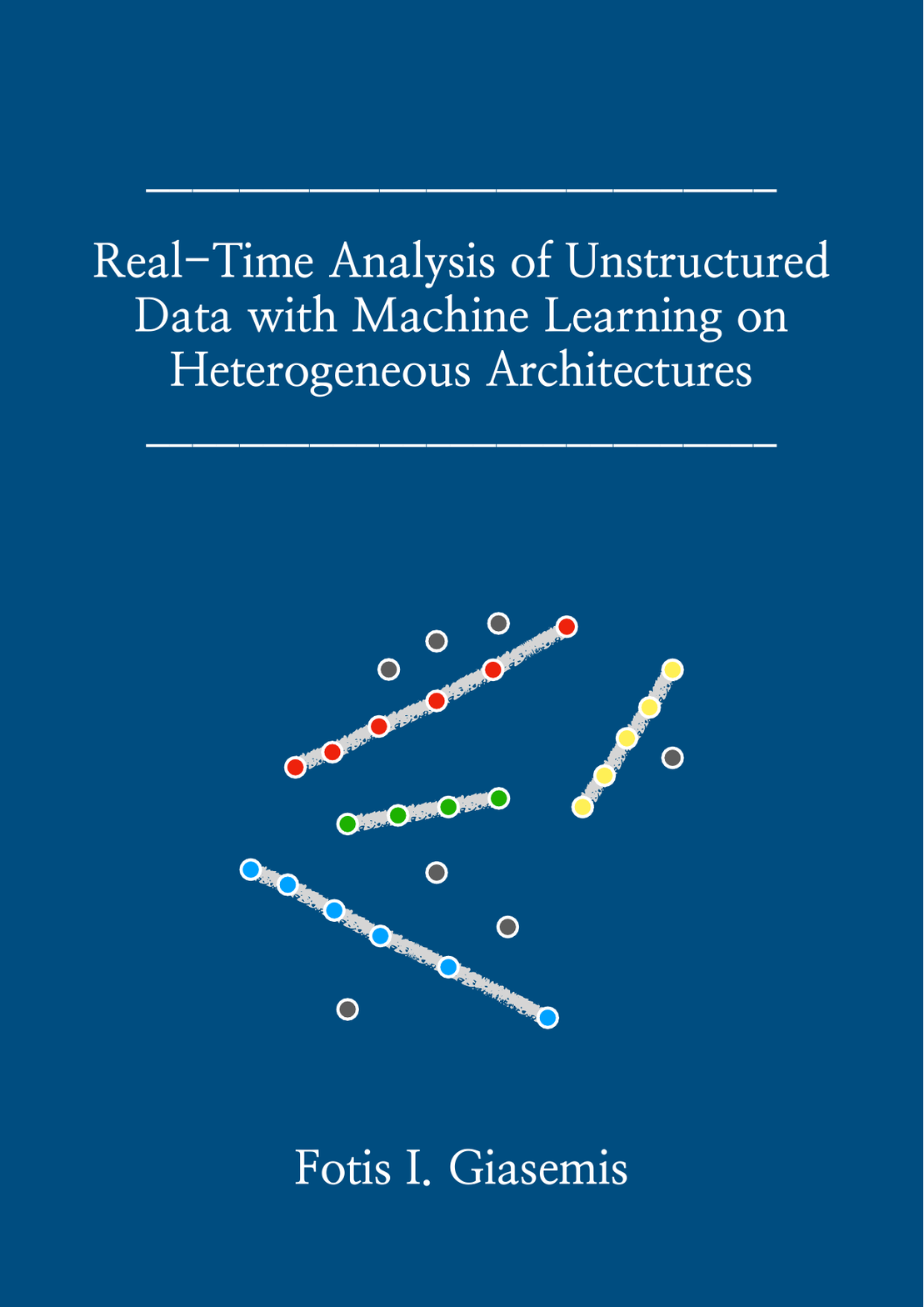}

\begingroup 
\let\newpage\relax
\let\clearpage\relax
\let\cleardoublepage\relax

\glsaddall 
\dominitoc 

\begin{titlepage} 
\begin{center}
{\scshape\LARGE Sorbonne Université\\}
{\large École Doctorale des Sciences de la Terre et de l'Environnement\\ et Physique de l'Univers, Paris\\}
\vspace{0.5cm}
\textsc{\Large Doctoral Thesis}\\

\rule{\linewidth}{0.5mm} \\[0.2cm]
{\huge \bfseries Real-Time Analysis of Unstructured Data with Machine Learning on Heterogeneous Architectures \\}\vspace{0.2cm}
\rule{\linewidth}{0.5mm} \\[0.2cm]
 
\begin{minipage}[t]{0.45\textwidth}
\centering
\emph{Author:}\\
Fotis I. Giasemis
\end{minipage}
\hfill
\begin{minipage}[t]{0.45\textwidth}
\centering
\emph{Supervisors:}\\
Vladimir Vava Gligorov\\
Bertrand Granado
\end{minipage}

\vspace{0.5cm}

\large \textit{Thesis submitted in partial fulfillment\\ of the requirements for the degree of}\\[0.1cm]
\textit{Doctor of Philosophy}\\[0.4cm]
Sorbonne Université, \large \today\\[0.4cm]

 \includegraphics[height=0.08\textheight]{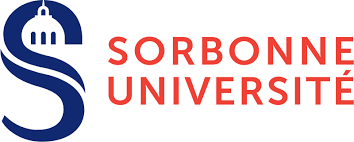}\\[0.2cm]

\vfill
\large Defended on September 5, 2025, before the committee below.
    
\vspace{0.5cm}
\begin{tabular}{l l l}
    Pierre & Astier & Jury president \\
    Jean Christophe & Prévotet & Reviewer \\
    David & Rousseau & Reviewer \\
    Eluned Anne & Smith  & Committee member \\
    Nicolas & Gac & Committee member \\
    Vladimir Vava & Gligorov & Supervisor \\
    Bertrand & Granado & Supervisor \\
\end{tabular}

\end{center}
\end{titlepage}

\endgroup 

\clearpage

\mbox{}
\vfill
\begin{wrapfigure}{L}{0.25\textwidth}
    \begin{center}
        \vspace{-\baselineskip}
        \includegraphics[width=0.25\textwidth]{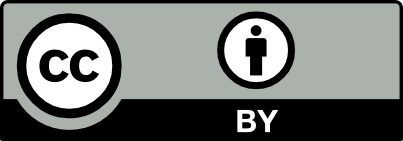}
    \end{center}
\end{wrapfigure}
\noindent
\textbf{Copyright}:\\
This thesis is licensed under a Creative Commons Attribution 4.0 International License. To view a copy of this license, visit \textbf{https://creativecommons.org/licenses/by/4.0/.}

\chapter*{Abstract}
\begin{center}
{\scshape \Large Sorbonne Université\\[1cm]}

{\Large \bfseries Real-Time Analysis of Unstructured Data with Machine Learning on Heterogeneous Architectures}\\[0.5cm]
    by\\[0.25cm]
    Fotis I. Giasemis\\[1cm]
    Doctor of Philosophy in High-Energy Physics\\

\noindent\textcolor{chapterblue}{\rule{0.6\textwidth}{0.4pt}}
\end{center}

\vspace{0.4cm}

As the particle physics community needs higher and higher precisions in order to test our current model of the subatomic world, larger and larger datasets are necessary. With upgrades scheduled for the detectors of colliding-beam experiments around the world, and specifically at the Large Hadron Collider (LHC) at CERN, more collisions and more complex interactions are expected. This directly implies an increase in data produced and consequently in the computational resources needed to process them.

In a world where the climate crisis becomes an ever more pressing concern, and with the ballooning electricity needs of artificial intelligence, developing new methods and algorithms in order to minimize the energy costs of compute becomes a priority. Along the new architectures and hardware available, algorithms need to be adapted to reduce compute waste.

At CERN, the amount of data produced is gargantuan: so big in fact that a year's worth of raw LHC data would roughly amount to the digital store capacity available in the entire world. This is why the data have to be heavily filtered and selected in real time before being permanently stored. This data can then be used to perform physics analyses, in order expand our current understanding of the universe and improve the Standard Model of physics.

This real-time filtering, known as triggering, involves complex processing happening often at frequencies as high as 40~MHz. This thesis contributes to understanding how machine learning models can be efficiently deployed in such environments, in order to maximize throughput and minimize energy consumption. Inevitably, modern hardware designed for such tasks and contemporary algorithms are needed in order to meet the challenges posed by the stringent, high-frequency data rates.

In this work, I present our graph neural network-based pipeline, developed for charged particle track reconstruction at the LHCb experiment at CERN. The pipeline was implemented end-to-end inside LHCb's first-level trigger, entirely on GPUs. Its performance was compared against the classical tracking algorithms currently in production at LHCb. The pipeline was also accelerated on the FPGA architecture, and its performance in terms of power consumption and processing speed was compared against the GPU implementation.

All in all, the work provides a thorough study of the nuances of deploying complex machine learning models in demanding, high-frequency data environments on heterogeneous computing architectures. Nonetheless, the field still has quite some progress to do in order to meet the challenges posed by the future accelerator experiments.

\chapter*{Résumé}
\begin{center}
{\scshape \Large Sorbonne Université\\[1cm]}

{\Large \bfseries Analyse en Temps Réel de Données Non Structurées à l'Aide de l'Apprentissage Automatique sur des Plateformes Hétérogènes}\\[0.5cm]
    par\\[0.25cm]
    Fotis I. Giasemis\\[1cm]
    Doctorat en Physique des Hautes Énergies\\

\noindent\textcolor{chapterblue}{\rule{0.6\textwidth}{0.4pt}}
\end{center}

\vspace{0.4cm}

La physique des particules nécessite des ensembles de données toujours plus volumineux pour atteindre une meilleure précision et tester notre modèle actuel du monde subatomique. Les expériences avec des accélérateurs de particules, notamment le Grand Collisionneur de Hadrons (LHC) au CERN, connaissent actuellement des améliorations majeures. Ces progrès génèrent un volume croissant de données, entraînant une hausse des besoins informatiques.

Face à la crise climatique et à l'augmentation rapide des besoins énergétiques de l'intelligence artificielle, il est essentiel de développer des méthodes et algorithmes réduisant la consommation énergétique des calculs. Ainsi, les algorithmes doivent être adaptés aux nouvelles architectures matérielles pour éviter le gaspillage énergétique.

Au CERN, les données produites par le LHC sont si volumineuses qu'une année de données brutes pourrait correspondre à la capacité mondiale totale de stockage numérique. Elles doivent donc être filtrées en temps réel avant stockage définitif, permettant des analyses approfondies pour affiner le Modèle Standard.

Ce filtrage en temps réel, ou « trigger », nécessite des traitements informatiques complexes à très haute fréquence (jusqu'à 40 MHz). Cette thèse explore l'efficacité du déploiement de modèles d'apprentissage automatique dans ces environnements exigeants, afin d'optimiser débit et consommation énergétique, en utilisant du matériel moderne et des algorithmes récents.

Je présente une chaîne de traitement basée sur des réseaux neuronaux à graphes, dédiée à la reconstruction des trajectoires de particules chargées pour l'expérience LHCb. Intégrée entièrement sur GPU dans le déclenchement de premier niveau, ses performances ont été comparées aux algorithmes traditionnels actuellement en production. Une implémentation accélérée sur FPGA a aussi été réalisée, permettant une comparaison de la consommation électrique et de la vitesse avec l'implémentation GPU.

En résumé, ce travail examine en profondeur les défis du déploiement d'apprentissage automatique dans des environnements à haut débit de données, utilisant différentes architectures de calcul, tout en identifiant les progrès nécessaires pour les expériences futures en physique des particules.

\cleardoublepage 
\thispagestyle{empty}
\vspace*{4cm}
\begin{center}
    \textit{
    To my parents\\[.5cm]
    and\\[.5cm]
    my sister.
    }
\end{center}

\chapter*{Acknowledgments}

I would like to thank my supervisors Vava and Bertrand for the support throughout these three years, and for helping me explore the vast domains tangent to this thesis's work. This interdisciplinary work, often more challenging because of communication barriers between scientific domains, proved fruitful and, most importantly, enjoyable. Vava led me to become an independent researcher by pushing me to ask the right physics questions, and Bertrand, with his LIP6 office two steps away, guided me into the exploration of a completely new field for me: FPGAs, GPUs, and parallelism. I gained so much knowledge over these three years---so much so that a 200-page thesis is not enough to contain it. The most important thing that I learned, however, is not to give up when things become difficult. For this, I would like to thank again Vava and Bertrand for listening to me complain repeatedly about problems which I was insisting were impossible to solve, and then proceeding to solve them one week later. Of course, I also have to thank Nabil Garroum for the many helpful and inspiring discussions we had at LPNHE. From LHCb, I would like to thank Roel Aaij, Dorothea vom Bruch, and Daniel Cámpora for helping me with the Allen framework.

I would like to thank the reviewers and all the jury members for taking the time to read this thesis and for the useful questions and feedback they gave me. I would also like to thank my \textit{comité de suivi}, Sophie Trincaz-Duvoid and Emmanuel Chailloux, for following me year after year and giving me helpful guidance on how to proceed at every stage of my PhD. I would also like to thank Sophie for giving me incredibly helpful advice on writing this thesis, for encouraging and supporting me during these three years, and of course for all the coffee she offered me at the LPNHE cafeteria during coffee breaks. I would also like to thank Julien Bolmont for helping me organize the defense in time.

I would also like to thank the PhD students before me---Alessandro Scarabotto, Lukas Calefice and Tommaso Fulghesu---for their advice in writing this thesis. I would like to thank all my friends from LPNHE, especially Pablo Correa, Laura Boggia, Paul Chabrillat, Gonzalo Diaz Lopez, Artur Oudot, Ugo Pensec, Anthony Correia, Lavinia Russo, Enya Van den Abeele, Aloïs Caillet, Line Delagrange, Marion Guelfand, Dylan Kuhn and Claudia de Dominicis. I would also like to thank my friends from the LIP6 office---Lokmane Demagh, Victor Enescu, Sylvain Takougang, Wenzheng Wang, Thomas Garbay---and Valentin Barbaza. This thesis would not have been possible without the lunches we shared and the countless coffee breaks. Thank you for all the support and the good times we had together. I would also like to wish the new LIP6 and LPNHE PhD students---Ghani Bourenane, Garance Lucas, Yuxuan Xi, Mounia Medjahed, Diego Mendoza---the very best for their future. From SMARTHEP, I would like to thank Caterina Doglioni for leading the project and all my friends from the network, especially Pratik Jawahar. I would also like to thank my other friends here in Paris, especially Vassilis and Christiana, my friends in Greece and the friends that I made in Oxford. 

\medskip
\noindent Finally, I would like to thank my family for supporting me throughout these years.

\bigskip
\noindent Fotis,\\
Paris, August 1, 2025

\chapter*{Funding Acknowledgment}
This work is part of the SMARTHEP network and it is funded by the European Union's Horizon 2020 research and innovation program, call H2020-MSCA-ITN-2020, under Grant Agreement n. 956086.

\vfill 

\begin{figure}[h!]
    \centering
    \includegraphics[height=1.5cm]{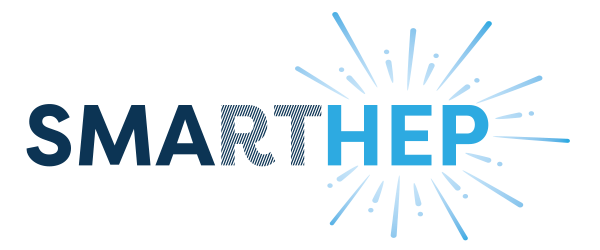} \hspace{0.2cm}
    \includegraphics[height=1.5cm]{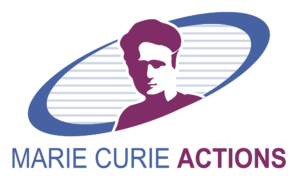} \hspace{0.2cm}
    \includegraphics[height=1.5cm]{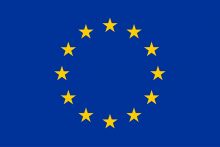}
\end{figure}

\tableofcontents

\cleardoublepage
\phantomsection
\addcontentsline{toc}{chapter}{\listfigurename}
\listoffigures

\cleardoublepage
\phantomsection
\addcontentsline{toc}{chapter}{\listtablename}
\listoftables

\printglossaries

\chapter{Research Context and Resources}

\section*{Research Context and Collaborations}

This interdisciplinary thesis, combining computer science and particle physics, was conducted as part of the Marie Sk\l{}odowska-Curie Actions (MSCA) SMARTHEP network~\cite{gooding_smarthep,smarthep_smarthep_nodate}, in the context of the Early-Stage Researcher (ESR) 5~\cite{smarthep_esr5_nodate} position. It took place at the LIP6 (computer science) and LPNHE (physics) laboratories and in collaboration with the LHCb experiment at CERN. During the course of the thesis, I undertook two secondments: one working on machine learning inference on FPGAs at CERN and the other working on traffic anomaly detection at the University of Lund and in collaboration with Ximantis~\cite{ximantis_ximantis_nodate}.

\section*{Digital Version}

The digital version of the thesis can be found on arXiv at \href{https://doi.org/10.48550/arXiv.2508.07423}{arXiv.2508.07423}.

\section*{Code Accessibility}

The code from all the projects is organized under the GitLab group Geometric Deep Learning for High-Energy Physics (GDL4HEP)~\cite{gdl4hep_geometric_nodate}.

\section*{Note on Typesetting}

This thesis was typeset using the \LaTeX{} document preparation system. The writing was carried out using Overleaf, an online collaborative \LaTeX{} editor, and compilation using pdfLaTeX. Plots were generated using the Matplotlib package in Python, while diagrams and illustrations using Keynote for macOS or the TikZ package, unless stated otherwise.

\section*{Feedback}

I welcome any feedback or corrections related to this work. My contact details can be found at \href{https://www.fotisgiasemis.com/about}{fotisgiasemis.com/about}.

\chapter{Publications and Original Work}

\section*{Original Work}

\noindent My own original contributions are described in Chapters~\ref{ch:etx4velo}, \ref{ch:etx4velo-gpu} and \ref{ch:etx4velo-fpga}. My work during the SMARTHEP secondment at Lund University can be found in \cite{giasemis_learning_2025}. The work in Chapters~\ref{ch:etx4velo} and \ref{ch:etx4velo-gpu}, the development of ETX4VELO and its implementation on GPU, was carried out in collaboration with my fellow PhD student, Anthony Correia, during the course of this thesis. In contrast, the work presented in Chapter~\ref{ch:etx4velo-fpga}, the partial implementation of the ETX4VELO pipeline on FPGA and the comparative studies between GPUs and FPGAs, was conducted solely by me. The work on traffic anomaly detection, in~\cite{giasemis_learning_2025}, was also done independently. Finally, I also contributed to the work presented in the SMARTHEP whitepapers~\cite{smarthep_network_review_2025}. A full list of my contributions is presented below.

\vspace{1em}
\noindent I confirm that this document is entirely my own work and that I did not use AI tools to generate any part of it.

\section*{Publications}

\begin{itemize}
    \item \textbf{Journals}
    \begin{itemize}
        \item ``Graph Neural Network-Based Track Finding in the LHCb Vertex Detector'', Journal of Instrumentation, 2024~\cite{correia_graph_2024}.
    \end{itemize}
    \item \textbf{International Conferences and Workshops}
    \begin{itemize}
        \item ``High-Throughout GNN Track Reconstruction at LHCb'', 8th International Connecting the Dots (CTD) Workshop, Toulouse, France, 2023~\cite{correia_ctd_2023,correia_graph_2023}.
        \item ``High-Throughput GNN-Based Track Reconstruction on GPUs at LHCb'', 42nd International Conference on High Energy Physics (ICHEP), Prague, Czech Republic, 2024~\cite{giasemis_ichep_2024,correia_graph_2024}.
        \item ``High-Throughput GNN-Based Track Reconstruction on GPUs at LHCb'', Machine Learning for Jet Physics (ML4Jets) 2024 Workshop, Paris, France, 2024~\cite{giasemis_ml4jets_2024,correia_graph_2024}.
        \item ``Comparative Analysis of FPGA and GPU Performance for Machine Learning-Based Track Reconstruction at LHCb'', 23rd IEEE International New Circuits and Systems (NEWCAS) Conference, Paris, France, 2025~\cite{giasemis_comparative_2025}.
        \item ``Learning Traffic Anomalies from Generative Models on Real-Time Observations'', 10th IEEE International Conference on Signal and Image Processing (ICSIP), Wuxi, China, 2025~\cite{giasemis_learning_2025}.
        \item ``Comparative Analysis of FPGA and GPU Performance for Machine Learning-Based Track Reconstruction at LHCb'', Machine Learning for Jet Physics (ML4Jets) 2025 Workshop, Caltech, Pasadena, California, USA, 2025~\cite{giasemis_comparative_2025,giasemis_ml4jets_2025}.
    \end{itemize}
    \item \textbf{National Conferences}
    \begin{itemize}
        \item ``Graph Neural Network for Track Finding at LHCb'',  Journées de Rencontres Jeunes Chercheurs (JRJC) 2023, Saint-Jean-de-Monts, France, 2023~\cite{giasemis_jrjc_2023, giasemis_graph_2024}.
    \end{itemize}
    \item \textbf{Communications and Seminars}
    \begin{itemize}
        \item ``Comparative Analysis of FPGA and GPU Performance for Machine Learning-Based Track Reconstruction at LHCb'', FastML Co-Processor Meeting, 2025~\cite{giasemis_comparative_2025,giasemis_co-processor_2025}.
    \end{itemize}

    \item \textbf{SMARTHEP Network Whitepapers}
    \begin{itemize}
        \item ``Review of Machine Learning for Real-Time Analysis at the Large Hadron Collider Experiments ALICE, ATLAS, CMS and LHCb'', arXiv, 2025~\cite{smarthep_network_review_2025}.
        \item \textit{Hybrid Architectures Whitepaper}, currently under internal review.
    \end{itemize}
\end{itemize}

\raggedbottom

\mainmatter

\chapter{Introduction}
\label{ch:intro}

The reconstruction of charged particle trajectories, or tracking, is an important step of the data processing involved in modern High-Energy Physics (HEP) collider experiments. It is often used as a tool to distinguish interesting physics processes from a background of uninteresting ones. With the colliders' granularity gradually increasing, and with an instantaneous luminosity rising after each upgrade of the detectors and the LHC machinery, the number of collisions, and subsequently the amount of data to be processed, is amplified. Specifically for tracking, this implies that each collision snapshot now contains a much larger and denser collection of detector hits, making it far more challenging to determine which hits belong to the same particle.

This increase in the data needed to be processed makes it more essential to perform this filtering at the earliest stages of the processing pipeline, including in real time. Already, two out of the four main experiments at the Large Hadron Collider (LHC), LHCb~\cite{aaij_allen_2020,lhcb_collaboration_lhcb_2020} and ALICE~\cite{gorbunov_alice_2011,rohr_gpu_2017,eulisse_o2_2024}, perform tracking in software at the full LHC collision rate. This filtering process, the so-called \textit{trigger}, in the case of LHCb, is performed by a two-stage real-time processing system, of which the first stage is implemented on Graphics Processing Units (GPUs) and is called Allen.

The ATLAS~\cite{atlas_collaboration_technical_2017} and CMS~\cite{cms_collaboration_phase-2_2021} collaborations, the other two main experiments at the LHC, are in the process of constructing upgraded detectors designed to operate at the High-Luminosity LHC (HL-LHC). These detectors will be capable of handling particle collision rates up to four times higher than the current ATLAS and CMS detectors and nearly forty times greater than the current LHCb detector. At the HL-LHC, both ATLAS and CMS aim to increase the rate of software-based track reconstruction by approximately an order of magnitude compared to current levels. Additionally, CMS plans to implement partial track reconstruction on Field-Programmable Gate Arrays (FPGAs) at the full LHC collision rate~\cite{cms_collaboration_phase-2_2020}.

The HEP community is therefore faced with a challenge. In general, for computational challenges similar to the one described, there are usually two approaches: hardware-driven and software-driven. In other words, when we want to speed up an algorithm, we can either use a faster hardware to run the algorithm on or, if possible, speed up the algorithm itself. On the one hand, the first approach would be relying on specialized hardware that is suitable to perform specific computations more efficiently and at a higher frequency, such as GPUs or FPGAs. On the other hand, with the second approach we would focus more on improving the algorithms from a computational complexity point of view. Finally, one could try a combination of the two approaches.

Indeed, this has already happened inside high-energy physics. Tracking constitutes a significant portion of the computational budget across all four main LHC experiments. The computational cost of classical tracking algorithms roughly scales with the number of hits raised to the power of 2~\cite{campora_perez_search_2021} in order to maintain the required physics performance. Meanwhile, advancements in computing architectures are increasingly driven by machine learning and artificial intelligence applications. This includes the development of hardware, such as Google's Tensor Processing Units (TPUs)~\cite{sato_-depth_2017} and Nvidia's tensor cores~\cite{markidis_nvidia_2018} integrated into GPUs, optimized for efficient deep learning computations. Over the past decade, major high-energy physics experiments have successfully re-optimized their classical tracking algorithms to leverage current parallel computing architectures effectively. However, it is worth considering whether tracking algorithms based on neural networks could offer a more suitable long-term solution for the hardware that supports our reconstruction processes and whether they could be capable of utilizing the hardware resources available better than classical algorithms.

This question is currently intensively explored inside the field~\cite{golling_trackml_2019,calafiura_trackml_2018,amrouche_tracking_2020,amrouche_tracking_2023,choma_track_2020,caillou_novel_2024}. In particular, the Exa.TrkX collaboration~\cite{exatrkx_exatrkx_nodate} developed a graph neural network-based pipeline for track finding~\cite{ju_performance_2021}. This pipeline was initially designed for tracking detectors similar to those used by the ATLAS and CMS experiments, specifically for the high-luminosity upgrade of the LHC. Using this pipeline as a starting point, we developed ``Exa.TrkX for VELO (ETX4VELO)'', our own pipeline for track finding at LHCb. The pipeline is specifically focused on the detector of the LHCb experiment known as the Vertex Locator (VELO). In Chapter~\ref{ch:etx4velo}, I present the pipeline, its development process, its early version and its final version. I also study its performance and compare it with LHCb's first-level trigger. In Chapter~\ref{ch:etx4velo-gpu}, the implementation of the pipeline inside the LHCb trigger on GPUs is presented. Finally in Chapter~\ref{ch:etx4velo-fpga}, I present the implementation of one of the ETX4VELO models on the FPGA architecture and I compare various aspects of the inference of the models on FPGAs and GPUs.

In Chapter~\ref{ch:motivation}, I start with the motivation behind doing real-time analysis with machine learning on heterogeneous architectures for high-energy physics. The physics background is given in Chapter~\ref{ch:physics}, machine learning is introduced in Chapter~\ref{ch:ml}, while computing methods are described in Chapter~\ref{ch:hpc}. Then, more details about the LHCb experiment and about particle track reconstruction are given in Chapters~\ref{ch:lhcb} and \ref{ch:tracking}, respectively. I finish with a conclusion and future work in Chapter~\ref{ch:conclusion}.

Finally, the notations, units and physical constants used throughput the thesis are summarized in Appendix~\ref{app:notations}. Details regarding the early development of the ETX4VELO pipeline can be found in Appendix~\ref{app:early-dev}. Further resources can be found in Appendix~\ref{app:further}.

\chapter{Motivation}
\label{ch:motivation}

\setcounter{mtc}{8} 
\minitoc

\section*{Introduction}

In this chapter, we explore the motivation behind Real-Time Analysis (RTA) in HEP, and why it is of interest to do it using Machine Learning (ML) on heterogeneous architectures.

\section{Real-Time Analysis}

Real-time analysis, as the name suggests, is the processing of data in ``real time''. However, its definition varies widely between disciplines and even within these disciplines and their specific use cases. In general, these systems are of interest in scenarios where making a decision is constrained in some way by time. The manner of this constraint varies between the different scenarios. For example, on the one hand, a system developed for the control of self-driving cars, in order to ensure the safety of the passengers, may need to have a reaction time below a specific threshold, for example lower than a human driver. This response time is known as \textit{latency} and we would say that this system for autonomous vehicles is latency-constrained. This latency constraint is known as a \textit{hard} real-time constraint. As defined in~\cite{kopetz_real-time_1997}, ``A real-time constraint is called hard, if not meeting that constraint could result in a catastrophe''---car accidents that may even be fatal.

On the other hand, for example for multimedia streaming applications, timely processing is preferable, but delayed processing does not cause system failure or result in catastrophic consequences. However, failing to meet time expectations can lead to reduced output quality, such as the freezing of the frames of the video, etc. This type of real-time constraint is known as \textit{soft}.

In HEP, the main field of focus of this text, real-time analysis has numerous applications. The amount of data produced in HEP experiments is often enormous. To put the size in scale, the amount of data produced by the four main experiments at CERN can reach up to several tens of terabytes per second. With the experiments running a significant amount during the year, this results into an amount of data that is impossible to store, even if all the storage available on earth was used. For this reason, numerous HEP experiments have a system that filters the data, which is known as the trigger. The trigger processes the incoming data in real time, keeping only the ones that are ``interesting'' enough. A lot of the data produced are not interesting given that they verify the knowledge that we already have about how the world works at subatomic scales, the so-called Standard Model. Instead, interesting physics analyses concern processes that are more rare, and the filtering system is designed in order to select, or trigger on, these rare events~\cite{aaij_lhcb_2013,atlas_collaboration_atlas_2008,schorner-sadenius_trigger_2003,head_lhcb_2014}.

Many of these data processing systems are indeed latency-constrained. Hardware triggers, specifically, have to decide whether to keep or discard the incoming data every few tens of nanoseconds: otherwise, the data are permanently lost. This hard time limit is what ultimately constrains how much processing you can pack into the trigger. By contrast, other architectures prioritize \textit{throughput}---the amount of data processed per unit time. In software triggers, for example, the incoming data from the detectors have to be processed by the system in order to make a decision about whether they should be saved or not, and, as before, a failure of this decision could result in the permanent loss of the data. Assuming the data come in streams of data packets, known in HEP as events, a constraint on throughput is equivalent to a constraint on the average processing time of each data packet. As long as the average processing time is below a specific threshold, the processing time of a single data packet can exceed that threshold. These systems are known as throughput-constrained.

The design of a system subject to these constraints would be significantly different between the two cases. For systems where the main constraint is latency, the processing is designed such that the processing time for each individual request is as small as possible. This may include techniques in order to ``simplify'' the computation and/or use less resources. When the main constraint is throughput, the focus of the design is on processing as much data as possible in a fixed amount of time. This will most likely include techniques in order to make the computations parallel. Contrary to sequential execution, where the different calculations are performed one after the other and each calculation has to finish before the next one is started, parallel execution is when one or multiple calculations are performed on one or multiple data at the same time. In this way, the available resources are better utilized. However, parallelizing a processing chain can be challenging, especially in cases when there is a data dependency between the different calculations. The two cases are summarized in Table~\ref{tab:latency-throughput}.

\begin{table}
    \centering
    \begin{tabular}{p{.3\textwidth}|p{.3\textwidth}|p{.3\textwidth}}
    \hline \hline
    \textbf{Constrain} & \textbf{Latency} & \textbf{Throughput} \\ \hline
    Primary Goal & Minimize delay per task/request & Maximize tasks handled per unit time\\ \hline
    Typical Design & Focused on low-latency execution & Focused on parallelism of execution and efficiency \\ \hline
    Resource Utilization & May under-utilize resources to reduce latency & Optimized for maximum resource usage \\ \hline \hline
    \end{tabular}
    \caption{Comparison of key characteristics and trade-offs between latency-constrained and throughput-constrained systems.}
    \label{tab:latency-throughput}
\end{table}

\begin{figure}
    \centering
    \includegraphics[width=0.9\linewidth]{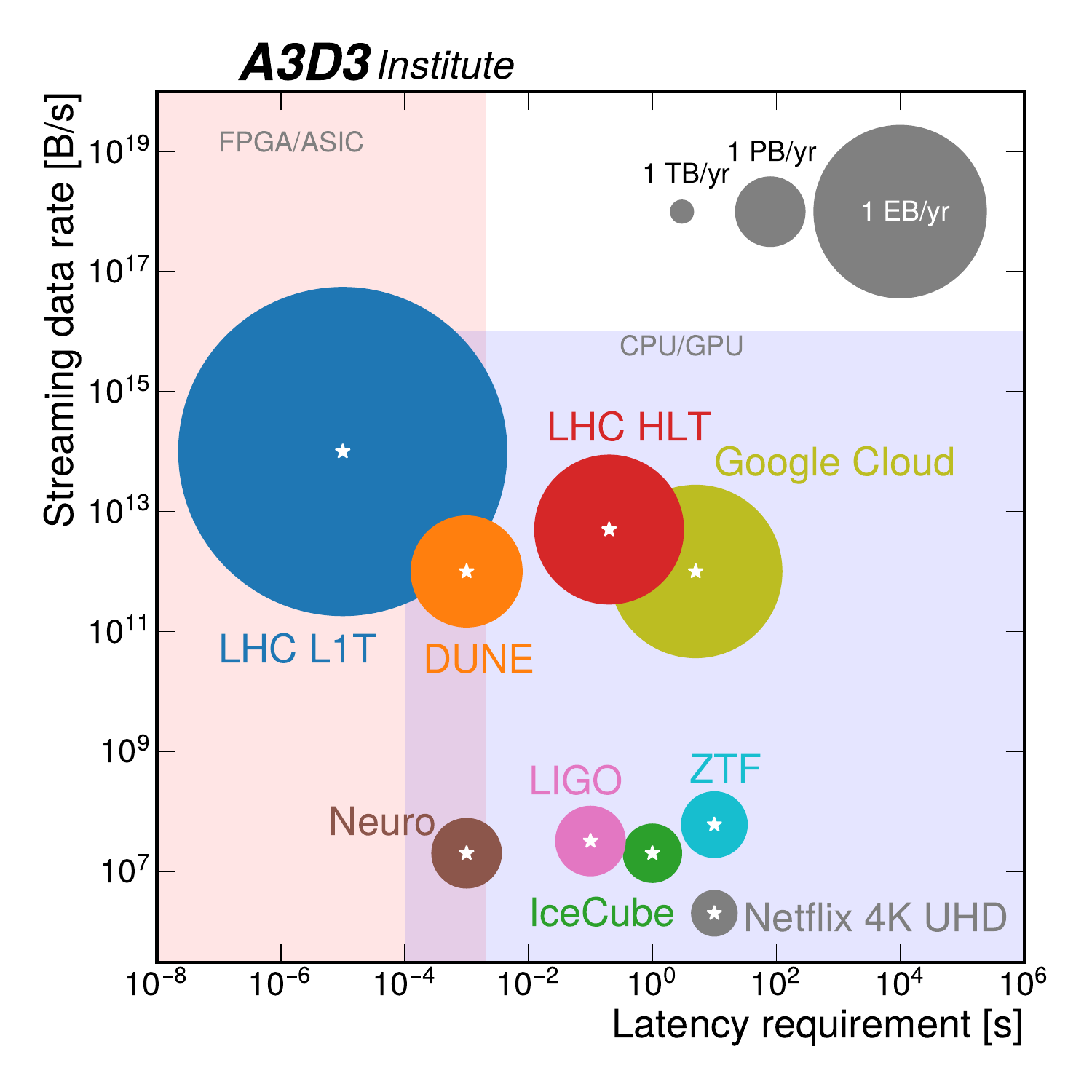}
    \caption{Comparison of the streaming data rates (in bytes per second) versus latency requirements (in seconds) across various experiments and domains, spanning high-energy physics to consumer-facing application such as Netflix. The traditional typical level-1 and high-level triggers at the LHC are labeled as ``LHC L1T'' and ``LHC HLT'', respectively. The area of markers is proportional to the total data volume. Figure from~\cite{a3d3_institute_accelerated_2025}.}
    \label{fig:a3d3}
\end{figure}

Real-time algorithms, also termed online, can have tremendously different time scales between different domains~\cite{harris_physics_2022}. For example, high frequency trading systems operate at sub-millisecond time scales while decision-making systems in autonomous vehicles operate on the order of a few hundred milliseconds. In HEP, real-time can mean anything from a few tens of nanoseconds all the way up to days. A comparison of various experiments is shown in Fig.~\ref{fig:a3d3}.

RTA has a wide variety of applications in science. Some of its applications are summarized below.

\begin{itemize}
    \item \textbf{Finance}
    \begin{itemize}
        \item Fraud Detection: Identifying anomalies in transactions, or other processes, in financial markets~\cite{borketey_real-time_2024}.
        \item Risk Management: Monitoring financial risk of positions in real time~\cite{abikoye_real-time_2024}.
        \item Stock Market Trading: Analyzing market data in real time, in order to execute trades~\cite{adlakha_real_2021,kalra_efficient_2024,jarunde_real-time_2023}.
    \end{itemize}

    \item \textbf{Healthcare}
    \begin{itemize}
        \item Patient Monitoring: Monitoring of vital signs of patients in real time in order to alert healthcare professionals~\cite{paganelli_real-time_2022}.
    \end{itemize}

    \item \textbf{Transportation}
    \begin{itemize}
        \item Fleet Management: Tracking vehicle locations and conditions in order to optimize their deployment and logistics~\cite{dinten_fleet_2023}.
        \item Traffic Management: Monitoring traffic using sensors and cameras in order to optimize the flow and reduce congestion~\cite{verma_real_2018,amini_big_2017}. In~\cite{giasemis_learning_2025}, traffic anomaly detection with a novel ML method is explored.
        \item Autonomous Vehicles: Processing navigation data from cameras and radars on the vehicle in order to make decisions about its movement~\cite{xie_edge_2024,kamel_real-time_2023}.
    \end{itemize}

    \item \textbf{Manufacturing and Industry}
    \begin{itemize}
        \item Predictive Maintenance: Analyzing data from sensors installed onto the machines and infrastructure in order to predict failures and schedule maintenance~\cite{belim_forecasting_2024,mahfoud_real-time_2024}.
    \end{itemize}

    \item \textbf{Energy and Utilities}
    \begin{itemize}
        \item Smart Grids: Managing the most efficient movement of energy from the production and storage grids to selling it, balancing electricity demand and supply~\cite{singh_deep_2025}.
        \item Renewable Energy Management: Adjusting operations based on variable weather conditions and predictions, and energy production~\cite{hadi_h_awaji_real-time_2024,elkazaz_optimization_2019}.
    \end{itemize}

    \item \textbf{Security and Surveillance}
    \begin{itemize}
        \item Threat Detection: Surveillance of network traffic in order to detect potential security threats~\cite{yarram_anomaly_2025}.
    \end{itemize}
    
\end{itemize}

\section{Real-Time Analysis in High-Energy Physics}

The HEP community, especially at the LHC, is preparing itself for a new era of unprecedented data rates~\cite{albrecht_hep_2018,shanahan_snowmass_2022}. In the wake of the High-Luminosity LHC~\cite{aberle_high-luminosity_2020,cern_high-luminosity_nodate}, increasing luminosity\footnote{Luminosity is discussed in Chapter~\ref{ch:physics}, Section~\ref{sec:accelerator}.}, increasing detector granularity and efficiency, and in general growing event complexity, the traditional methods will soon be outdated. The computational costs of the current reconstruction algorithms will skyrocket unless they are optimized or even redeveloped from scratch. 

In particular, for the duration of Run~2, the average number of proton--proton interactions per bunch crossing\footnote{Bunch crossing is discussed in Chapter~\ref{ch:physics}, Section~\ref{sec:accelerator}.}---what is known as pile-up---was at $\langle \mu \rangle \approx 30$, and the peak instantaneous luminosity recorded was $L = 2 \times 10^{34}~\text{cm}^{-2} \, \text{s}^{-1} $. The integrated luminosity of Run~2 was at around 190~fb$^{-1}$, while by the end of the current run, Run~3, the goal is to reach 350~fb$^{-1}$. 

\begin{sidewaysfigure}
    \centering
    \includegraphics[width=1\linewidth]{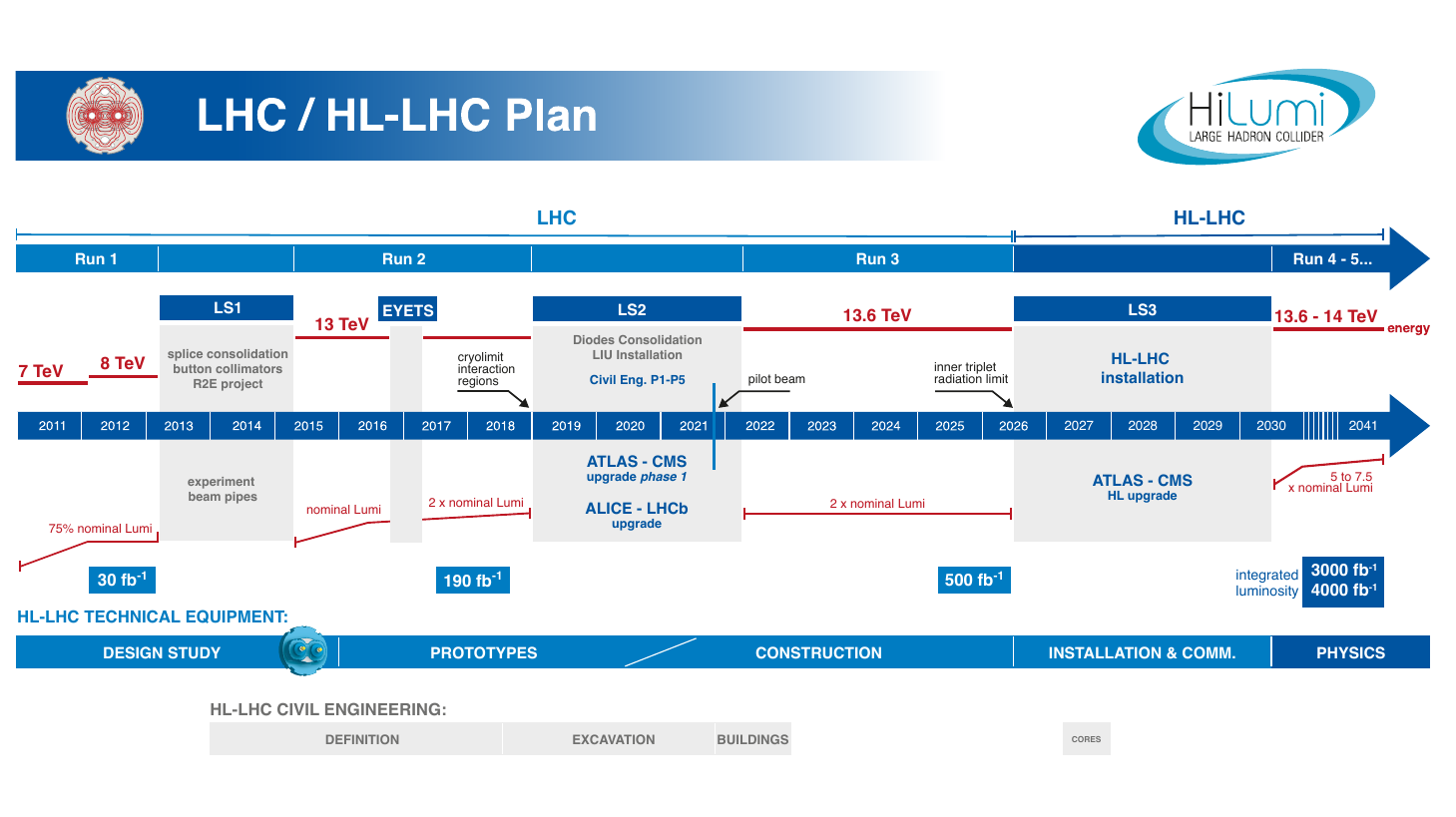}
    \caption{Plan for the LHC/High-Luminosity LHC (updated in January 2025). Figure from~\cite{cern_high-luminosity_nodate}.}
    \label{fig:hl-lhc}
\end{sidewaysfigure}

By contrast, starting from Run~4, the HL-LHC project has been engineered with the ambitious goal of achieving a peak instantaneous luminosity of $L = 7.5 \times 10^{34}~\text{cm}^{-2} \, \text{s}^{-1}$. This will correspond to an average pile-up of $\langle \mu \rangle \approx 200$. The ultimate goal is to achieve a per-year integrated luminosity of 250~fb$^{-1}$, with the goal of 3000~fb$^{-1}$ in the 12 years or so following the HL-LHC installation. Fig.~\ref{fig:hl-lhc} summarizes these plans for the LHC/HL-LHC.

\begin{figure}
    \centering
    \includegraphics[width=0.95\linewidth]{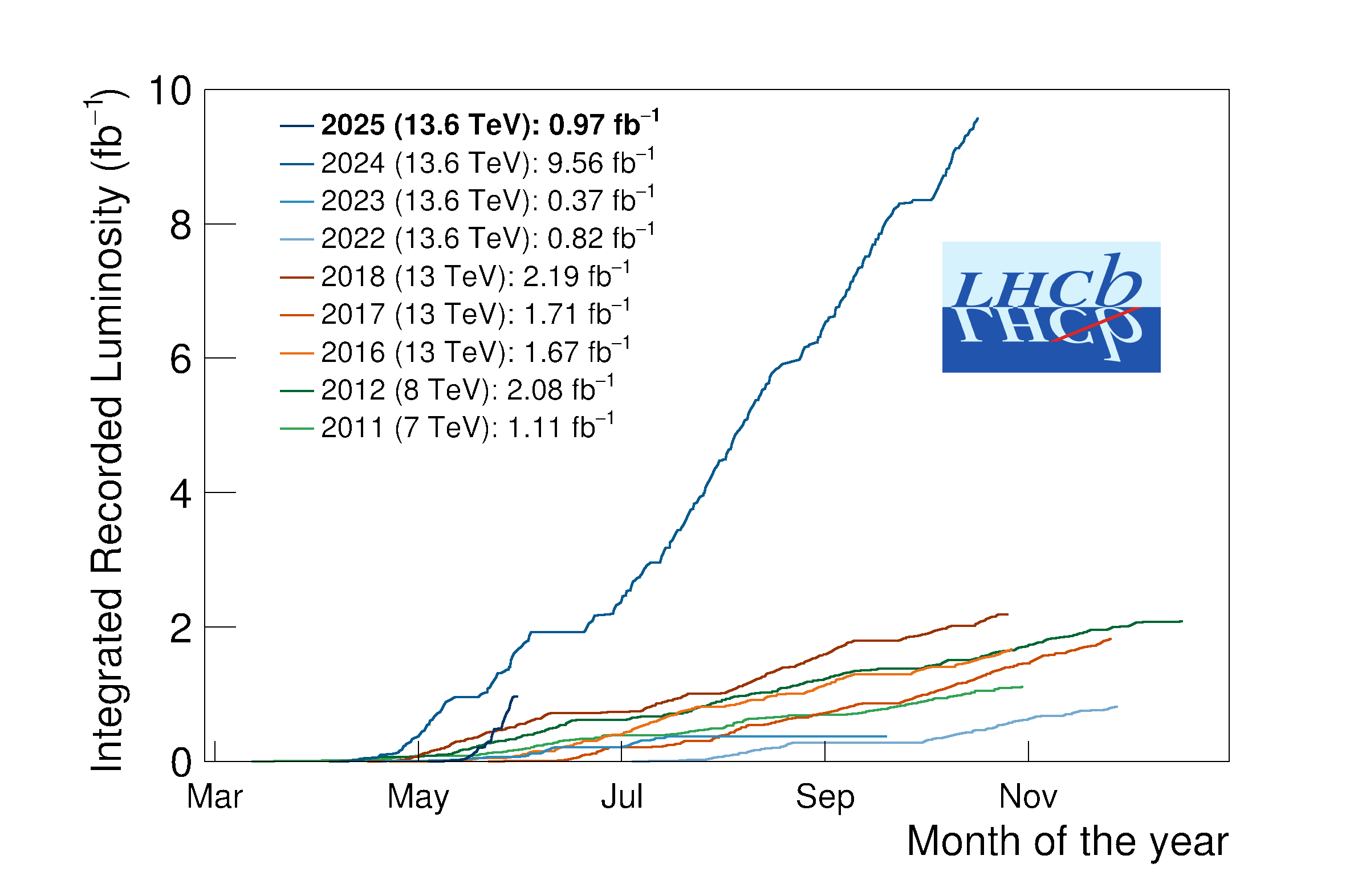}
    \caption{LHCb luminosity results over the various data-taking periods at the LHC. Figure from~\cite{cavallero_navigating_2024}.}
    \label{fig:lumiperyear}
\end{figure}

\begin{figure}
    \centering
    \includegraphics[width=1\linewidth]{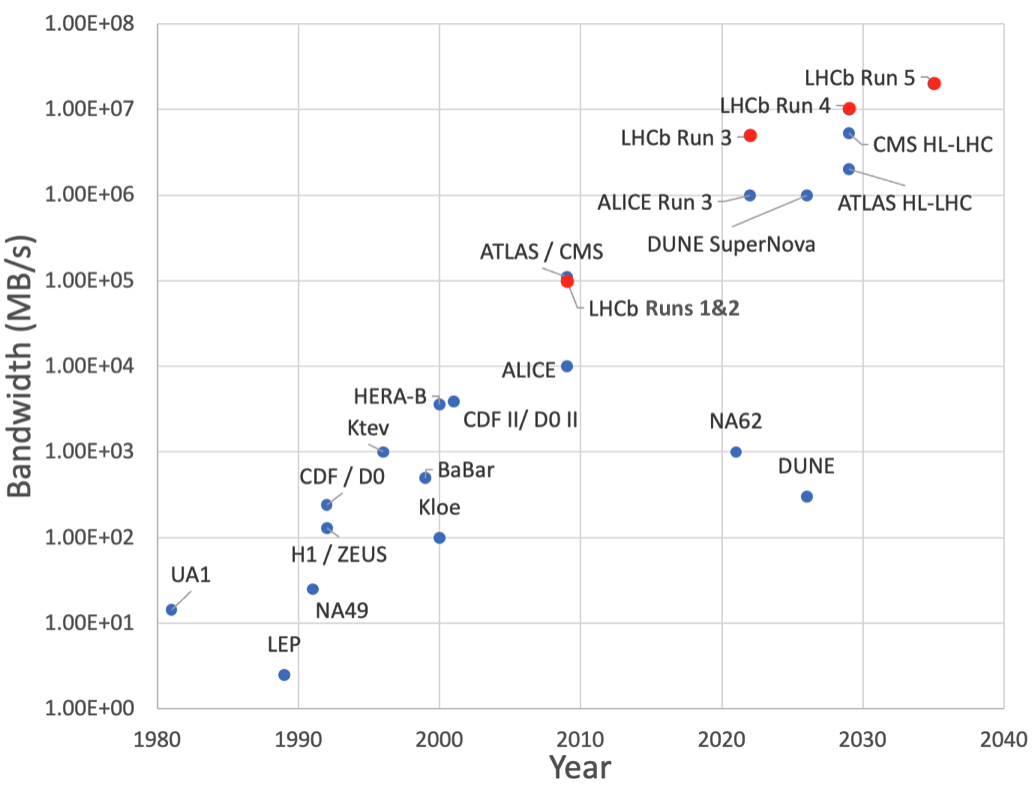}
    \caption{The evolution of the data bandwidth as a function of time for past and planned experiments. The LHCb runs are highlighted in red. Figure by Alessandro Cerri, University of Sienna, from~\cite{lhcb_collaboration_computing_2025}.}
    \label{fig:data-rates}
\end{figure}

Specifically for LHCb, the experiment will operate at its current configuration until the end of Run~4 (2033) reaching a maximum instantaneous luminosity of $L = 2 \times 10^{33}~\text{cm}^{-2} \, \text{s}^{-1}$~\cite{atlasbelle_iicmslhcb_collaborations_projections_2025}. By the end of Run~4 it will have recorded 50~fb$^{-1}$ of high-energy $p$--$p$ collision data. In contrast, after the upgrade during Long Shutdown 4 (LS4), the detector will operate at $L = 1 \times 10^{34}~\text{cm}^{-2} \, \text{s}^{-1}$, corresponding to $\langle \mu \rangle \approx $ 28--42 interactions per bunch crossing, compared to the current $\langle \mu \rangle \approx 5$. By the end of the HL-LHC operation, the detector will have recorded at least 300~fb$^{-1}$. A summary of the integrated luminosities is shown in Fig.~\ref{fig:lumiperyear}. Furthermore, Fig.~\ref{fig:data-rates} illustrates how data bandwidth has evolved over time across past and upcoming experiments. This demonstrates the scale of the challenge currently facing the LHC community.

In order for the HL-LHC project to be successfully completed, a massive effort is underway. The detectors and all the infrastructure at the LHC---including magnets, cryogenics, vacuum systems, and beam instrumentation---has to be modernized. 

\begin{figure}
    \centering
    \includegraphics[width=0.9\linewidth]{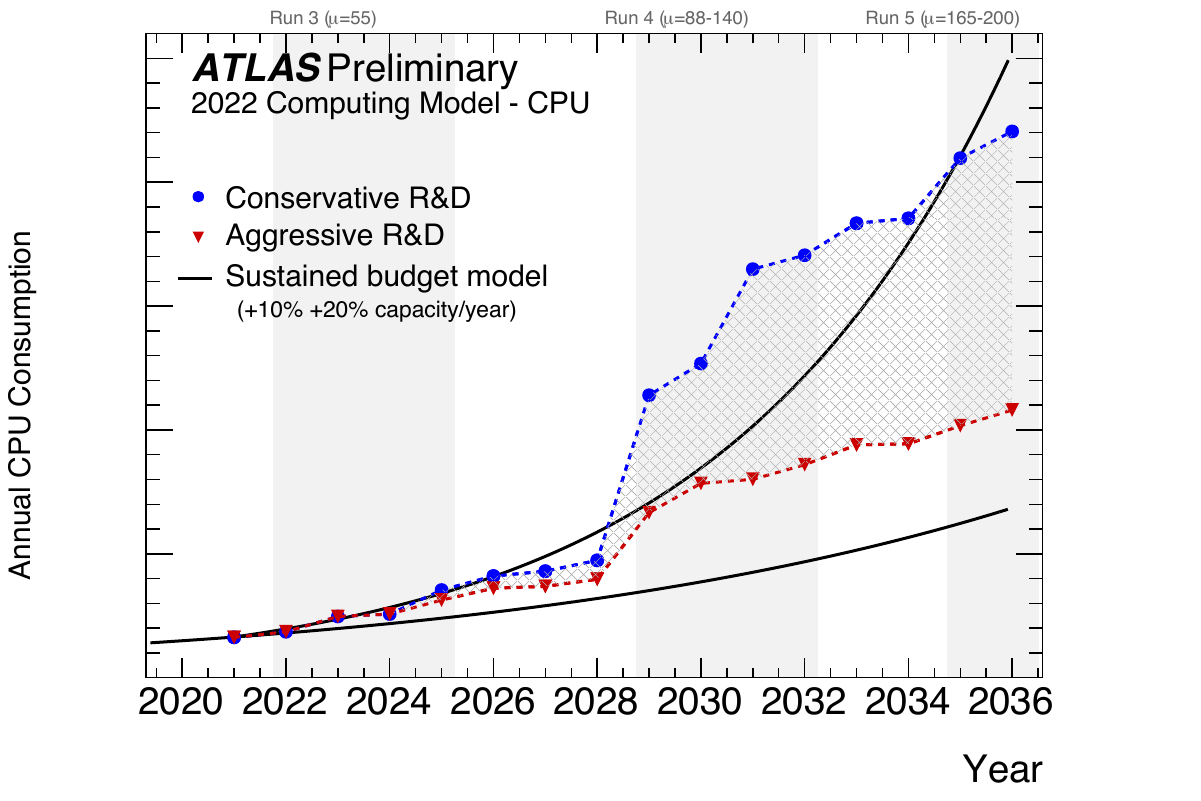}
    \caption{Projected evolution of the ATLAS compute usage from 2020 until 2036, in arbitrary units, under conservative (blue) and aggressive (red) Research and Development (R\&D) scenarios. The gray-hatched shading between the red and blue lines represents the range of resource consumption that would occur if the aggressive scenario were only partially implemented. The black lines represent the effects of consistent annual budget increases and advancements in new hardware, resulting in overall capacity increases of 10\% (lower line) and 20\% (upper line). The vertical shaded bands indicate the LHC runs during which ATLAS will be gathering data. Adapted from~\cite{atlas_collaboration_atlas_2022}.}
    \label{fig:aggressive-rd}
\end{figure}

Along with the upgrade of the infrastructure, a refinement of the computational and software tools at the disposal of the LHC collaborations must also be undertaken. Computing is a crucial component of all the experiments, encompassing operation, calibration and monitoring of the detectors. Furthermore, the trigger is an integral part of the processing pipelines leading to the physics analyses conducted at the LHC. The steep increase in pile-up puts significant strain on the computational systems in place, and depletes the limited resources needed for meaningful computations aimed at identifying rare decays and interesting signals within massive datasets. This increasing demand for computation, as illustrated in Fig.~\ref{fig:aggressive-rd} for ATLAS, necessitates the upgrading and redevelopment of the computing infrastructure at the LHC experiments. 

One example of particular interest is track reconstruction. The computational cost of classical tracking algorithms scales with the number of hits roughly quadratically~\cite{campora_perez_search_2021,fruhwirth_data_2000}. The ATLAS~\cite{atlas_collaboration_technical_2017} and CMS~\cite{cms_collaboration_phase-2_2021} collaborations are in the process of upgrading their detectors in order to operate in the HL-LHC environment, where instantaneous luminosities will be up to four times the current ones at ATLAS and CMS, and almost forty times the current one at the LHCb detector. Therefore, if nothing is done to improve the processing infrastructure of these experiments, the maximum potential of the HL-LHC will not be achieved.

At the same time, this increasing event complexity and immense data volume makes it more essential to perform tracking at the earliest possible stages of the pipeline, in order to reduce the memory and computational footprint of these algorithms while improving their efficiency~\cite{gligorov_real-time_2015}. The LHCb~\cite{aaij_allen_2020,lhcb_collaboration_lhcb_2020} and ALICE~\cite{gorbunov_alice_2011,rohr_gpu_2017,eulisse_o2_2024,alice_collaboration_real-time_2019} collaborations at CERN already perform track reconstruction at the full LHC collision rate. 

Firstly, hardware triggers are fast and simple, but operate on coarse detector data and hence on a rudimentary representation of the collision events. On the other hand, RTA allows a more sophisticated and physics-rich event selection using information from various subdetectors. In this way, interesting events that would otherwise have been discarded by the coarse filtering of the hardware trigger, can still be kept. For example, rare processes traditionally studied by the LHCb experiment, such as Charge-Conjugation Parity (CP) violation and heavy flavor decays, may not trigger efficiently on hardware but can be identified well using full tracking information in real time.

Secondly, with RTA, the online data reduction performed by the trigger becomes more efficient, since events are better ``understood'' before they are discarded. In addition, due to technology improvements in hardware, such as GPUs and communication, RTA systems are able to match or even exceed the capabilities of older hardware triggers, while being more flexible and while offering more maintainability and upgradability.

\section{RTA using ML on Heterogeneous Architectures}

The interest in ML for HEP~\cite{albertsson_machine_2018,moreno_jedi-net_2020,bakina_deep_2022,savard_emerging_2024} lies in the same reasons that make ML algorithms interesting in the first place. Artificial Intelligence (AI) and ML methods have proven to be powerful tools that can surpass, at times, classical algorithms in terms of various metrics, and across a diverse set of tasks~\cite{lecun_deep_2015,goodfellow_deep_2016}. Their ability to adjust to each problem and to extract useful information out of raw data have made them ubiquitous in many industries. Moreover, the potential to optimize the usage of the computational resources available makes them particularly attractive from a energy/cost perspective~\cite{albrecht_roadmap_2019}.

Furthermore, ML applications ``on the edge'', i.e., ultra-low latency and on-detector/sensor, are growing day by day. Examples include $\tau$~\cite{cms_collaboration_phase-2_2020}, $b$-quark~\cite{cms_collaboration_neural_2022}, and electron~\cite{cms_collaboration_electron_2023} identification, anomaly detection~\cite{cms_collaboration_anomaly_2023,cms_collaboration_2024_2024,cms_collaboration_level-1_2023}, data compression~\cite{guglielmo_reconfigurable_2021}, and continual learning~\cite{cms_collaboration_continual_2023} in the CMS trigger. Other examples include calorimeter peak finding~\cite{chiedde_machine_2022} in the ATLAS trigger.

Lipschitz neural networks~\cite{schulte_development_2023,delaney_applications_2023,kitouni_robust_2023} have also been developed for the LHCb topological trigger~\cite{likhomanenko_lhcb_2015,gligorov_efficient_2013,aaij_lhcb_2013}, but it should be noted that these models are not deployed on the edge in the sense of being on-detector, neither are they used in a latency-bound environment. The LHCb architecture is explicitly not latency bound so it can process at the full LHC collision rate while being throughput-constrained. Apart from HEP, examples in other sectors include healthcare~\cite{amin_edge_2021}, autonomous driving~\cite{yang_edge_2021}, industrial predictive maintenance~\cite{tang_computational_2023} and smart cities~\cite{alnoman_edge_2021}.

Performing ML in real time is challenging, especially in scientific applications, a domain referred to as the ``FastML Science domain'' in~\cite{duarte_fastml_2022}. The domain generates an immense volume of data, with inference latency requirements that are several orders of magnitude more stringent than those typically found in traditional consumer-facing applications. Therefore, on the one hand, real-time processing is by itself a significant challenge in many scenarios. On the other hand, ML is a computationally intensive process, making it more of an issue in a constrained environment~\cite{strubell_energy_2020,thompson_computational_2022}. One approach, in order to mitigate these computing challenges, is the use of \textit{heterogeneous} computational architectures. Instead of performing all the treatment of the data on a traditional processor, the tasks can be split and distributed between various processors, each of which is specialized to do a specific family of tasks. This system will contain processors of different type, and can therefore be called heterogeneous.

\section*{Conclusion}

We have now seen the motivation behind doing real-time analysis for high-energy physics using machine-learning methods on heterogeneous architectures. Next, we turn to the background, Part~\ref{part:background}, essential in understanding the work presented in Part~\ref{part:main}: the main results of this thesis.

\part{Background}
\label{part:background}

\chapter{Physics Background}
\label{ch:physics}

\minitoc

\section*{Introduction}

In this chapter, we delve into the primary field of focus of this text: high-energy particle physics. We begin by introducing fundamental concepts in accelerator physics, followed by an overview of the Standard Model (SM) and some key open questions in the field. Finally, we touch on heavy flavor physics in a bit more detail. This background will be necessary to understand and precisely describe the work from the physics point of view.

\section{Accelerator Physics}
\label{sec:accelerator}

\subsection*{Cylindrical Coordinates}

\begin{figure}
    \centering
    \includegraphics[width=0.6\linewidth]{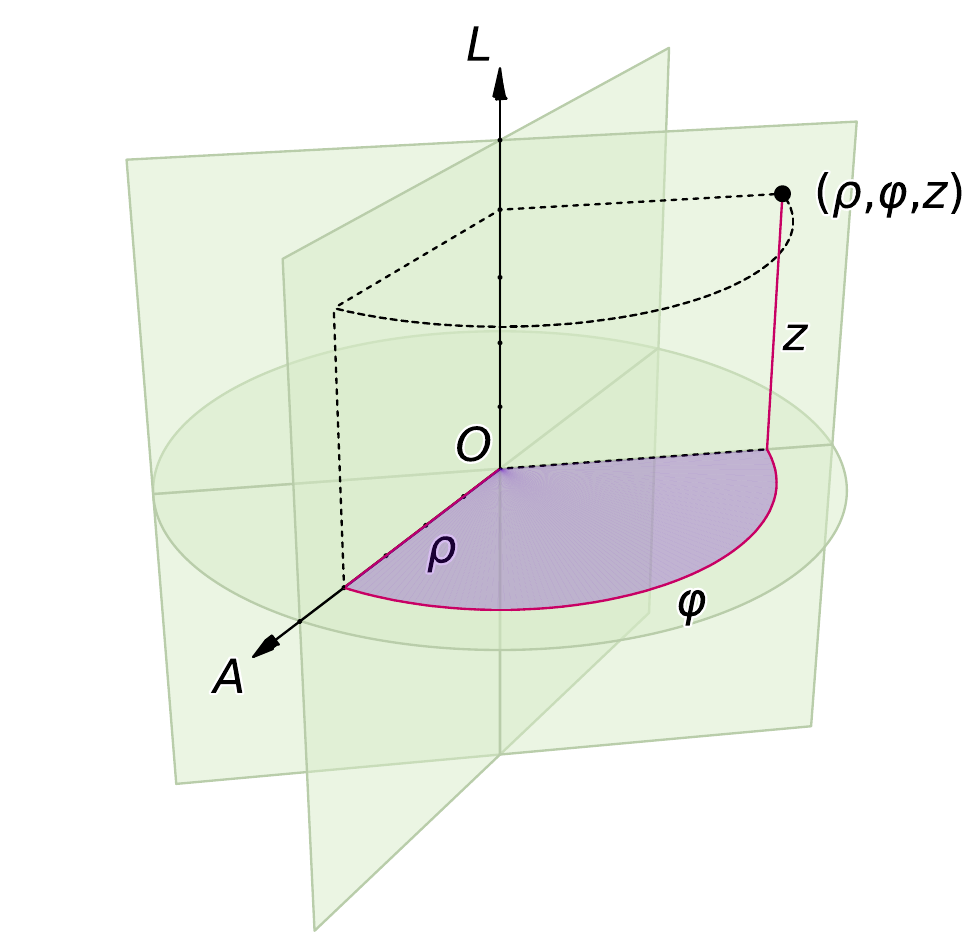}
    \caption{A cylindrical coordinate system defined by an origin $O$, a polar (radial) axis $A$, and a longitudinal (axial) axis $L$. Figure from~\cite{stolfi_cylindrical_2009}.}
    \label{fig:cylindrical}
\end{figure}

In accelerator physics, cylindrical coordinates $(\rho, \varphi, z)$~\cite{riley_mathematical_2006} are often used, instead of Cartesian coordinates $(x,y,z)$. In this configuration, points are identified with respect to the main axis called cylindrical or longitudinal axis, and an auxiliary axis called the polar axis, as shown in Fig.~\ref{fig:cylindrical}. $\rho$ denotes the perpendicular distance from the main axis, $z$ denotes the distance along the main axis, and $\varphi$ is the plane (or azimuthal) angle of the point of projection on the transverse plane. The beamline is naturally identified with the cylindrical axis of the coordinate system. 

\subsection*{Pseudorapidity}

\begin{figure}
    \centering
    \includegraphics[width=0.5\linewidth]{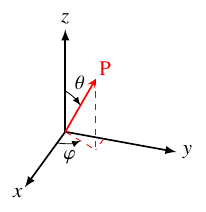}
    \caption{The polar ($\theta$) and azimuthal ($\varphi$) angles. Adapted from~\cite{neutelings_3d_2021}.}
    \label{fig:angles}
\end{figure}

In experimental particle physics, another frequently used spatial coordinate is the pseudorapidity $\eta$ . It describes the angle between a particle's momentum $\mathbf{p}$ and the positive direction of the beam axis---identified with the $z$-direction. This angle is referred to as the polar angle $\theta$, as shown in Fig.~\ref{fig:angles}.

Pseudorapidity is defined as~\cite{wong_introduction_1994}:

\begin{equation}
    \eta = - \ln \left[ \tan \left( \frac{\theta}{2} \right) \right]\,,
\end{equation}
or inversely

\begin{equation}
    \theta = 2 \arctan \left( e^{-\eta}\right) \,.
\end{equation}
As a function of the three-momentum $\mathbf{p}$, pseudorapidity can be expressed as

\begin{equation} \label{eq:pseudorapidity}
    \eta = \frac{1}{2} \ln \left( \frac{|\mathbf{p}| + p_L}{|\mathbf{p}| - p_L} \right)\,
\end{equation}
where $p_L$ is the longitudinal component of the momentum, along the beam axis. Due to its desirable physical properties, this definition is highly favored in experimental particle physics.

\begin{figure}
    \centering
    \includegraphics[width=1\linewidth]{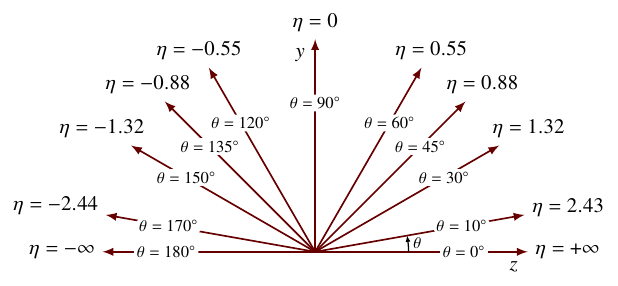}
    \caption{Values of pseudorapidity $\eta$ versus polar angle $\theta$. Figure from~\cite{neutelings_pseudorapidity_2021}.}
    \label{fig:pseudorapidity}
\end{figure}

From Eq.~\eqref{eq:pseudorapidity}, we can see that when the momentum tends to be all along the beamline, i.e., $p_L \rightarrow |\mathbf{p}| $ ($\theta \rightarrow 0 $), pseudorapidity blows up $\eta \rightarrow \infty $. On the other hand, when most of the momentum is in transverse directions, $p_L \rightarrow 0 $ ($\theta \rightarrow 90^{\circ} $), then $\eta \rightarrow 0 $, as shown in Fig.~\ref{fig:pseudorapidity}. 

\subsection*{Beam Bunching}

In particle beams, in many modern experiments including the LHC, particles are distributed into pulses, or \textit{bunches}. Bunched beams are common because most modern accelerators require bunching for acceleration~\cite{edwards_introduction_1993}.

\begin{figure}
    \centering
    \includegraphics[width=1\linewidth]{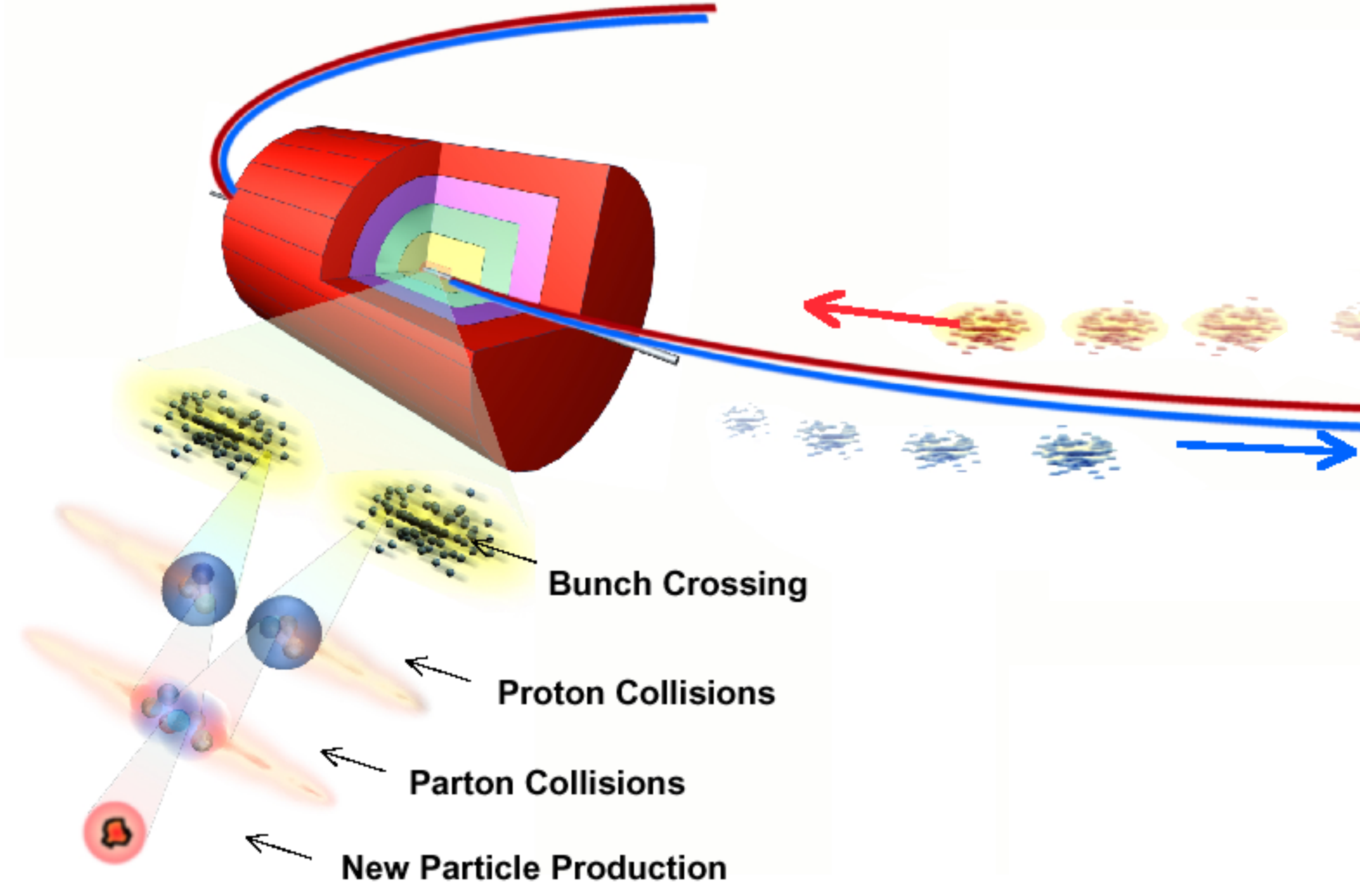}
    \caption{Illustration of beam bunching utilized at the Large Hadron Collider at CERN. Adapted from~\cite{cern_large_2013}.}
    \label{fig:bunches}
\end{figure}

At the LHC, after accelerating the particles in bunches, the two beams are focused resulting in the crossing of these bunches---the so-called \textit{bunch crossing}, as shown in Fig.~\ref{fig:bunches}. These bunch crossings, also known as \textit{events}, may result in one or multiple collisions between protons and consequently in the production of new particles. The number of these collisions during a bunch crossing is known as pile-up.

\subsection*{Primary and Secondary Vertices}

\begin{figure}
    \centering
    \includegraphics[width=0.75\linewidth]{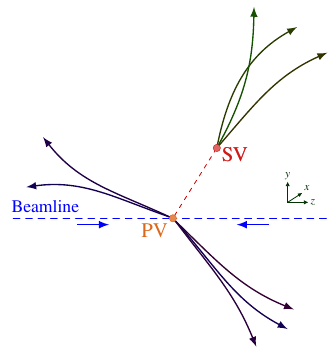}
    \caption{Illustration of Primary Vertices (PVs) and Secondary Vertices (SVs) in colliding-beam experiments. PVs are points in space where a primary particle collision occurred, and can reconstructed from the tracks of particles emerging directly from the collision. SVs, on the other hand, are points displaced from the PV where the decay of a long-lived particle occurred. They can be reconstructed from the tracks of decay products that do not originate from the primary interaction. Adapted from~\cite{neutelings_b_2021}.}
    \label{fig:vertices}
\end{figure}

Primary vertices are points in space where a particle collision occurred, resulting in the generation of other particles at this point, as shown in Fig.~\ref{fig:vertices}. The location of this point can be reconstructed from the tracks of particles emerging directly from the collision. Secondary (or displaced) vertices are points displaced from the primary vertex, where the decay of a long-lived particle occurred. These points can be reconstructed from the tracks of decay products that do not originate from the primary interaction.

Primary vertices are a crucial element of many physics analyses~\cite{piacquadio_primary_2008}. The precise reconstruction of many processes, the identification of $b$- or $\tau$-jets, the reconstruction of exclusive
$b$-decays and the measurement of lifetimes of long-lived particles are all dependent upon the precise knowledge of the location of the primary vertex. Secondary vertices, on the other hand, are tools for identifying heavy flavor hadrons and $\tau$ leptons~\cite{erdmann_vertex_2008}.

\subsection*{Luminosity}

\textit{Luminosity} $L$ is defined as the ratio of the number of events detected $dN$ in a certain period of time $dt$ and across a cross section $\sigma$~\cite{herr_concept_2006,martin_particle_2008,myers_particle_2020}:

\begin{equation}
    L = \frac{1}{\sigma} \frac{dN}{dt} \,,
\end{equation}
and is often given units of $\text{cm}^{-2} \cdot \text{s}^{-1}$. In practice, the luminosity depends on the parameters of the particle beam, such as the beam width and particle flow rate.

\textit{Integrated luminosity} $L_{\text{int}}$ is defined as the integral of the luminosity with respect to time:

\begin{equation}
    L_{\text{int}} = \int L \,dt = \frac{N}{\sigma} \,,
\end{equation}
where $N$ is now the total number of collision events produced. $L$ is frequently referred to as instantaneous luminosity, in order to emphasize the distinction between its integrated-over-time counterpart $L_{\text{int}}$. Integrated luminosity, having units of $1/\sigma$, is sometimes measured in inverse femtobarns $\text{fb}^{-1}$. It measures the number of collisions produced per femtobarn of cross section. 

These variables are useful quantities to evaluate the performance of a particle accelerator. In particular, most HEP collision experiments aim to maximize their luminosity, since a higher luminosity means more collisions and consequently a higher integrated luminosity means a larger volume of data available to be analyzed. 

For beam-to-beam experiments, where the particles are accelerated in opposite directions before collided, like the majority of the time at the LHC, the instantaneous luminosity can be calculated as~\cite{herr_concept_2006}:

\begin{equation}
    L = \frac{N^2 f N_b}{4 \pi \sigma_x \sigma_y} \,,
\end{equation}
where $N$ denotes the number of particles per bunch, $f$ is the revolution frequency, and $N_b$ is the number of bunches in each beam. The transverse dimensions of the beam, assuming a Gaussian profile, are described by $\sigma_x$ and $\sigma_y$.

\subsection*{Impact Parameter}

\begin{figure}
    \centering
    \includegraphics[width=0.5\linewidth]{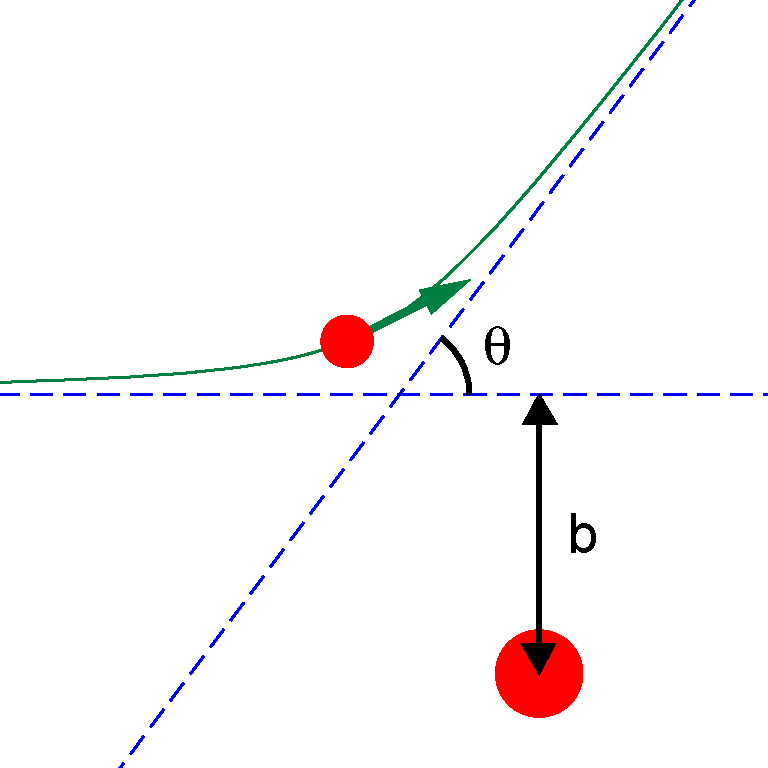}
    \caption{A projectile scattering off a target particle. The impact parameter $b$ and the scattering angle $\theta$ are shown. Figure from~\cite{tonatsu_impact_2007}.}
    \label{fig:impact}
\end{figure}

The impact parameter $b$ represents the shortest, perpendicular distance between the trajectory of a projectile and the center of the potential field generated by the target particle, as shown in Fig.~\ref{fig:impact}. In accelerator experiments, collisions can be classified based on the value of the impact parameter. Central collisions have $b \approx 0$, while peripheral collisions have impact parameters comparable to the radii of the colliding nuclei.

\subsection*{Detector Acceptance}

In particle collider experiments, the location of the collisions is predetermined. However, the direction of the produced particles due to the interactions is not predetermined, i.e., the products can fly in every possible direction. However, depending on the geometry of the experiment or its physics program, detecting all the products is not feasible or desirable. The region of the detector where the particles are in fact detectable is referred to as the \textit{acceptance}. In some cases, detection depends also on the energy, or other characteristics of the particle, meaning that the acceptance is not only a function of the particle's direction, but also of those extra characteristics.

\section{The Standard Model of Particle Physics}

The SM is a relativistic quantum field theory classifying all known elementary particles and describing three out of the four fundamental forces: the electromagnetic, weak nuclear and strong nuclear interactions, excluding gravity. It was developed progressively during the latter half of the 20th century through the contributions of numerous scientists worldwide~\cite{oerter_theory_2006}. Its current form was established in the mid-1970s following the experimental confirmation of quarks. Subsequent discoveries, including the top quark in 1995~\cite{cdf_collaboration_observation_1995}, the tau neutrino in 2000~\cite{donut_collaboration_observation_2001}, and the Higgs boson in 2012~\cite{atlas_collaboration_observation_2012, cms_collaboration_observation_2012}, have further reinforced the validity of the Standard Model.

\begin{figure}
    \centering
    \includegraphics[width=1\linewidth]{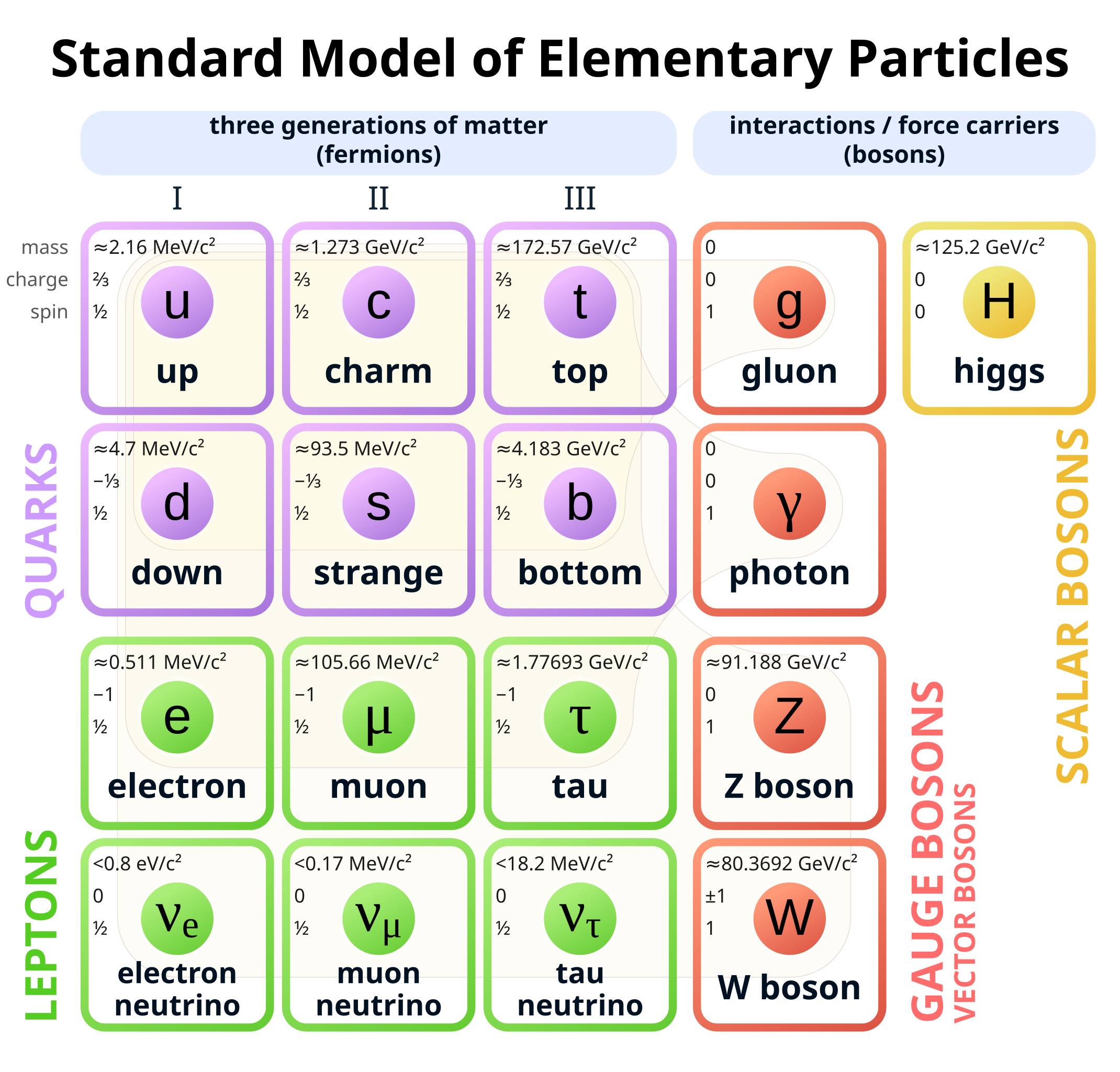}
    \caption{The Standard Model of elementary particles including twelve fundamental fermions and five fundamental bosons. Brown loops indicate the interactions between the bosons (red) and the fermions (purple and green). Please note that the masses of some particles are periodically reviewed and updated by the scientific community. The values shown in this graphic are taken from~\cite{particle_data_group_review_2024}. Figure from~\cite{missmj_standard_2019}.}
    \label{fig:sm}
\end{figure}

Fig.~\ref{fig:sm} depicts the elementary particles of the SM and their interactions. They can be divided into twelve \textit{fermions} with spin-$1/2$, five spin-1 gauge \textit{bosons} ($\gamma, g^a, W^{\pm}, Z^0$), carriers of the electromagnetic, weak and strong interactions, and the spin-0 (scalar) Higgs boson ($H$). 

The fermions are further grouped into six \textit{quarks} and six \textit{leptons}. The main difference is that quarks interact with all three fundamental forces of the SM, while leptons only interact with the weak and electromagnetic interactions. Quarks appear in six different flavors. In increasing order of quark masses they are called: up ($u$), down ($d$), strange ($s$), charm ($c$), bottom or beauty ($b$) and top ($t$) quarks. The quarks are further grouped into three generations of increasing masses. Up-type quarks ($u$, $c$, $t$) have an electric charge $q=+(2/3)e$ while down-type quarks ($d$, $s$, $b$) have $q=-(1/3)e$, where $e$ is the elementary charge.

Quarks possess a property known as color charge, which causes them to interact through the strong force. Due to color confinement, quarks are tightly bound together, forming color-neutral composite particles called \textit{hadrons}. As a result, quarks cannot exist in isolation and must always combine with other quarks. Hadrons are classified into two types: \textit{mesons}, which consist of a quark-antiquark pair, such as the pion ($\pi$), the kaon ($K$), the $B$, $D$ and $J/\psi$ mesons, and \textit{baryons}, which are made up of three quarks. The lightest baryons are the nucleons: the proton and the neutron.

Furthermore, the solutions of the Dirac equation~\cite{dirac_quantum_1997} predict that each of the twelve SM fermions has a corresponding counterpart, known as its antiparticle, which possesses the same mass but opposite charge.

Similarly, the leptons are also grouped into three generations. Each generation contains a charged lepton and its corresponding uncharged neutrino. The charged leptons are the electron ($e^-$), the muon ($\mu^-$) and the tau ($\tau^-$). Their uncharged partners are the electron, muon and tau neutrinos ($\nu_e$, $\nu_{\mu}$, $\nu_{\tau}$). Being chargeless, they are not sensitive to the electromagnetic interaction and moreover, they are considered massless in the SM. The observation of neutrino oscillations~\cite{super-kamiokande_collaboration_evidence_1998} requires that neutrinos have small but non-zero masses and thus implies physics beyond the SM.

The five types of gauge bosons mediate the interactions between the fermions. The electromagnetic is mediated by the photon $\gamma$, the strong by eight distinct gluons $g^a$, and the weak by the W$^{\pm}$ and Z$^0$ bosons. The Higgs boson plays a special role in the Standard Model by providing an explanation for why elementary particles, except for the photon and gluon, have mass. Specifically, the Higgs mechanism is responsible for the generation of the gauge boson masses while the fermion masses result from Yukawa-type interactions with the Higgs field.

Table~\ref{tab:fermions} summarizes the masses $m$ and electric charges $q$ of the fermionic elementary particles of the SM, while in Table~\ref{tab:bosons}, the masses, charges and spins of the elementary bosons are shown.

\begin{table}
    \centering
    \begin{tabular}{c|ccc|ccc} \hline \hline
        \multirow{2}{*}{Generation} & \multicolumn{3}{c|}{Quarks} & \multicolumn{3}{c}{Leptons} \\
         & Flavor & $m$ (MeV/$c^2$) & $q$ ($e$) & Flavor & $m$ (MeV/$c^2$) & $q$ ($e$) \\ \hline
         \multirow{2}{*}{1}& $u$ & $2.16 \pm 0.07$ & \nicefrac{+2}{3} & $\nu_e$ & $<2 \times 10^{-6}$ & 0 \\
         & $d$ & $4.70 \pm 0.07$ & \nicefrac{-1}{3} & $e^-$ & 0.511 & $-1$ \\ \hline
         \multirow{2}{*}{2} & $c$ & $1273.0 \pm 4.6$ & \nicefrac{+2}{3} & $\nu_{\mu}$ & $<0.19$ & 0 \\
         & $s$ & $93.5 \pm 0.8$ & \nicefrac{-1}{3} & $\mu^-$ & 105.66 & $-1$ \\ \hline
         \multirow{2}{*}{3} & $t$ & $172\,570\pm 290$ & \nicefrac{+2}{3} & $\nu_{\tau}$ & $<18.2$ & 0 \\
         & $b$ & $4183 \pm 7$ & \nicefrac{-1}{3} & $\tau^-$ & $1777$ & $-1$ \\ \hline \hline
    \end{tabular}
    \caption{Summary of the masses and charges of the elementary fermions in the SM. Mass values taken from~\cite{particle_data_group_review_2024}. Uncertainties are not displayed for masses if they are smaller than the last digit of the value.}
    \label{tab:fermions}
\end{table}

\begin{table}
    \centering
    \begin{tabular}{c|c|c|c|c} \hline \hline
         Boson & Type & Spin & $m$ (GeV/$c^{\rule{0pt}{1.5ex}2}$) & $q$ ($e$) \\ \hline
         Photon & \multirow{4}{*}{Gauge} & \multirow{4}{*}{1}  & 0 & 0 \\
         Gluon & & & 0 & 0 \\
         Z$^0$ & & & $91.1880 \pm 0.0020$ & 0 \\
         W$^{\pm}$ & & & $80.3692 \pm 0.0133$  & $\pm 1$ \\ \hline
         Higgs & Scalar & 0 & $125.20 \pm 0.11 $ & 0 \\ \hline \hline
    \end{tabular}
    \caption{Summary of the masses, charges and spins of the elementary bosons of the SM. Mass values taken from~\cite{particle_data_group_review_2024}. The masses of the photon and the gluon are the theoretical values.}
    \label{tab:bosons}
\end{table}

\section{Open Questions}

Despite the successes of the Standard Model, it is not a complete theory of fundamental interactions and several questions in physics remain open~\cite{martin_particle_2008}. For example, even though the three out of the four fundamental forces have been combined into the same theory, gravity, described by the general theory of relativity, cannot be integrated into the SM. The problem remains elusive, and theories Beyond the Standard Model (BSM) are needed, such as string theory or quantum gravity. In addition, the question of why there is more matter in the universe than antimatter, remains an open question. This problem is known as the matter-antimatter asymmetry and is a core question in the LHCb physics program. Furthermore, this question is related to CP violation, the violation of the charge-conjugation parity symmetry in particle interactions. This is one of the reasons why CP violation is heavily studied at LHCb. Moreover, it does not account for the accelerating expansion of the universe, and how it is possibly described by dark energy. Finally, the origin of dark matter remains to be understood as well as the explanation for neutrino oscillations and their non-zero masses.

\section{Heavy Flavor Physics}

Going into more detail, the gigantic datasets being collected by the various accelerator experiments---and specifically by the Large Hadron Collider beauty (LHCb) experiment---are crucial to shed light on many of the open questions in particle physics~\cite{atlasbelle_iicmslhcb_collaborations_projections_2025}, and in particular in heavy flavor physics.

An important matrix in flavor physics is the so-called Cabibbo--Kobayashi--Maskawa (CKM) matrix~\cite{cabibbo_unitary_1963,kobayashi_cp-violation_1973}, and is of the form:

\begin{equation}
    V_{CKM} =
    \begin{pmatrix}
        V_{ud} & V_{us} & V_{ub} \\
        V_{cd} & V_{cs} & V_{cb}\\
        V_{td} & V_{ts} & V_{tb}\\
    \end{pmatrix} \,.
\end{equation}
It is a unitary matrix that dictates the quark mixing strengths of the flavor-changing weak interaction, and is crucial in understanding CP violation. The unitarity of the CKM matrix imposes constraints on its elements, which can be visualized geometrically through the construction of so-called unitarity triangles. Unitarity triangles have angles conventionally labeled as $\alpha$, $\beta$ and $\gamma$. The angle $\beta$ is conventionally measured from the mixing-induced CP violation in $B^0 \to J/\psi K^0_S$ decays. The angle $\alpha$ is determined using the $B\to \pi \pi$, $\pi \rho$ and $\rho \rho$ decays, while $\gamma$ is inferred from CP violation effects in $B^+ \to D K^+$~\cite{atlasbelle_iicmslhcb_collaborations_projections_2025}. The angles above are related to the unitarity relation between the rows of the CKM matrix corresponding to the couplings of the $b$ and $d$ quarks to $u$ quarks. The current uncertainties, measured by LHCb, are $0.57^{\circ}$~\cite{lhcb_collaboration_measurement_2024} and $2.8^{\circ}$~\cite{lhcb_collaboration_simultaneous_2024} for $\beta$ and $\gamma$, respectively. These sensitivities have been achieved using data samples of integrated luminosity 2--9~fb$^{-1}$. These values are projected to be reduced to $0.20^{\circ}$ and $0.8^{\circ}$, respectively, with 50~fb$^{-1}$ of data recorded by the early 2030s, and even to $0.08^\circ$ and $0.3^\circ$, respectively, with 300~fb$^{-1}$ of data recorded by the early 2040s.

Improving our understanding of the CKM matrix through global fits requires more precise knowledge of the magnitudes of the $|V_{ub}|$ and $|V_{cb}|$ CKM matrix elements. We can determine these magnitudes by studying semileptonic decays like $b \to u l \nu$ and $b \to cl \nu$, where $l$ denotes a charged lepton. Semileptonic decays can also be utilized to test the SM predictions on universality between the charged current weak interactions with different lepton flavors. This can be done using observables such as $R(D^{(*)})$, which are the branching fraction ratios

\begin{equation}
    \frac{B \to D^{(*)} \tau \nu}{B \to D^{(*)} e \nu}
\end{equation}
or
\begin{equation}
    \frac{B \to D^{(*)} \tau \nu}{B \to D^{(*)} \mu \nu} \,.
\end{equation}
The current values of these quantities suggest possible discrepancies with the SM. In order to further explore these discrepancies, the measured uncertainties on these values have to be reduced. Currently, the uncertainty on both $|V_{ub}|$~\cite{lhcb_collaboration_determination_2015} and $R(D^{(*)})$~\cite{lhcb_collaboration_test_2023} is at 6\%, from LHCb measurements. These uncertainties are projected to be reduced down to 1\% and 3\%, for $|V_{ub}|$ and $R(D^{(*)})$, respectively, with the increased number of collisions expected until the early 2040s.

Moreover, even though all CP violation in the charm sector is suppressed in the SM, CP violation in $D^0$-meson decays has been observed through asymmetries in $D^0 \to K^+ K^-$ and $D^0 \to \pi^+ \pi^-$ decays, captured by the observable $\Delta A_{CP} = A_{CP}\left(D^0 \to K^+ K^- \right) - A_{CP}\left(D^0 \to \pi^+ \pi^- \right)$. $A_{CP}(D^0 \to f)$ denotes the asymmetry between the $D^0 \to f$ and $\bar{D}^0 \to f$ decay rates to a final state  $f$. With a sample of 5.9~fb$^{-1}$, LHCb quoted an uncertainty of $29 \times 10^{-5}$~\cite{lhcb_collaboration_observation_2019}. This uncertainty can be potentially reduced almost by a factor of 10, down to $3.3 \times 10^{-5}$, given the expected integrated luminosities of 300~fb$^{-1}$. Furthermore, the charm samples essential to these measurements are produced at very large signal rates. Without real-time processing at the full collision rate these samples would be impossible to collect. The need for an RTA trigger at LHCb is further discussed in Chapter~\ref{ch:lhcb}, Section~\ref{sec:trigger}.

Beyond CP violation, the study of lepton flavor violation offers another compelling avenue for discovering BSM physics. While lepton flavor violation occurs in neutrino oscillations, any related effect in charged leptons is unobservably small within the SM framework. Consequently, observing any non-zero effect would be an unambiguous sign of BSM physics. Similarly, stringent upper limits on branching fractions, like $\mathcal{B}(\tau^+ \to \mu^+ \gamma)$ and $\mathcal{B}(\tau^+ \to \mu^+ \mu^+ \mu^-)$, tightly constrain potential BSM extensions of the Standard Model. For example, with a data sample of 424~fb$^{-1}$, the Belle II collaboration has constrained $\mathcal{B}(\tau^+ \to \mu^+ \mu^+ \mu^-)$ down to $<1.8 \times 10^{-8}$~\cite{belle_ii_collaboration_search_2024}. This uncertainty, using 50~ab$^{-1}$ instead, is projected to be reduced down to $<0.02 \times 10^{-8}$ until the early 2040s.

Heavy flavor physics remains a vital part of the global particle physics program. While experiments including ATLAS, CMS, LHCb and Belle II offer complementary strengths, they will also compete for the best precision on certain observables. This competion will allow for crucial consistency checks and ultimately lead to even more precise world average combinations. Collectively, these experiments can significantly advance the experimental precisions of all the key observables in $b$, $c$ and $\tau$ physics, with an expected improvement of typically one order of magnitude from what is available today. Nonetheless, this represents only a partial evaluation of the true physics reach, suggesting the impact will probably be even more significant. The precision currently at reach with these experiments, including their upgrades, provides an unprecedented capability to probe the flavor sector of the Standard Model.

\section*{Conclusion}

In this chapter, I started by introducing fundamental concepts in accelerator physics, necessary to understand the technical aspects related to the detector physics of this work. I also described the Standard Model of particle physics, the open questions in the field, and finally the research outlook and expected impact of heavy flavor physics research.

\chapter{Machine Learning Background}
\label{ch:ml}

\minitoc
\noindent Parts of this chapter were inspired by~\cite{goodfellow_deep_2016,burkov_hundred-page_2019}.

\section*{Introduction}

This chapter is a short and pedagogical introduction to the field of machine learning and its brief history, its subfields Deep Learning (DL) and Graph Neural Networks (GNNs), as well as some important techniques highly relevant to the field of ML and the work undertaken during this thesis.

\section{Machine Learning}

\begin{wrapfigure}{L}{0.5\textwidth}
    \centering
    \includegraphics[width=0.5\textwidth]{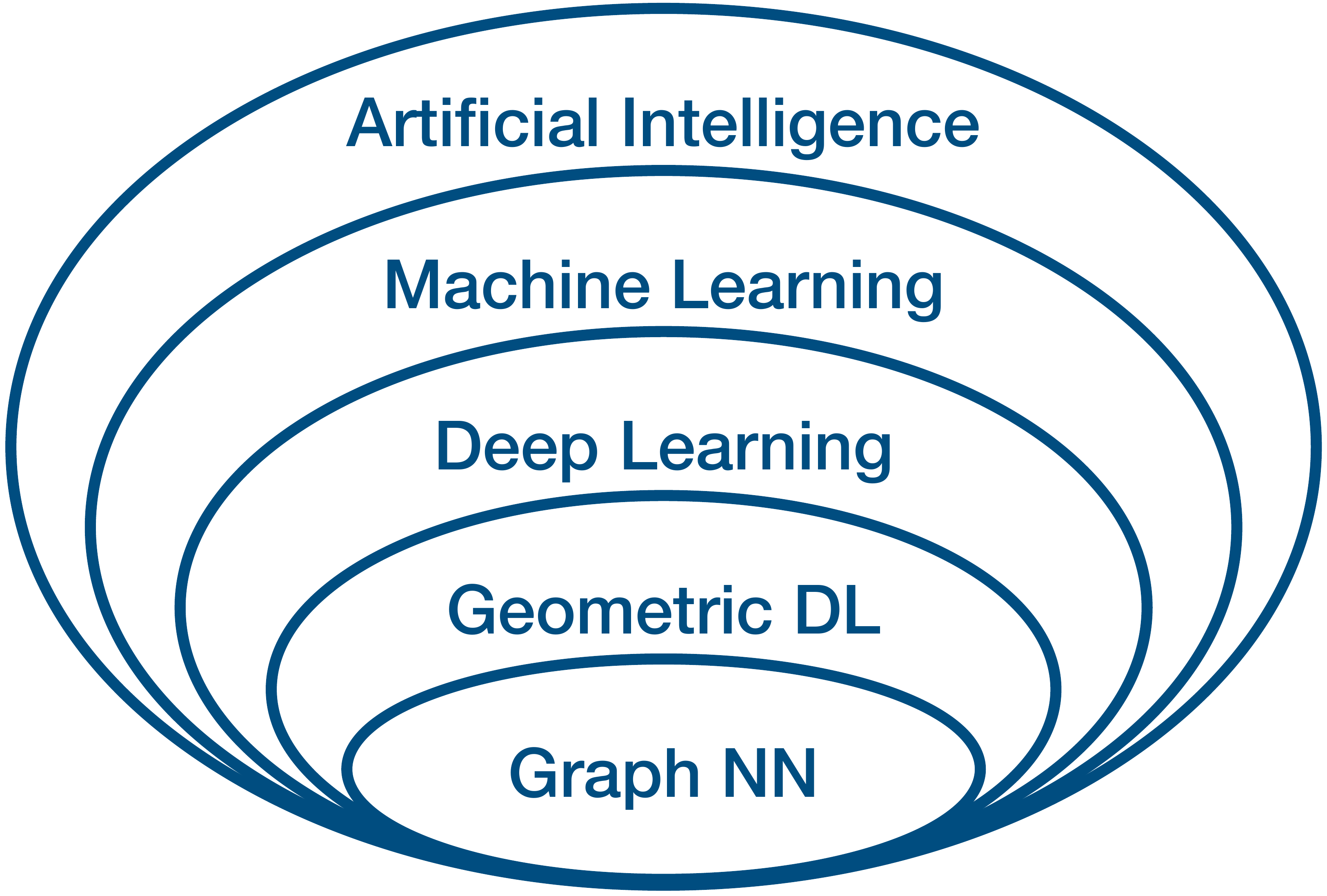}
    \caption{Euler diagram of AI and its subfields as relevant to this thesis.}
    \label{fig:ai-ml-dl}
\end{wrapfigure}

Machine learning is the field of how machines---specifically computers---can ``learn''. Although ``learn'' is perhaps a generous term, it refers to how computers manage to do specific tasks without being explicitly programmed to do them. Unlike classical algorithms, which follow hand-crafted rules defined by developers, ML algorithms, and by consequence ML models, are data-driven: By an iterative process of providing data to the ML model, the model is \textit{trained} and progressively learns to perform a task solely based on the data it has been given. At the end of this process, without the need for the developer to describe the logic of the algorithm itself, the model can carry out the task effectively without the developer needing to explicitly define how it should be done.

The term machine learning is believed to have been coined by Arthur Samuel in 1959 for his work on programming a computer to play checkers~\cite{samuel_studies_1959}. In general, AI is considered as a more general term than ML, as shown in Fig.~\ref{fig:ai-ml-dl}. Strictly speaking it refers to the capability of computational systems to mimic tasks which normally require human intelligence, such as learning, reasoning, decision-making, and problem solving. However, the two terms ML and AI are often used interchangeably.

Classical, or probabilistic, ML has been in use long before the term ML came into existence. These algorithms are statistical models that try to capture relationships between various variables. Arguably, the most famous example is linear regression, originally developed by Isaac Newton for his work on the equinoxes around 1700~\cite{belenkiy_history_2005}, and later formalized by Legendre and Gauss in the early 19th century~\cite{stigler_history_1986}. 

\begin{figure}
    \centering
    \includegraphics[width=1\linewidth]{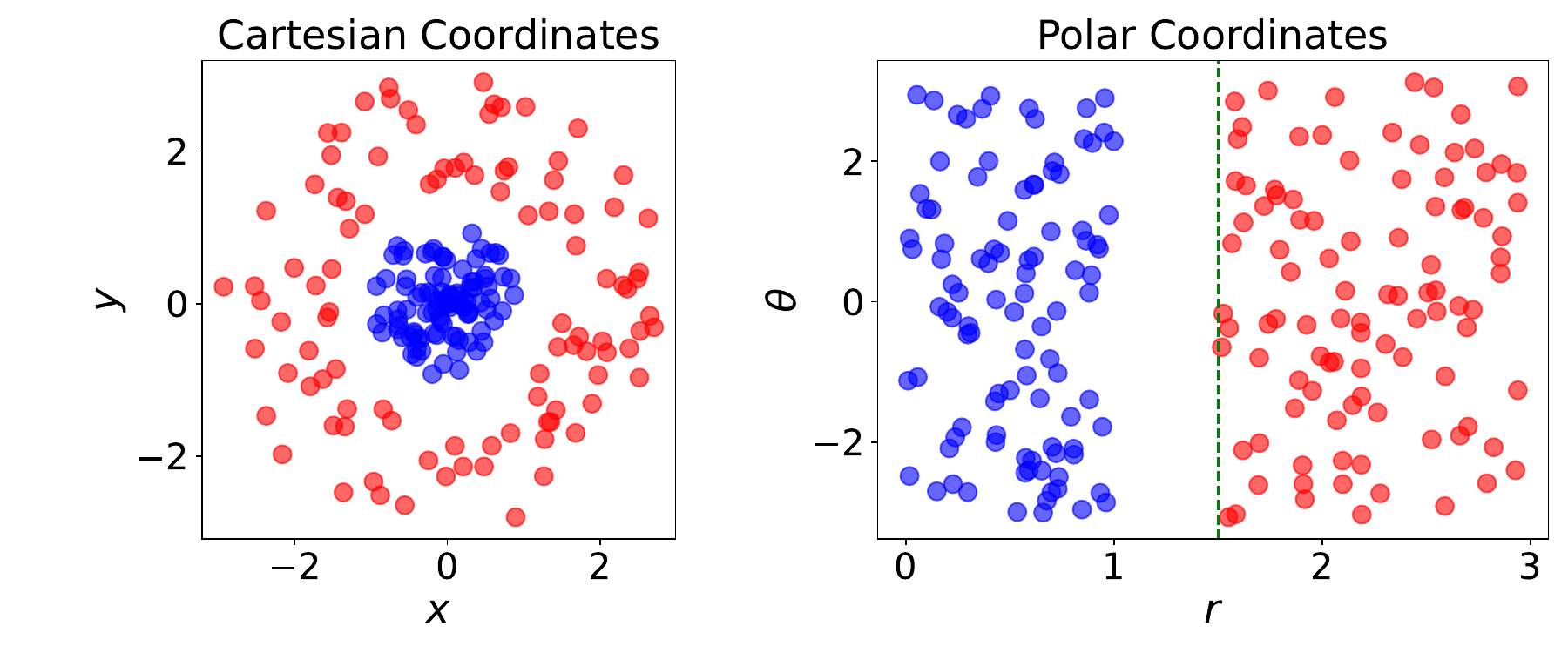}
    \caption{Example of different representations: Suppose we want to separate two classes of data by drawing a line between them. If the data are represented in Cartesian coordinates (left) the task is impossible. On the other hand, when the same points are represented in polar coordinates (right), the task becomes very simple to solve with a vertical separator.}
    \label{fig:coordinate_transform}
\end{figure}

The performance of these simple ML algorithms strongly depends on the \textit{representation} of the data they are given. For example, as illustrated in Fig.~\ref{fig:coordinate_transform}, the coordinate system used: Switching from Cartesian to polar coordinates might have a dramatic impact on the performance of an algorithm in solving a specific task. Each piece of information included in the representation of a data class, coordinates $x$, $y$ and $r$, $\theta$ in our example in Fig.~\ref{fig:coordinate_transform}, is known as a \textit{feature}. Linear regression tries to capture the relationship between these features, the independent variables, and the dependent variables. However, it cannot influence our choice for the definition of the features to be used.

Many ML tasks can be efficiently solved by designing the right set of features for that task, and then providing these features to a simple machine learning algorithm. As an example, imagine we have a set of images of either grass fields or the sea. What feature can we design to separate the two groups of images? We could find the average color of all the pixels and if the average is close to green then we would label the photo as ``grass'', while if it is close to blue as ``sea''. We can be confident that with this simple feature we have extracted, the performance of our classification algorithm is likely to be adequate for this task.

However, what happens if we pass each photo through a color filter, changing the color of the pixels? In this case, the algorithm breaks down completely. However, to a human eye, the classification task remains identically easy. So, how do we capture the ``seaness'' of the sea and the ``grassness'' of the grass? This is exactly where things get difficult. It is not obvious how to design a feature exactly in order to capture, for example, the texture of the grass in terms of pixel values. This is where \textit{representation learning}, also known as feature learning, comes in. It is a set of techniques that allows a system to automatically discover the representation needed for a specific problem, completely bypassing the need for hand-designing. And as it turns out, learned representations often result in much better performance than hand-designed ones~\cite{goodfellow_deep_2016}.

Deep learning is a form of representation learning and involves Neural Networks (NNs) with multiple layers. The NN learns hierarchical representations of data, i.e from low-level (e.g., edges in images) to high-level features (e.g., faces, objects). Frank Rosenblatt is attributed with introducing the \textit{perceptron} in 1958~\cite{rosenblatt_perceptron_1958}. Combining multiple of these perceptrons arranged in layers results in the so-called Multilayer Perceptron (MLP), also known as a Feedforward Neural Network (FNN). The first MLP trained by stochastic gradient descent~\cite{robbins_stochastic_1951} was published by Shun'ichi Amari in 1967~\cite{amari_theory_1967}. The ReLU (Rectified Linear Unit) activation function, introduced in 1969 by Kunihiko Fukushima~\cite{fukushima_visual_1969}, has now become the most popular activation function for deep learning~\cite{ramachandran_searching_2017}. Finally, the modern form of backpropagation was first published in 1970 by Seppo Linnainmaa~\cite{linnainmaa_representation_1970,linnainmaa_taylor_1976}. The method applied to neural networks was popularized by David E. Rumelhart et al. in 1986~\cite{rumelhart_learning_1986}.

During the 1990s, introduced by Yann LeCun~\cite{lecun_gradient-based_1998}, Convolutional Neural Networks (CNNs) marked a major breakthrough. In his seminal work, he proposed the LeNet-5 architecture, which utilized convolutional layers to recognize hand-written digits from the MNIST database---a significant shift from traditional fully connected layers.

\subsection*{The Revolution}

The ML/DL revolution was kick-started by CNN-based computer vision in 2012~\cite{krizhevsky_imagenet_2012}, driven by advancements in computation, particularly the graphics processing unit. Although CNNs trained via backpropagation had existed for decades, and neural networks---including CNNs---had already been implemented on GPUs for years~\cite{oh_gpu_2004, chellapilla_high_2006}, advancements in computer vision required faster GPU implementations. At the same time, in 2006, GPUs became programmable with Nvidia's CUDA framework~\cite{sanders_cuda_2010}. As deep learning gained widespread adoption, specialized hardware and optimized algorithms were subsequently developed to meet its growing demands~\cite{sze_efficient_2017}. In 2009, Rajat Raina et al. demonstrated an early example of GPU-accelerated deep learning by training a 100-million-parameter deep belief network using 30 Nvidia GeForce GTX 280 GPUs~\cite{raina_large-scale_2009}. Their approach achieved training speeds up to 70 times faster than traditional CPU-based methods.

Another reason why deep learning has only recently gained such traction is the availability of data in the era of ``big data''. ML algorithms are data-driven and in fact need a large amount of data in order to be able to be trained and to generalize well on unseen data. With the increasing digitization of society, data became abundant. Furthermore, it was possible to gather all these records and curate them into large datasets appropriate for training ML models.

Finally, even more recently, advances in Natural Language Processing (NLP) are beginning to transform our everyday lives. This was largely initiated by a novel architecture called \textit{transformer}, introduced by Google researchers in 2017~\cite{vaswani_attention_2017}, which was based mainly on the attention mechanism developed by Bahdanau et al.~\cite{bahdanau_neural_2016}. Based on the transformer architecture, Large Language Models (LLMs) can be constructed, containing billions of trainable parameters. One popular example is the chatbot ``ChatGPT''~\cite{openai_introducing_2022} which has an impressive ability to respond to various questions, and in diverse contexts, in a remarkably human-like manner. Ever since the introduction of the chatbot, the field of AI has been increasingly becoming the spotlight of attention, driving advancements and drawing the interest of academia, industry, and the public. However, the true capabilities of LLMs remain insufficiently understood~\cite{shojaee_illusion_2025}.

\subsection*{The Learning Procedure}

We now turn to the fundamental concepts related to the process of training a machine learning model. ML has a diverse set of application tasks including classification, regression, clustering, anomaly detection, transcription, denoising, density estimation and more. Each of these tasks has different specific requirements and objectives and hence the training procedure is different and focuses on optimizing different evaluation metrics. However, in general, ML algorithms can be broadly categorized as unsupervised or supervised based on their learning process.

\textbf{Unsupervised learning algorithms} have access to the entirety of a dataset containing various features, and learn useful properties and characteristics of the structure of this dataset. Clustering, for example, is possibly the most important unsupervised learning problem. It attempts to organize the elements of a dataset into groups which are similar in some way.

In high-energy physics, clustering plays a central role across many stages of data processing. For example, in pixel detectors, clustering is used to group adjacent hits in the sensor planes that are likely to have originated from the same charged particle, forming the basis for subsequent track reconstruction. Similar techniques are applied in calorimetry to group energy deposits and in jet reconstruction to cluster final-state particles.

While clustering is commonly framed as an unsupervised learning task, it can also appear in supervised or semi-supervised contexts, especially when the goal is to learn a model that mimics or improves upon a known clustering procedure, such as in learned jet tagging.

\textbf{Supervised learning algorithms}, on the other hand, have access to a dataset but each element of that set has an associated \textit{label}. For example, for a simple image classification task of animals, each image needs to have a label which specifies the animal that is the target of the classification.

Other learning paradigms exist such as semi-supervised learning and \textit{reinforcement learning}. The former is when some examples in the dataset include supervision targets while others do not, while the latter is when the learning algorithm interacts with an environment, so there is a feedback loop between the learning system and its actions.

\subsubsection{Example: Linear Regression}

To give an example of how a learning algorithm works we walk through possibly the simplest learning algorithm: linear regression. 

The goal of linear regression is to build a system that takes in a vector $\mathbf{x} \in \mathbb{R}^n$ as input and predict the value of a scalar $y \in \mathbb{R}$ as its output. Let $\hat{y}(\mathbf{x}_i)$ denote the value that our model predicts $y$ should be for example $\mathbf{x}_i$. We define the output to be 

\begin{equation}
    \hat{y}_i = \mathbf{w}^\top \mathbf{x}_i + b\,,
    \label{eq:linear-regression}
\end{equation}
where $\mathbf{w} \in \mathbb{R}^n$ and the scalar $b$ are the parameters we are trying to learn. We can think of $\mathbf{w}$ as the \textit{weights} and $b$ as the $bias$. We can further organize our dataset into a \textit{design matrix} $\mathbf{X}$, where the different examples $\mathbf{x}_i$ are organized in the rows of the matrix, and each column corresponds to a different feature. For simplicity, we can set $b=0$. In terms of the design matrix, $\hat{y}$ becomes a vector $(\hat{\mathbf{y}})_i = \hat{y}_i$ $\forall i $, and:

\begin{equation}
    \hat{\mathbf{y}} = \mathbf{X}\mathbf{w}\,.
\end{equation}

To make a learning algorithm we need to create an algorithm that can improve the weights $\mathbf{w}$ in order to improve the performance of the model, when the algorithm is allowed to gain experience by observing the dataset. However, how do you evaluate the performance of the model? One way of doing this is to compute the Mean Square Error (MSE) between the predictions and the actual values:

\begin{align}
    \text{MSE} &= \frac{1}{m} ||\hat{\mathbf{y}} - \mathbf{y}||^2 \\
    &= \frac{1}{m} \sum_{i=1}^m (\hat{\mathbf{y}} - \mathbf{y})_i^2\,,
    \label{eq:mse}
\end{align}
where $\mathbf{y}$ are the regression targets, and $m$ is the size of the set over which we are doing this evaluation. Furthermore, because we want to do a fair evaluation, we want to evaluate our model on examples it has never seen before. This can be achieved by splitting the dataset into a \textit{test} and a \textit{train} set. During the learning procedure the algorithm only has access to the training set, and after the end, the model is evaluated solely on the test set.

Therefore, in order to now minimize $\text{MSE}_{\text{train}}$, known as the \textit{loss function}, we can simply solve for where its gradient is $\mathbf{0}$:

\begin{equation}
    \nabla_{\mathbf{w}} \text{MSE}_{\text{train}} = \mathbf{0}
\end{equation}

\begin{equation}
    \Rightarrow \nabla_{\mathbf{w}} ||\hat{\mathbf{y}}^{\text{(train)}} - \mathbf{y}^{\text{(train)}}||^2 = \mathbf{0}
\end{equation}

\begin{equation}
    \Rightarrow \nabla_{\mathbf{w}} ||\mathbf{X}^{\text{(train)}} \mathbf{w} - \mathbf{y}^{\text{(train)}}||^2 = \mathbf{0}
\end{equation}

\begin{equation}
    \Rightarrow \nabla_{\mathbf{w}} (\mathbf{X}^{\text{(train)}} \mathbf{w} - \mathbf{y}^{\text{(train)}})^\top (\mathbf{X}^{\text{(train)}} \mathbf{w} - \mathbf{y}^{\text{(train)}}) = \mathbf{0}
\end{equation}

\begin{equation}
    \Rightarrow \nabla_{\mathbf{w}} \left( \mathbf{w}^\top \mathbf{X}^{\text{(train)}\top} \mathbf{X}^{\text{(train)}} \mathbf{w} - 2 \mathbf{w}^\top \mathbf{X}^{\text{(train)}\top}\mathbf{y}^{\text{(train)}} + \mathbf{y}^{\text{(train)}\top}\mathbf{y}^{\text{(train)}} \right)= \mathbf{0}
\end{equation}

\begin{equation}
    \Rightarrow 2 \mathbf{X}^{\text{(train)}\top} \mathbf{X}^{\text{(train)}} \mathbf{w} - 2 \mathbf{X}^{\text{(train)}\top}\mathbf{y}^{\text{(train)}} = \mathbf{0}
\end{equation}

\begin{equation}
    \Rightarrow \mathbf{w}  = \left( \mathbf{X}^{\text{(train)}\top} \mathbf{X}^{\text{(train)}} \right)^{-1} \mathbf{X}^{\text{(train)}\top}\mathbf{y}^{\text{(train)}}\,,
    \label{eq:normal-eqns}
\end{equation}
assuming that $\mathbf{X}^{\text{(train)}\top} \mathbf{X}^{\text{(train)}}$ is invertible. Evaluating Eq.~\eqref{eq:normal-eqns} constitutes a simple learning algorithm. However simple and limited this algorithm may be, it provides a good example of how a classical learning algorithm works.

From the previous example, certainly one question arises: Why did we choose to minimize MSE and not some other function? For each problem, rather than guessing that some function may be appropriate as an estimator, we would like to have a systematic way of deciding its form. The most common such principle is the principle of maximum likelihood, and the method is known as Maximum Likelihood Estimation (MLE).

\subsubsection*{Maximum Likelihood Estimation}

We demonstrate the MLE method and give the set of probabilistic assumptions under which least-squares regression is derived as a very natural algorithm~\cite{ng_cs229_nodate}.

Let us assume that, in line with Eq.~\eqref{eq:linear-regression}, the target variables and the input variables are related via the equation

\begin{equation}
    y_i = \mathbf{w}^\top \mathbf{x}_i + \epsilon_i\,,
    \label{eq:mle-error}
\end{equation}
where $\epsilon_i$ is the error term that captures random noise, or unmodeled effects. Let us further assume that these terms $ \left\{ \epsilon_i \right\}_{i=1}^{m} $, given $m$ observations, are independent and identically distributed (IID) random variables, and that they follow the Gaussian (or normal) distribution $\epsilon_i \sim \mathcal{N}(0,\sigma^2)$. The probability density function is therefore as follows

\begin{equation}
    p(\epsilon_i) = \frac{1}{\sqrt{2 \pi} \sigma} \exp(- \frac{\epsilon_i^2}{2 \sigma^2}) \,.
\end{equation}
This, given that $ \epsilon_i = y_i - \mathbf{w}^\top \mathbf{x}_i $ from Eq.~\eqref{eq:mle-error}, implies that

\begin{equation}
    p(y_i | \mathbf{x}_i ; \mathbf{w}) =  \frac{1}{\sqrt{2 \pi} \sigma} \exp(- \frac{(y_i - \mathbf{w}^\top \mathbf{x}_i)^2}{2 \sigma^2}) \,,
\end{equation}
the probability that $y_i$ will take a specific value, given the measurement of an example $\mathbf{x}_i$ and parametrized by $\mathbf{w}$.

Now, if we take into account all the measurements $\mathbf{x}_i$, in other words given the design matrix $\mathbf{X}$, what is the distribution of the $y_i$'s? Since we assumed independence, the probability will be a simple product of the respective probabilities for each observation:

\begin{align}
    p(\mathbf{y} | \mathbf{X}; \mathbf{w}) &= \prod_{i=1}^m p(y_i | \mathbf{x}_i ; \mathbf{w}) \\
    &= \prod_{i=1}^m \frac{1}{\sqrt{2 \pi} \sigma} \exp \left( - \frac{(y_i - \mathbf{w}^\top \mathbf{x}_i)^2}{2 \sigma^2} \right ) \,,
\end{align}
for $m$ measurements $\left\{ \mathbf{x}_i \right\}_{i=1}^m $. We can view this function as a function of $\mathbf{w}$, and in this case this function is known as the likelihood:

\begin{equation}
    L (\mathbf{w}) = L (\mathbf{w}; \mathbf{X}, \mathbf{y}) = \prod_{i=1}^m \frac{1}{\sqrt{2 \pi} \sigma} \exp \left( - \frac{(y_i - \mathbf{w}^\top \mathbf{x}_i)^2}{2 \sigma^2} \right) \,.
\end{equation}
Given this probabilistic model for the $y_i$'s based on the data points $ \left\{ \mathbf{x}_i \right\}_{i=1}^m$, what is the best way to choose the values for the parameters $\mathbf{w}$? The \textit{principle of maximum likelihood} states that the parameters for which the observations are as highly probable as possible should be chosen. This is equivalent to maximizing the likelihood function $L(\mathbf{w})$. 

The maximization of $L(\mathbf{w})$ is equivalent to the maximization of the logarithm of $L(\mathbf{w})$, since the logarithmic function is strictly increasing. Hence, we want to maximize the log likelihood $l(\mathbf{w})$:

\begin{align}
    l(\mathbf{w}) &= \log L(\mathbf{w}) \\
    &= \log \prod_{i=1}^m \frac{1}{\sqrt{2 \pi} \sigma} \exp \left( - \frac{(y_i - \mathbf{w}^\top \mathbf{x}_i)^2}{2 \sigma^2} \right) \\
    &= \sum_{i=1}^m \log \frac{1}{\sqrt{2 \pi} \sigma} \exp \left( - \frac{(y_i - \mathbf{w}^\top \mathbf{x}_i)^2}{2 \sigma^2} \right) \\
    &= m \log \frac{1}{\sqrt{2\pi} \sigma} - \frac{1}{\sigma^2} \frac{1}{2} \sum_{i=1}^m (y_i - \mathbf{w}^\top \mathbf{x}_i)^2 \,.
\end{align}
Hence, maximizing $l(\mathbf{w})$, is equivalent to minimizing

\begin{equation}
    \sum_{i=1}^m (y_i - \mathbf{w}^\top \mathbf{x}_i)^2 \,,
\end{equation}
which we recognize to be our original least-squares (MSE) cost function of Eq.~\eqref{eq:mse}.

Therefore, under the assumptions of Gaussian IID errors, the least-squares linear regression algorithm corresponds to the maximization of the likelihood function. Depending on the problem at hand, by a similar approach, one can prove that, for example, for a binary classification task, the most appropriate cost function is given by the binary cross entropy~\cite{rumelhart_learning_1986, goodfellow_deep_2016, shannon_mathematical_1948}.

\subsubsection*{Generalization, Overfitting, and Underfitting}

\begin{figure}
    \centering
    \includegraphics[width=1\linewidth]{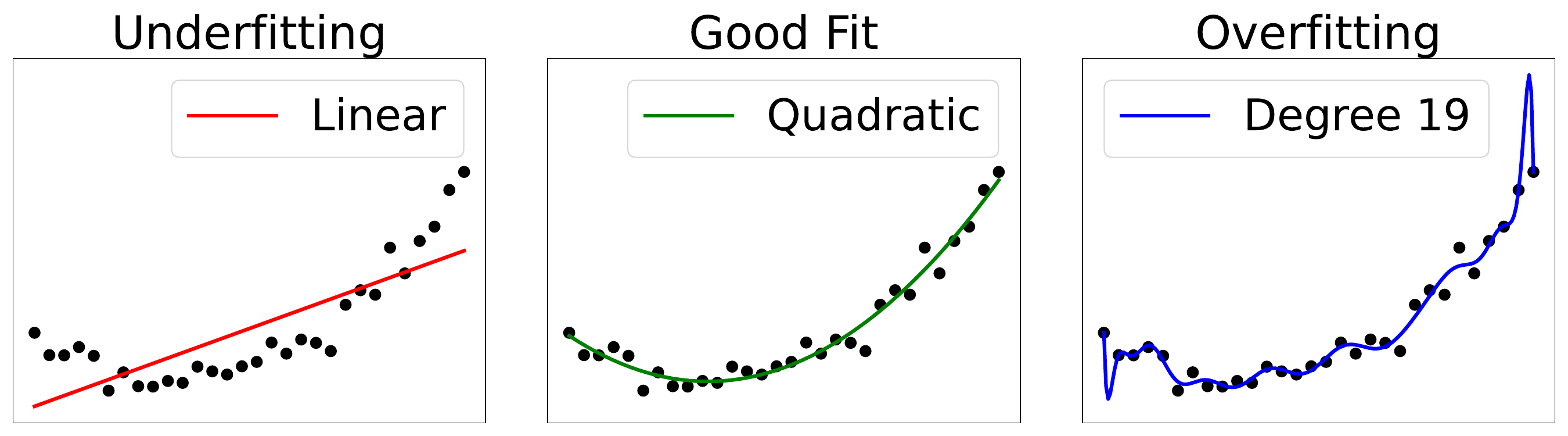}
    \caption{Examples of underfitting and overfitting on a synthetically generated dataset with quadratic structure. Left: A linear fit cannot capture the curvature present in the data. Center: A quadratic fit generalizes well to unseen points and hence does not suffer from a significant amount of either underfitting or overfitting. Right: A polynomial fit of degree 19 suffers from strong overfitting. The solution passes exactly through many points in the dataset, however, the structure has not been correctly extracted, and the performance on unseen data will be poor.}
    \label{fig:underfitting-overfitting}
\end{figure}

Another important challenge in this process, one of the most central ones, is to further make the learning algorithm perform well on the test set, on \textit{new, unseen} inputs, not only on the dataset that the model was trained on. In other words, we want the model to be able to \textit{generalize}. In order to decide, whether a model is doing this well, we have to compare the loss on the test set, $\text{MSE}_{\text{test}}$ in our example, with the loss on the training set $\text{MSE}_{\text{train}}$. If the model is generalizing well, we expect the error on the test set to be roughly the same as the error on the training set. If the model is not generalizing well, we talk about overfitting or underfitting. The former refers to the case where a model corresponds too closely to the dataset it was trained on, and hence performs poorly on new unseen data. The latter refers to the case where a model cannot adequately capture the underlying structure of the data. In Fig.~\ref{fig:underfitting-overfitting}, examples of underfitting and overfitting are compared.

Furthermore, if the model's deviations from the data are, on average, roughly the same size as the measurement uncertainties of the data points, that means the ML model is doing a ``good-enough'' fit of the data---i.e., it's actually fitting the signal and not the noise. On the other hand, if the residuals are significantly smaller than the measurement uncertainties, this indicates that the model is also fitting random fluctuations and thus overfitting.

\section{Deep Learning}
\label{sec:dl}

Deep feedforward networks, also known as MLPs, are the archetype of deep learning models. They are called deep because they have several layers and feedforward because of how the information is progressively fed into the successive layers, flowing towards the output. The term neural is a remnant of the models' origins in neuroscience, specifically the McCulloch-Pitts neuron~\cite{mcculloch_logical_1943}, a simplified model of the biological neuron that can be used as a form of computing element. However, the modern use in deep learning no longer draws these parallels from biology. Finally, these models are called networks because they are typically represented by combining and chaining various neurons together.

\begin{figure}
    \centering
    \includegraphics[width=0.85\linewidth]{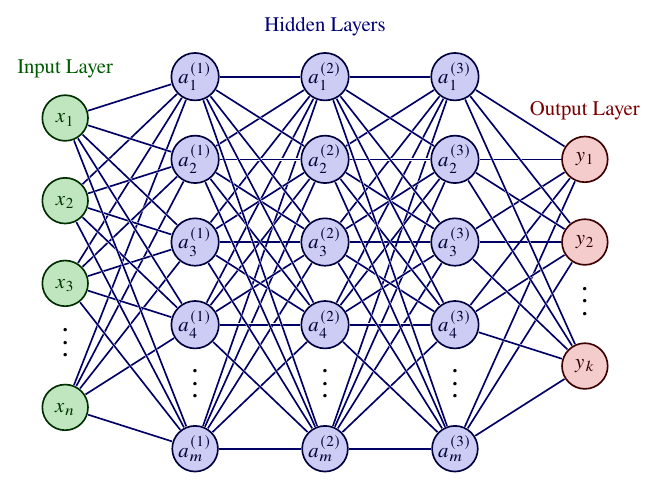}
    \caption{Illustration of a deep feedforward neural network, highlighting its input, output and hidden layers. Adapted from~\cite{neutelings_neural_2024}.}
    \label{fig:nns}
\end{figure}

A feedforward neural network with three hidden layers is shown in Fig.~\ref{fig:nns}. In our example, input, hidden and output layers have $n$, $m$ and $k$ units, respectively. Moreover, we can see that the network is fully-connected since every neuron of a layer is connected to every neuron in neighboring layers.

One way to understand neural networks is to consider the limitations of linear models. The obvious problem with linear models is that they are limited to linear functions. In order to extend linear models to approximate nonlinear functions of $x$, we can apply the linear model not to $\mathbf{x}$ itself but to a transformed input $\phi(\mathbf{x})$, where $\phi$ is a nonlinear transformation. We can think of this function $\phi$ as providing a new representation of $\mathbf{x}$. 

So how can this nonlinear transformation $\phi$ be chosen? We already saw that in classical ML approaches, this is hand-crafted by the engineer. However, here, since deep learning is a type of representation learning, the goal is to learn this transformation $\phi$. If we assume that this transformation depends on some set of parameters $\mathbf{w}$, then we can learn what these parameters have to be for a good representation.

So how do we do this? We start from our input say $\mathbf{x}$. For linear regression, we had:

\begin{equation}
    f(\mathbf{x}; \mathbf{w},b) = \mathbf{x}^\top \mathbf{w} + b\,.
\end{equation}
The output of this model is a scalar even though the input is a vector. However, if we wanted a multidimensional output, where the linear parameters $\mathbf{w}$ are different for each dimension, we can organize the parameters in a matrix $\mathbf{W}$ such that:

\begin{equation}
    \mathbf{h}(\mathbf{x}; \mathbf{W}, \mathbf{b}) = \mathbf{W} \mathbf{x} + \mathbf{b}\,,
\end{equation}
where now we have a different bias, i.e., additive constant, $(\mathbf{b})_i$ for each output dimension.

\begin{figure}
    \centering
    \includegraphics[width=0.95\linewidth]{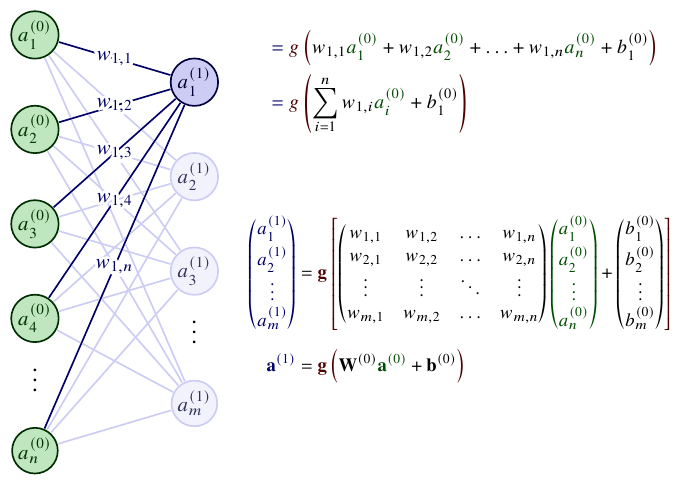}
    \caption{The operations between the input and the first hidden layer. Weights are denoted as $w$, biases as $b$, and the activation function as $g$. The element-wise, vector version of the activation is denoted by $\mathbf{g}$. Adapted from~\cite{neutelings_neural_2024}.}
    \label{fig:nn-operations}
\end{figure}

Finally, to overcome the defect of linear models, we use a nonlinear function after this affine transformation. This nonlinear function is known as the \textit{activation function} and can be denoted by $\mathbf{g}$. Therefore, our model now is as follows:

\begin{equation}
    \mathbf{h}(\mathbf{x}; \mathbf{W}, \mathbf{b}) = \mathbf{g} (\mathbf{W} \mathbf{x} + \mathbf{b} ) \,,
\end{equation}
where $\mathbf{g}$ is element-wise. The nonlinear function $\phi$ now comprises an affine transformation based on the learnable parameters $\mathbf{W}$ and $\mathbf{b}$, and a fixed nonlinear function $\mathbf{g}$. The parameters are adjusted during training, while the form of the activation $\mathbf{g}$ is chosen beforehand. These operations are also summarized in Fig.~\ref{fig:nn-operations}.

\begin{figure}
    \centering
    \includegraphics[width=1\linewidth]{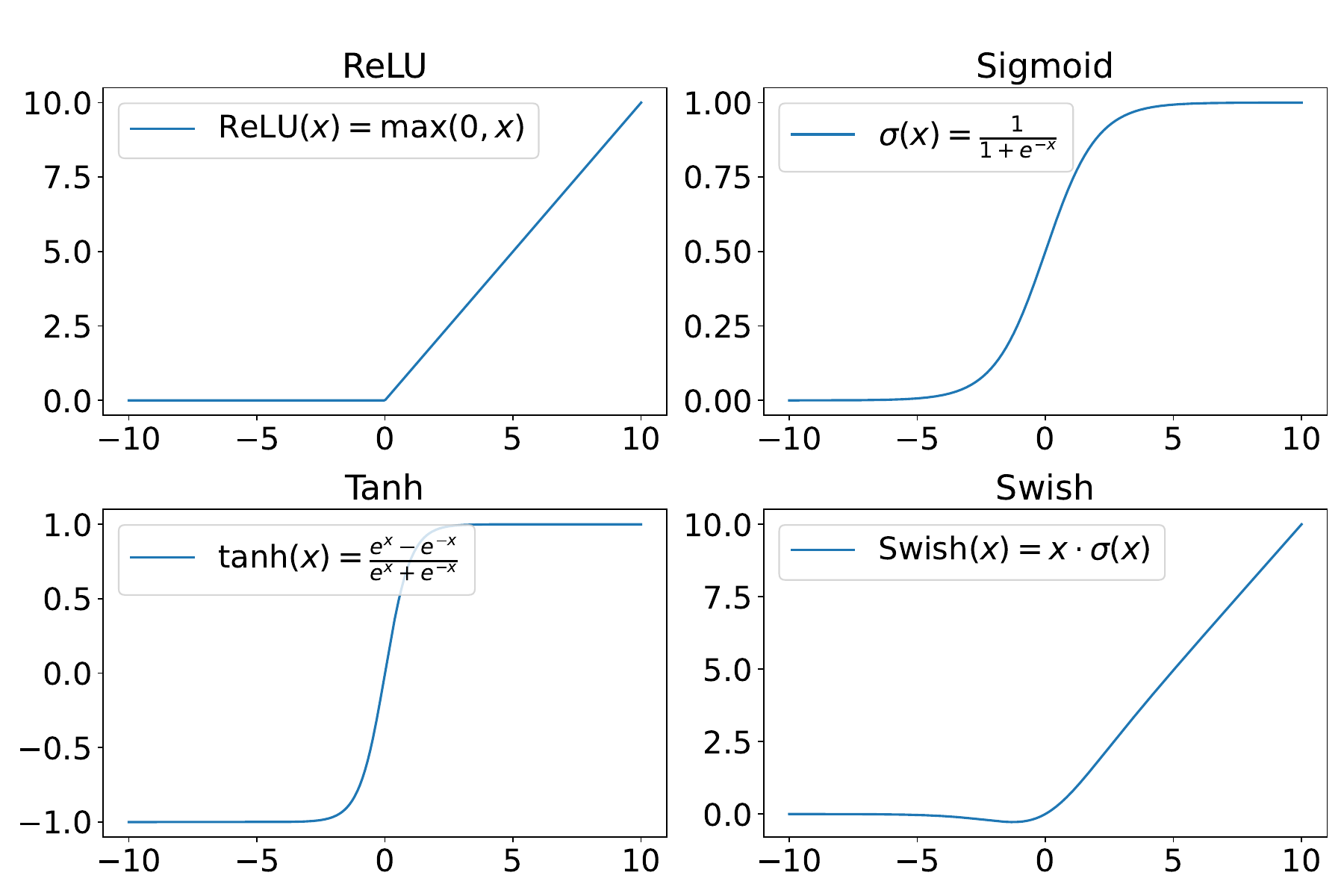}
    \caption{Popular activation functions. }
    \label{fig:activations}
\end{figure}

Various popular activations are plotted in Fig.~\ref{fig:activations}. ReLU has only nonnegative values and is defined as $\text{ReLU}(x) = \max(0,x)$. It is computationally efficient and mitigates the vanishing gradient problem, making it the default activation for various deep learning architectures. However, it suffers from the so-called ``dying ReLU'' problem, where neurons can become completely inactive and only output zero for all inputs.

The sigmoid function is defined as $\sigma (x) = 1/(1+e^{-x})$, taking values between 0 and 1. While historically important, sigmoid activations are prone to the vanishing gradient problem for large absolute values of the input, which can hamper the training of deep networks, unless intermediate layers designed to avoid this are introduced.

The hyperbolic tangent is defined as $\tanh (x) = (e^x - e^{-x} )/(e^x + e^{-x})$ so the function takes values between $-1$ and 1. The function is zero-centered which can help with convergence compared to the sigmoid. Nonetheless, it still suffers from vanishing gradients for large inputs.

Finally, the swish function $\text{swish} (x) = x/(1+e^{-x})$~\cite{ramachandran_searching_2017} is an attempt to interpolate between the linear function and the ReLU function. Swish has been shown to outperform ReLU in some deep architectures, especially in deeper models. However, it is computationally more expensive, which can be a serious drawback in resource-constrained settings.

A neural network is nothing more than a chain function of these successive transformations. So, for a $k$-layer neural network that returns a scalar, the combined action of the neural network $f_{\text{NN}}$ on an input $\mathbf{x}$ is simply:

\begin{equation}
    y = f_{\text{NN}} (\mathbf{x}) = f_k ( \boldsymbol{f}_{k-1} ( \cdots \boldsymbol{f}_2 ( \boldsymbol{f}_1 ( \mathbf{x})))) \,,
    \label{eq:nested_function}
\end{equation}
where $\boldsymbol{f}_l$,  for the layer index $l = 1,...,k-1$, are functions with vector output of the form:

\begin{equation}
    \boldsymbol{f}_l (\mathbf{z}) = \mathbf{g_l} (\mathbf{W}_l \mathbf{z} + \mathbf{b}_l) \,,
\end{equation}
where $\mathbf{W}_l$ are the weights between layers $l$ and $l-1$, $\mathbf{g_l}$ and $\mathbf{b}_l$ are the activation and biases, respectively, of layer $l$,
while $f_k$ returns a scalar.

The remarkable result of the universal approximation theorem~\cite{hornik_multilayer_1989} states that, under mild assumptions on the activation functions used for the neural network, any continuous function $f : [0, 1]^n \rightarrow [0, 1]$ can be in fact approximated arbitrarily well by a neural network with \textit{as few as one} hidden layer and with a finite number of weights. By adding more layers, we are increasing the complexity of the model and hence its capacity to approximate a complex function, as well as to generalize. At the same time, however, we are increasing the computational cost of the algorithm, and therefore, the development of DL models is always a trade-off between these two aspects. By learning the parameters of these models, we essentially can learn how to solve any task, along the representations needed for this specific task. 

In order for the learning process to happen, a loss function, similarly to the loss in Eq.~\eqref{eq:mse} of our linear regression example, is needed. Depending on the problem, a suitable form can be chosen using the MLE method. The weights and biases have then to be chosen such that this function is minimized. This is most frequently done using a form of gradient-based optimization.

\subsubsection{Gradient-Based Optimization}

Optimization, in general, refers to the minimization or maximization of an \textit{objective function} $J$, a more general term for what we have been calling the loss function so far. In more general optimization problems---including reinforcement learning and economic modeling---the objective function may take a different form from the loss functions encountered previously, and the goal may instead be to maximize it, such as maximizing a reward signal or economic profit.

For the case of neural networks, we are minimizing the prediction error of the model and this objective function is called a loss function. It is a smooth differentiable function of the parameters $\boldsymbol{\theta}$ of the model. In addition, even though it has multiple inputs, for the concept of ``minimization'' to make sense, there must be only one output, i.e., $J : \mathbb{R}^n \rightarrow \mathbb{R}$. In order to minimize $J(\boldsymbol{\theta})$, we need to find the direction, in the $n$-dimensional parameter space, that $J$ decreases the fastest and move in this direction. Since, by the definition of the gradient, $\nabla_{\boldsymbol{\theta}} J (\boldsymbol{\theta})$ gives the direction in which $J$ increases the fastest, we have to update $\boldsymbol{\theta}$ by going in the opposite direction:

\begin{equation}
    \boldsymbol{\theta} \leftarrow \boldsymbol{\theta} - \alpha \nabla_{\boldsymbol{\theta}} J(\boldsymbol{\theta}) \,,
    \label{eq:param_updates}
\end{equation}
where $\alpha$ controls the size of the step in this direction and is known as the learning rate. This method proceeds in \textit{epochs}. An epoch consists of using the entire training dataset to update each parameter. This iterative optimization algorithm is known as \textit{gradient descent}.

Depending on the size of the dataset, one epoch could be too time consuming for the purposes of developing an ML model. In that case, a family of methods known as Stochastic Gradient Descent (SGD) can be used. For example, instead of using the entire dataset for the parameters updates in Eq.~\eqref{eq:param_updates}, we can sample a \textit{mini-batch} of data drawn uniformly from the training set. The convergence to a local minimum is thus noisier but significantly faster. At the same time, using this method during training, non-optimal local minima can be avoided.

\begin{figure}
    \centering
    \includegraphics[width=0.75\linewidth]{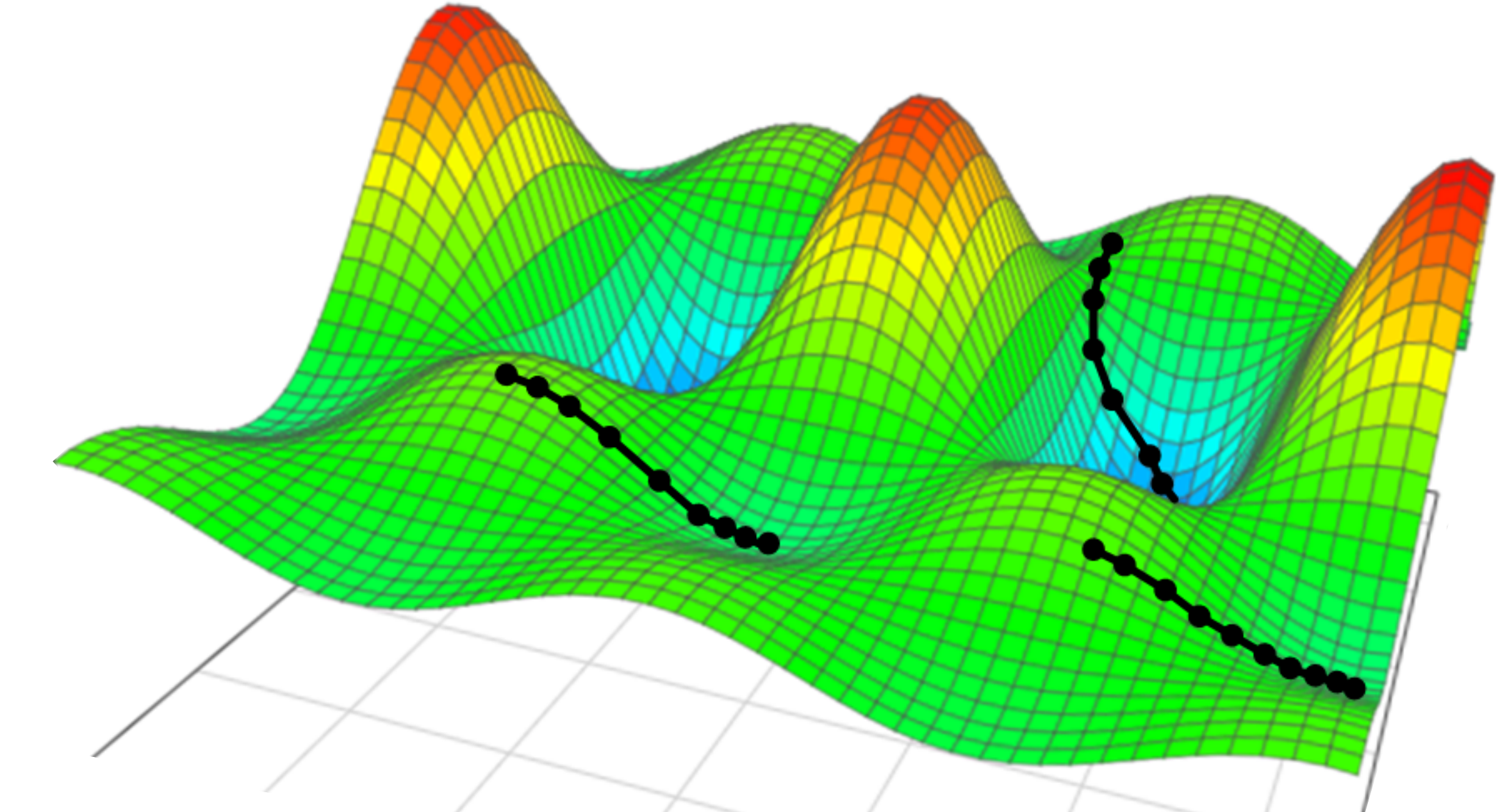}
    \caption{Illustration of gradient descent in a two-dimensional parameter space. Different trajectories may lead to different local minima, and hence may give qualitatively different results. Figure from~\cite{cohen_understanding_2018}.}
    \label{fig:sgd}
\end{figure}

The process for two learnable parameters is visualized in Fig.~\ref{fig:sgd}. Different trajectories can lead to different local minima, potentially resulting in qualitatively distinct outcomes. This problem can be mitigated using optimized versions of these algorithms, with, for example, a variable learning rate. A frequently used example, is the Adam optimizer~\cite{kingma_adam_2017}. It combines an adaptive learning rate with momentum, which accumulates a moving average of past gradients to sustain optimization in consistent directions, thereby reducing the risk of stalling in small local minima or flat regions (plateaus) of the loss landscape. In this way, convergence is accelerated and robustness is improved across a wide range of tasks.

The next question that arises is the following. Since we said our neural network is essentially a complex nested function of these combinations of nonlinear activations and affine transformations, as in Eq.~\eqref{eq:nested_function}, that means that the loss function is going to have a similar structure. So, how do we know how to update the individual parameters of each layer of the neural network, in order to minimize this objective function?

\subsubsection*{Backpropagation}

When we use a feedforward neural network that accepts an input $\mathbf{x}$ and produces an output $\mathbf{y}$, information flows ``forward'' through the network, as in, from left to right in Fig.~\ref{fig:nns}. The input vector $\mathbf{x}$ provides the initial information that propagates, layer by layer and finally results in $\mathbf{y}$. This vector $\mathbf{y}$ is a function of all the weights and biases of all the layers of the neural network, denoted collectively as $\boldsymbol{\theta}$. This process is known as forward propagation. A scalar cost function $J(\boldsymbol{\theta})$ can then be formed using the output $\mathbf{y}$.

The backpropagation algorithm, is the reverse process where the information from the cost $J(\boldsymbol{\theta})$ flows ``backward'', i.e., from right to left in Fig.~\ref{fig:nns}, through the network in order to compute the gradients needed for the updates in Eq.~\eqref{eq:param_updates}. Essentially, it is an efficient application of the chain rule to neural networks. Backpropagation computes the gradient of a loss function with respect to the parameters of the network for a single input-output example by applying the chain rule layer by layer in reverse order. This backward iteration avoids redundant calculations of intermediate derivatives and is related to dynamic programming, as it reuses intermediate results in order to improve efficiency~\cite{goodfellow_deep_2016}.

Strictly speaking, the term backpropagation refers only to the algorithm used for this computation and does not include how the computed gradients are used. The term however, is often used loosely to refer to the entire learning algorithm, including the parameter updates in Eq.~\eqref{eq:param_updates}.

\section{Convolutional Neural Networks}

Convolutional neural networks are a special kind of deep learning model, especially suited to image data. When the training data are images, the input is high-dimensional. Even for a low resolution image of 256 by 256, the input would have to be of size $256 \times 256 = 65\,536$. At this size, using a fully connected feedforward neural network to process the input starts being problematic. In addition, by treating the pixels essentially as a vector, we lose information about the ``local structure'' of the image. Apart from the value of the pixel itself, there is a significant amount of information in the placement of the pixels relative to each other. Going back to our earlier example, even by changing the value of the pixel colors, one could still understand whether a photo depicts a grass field or not. The information of the texture of the grass is somehow encoded in how the relative values of the pixels are arranged together to form the edges that correspond to the grass blades, and the patterns in general, which together convey the texture and structure typical of a grass field.

In order to capture this local structure of the image, the idea is to instead of flattening the input into a vector, to process it in its original, matrix-like, form. To make this easier, we can split the image into small square patches, of equal size. Each patch can then be processed to extract meaningful local features. In practice, this is done using shared filters, also known as kernels, that learn to detect patterns relevant to the task. How can this really be done?

\begin{figure}
    \centering
    \includegraphics[width=0.55\linewidth]{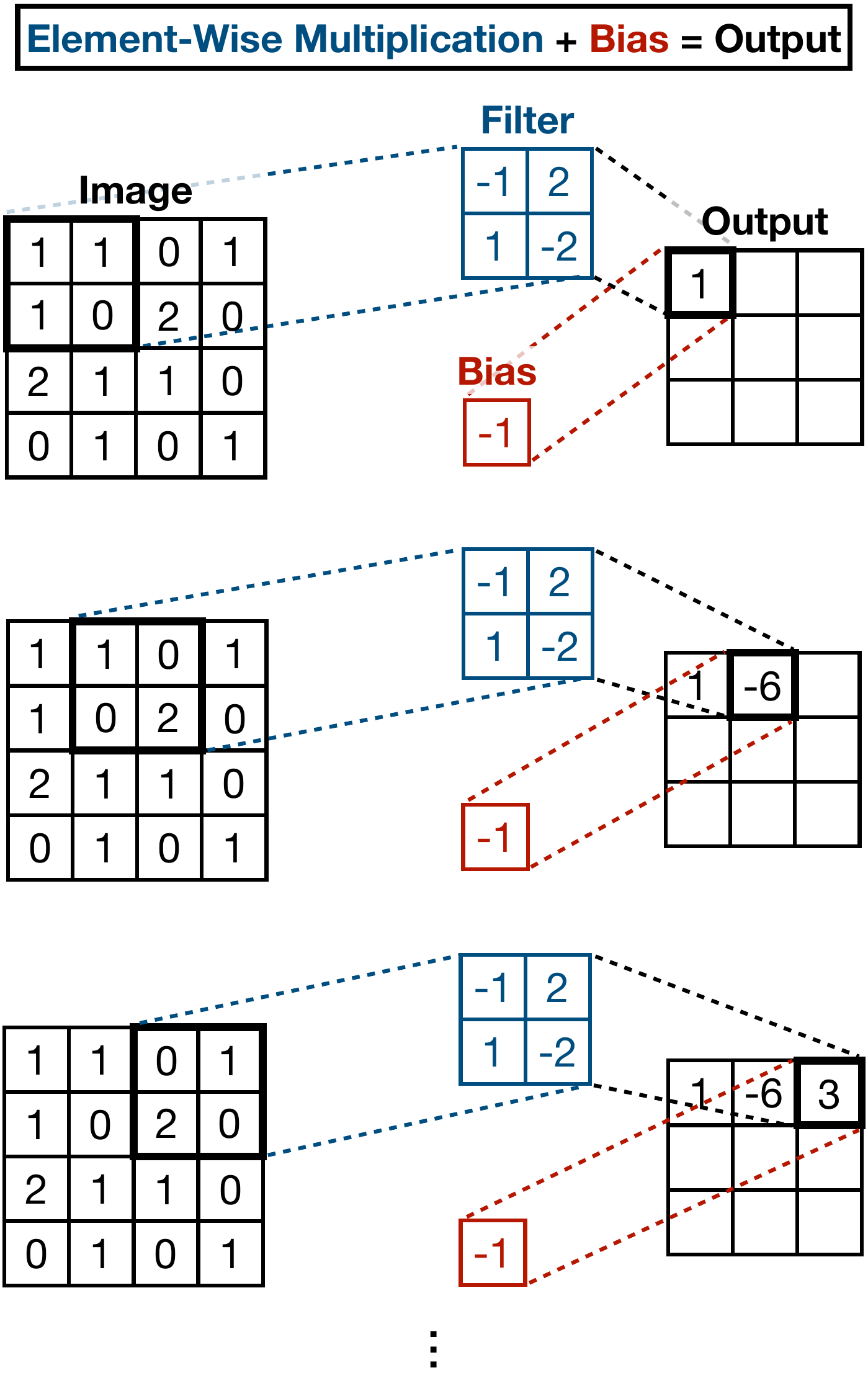}
    \caption{Illustration of the process of convolving a filter across an image using a sliding window approach. Inspired by~\cite{burkov_hundred-page_2019}.}
    \label{fig:conv}
\end{figure}

In order to preserve the local structure, we organize the learnable parameters of the model in a matrix $\mathbf{F}$, for ``filter'', of size equal to the size of the patches. We then perform the \textit{convolution} of the filter matrix $\mathbf{F}$, across the original image using a moving window approach, as illustrated in Fig.~\ref{fig:conv}. The pixels of the patch are multiplied element-wise into a scalar, and then the bias is added. The output of this operation is sometimes referred to as the feature map. Like before, a nonlinearity is applied to the output, typically the ReLU activation. The learnable parameters of this algorithm are the values of the matrix ``filter'' as well as the value of the ``bias''.

This operation is performed for a number $k$ of filters, in order to extract various features in the image, and each filter's parameters are completely independent. The output for each filter is different, and hence the operation of this convolutional layer results in a collection of $k$ feature maps. This collection can be thought of as a higher-dimensional tensor and is called a volume. For color images, the input is actually also a volume, since the image is usually represented by three channels: R (red), G (green), and B (blue), where each channel is a monochrome picture.

In a convolutional layer with a multi-channel input volume, the operation is similar to the single-channel case. The convolution of a patch from a multi-channel volume is equal to the sum of the convolutions of the corresponding patches from each individual channel.

By applying various convolutional layers in sequence, the model can learn hierarchical representations of data, starting from low-level representations such as edges in images, all the way to high-level features such as faces and objects. 

Another operation frequently used in CNNs is \textit{pooling}. It works in a similar way to the convolution, as a filter is applied using a sliding window approach. However, instead of applying a trainable filter, a fixed operation is applied: commonly max pooling (which selects the maximum value) or average pooling (which computes the mean value) within each window. Pooling is used to reduce the spatial dimensions of feature maps, helping to retain the most significant features from the input. This subsampling process lowers the number of parameters, decreases computation time, and helps prevent overfitting, ultimately improving model performance.

\begin{sidewaysfigure}
    \centering
    \includegraphics[width=1\linewidth]{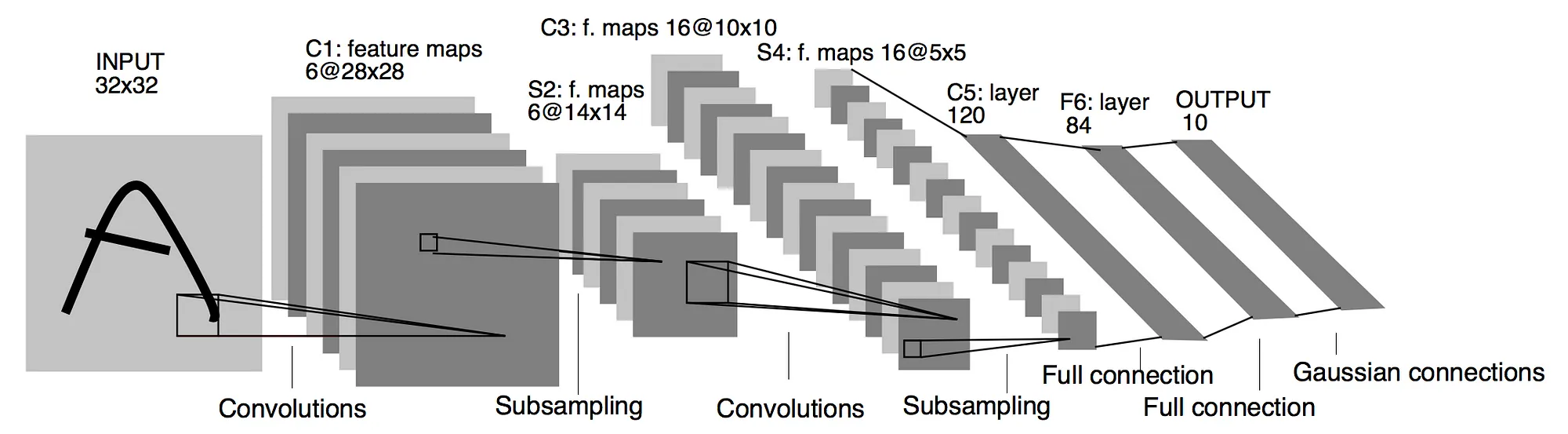}
    \caption{The architecture of LeNet-5, a convolutional neural network for digits recognition, as depicted in the original paper~\cite{lecun_gradient-based_1998}. The feature extraction module is illustrated using convolution and pooling operations. The classification is performed in the fully connected layers. The input is images of size $32 \times 32$. Layer C1 has 6 feature maps of size $28 \times 28$, while layer C3 has 16 feature maps of size $10\times10$. After subsampling, layers S2 and S4 reduce the size of the maps by one half. The output is then fed into the fully connected network of layers with 120 and 84 units. Finally, the output of the network is a vector of dimension 10.}
    \label{fig:lenet}
\end{sidewaysfigure}

A famous and illustrative example of the CNN architecture is shown in Fig.~\ref{fig:lenet}. The LeNet-5 architecture~\cite{lecun_gradient-based_1998}, designed for digits recognition, is split into two modules: the feature extraction module and the trainable classifier module. For the former, a convolutional layer is combined with a subsampling layer twice, C1-S2 and C3-S4, and then layer C5 creates 120 feature maps of size $1\times 1$. These feature maps are then ``flattened'' into a 1-dimensional vector of size 120. For the classification part, this vector is then fed into the feedforward fully connected layers.

\section{Graph Neural Networks}

What happens when the data that we have are not structured in the traditional tabular manner, such as vectors in the case of series, or matrices in the case of images? Furthermore, what happens when our data possess an inherent network structure which we would like to take into account, or even learn about directly?

Networks are ubiquitous---and so are graphs. In many real-world scenarios, it is beneficial to think of data points not in isolation but as part of a web of complex connections: people connected through social interactions, proteins by biochemical interactions, or web pages by hyperlinks. Capturing and using this connectivity is crucial for understanding the underlying relationships and dynamics~\cite{hamilton_graph_2020,leskovec_cs224w_nodate,battaglia_relational_2018,bacciu_gentle_2020}. 

\begin{figure}
    \centering
    \begin{minipage}[t]{0.48\linewidth}
        \centering
        \includegraphics[width=\linewidth]{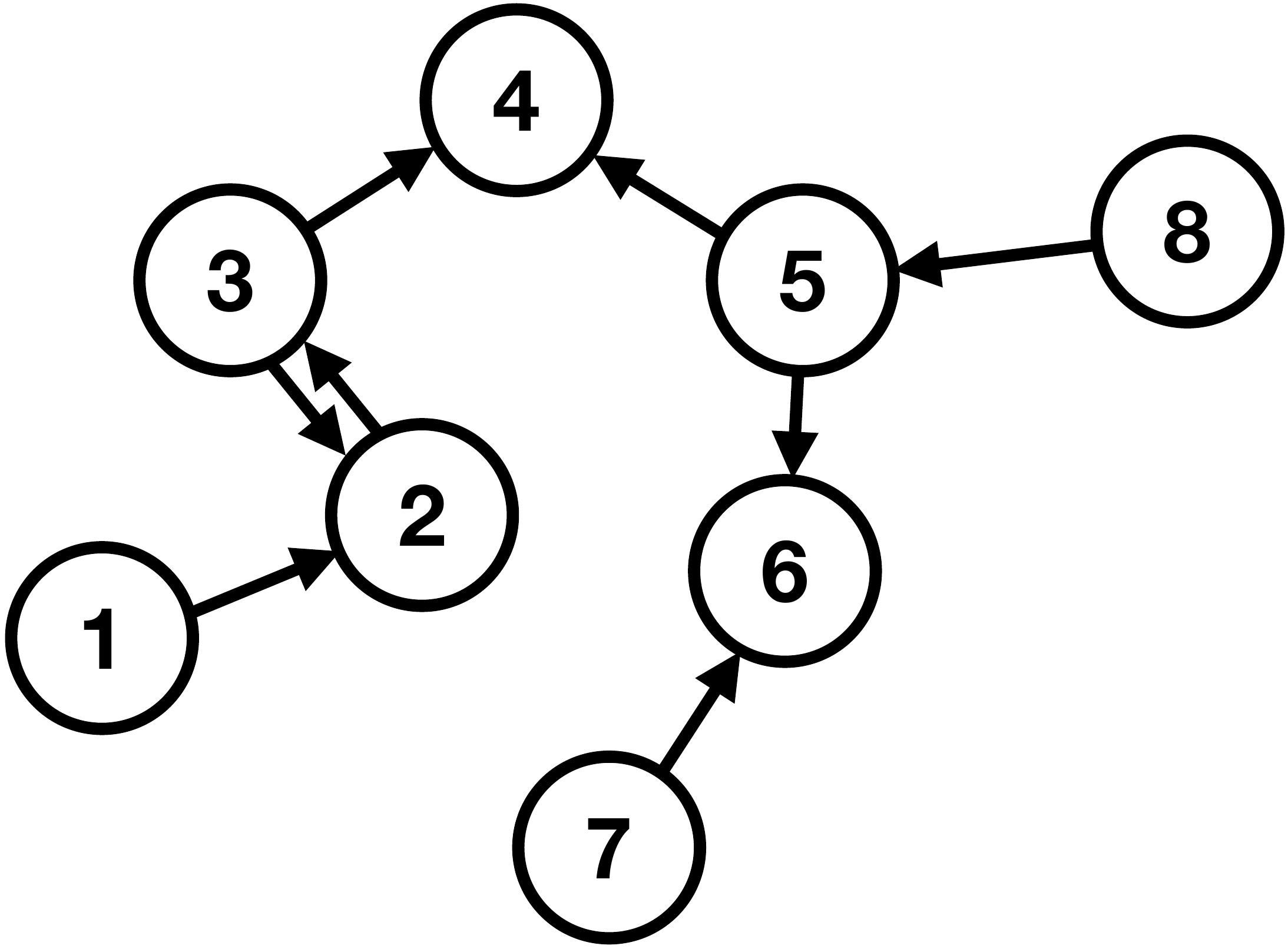}
        \caption{A directed graph with eight vertices and seven edges.}
        \label{fig:graph-dir}
    \end{minipage}
    \hfill
        \begin{minipage}[t]{0.48\linewidth}
        \centering
        \includegraphics[width=\linewidth]{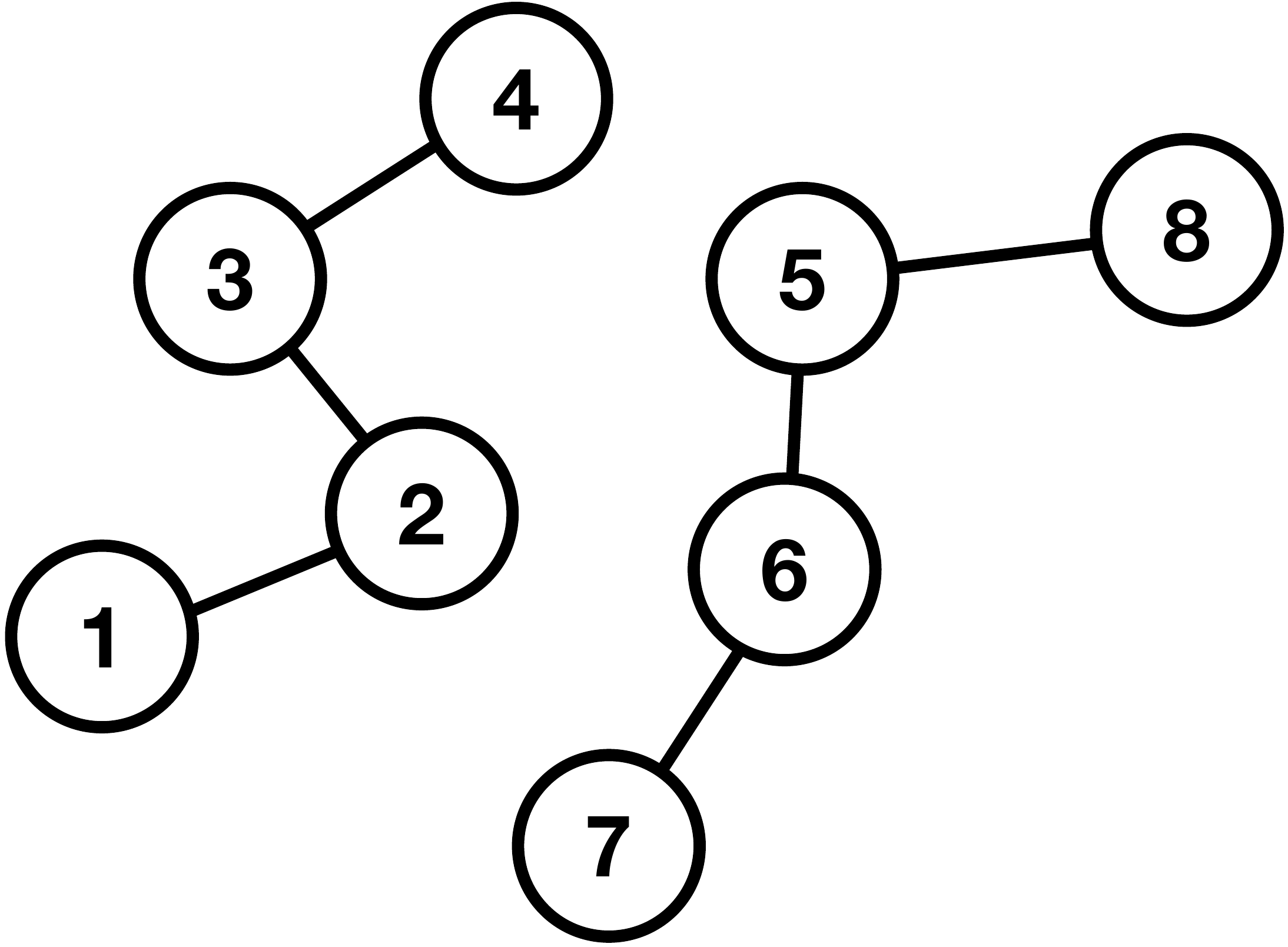}
        \caption{An undirected graph with eight vertices and seven edges, and two connected components.}
        \label{fig:graph}
    \end{minipage}
\end{figure}

Similarly to images being processed by CNNs, we would like to have an algorithm that can have these complex network structures as input. These structures are known as graphs. In general, a graph is a pair $G = (V, E)$, where $V$ is a finite set of vertices (or nodes), and $E$ is the set of connections (known as edges) between these nodes. Graphs can be further classified into directed and undirected. The former means that the edges have a certain direction, for example, we can go from node 5 to 6, but not the other way around, as illustrated in Fig.~\ref{fig:graph-dir}. The latter means that the connections are only symmetrical and mutual, as illustrated in Fig.~\ref{fig:graph}. In addition, in Fig.~\ref{fig:graph}, we can see that the graph comprises two so-called connected components, i.e., maximally connected subgraphs which are disconnected with each other.

Graphs can be represented in various ways. A frequently used representation is the so-called \textit{adjacency matrix}. The elements of the adjacency matrix $\mathbf{A}$ are given simply by
\begin{equation} \label{eq:adj}
\mathbf{A}_{ij} =
\begin{cases}
1, & \text{if there is a link from node $i$ to node $j$}, \\
0, & \text{otherwise}.
\end{cases}
\end{equation}
The adjacency matrix $\mathbf{A}$ is therefore symmetric for undirected graphs but not necessarily symmetric for directed ones. Furthermore, the edges themselves, may possess some value based on some characteristic, instead of simply 0 and 1, as in Eq.~\eqref{eq:adj}. In this case, the graph is called weighted. Finally, the information associated with the nodes is referred to as \textit{node features}, while the information associated with the edges is known as \textit{edge features}.

The question now is the following: How do we take advantage of the relational structure of graphs, in order to achieve better predictions? Drawing inspiration from CNNs, where we wanted to capture the local structure of the pixels in the images, we will try to do something similar. The idea is to do a series of ``convolutions'', similar to the ones for images, but this time suited for data with a network structure.

\subsection*{Node Embeddings}

Similarly to deep learning, we wanted to avoid hand-designing the representations of the problem, and we tried to learn them, in a process that we called representation learning. In the same vein, we will use the same method for our graphs. We will learn node representations, which we will call node embeddings, that will contain information about any node and its connections to neighboring nodes. In this mapping, that can be learned using a neural network, similar nodes in the network are embedded close to each other.

\subsection*{Message Passing}

In order to capture and encode inside the node embeddings the connectivity of the network, for each node in the graph, the process is as follows~\cite{scarselli_graph_2009,hamilton_inductive_2017}. 

\begin{figure}
    \centering
    \includegraphics[width=1\linewidth]{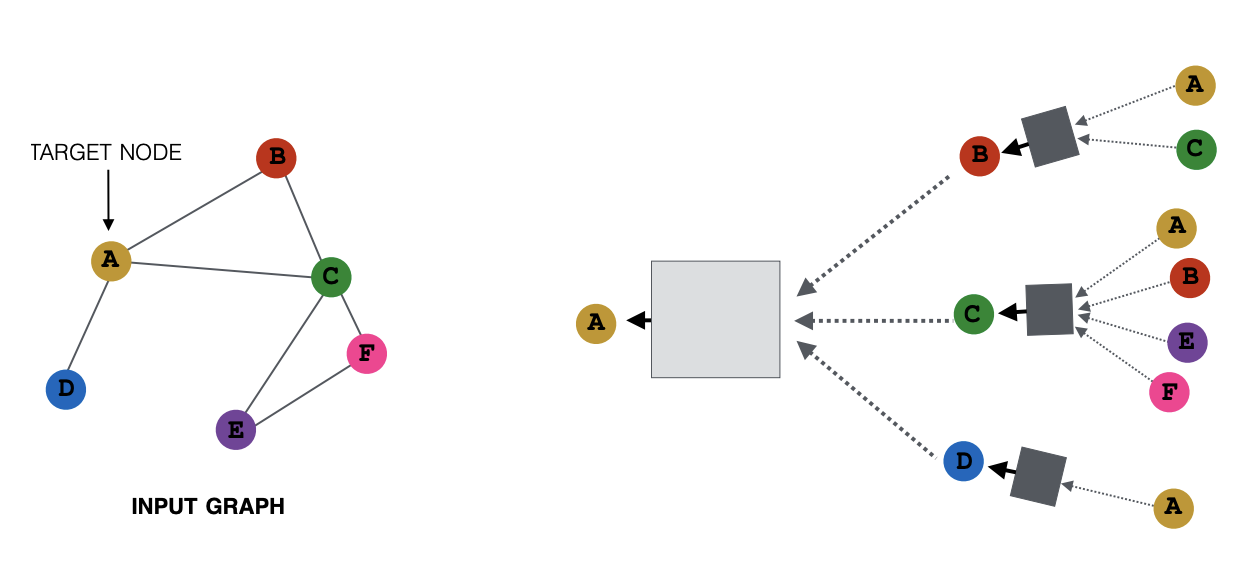}
    \caption{Illustration of the process of message passing. Every node defines its own computation graph based on its neighborhood. Left: The input graph and the target node based on which the series of computations is defined. Right: The message passing steps for two hops away from the target node. Gray rectangles represent neural networks. Figure from~\cite{leskovec_cs224w_nodate}.}
    \label{fig:aggregate}
\end{figure}

\begin{enumerate}
    \item The embeddings of neighboring nodes are aggregated using a permutation invariant function. This is justified because a permutation of the graph nodes should not give a different result. Examples of these aggregating functions include the max, sum or average functions. This process is referred to as the aggregation of the \textit{messages} received from the immediate neighbors.
    \item This aggregated information is then passed through a neural network.
    \item Finally, the node embedding of the target node is updated based on the aggregated messages from its neighbors. This iterative process of updating the node representations by exchanging information between neighbors is known as \textit{message passing}.
\end{enumerate}
In this way, after each message passing step, the receptive field of the GNN increases by one hop. Hop, here, refers to a traversal from one node of a graph to a neighboring node via a connecting edge. The process is summarized in Fig.~\ref{fig:aggregate}. 

For a graph $G = (V,E)$, the message passing layer can also be expressed as:
\begin{equation}
    \mathbf{h}_u = \phi \left( \mathbf{x}_u, \bigoplus_{v \in \text{Adj}[u]} \psi (\mathbf{x}_u, \mathbf{x}_v,\mathbf{e}_{uv}) \right) \,,
\end{equation}
where $\phi$ and $\psi$ are differentiable functions representing neural networks, $\text{Adj}[u]$ is the immediate neighborhood of node $u \in V$, $\mathbf{x}_u$ represents the node features of node $u \in V$, and $\mathbf{e}_{uv}$ represents the edge features of edge $(u,v) \in E$. Finally, $\bigoplus$ is a permutation invariant aggregation operator (e.g., element-wise sum, mean) accepting an arbitrary number of inputs. Functions $\phi$ and $\psi$ are referred to as the update and message functions, respectively. 

Other ``flavors'' of this message passing process have been developed, such as the famous graph convolution networks~\cite{kipf_semi-supervised_2016,kipf_deep_2020} and interaction networks~\cite{battaglia_interaction_2016}.

Having presented these ML models, we now move on to an important technique used in this thesis: quantization.

\section{Quantization}

Quantization, in signal processing in general, is the process of mapping a set of values from a continuous set to a finite set. Examples of this include rounding and truncation. In this form, quantization is involved to some extent in nearly all digital signal processing, because the continuous analog signal of any quantity has to be digitized, to discrete values.

In the context of ML/DL~\cite{hugging_face_quantization_nodate,jacob_quantization_2017,wu_integer_2020}, quantization refers to a process of reducing the size of the models, by representing their weights and activations using numbers with less bits than standard floating-point systems, where 32 or 64 bits are typical. In this way, the computational and memory costs of inference can be reduced significantly. On the one hand the required memory is reduced because simply the space required by each weight is reduced. On the other hand, the operations happen between low-precision data types and hence are considerably less computationally expensive.

As a simple example, let's consider a symmetric quantization scheme, from 32-bit float to 8-bit integer precision. With 8 bits, only $2^8 = 256$ numbers can be represented, while using 32-bit floats, a wide range of values is possible. Let's consider a float $x \in [-\alpha,\alpha]$, where $\alpha$ is a real number with $\alpha>0$. How do we best project this symmetric interval $[-\alpha,\alpha]$ of floats onto the space of 8-bit integers? We can write the following quantization scheme:

\begin{equation}  \label{eq:dequant}
    x = S \times x_q \,,
\end{equation}
where $x_q$ is the quantized representation of float $x$, and float $S$ is the scale quantization parameter. The quantized value can then be calculated as follows:

\begin{equation} \label{eq:quant}
    x_q = \text{round}(x/S) \,.
\end{equation}
Finally, any float values outside interval $[-\alpha,\alpha]$ are clipped, so for any float $x$:

\begin{equation}
    x_q = \text{clip}(\text{round}(x/S), -\alpha_q, \alpha_q) \,,
\end{equation}
where $\alpha_q = \text{round}(\alpha/S)$, and $\text{clip}(x, x_{\text{min}}, x_{\text{max}})$ denotes the clamp (or clipping) function between $x_{\text{min}}$ and $x_{\text{max}}$.

\subsection*{Calibration and Quantization Types}

\begin{figure}
    \centering
    \includegraphics[width=1\linewidth]{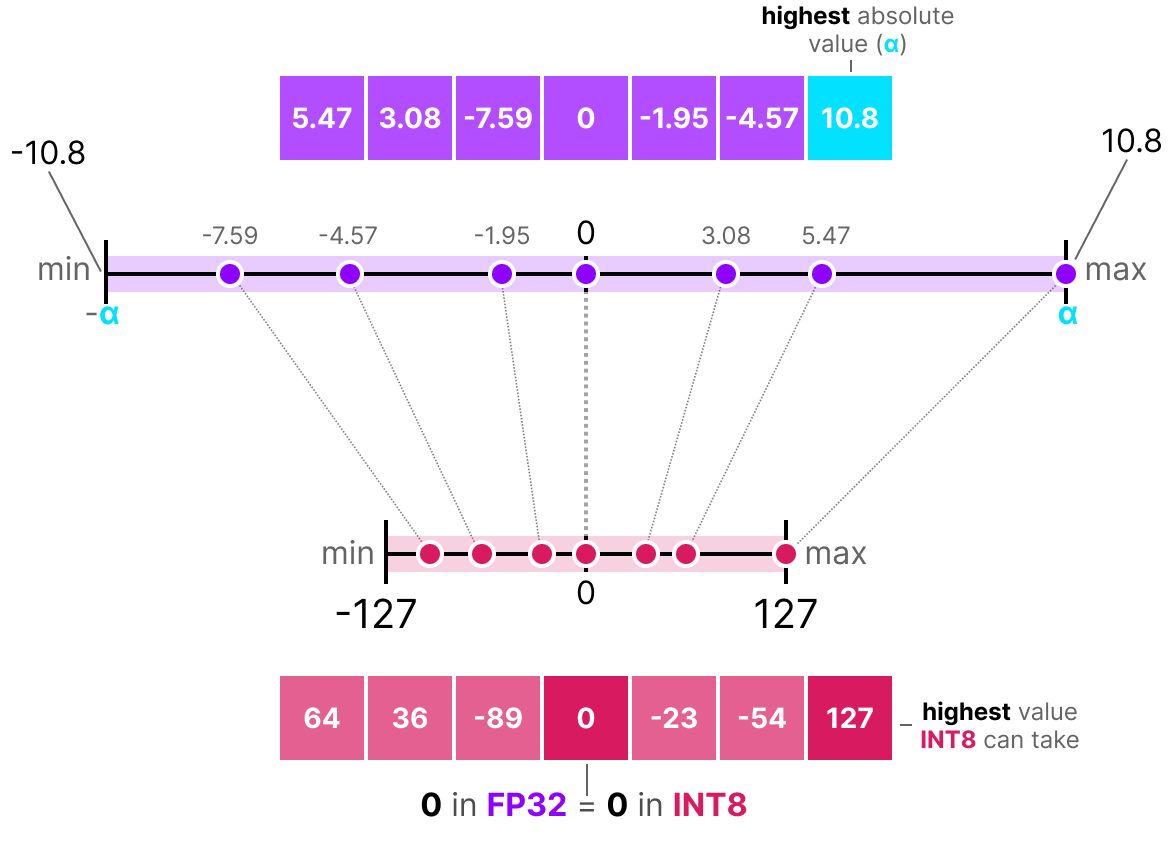}
    \caption{Illustration of the process of symmetric quantization. The scale is chosen to best fit the input values to be quantized. Figure from~\cite{grootendorst_visual_2024}.}
    \label{fig:quant}
\end{figure}

Calibration is the process during which the ideal values for the quantization parameters, the scale $S$ in our example, are chosen based on the distribution of the input values. For example, as shown in Fig.~\ref{fig:quant}, based on the range of the input values, the interval limits $[-\alpha,\alpha]$ are chosen, and the value of $S$ is chosen such that $\alpha$ is mapped to the highest value the quantized type can take. For the values shown, and according to Eq.~\eqref{eq:dequant}, the scale will have to be $S = 10.8 / 127$. Due to the interval being symmetric, from the 256 available values in INT8, we effectively only have half the numbers to represent positive values, while the rest are reserved for the zero point and the negative values.

For the case of neural networks, the input values of the quantization are the weights and the activations of the model. For weights, the process is quite easy since the actual range can be easily calculated at the time of quantization. For activations, however, things are a bit more complicated, and the approaches are different depending on the type of quantization pursued: 

\begin{itemize}
    \item \textbf{Post-Training Quantization (PTQ):} The quantization of the weights and activations is performed after the training of the model in full precision.
    \item \textbf{Quantization-Aware Training (QAT):} The quantization is performed during the training process.
\end{itemize}
Depending on the type of quantization, a different method for the calibration of the activations is used~\cite{hugging_face_quantization_nodate}:

\begin{itemize}
    \item[] Static PTQ: At the time of quantization, a representative sample of the data is passed through the model and the activation values are recorded, using ``observers'' placed at the activations. After several forward passes, the ranges of the computations can be deduced using some calibration technique.
    \item[] Dynamic PTQ: For each activation, the range is computed at runtime. However, this can prove slow and even not an option on several types of hardware.
    \item[] QAT: The ranges of computations are computed during training. ``Fake quantize'' operators simulate the effects of quantization during training, enabling the model to adjust and become robust to the errors introduced by the quantization process.
\end{itemize}

Quantization to integer precision was used when porting ETX4VELO models on GPUs and FPGAs, as explained in Chapters~\ref{ch:etx4velo-gpu} and \ref{ch:etx4velo-fpga}.

\section*{Conclusion}

In this chapter, I presented a brief history of machine learning, and sketched, from the ground up, the inner workings of graph neural networks. Quantization was also introduced. In the main results part of this work, Chapter~\ref{ch:etx4velo} onwards, GNNs were used to perform the task of track reconstruction, which is introduced in Chapter~\ref{ch:tracking}.

\chapter{High Performance Computing}
\label{ch:hpc}

\minitoc

\noindent Parts of this chapter were inspired by~\cite{calefice_standalone_2022,scarabotto_search_2023,hennequin_performance_2022,boettcher_lhcb_2021}.

\section*{Introduction}

In this chapter, we look into parallel, as opposed to sequential, computation, specialized hardware, and High Performance Computing (HPC). This background is crucial in understanding the computational aspects of my work as well as the motivations behind it. HPC is particularly motivated by the need to perform RTA, which requires specific hardware and computing paradigms---such as parallel programming---in order to meet the strict latency and throughput constraints imposed by the extreme data rate environments at LHC experiments.

\section{Parallelism}

Traditionally, computer software has been sequential. A computer program was constructed as a series of instructions to be executed one after the other on the Central Processing Unit (CPU) of the computer. Parallel computing~\cite{kumar_introduction_1994,almasi_highly_1989,czech_introduction_2017}, on the other hand, uses multiple processing elements in order to tackle a problem simultaneously. Many tasks are essentially a repetition of the same calculation a large number of times. So, if these calculations are independent from each other, why wait for each one to finish before proceeding to the next one? The execution can be performed in parallel and thus the routine can be sped up. Historically, parallel computing was used for scientific problems and simulations, such as meteorology. This led to the design of parallel hardware architectures and the development of software needed to program these architectures, as well as HPC~\cite{rauber_parallel_2013}.

\subsection*{Amdahl's Law}

\begin{figure}
    \centering
    \includegraphics[width=0.65\linewidth]{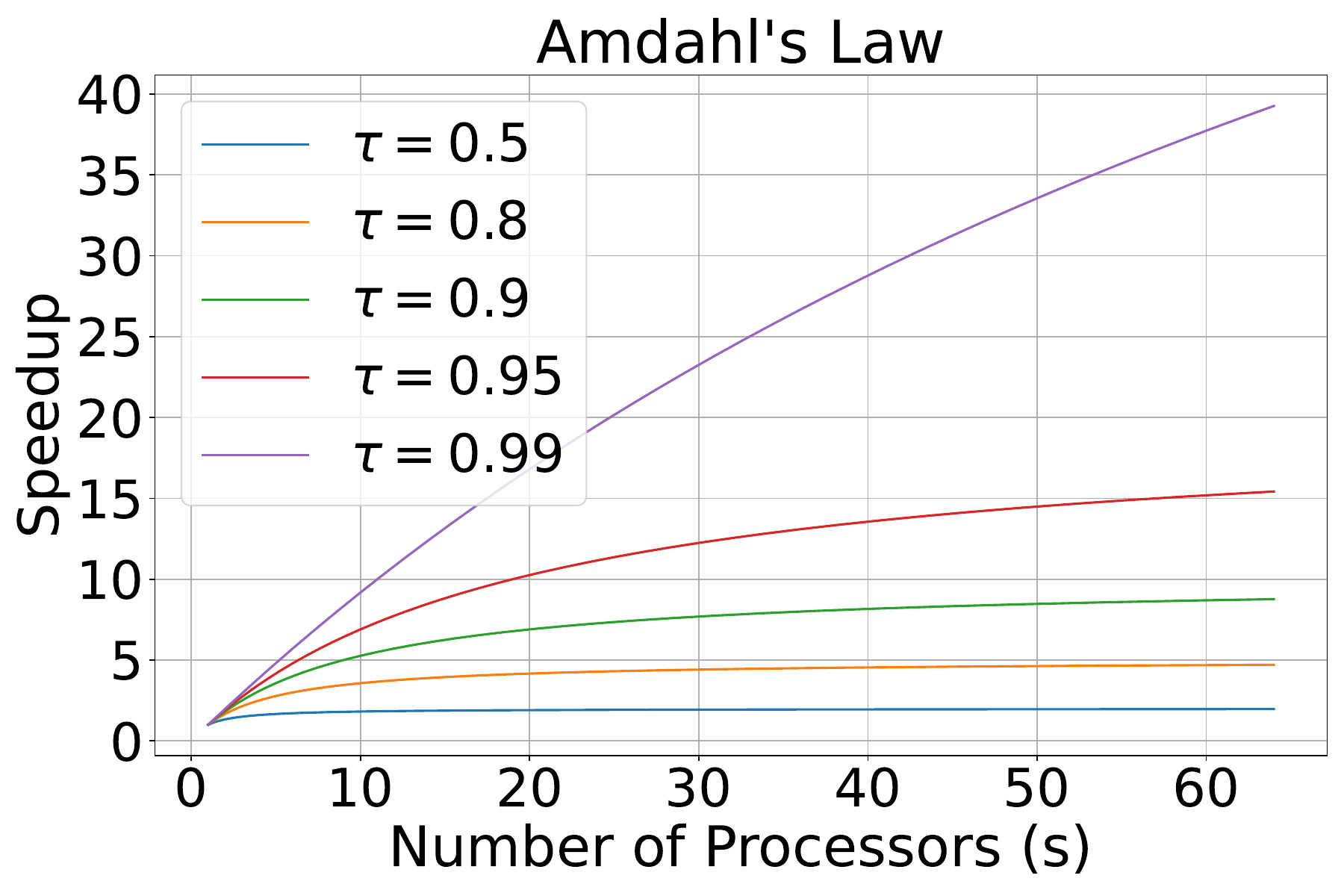}
    \caption{Demonstration of Amdahl's law for the theoretical maximum  speedup of a computational system, as a function of the fraction of the parallelizable code $\tau$, and the speedup factor $s$ that the parallelization results in.}
    \label{fig:amdahl}
\end{figure}

Ideally, doubling the number of processors would result in the halving of the runtime. However, in practice, very few algorithms achieve optimal speedup. The maximum potential speedup is given by Amdahl's law~\cite{amdahl_validity_1967}. A task executed on a multicore system can be categorized into two parts: a part that does not benefit from the usage of multiple cores, and a part that does benefit. Assuming that the latter is a fraction $\tau$ of the task, and that it benefits from an acceleration by a factor $s$ compared to single core execution then, the maximum speedup is given by:

\begin{equation}
    \text{Speedup}(s) = \frac{1}{1 - \tau + \frac{\tau}{s}} \,.
\end{equation}
The relationship is illustrated in Fig.~\ref{fig:amdahl}. Interestingly, this law reveals that increasing the number of processors yields diminishing returns past a certain point. In addition, it demonstrates that the enhancements of the code have to be focused on both the parallelizable and non-parallelizable components. Of course, the simplistic view of computation needed for the derivation of Amdahl's law neglects various aspects of inter-process communication, synchronization and memory access overheads. A more complete assessment is given by Gustafson's law~\cite{gustafson_reevaluating_1988}.

\subsection*{The CPU as a Parallel Processor}

During the 1980s until the early 2000s, various methods were developed for increasing the computational performance of the CPU. A crucial method was frequency scaling: By increasing the clock frequency of the CPU, more instructions can be executed in the same amount of time. Other methods included the use of reduced instruction sets, out-of-order execution, memory hierarchy or vector processing. 

\begin{figure}
    \centering
    \includegraphics[width=1\linewidth]{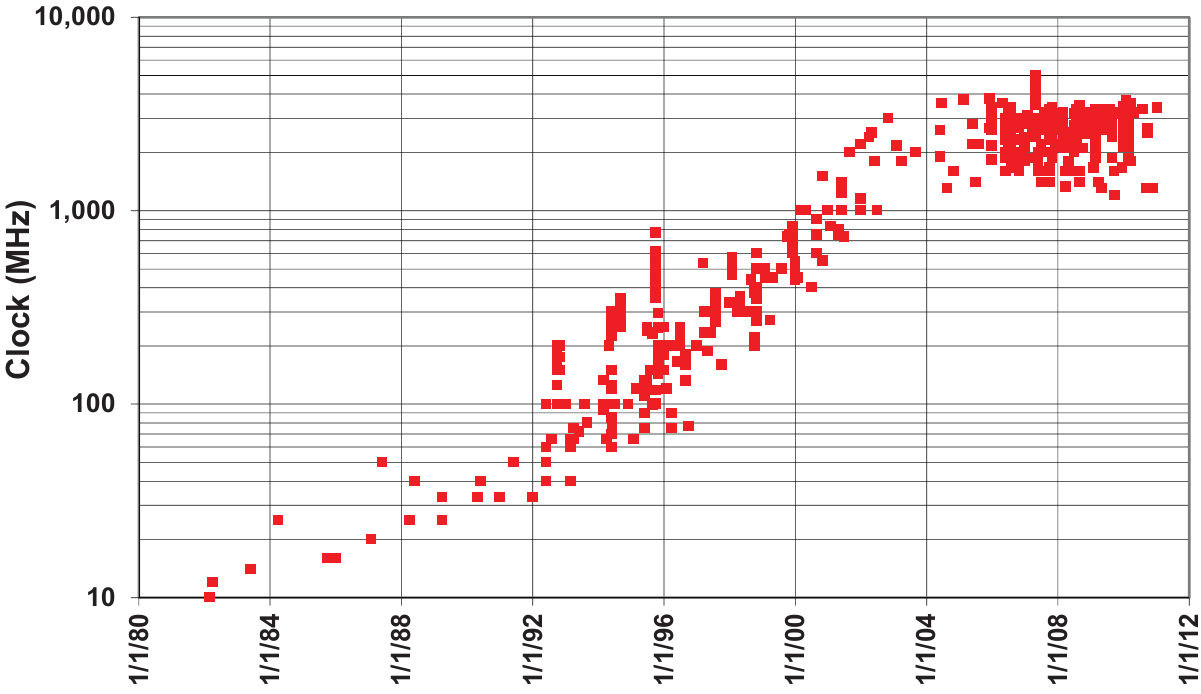}
    \caption{Historical evolution of microprocessor clock rates from 1980 to 2012, illustrating the scaling plateau beginning in 2004. This effect demonstrates the breakdown of Dennard scaling and the so-called ``power wall'', limiting further gains through increased frequency due to thermal and energy constraints. Figure from~\cite{gropp_designing_2015}.}
    \label{fig:power-wall}
\end{figure}

The Dennard scaling law was introduced in 1974~\cite{dennard_design_1974} and it stated that as transistors get smaller the power consumption of a chip of constant size stays the same even if the number of transistors increases. As transistors became smaller and operating voltages decreased, circuits were able to run at higher frequencies without increasing power consumption. However, this scaling is considered to have broken down around 2006. Dennard scaling overlooked factors like the ``leakage current'' and the ``threshold voltage'', which set a minimum power requirement per transistor. As transistors shrink, these parameters don't scale proportionally, leading to an increase in power density. This created a so-called ``power wall'', as shown in Fig.~\ref{fig:power-wall}, that practically limited processor frequency to around 4~GHz~\cite{gropp_designing_2015}, and which eventually led to Intel canceling the Tejas and Jayhawk microprocessors in 2004~\cite{flynn_intel_2004}.

In order to address the problem of power consumption, manufacturers turned to producing power efficient processors that have multiple cores. Each core is independent and can access the same memory concurrently. This design principle brought multi-core processors to the mainstream. By early 2010s, computers by default had multiple cores, while servers had more than ten core processors. By contrast, in early 2020s, some processors had over one hundred cores~\cite{rauber_parallel_2013}. Moore's law~\cite{moore_cramming_1998}, that predicts that the number of transistors in an integrated circuit will double every roughly two years, can be extrapolated to the doubling of the number of cores per processor.

The operating system of the CPU ensures that the different tasks are performed concurrently using the resources of the processor by distributing them across the free cores. However, in order to unlock the full capacity of the processing unit, the code itself has to be designed in a way that leverages the new computational capabilities of multicore architectures~\cite{rauber_parallel_2013}.

\subsection*{Flynn's Taxonomy}

One of the earliest classifications of parallel computers and programs is the so-called Flynn's taxonomy~\cite{flynn_very_1966,flynn_computer_1972}. It categorizes programs based on whether they are operating using a single instruction or multiple instructions, and whether these instructions are executed on one or multiple data.

\begin{figure}
\begin{center}
  \subfloat[][]{\includegraphics[width=0.45\textwidth]{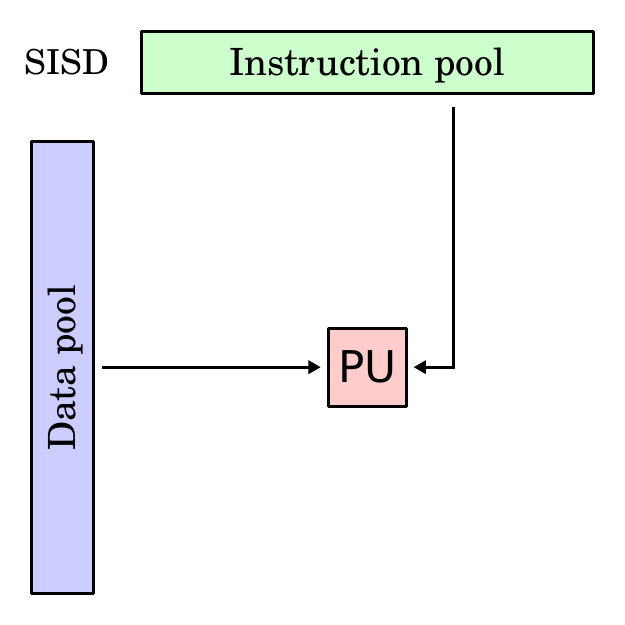}\label{fig:sisd}}\quad
  \subfloat[][]{\includegraphics[width=0.45\textwidth]{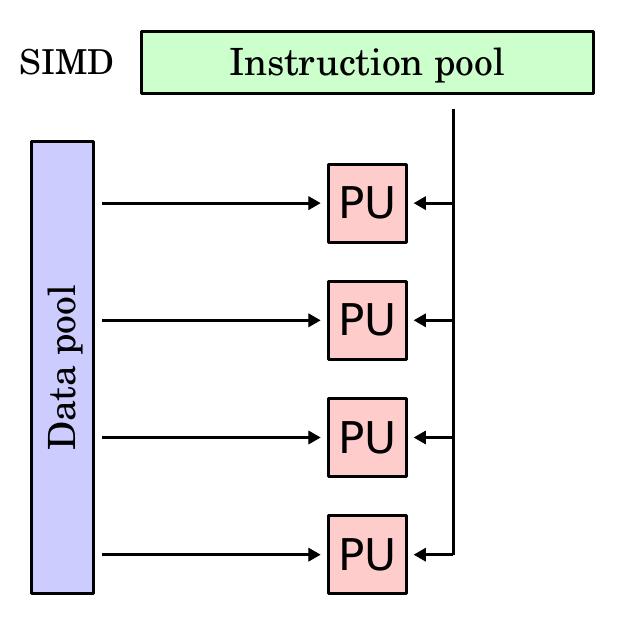}\label{fig:simd}}\\\vspace{4pt}
  \subfloat[][]{\includegraphics[width=0.45\textwidth]{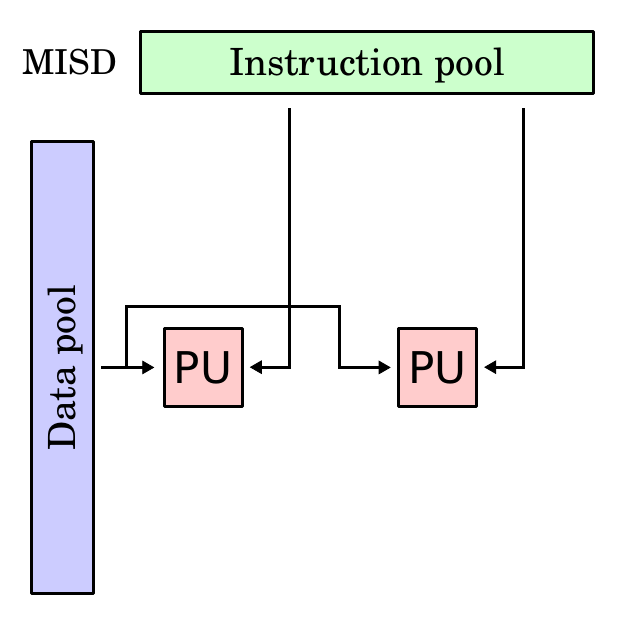}\label{fig:misd}}\quad
  \subfloat[][]{\includegraphics[width=0.45\textwidth]{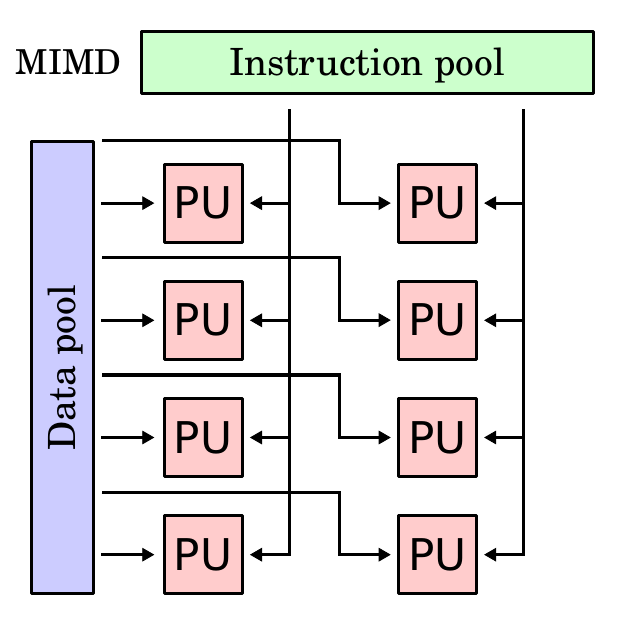}\label{fig:mimd}}
\end{center}
\caption{Flynn's Taxonomy. \protect\subref{fig:sisd} Single Instruction Stream, Single Data Stream (SISD), \protect\subref{fig:simd} Single Instruction Stream, Multiple Data Stream (SIMD), \protect\subref{fig:misd} Multiple Instruction Stream, Single Data Stream (MISD), \protect\subref{fig:mimd} Multiple Instruction Stream, Multiple Data Stream (MIMD). The instruction and data pools are shown, as well as the Processing Units (PUs). Figures from~\cite{cburnett_sisd_2007,cburnett_misd_2007,cburnett_simd_2007,cburnett_mimd_2007}.}
\label{fig:flynn}
\end{figure}

An entirely sequential program is equivalent to the Single Instruction Stream, Single Data Stream (SISD) classification. When the operation is repeated over multiple data, it corresponds to the Single Instruction Stream, Multiple Data Stream (SIMD) class, a form of data parallelism. On the other hand, when multiple instructions are performed on a single data, a form of dataflow parallelism, the program is classified as Multiple Instruction Stream, Single Data Stream (MISD). While systolic arrays are sometimes put in this category, the class is rather rare in practice. Multiple Instruction Stream, Multiple Data Stream (MIMD) is by far the most common type of modern programs, and is known as control parallelism. The taxonomy is summarized in Fig.~\ref{fig:flynn}.

In this context, data dependencies are a crucial aspect of implementing parallel code. If we have a sequence of steps, and each step depends on the result of the previous step then this sequence is not parallelizable since it must be executed in order. However, most algorithms contain opportunities where the execution can be parallelized. Notable examples of this are deep learning algorithms. The operations we saw in Chapter~\ref{ch:ml}, Section~\ref{sec:dl}, such as the propagation of neuron activations in the feedforward layers, are essentially matrix multiplication operations that can be performed in parallel.

\subsection*{Parallelism for RTA in High-Energy Physics}

HPC and parallelism have emerged as essential components of the processing infrastructure at LHC experiments. This development is largely driven by the need for RTA at increasingly higher data rates. Meeting the stringent requirements for latency and throughput in such environments demands both specialized hardware and modern computing paradigms.

Having introduced the concept of parallelism in general terms, we now turn to the specific types of hardware architectures particularly interesting for exploiting parallelism in order to perform real-time analysis in high-energy physics. Specifically, the GPU and FPGA architectures are outlined.

\section[From Video Games to the GPU]{From Video Games to the GPU Architecture}

Early arcade video games used specialized video hardware to handle graphics due to expensive memory units since the 1970s. The first integrated graphics processing unit, NEC's $\upmu$PD7220, was the most well known GPU until the mid-1980s. It supported graphics display monitors of $1024 \times 1024$ resolution, and laid the foundations for the GPU market~\cite{hopgood_advances_1986}.

Early 3D graphics emerged in the 1990s in arcades and consoles and GPUs started integrating 3D functions. The term GPU was coined by Sony in reference to their 32-bit Sony GPU used in the PlayStation 1 video game console, released in 1994~\cite{anderson_is_2018}. Nvidia and ATI started creating consumer graphics accelerators, leading to the release of GeForce 256. This GPU was marketed as the world's first GPU capable of performing advanced graphics rendering. These capabilities included tasks such as rasterization, where an image described in a vector graphics format is translated into an array of pixels that best represents this vector description in the available screen granularity. Shading, another essential task for a graphics processor, is the process through which a GPU calculates the appropriate levels of light and color, in order to render a 3D scene more realistically. The first GPU capable of shading was the GeForce 3, used in the Xbox console, competing with the chip used in PlayStation 2.

Nvidia introduced the Compute Unified Device Architecture (CUDA) in 2006, sparking what is now known as General-Purpose Graphics Processing Unit (GPGPU) computing~\cite{sanders_cuda_2010}. This marked a revolution in computing: previously, GPUs were dedicated chips designed to accelerate 3D rendering tasks for gaming and graphics applications. With CUDA, GPUs became programmable parallel processors equipped with hundreds of processing elements, enabling them to perform a broad range of tasks traditionally tackled using CPUs. This can include scientific computing (simulations, climate, etc.), financial modeling, signal processing, machine learning and deep learning. For the first time, Nvidia provided a dedicated programming model and language for its GPUs, enabling developers to write general-purpose code that could run directly on the GPU---something that was previously not possible with such flexibility and ease.

CUDA is a proprietary language, which led to the need for a standardized parallel programming language that could be used across GPUs from different manufacturers. In response, OpenCL~\cite{khronos_group_opencl_2013,stone_opencl_2010} was defined by Khronos Group as an open standard. It allows the development of code compatible with both GPU and CPU. This emphasis on portability---the ability to write a single kernel that can run across heterogeneous platforms---made OpenCL the second most popular HPC tool at the time~\cite{handy_amd_2011}.

In the 2010s, GPUs were used in consoles such as the PlayStation 4 and the Xbox One~\cite{lendino_xbox_2015}, and on automotive systems, after Nvidia partnered with Audi to power car dashboard displays~\cite{teglet_nvidia_2010}. Nvidia architectures developed further, increasing the number of CUDA cores and further adding the new technology of the so-called tensor cores~\cite{sarkar_nvidia_2018}. Tensor cores were designed to bring better performance to deep learning operations. Real-time ray tracing---simulation of reflections, shadows, depth of field, etc.---debuted with Nvidia RTX 20 series in 2018~\cite{burnes_nvidia_2020}.

In 2020s, after the deep learning explosion we described in Chapter~\ref{ch:ml}, GPUs are heavily used in the training and inference of large language models, such as the ChatGPT~\cite{openai_introducing_2022} chatbot by OpenAI. This surge in interest of dedicated hardware, infrastructure and electricity to support these heavy models has created a booming artificial intelligence ecosystem. It is further fueling a re-evaluation of our electricity needs, infrastructure organization, and the direction of hardware development, while also raising questions about the feasibility of continued scaling.

\section{CUDA Programming Model}

Introduced in 2006 by Nvidia~\cite{sanders_cuda_2010}, CUDA is a parallel programming model designed for developing general purpose applications that leverage the parallelization capabilities and architecture of Nvidia GPUs. It can be thought of as an Application Programming Interface (API) that allows software to access the GPU's virtual instruction set and parallel computation elements for the execution of compute kernels.

The C++ version of CUDA is a language extension of C++ that allows the programmer to define specific parallel functions called kernels, and run code on CPU and GPU using a single language~\cite{nvidia_cuda_2025}. By splitting the code into a \textit{host} (traditional CPU) and a \textit{device} (GPU) part, the instructions dictated by the CPU are executed on the GPU. The device code is organized into kernels, and kernels are executed by the threads available on the GPU. Multiple threads execute the same kernel simultaneously, in the so-called Single Instruction, Multiple Threads (SIMT) execution model. SIMT can be thought of as a subcategory of SIMD. In SIMD, a single thread executes an instruction on multiple data. On the other hand, in SIMT, a small group of threads called a warp executes the same instruction on multiple data, but each thread has its own independent program counter, stack and registers, so threads can have divergent execution. This per-thread autonomy gives more flexibility to the SIMT execution model.

\subsection*{Memory Hierarchy}

\begin{figure}
    \centering
    \includegraphics[width=1\linewidth]{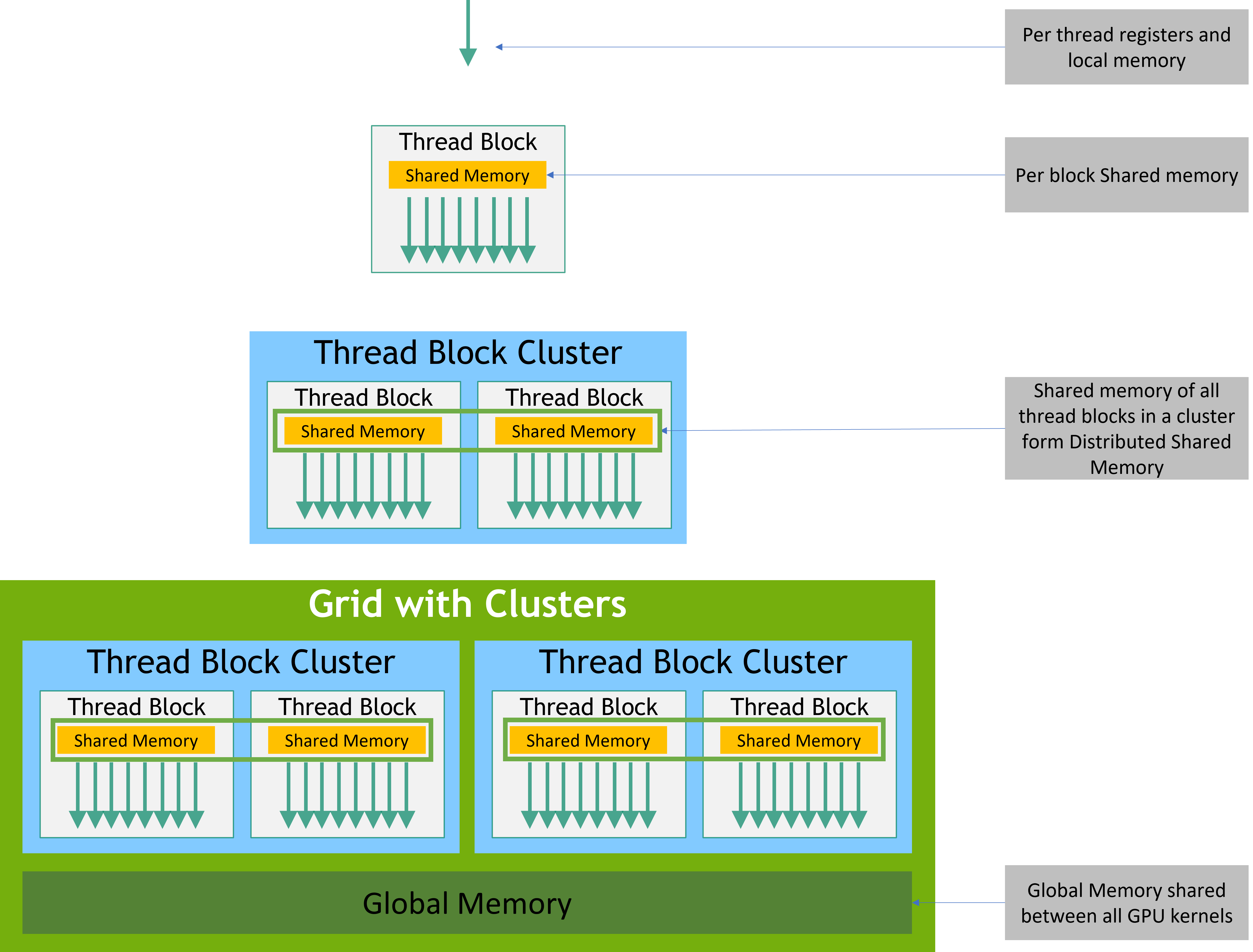}
    \caption{CUDA thread and memory hierarchy. Figure from~\cite{nvidia_cuda_2025}.}
    \label{fig:threads-blocks}
\end{figure}

In the CUDA programming model, threads are organized into blocks. In particular, threads that execute the same instruction are grouped into warps and several warps constitute a thread block. Blocks of threads are further organized into grids. These two levels---blocks and grids---correspond to different communication bandwidths and shared memory capacities. Blocks have shared memory that is accessible to all threads in the block, while threads from the different blocks only share the view of the device memory. The model is summarized in Fig.~\ref{fig:threads-blocks}.

\begin{figure}
    \centering
    \includegraphics[width=0.65\linewidth]{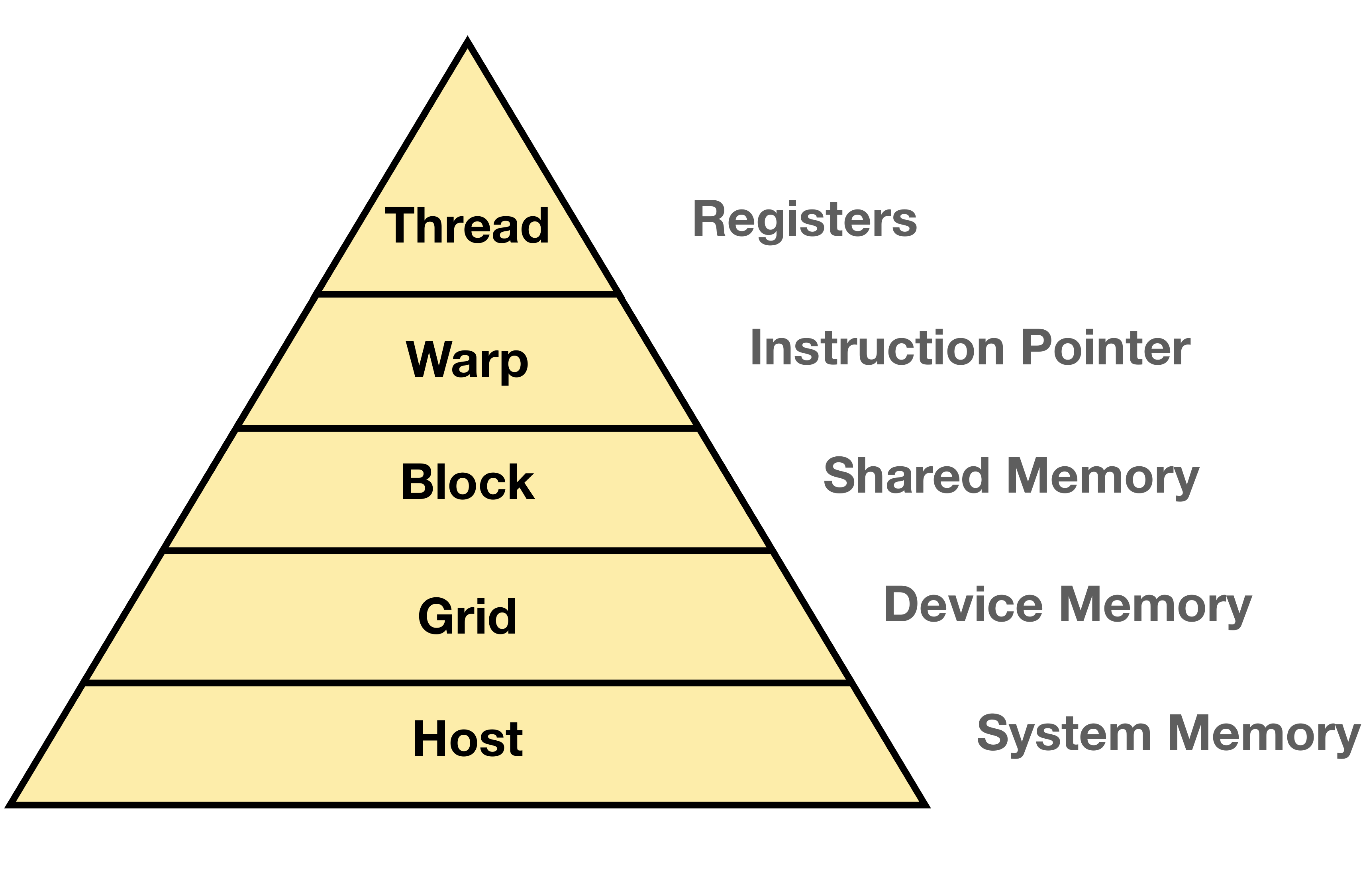}
    \caption{Illustration of the memory hierarchy for a Single Instruction, Multiple Threads (SIMT) program. Inspired by~\cite{bensdorp_gpu_2024}.}
    \label{fig:memory}
\end{figure}

Register memory, is the fastest kind of memory but is of the smallest size, usually around 1~KB per thread. Shared memory, on the other hand, is slower, accessible by all the threads within a block, and is usually on the order of hundreds of kilobytes. The device memory, even slower, is accessible by all the threads of the device and is what is commonly known as Random Access Memory (RAM). As of 2025, most modern GPUs do not go over 80~GB of RAM. Finally, the host RAM is the most costly, in terms of access latency. The memory hierarchy is illustrated in Fig.~\ref{fig:memory}, along with Fig.~\ref{fig:threads-blocks}.

\subsection*{Architecture}

The GPU delivers significantly higher instruction throughput and memory bandwidth than the CPU, all with similar cost and power range. Various applications take advantage of these enhanced capabilities compared to the CPU, such as GPGPU programming. While FPGAs are also energy-efficient, GPUs offer unmatched programming flexibility.

This difference stems from fundamental design differences. The CPU is optimized to execute a series of operations, by a single thread, at the highest clock frequency possible, and can handle a few dozen concurrent threads. In contrast, GPUs are designed to run thousands of threads in parallel, exploiting data parallelism, but at a lower frequency. However, by trading off individual speed, a much higher overall throughput can be achieved.

\begin{figure}
    \centering
    \includegraphics[width=1\linewidth]{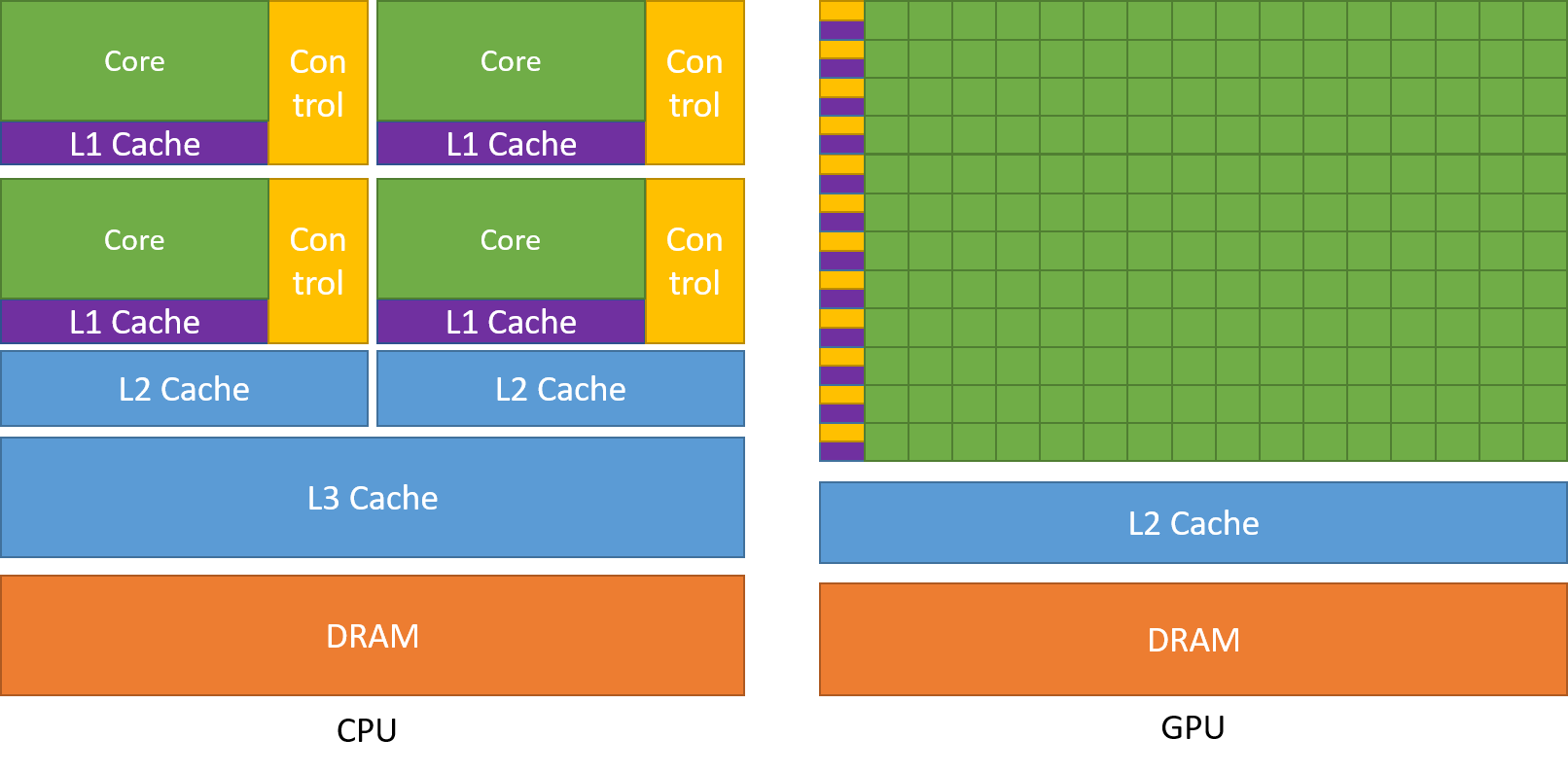}
    \caption{Comparison of the allocation of resources between a CPU and a GPU. Figure from~\cite{nvidia_cuda_2025}.}
    \label{fig:cpu-gpu}
\end{figure}

To support this level of parallelism, GPUs devote more transistors to data processing rather than to data caching and control logic. This design philosophy is illustrated in Fig.~\ref{fig:cpu-gpu}, which compares the typical allocation of resources between a CPU and a GPU.

Nvidia's GPU architecture is an array of the so-called Streaming Multiprocessors (SMs). A multithreaded program is divided into thread blocks that run independently of one another. When a kernel is launched over several blocks, the blocks are distributed across the available SMs for execution. An SM can execute multiple blocks simultaneously. On a GPU with more SMs, the program will be executed automatically in less time than a GPU with fewer multiprocessors. In this way, scaling is automatically guaranteed.

\subsection*{C++ Extension}

In the C++ version of CUDA, compute kernels are defined as C++ functions using the \texttt{\_\_global\_\_} declaration specifier. The launch of the kernel is defined using the CUDA execution configuration syntax \texttt{\symbol{60}\symbol{60}\symbol{60}K, M\symbol{62}\symbol{62}\symbol{62}(...)}. In this way, a kernel is launched on \texttt{K} blocks per grid, each with \texttt{M} threads, and is executed in parallel by the active threads. Furthermore, CUDA exposes built-in variables that can be accessed by the developer. In particular, \texttt{threadIdx} gives the identifier of the thread currently executing and \texttt{blockDim} gives the block dimension, i.e., the number of threads in each block---\texttt{M} above. Finally, \texttt{blockIdx} gives the identifier of the block currently in execution. These three variables are 3-component vectors, providing a natural way to invoke computations on vectors, matrices and volumes.

As an example, in Listing~\ref{lst:saxpy}, an implementation of ``Single-precision A*X Plus Y (SAXPY)''~\cite{harris_easy_2012} is presented, a basic function of the Basic Linear Algebra Subroutines (BLAS) library, in CUDA/C++. The saxpy function takes two $n$-dimensional input vectors, $\mathbf{x}$ and $\mathbf{y}$, as well as a scalar $a$. It then computes the expression $a \times (\mathbf{x})_i + (\mathbf{y})_i$, and stores the result in $\mathbf{y}$. In the host code, we start by moving the prepared data of $\mathbf{x}$ and $\mathbf{y}$ from the host to the device. We then invoke the kernel with 4096 blocks, of 256 threads each, for a total of 1\,048\,576 active threads (line 21). In this way we launch exactly the number of threads we need to perform the calculation on the number of elements $N=1\,048\,576$. Each thread is supposed to perform the calculation of each element independently, so in the device code, threads first calculate the index of the element they need to calculate (line 4). After checking that this index does not exceed the length of the vector $n$ (line 5), they then perform the calculation (line 6). The data are moved from the host to the device and back using API calls (lines 16, 17, 24).

\begin{lstlisting}[caption={Saxpy implementation in CUDA C++. Adapted from~\cite{harris_easy_2012}.}, label={lst:saxpy}]
// Device code (kernel definition)
__global__ void saxpy(int n, float a, float *x, float *y)
{
  int i = blockIdx.x*blockDim.x + threadIdx.x;
  if (i < n) {
    y[i] = a*x[i] + y[i];
  }
}

int main(void)
{
  // ...
  int N = 1<<20; // 2^20 = 1048576

  // Copy data from host to device
  cudaMemcpy(x_device, x_host, N*sizeof(float), cudaMemcpyHostToDevice);
  cudaMemcpy(y_device, y_host, N*sizeof(float), cudaMemcpyHostToDevice);

  // Perform SAXPY on 1M elements
  // Invoke kernel with 4096 blocks of 256 threads each
  saxpy<<<4096, 256>>>(N, 2.0f, x_device, y_device);

  // Transfer result back to the host
  cudaMemcpy(y_host, y_device, N*sizeof(float), cudaMemcpyDeviceToHost);

  // ...
}
\end{lstlisting}

\begin{figure}
    \centering
    \includegraphics[width=0.75\linewidth]{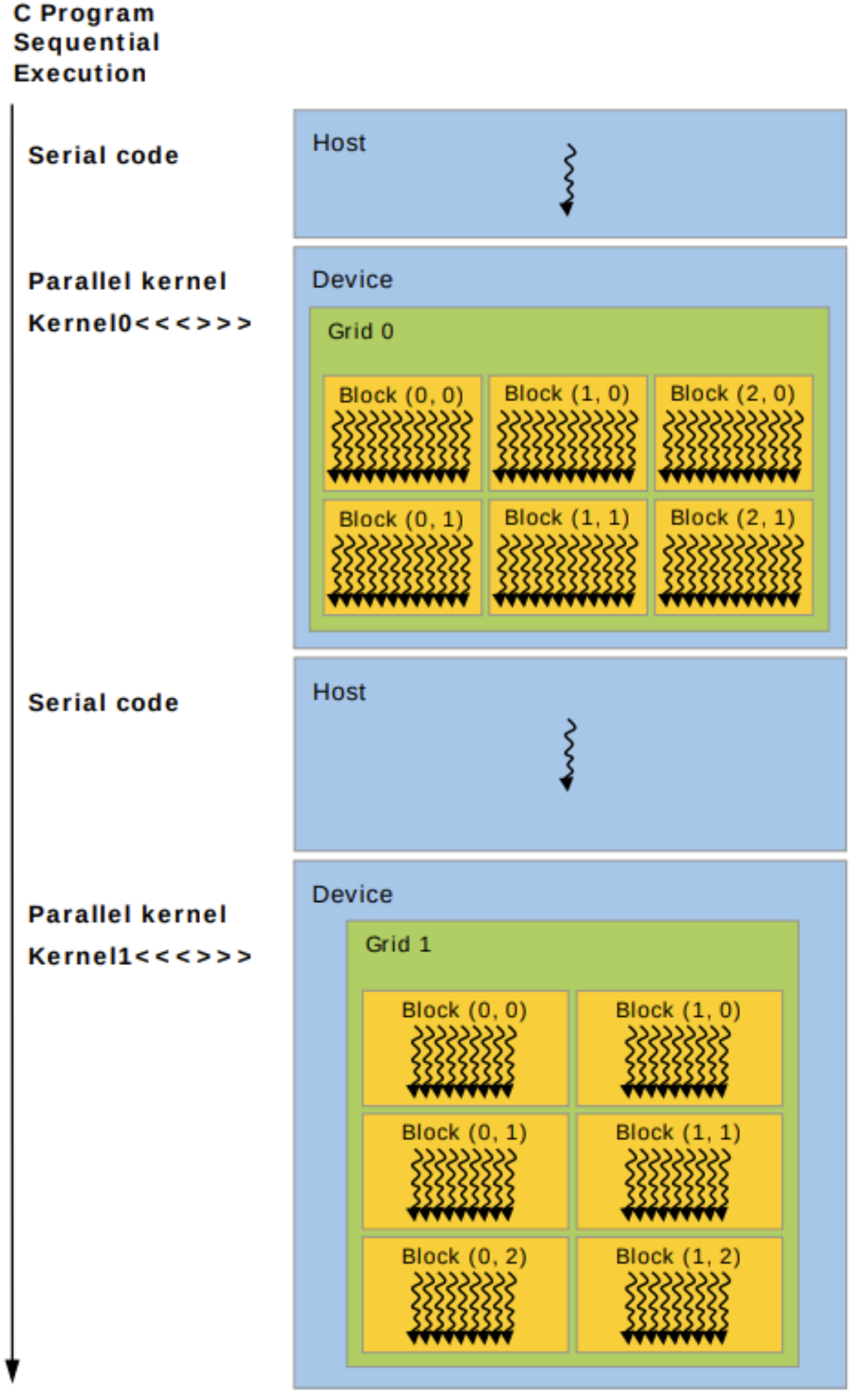}
    \caption{Illustration of heterogeneous programming using the CUDA programming model. Adapted from~\cite{nvidia_cuda_2025}.}
    \label{fig:hetero}
\end{figure}

CUDA threads operate on a physically separate device to the host running the C++ script. The kernel is invoked by the host, but it runs on the device. The execution model is illustrated in Fig.~\ref{fig:hetero}. 

\section{Programmable Logic}

While GPUs are programmable parallel processors designed for general-purpose computing, FPGAs are electronic chips that enable the integration of dedicated parallel architectures. The FPGA sprouted from developments in technology around programmable logic, and in particular from Programmable Read-Only Memory (PROM) and Programmable Logic Devices (PLDs). Both PROMs and PLDs could be programmed outside the factory, i.e., in the field, which explains the ``field-programmable'' part of the abbreviation~\cite{franz_history_2016,roelandts_15_1999,mencer_history_2020,arm_what_nodate}.

Altera, founded in 1983, produced the first erasable programmable ROM circuit in 1984. However, Xilinx delivered the first commercial field-programmable gate array in 1985, the XC2064. Until the mid-1980s, FPGAs were only used in networking and telecommunications. However, by the end of the decade, FPGAs had been adopted across consumer, automotive, and industrial applications~\cite{maxfield_design_2004}. With the AI boom around the 2010s, FPGAs are increasingly being used for applications in constrained environments and for prototyping.

FPGAs are extremely versatile due to the fact that they are reconfigurable. This allows developers to test numerous designs after the board has been built. When changes to the design are required, the device is simply restarted and the configuration file, usually called the bitstream, is transferred onto the device.

In particular, FPGAs are crucial for designing Application-Specific Integrated Circuits (ASICs).
The manufacture of ASICs is extremely costly, so before a design is decided and put into production, it has to be prototyped. The digital hardware design is then verified and finalized.

\section{Field-Programmable Gate Arrays}
\label{sec:intro-fpga}

\begin{figure}
    \centering
    \includegraphics[width=0.7\linewidth]{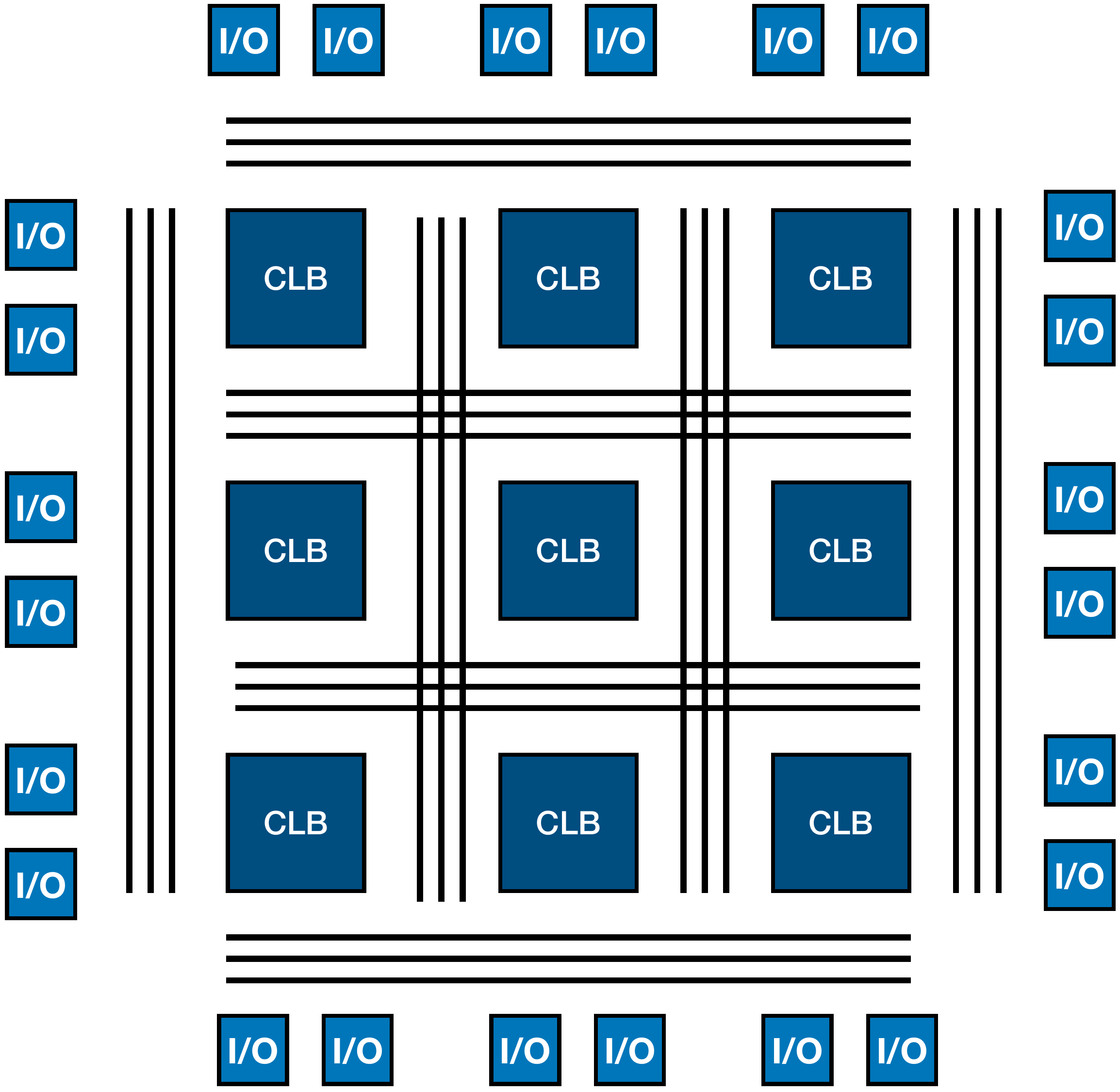}
    \caption{Illustration of the structure of an FPGA, highlighting its three fundamental digital logic components: Configurable Logic Blocks (CLBs), Input/Output (I/O) pads, and routing channels. Inspired by~\cite{university_of_toronto_fpga_nodate}.}
    \label{fig:fpga}
\end{figure}

The most common FPGA architecture includes an array of Configurable Logic Blocks (CLBs), Input/Output (I/O) cells, and routing channels~\cite{ahmed_multi-tenant_2022, university_of_toronto_fpga_nodate}, as illustrated in Fig.~\ref{fig:fpga}. The CLB typically consists of a Lookup Table (LUT) and a clocked Flip-Flop (FF). An LUT of $n$-bit input can encode any Boolean function of $n$ inputs by simply storing the value of the function for each input, i.e., by storing its truth table. FFs on the other hand, are used to register the value of the output of the logic function and to synchronize the data with the system clock. In this way, by storing the value of a state, sequential logic can be implemented. The routing channels are used to interconnect the logic blocks, and the I/O pads are used for interfacing with external signals. By ``configuring'' an FPGA, the developer can define the arrangement of these logic gates and their connections, in order to implement a series of operations such as additions, subtractions and logical operations.

FPGAs are often also equipped with Digital Signal Processing (DSP) blocks, responsible for performing more complex operations such as multiplications and divisions. These operations become more and more complex as the bit width of the operands increases. Furthermore, Block RAM (BRAM) is often added on the CLB grid, to enable the storage of large amounts of data inside the FPGA.

\subsection*{System on a Chip FPGAs}

\begin{figure}
    \centering
    \includegraphics[width=0.5\linewidth]{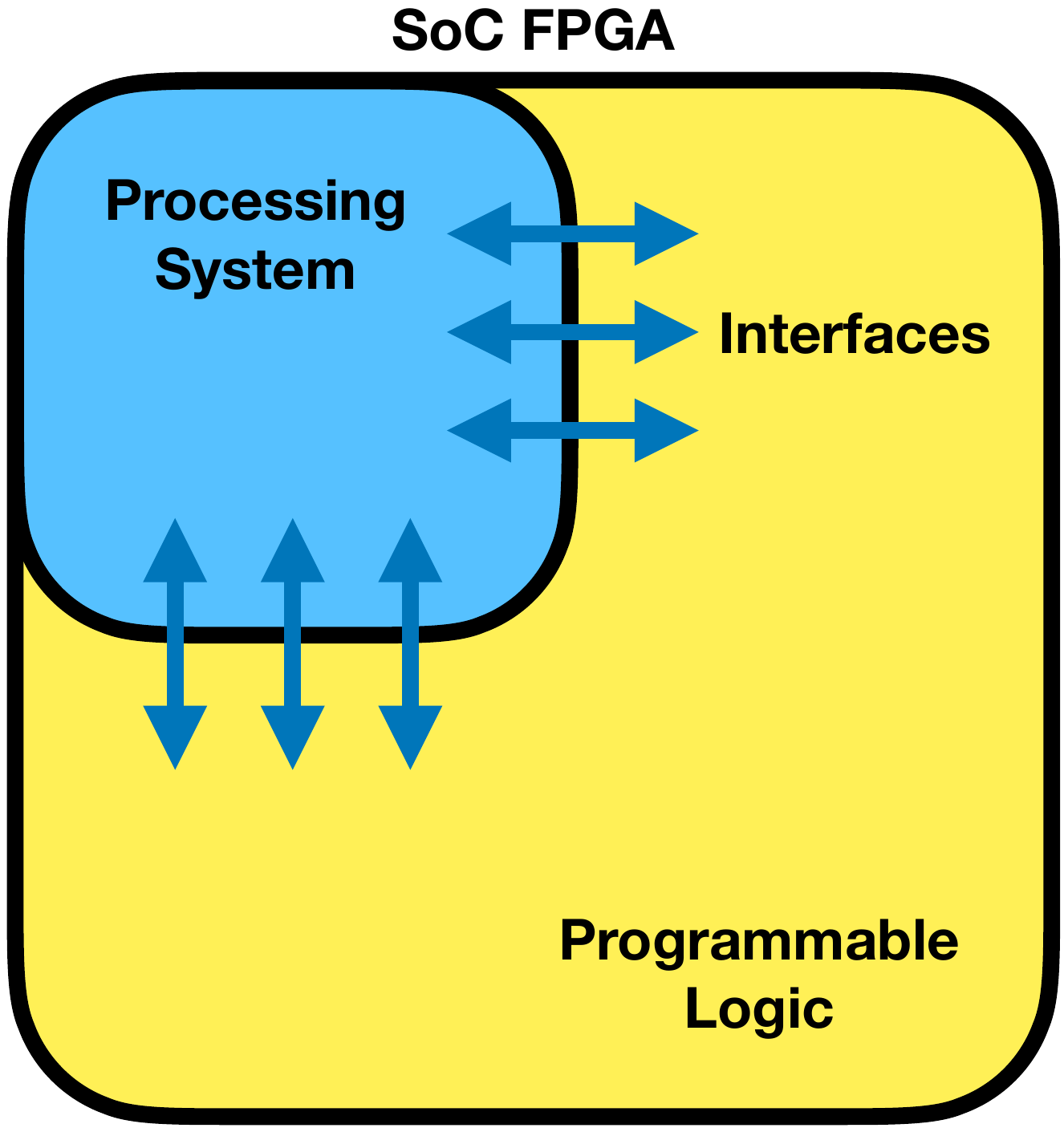}
    \caption{Block diagram illustration of a System on a Chip (SoC) FPGA, highlighting the division between the processing system and the programmable logic part, as well as the communication between them.}
    \label{fig:ps-pl-soc}
\end{figure}

Often, FPGAs are sold as a System on a Chip (SoC). The SoC board is divided into two parts, the Processing System (PS) and the Programmable Logic (PL), as shown in the block diagram in Fig.~\ref{fig:ps-pl-soc}. This type of diagram is a high-level representation showing the main functional components of the FPGA and how these are connected. It is used to understand the internal organization of the chip.

The PS is a traditional CPU, while the PL is the traditional reconfigurable FPGA part. SoCs comprise many execution units. These units communicate by sending data and instructions between them. A very common data bus for SoCs is ARM's Advanced Microcontroller Bus Architecture (AMBA) standard. Direct memory access controllers transfer data directly between external interfaces and the SoC memory, bypassing the CPU or control unit, which enhances the overall data throughput of the SoC.

\subsection*{Development}

In order to configure FPGAs, a developer needs to use a specialized computer language called Hardware Description Language (HDL). This type of language is used for describing the structure and behavior of electronic circuits, usually for ASICs and FPGAs. This design abstraction is known as Register-Transfer Level (RTL), modeling the digital logic circuit in terms of the flow of signals between the registers~\cite{vahid_digital_2010}. HDLs differ from normal programming languages because they describe concurrent hardware operations and timing behavior rather than sequential instruction execution. Because of this particularity, FPGA programming is notoriously difficult and comes with a high resource cost.

After the RTL description has been validated with test benches, the design is synthesized and the RTL description is translated to the gate-level description of the circuit. Finally, the design is laid out and routed on the FPGA.

\subsection*{High-Level Synthesis}

In order to avoid the cost related to developing FPGAs, various tools have been designed to abstract out the complexity in configuring FPGAs. One particularly well-known tool is High-Level Synthesis (HLS)~\cite{coussy_high-level_2008,mcfarland_high-level_1990}. It is an automated process that takes an abstract high-level description, in languages such as C, C++ and MATLAB, of a digital system and produces the RTL architecture that realizes the given behavior. The code at the algorithmic level is analyzed, architecturally constrained, and scheduled for transcompilation into an RTL design in HDL, which is then typically synthesized to the gate level using a logic synthesis tool.

\section*{Conclusion}

\begin{figure}
    \centering
    \includegraphics[width=0.85\linewidth]{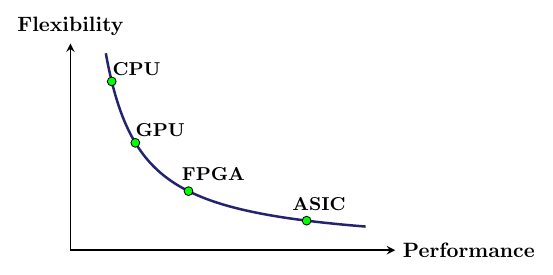}
    \caption{Illustration of a comparison of different processor architectures based on their flexibility and their performance potential.}
    \label{fig:flexibility-performance}
\end{figure}

In this chapter, I introduced parallelism, briefly summarized the histories of GPUs and FPGAs, and presented the CUDA programming model. I also described the architecture of FPGAs and touched upon the nuances of their design. While CPU remains the strongest candidate for general-purpose, control-intensive, and sequential tasks, offering flexibility and ease of programming, they lack in ability to parallelize at large scale. GPUs on the other hand are well-suited for highly parallel, throughput-oriented tasks, particularly those with structured, data-parallel workloads. FPGAs provide customizable hardware-level parallelism with low latency and high energy efficiency, ideal for real-time and resource-constrained applications. However, their programming complexity remain significant barriers. This comparison is illustrated in Fig.~\ref{fig:flexibility-performance}. The choice between the different architectures presented depends on many factors, including performance, energy efficiency, flexibility and cost.

Understanding the trade-offs between these architectures is crucial for designing optimized pipelines that meet specific requirements on throughput, latency or power consumption. This motivates the hardware choices made in the course of this thesis. Our graph neural network-based pipeline, is accelerated on the GPU architecture in Chapter~\ref{ch:etx4velo-gpu}. In Chapter~\ref{ch:etx4velo-fpga}, the pipeline is partially accelerated on FPGAs, and a comparison between the two architectures is performed.

\chapter{The LHCb Experiment at CERN}
\label{ch:lhcb}

\minitoc

\noindent Parts of this chapter were inspired by~\cite{scarabotto_search_2023,valls_canudas_calorimeter_2023,gunther_track_2023,calefice_standalone_2022}.

\section*{Introduction}

In this chapter, we look at high-energy particle physics at the LHC, specifically through the lens of the LHCb experiment, with which this work is associated. The detector of the experiment, the dataflow and its trigger system are described. Only the Upgrade~I detector for Run~3~\cite{lhcb_collaboration_framework_2012} of the LHC is discussed. Information about the previous LHCb configuration can be found in~\cite{lhcb_collaboration_lhcb_2008,aaij_design_2019}.

\section{The Large Hadron Collider at CERN}

\begin{figure}
    \centering
    \includegraphics[width=1\linewidth]{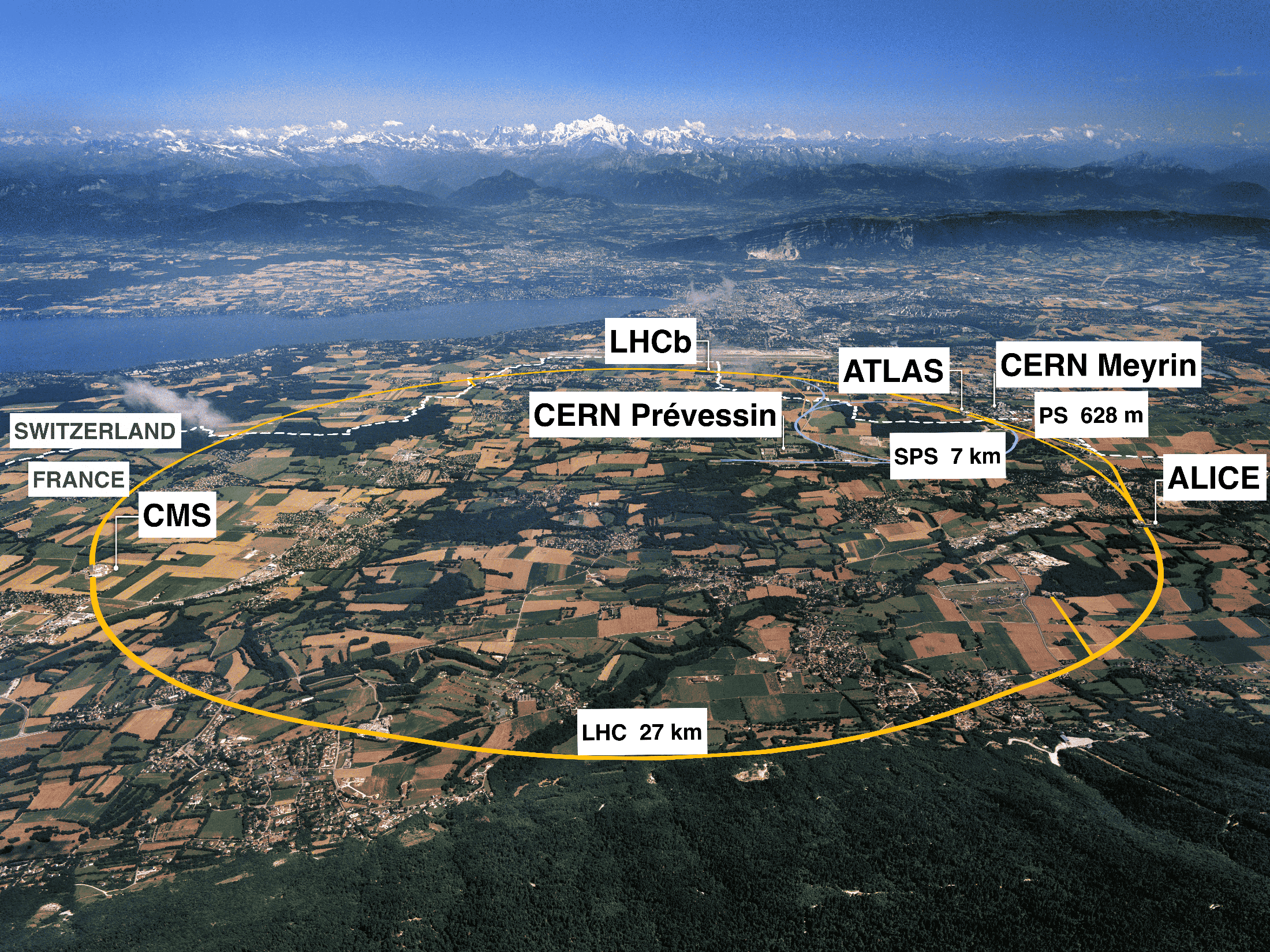}
    \caption{Aerial view of the European Organization for Nuclear Research (CERN), showing the main sites at Meyrin at Prévessin, operating the largest particle physics accelerator in the world: the Large Hadron Collider (LHC). The LHC lies in an underground tunnel 27 kilometers in circumference beneath the French--Swiss border near Geneva. The position of the main experiments, ATLAS, CMS, ALICE and LHCb, are shown. The Proton Synchrotron (PS) and the Super Proton Synchrotron (SPS) can also be seen. Adapted from~\cite{brice_aerial_nodate}.}
    \label{fig:cern-aerial}
\end{figure}

The \textit{Conseil Européen pour la Recherche Nucléaire} (CERN) is the European Organization for Nuclear Research. It is an intergovernmental organization, comprising 24 member states, that operates the largest particle physics laboratory in the world. Established in 1954, it is based in Meyrin, a suburb of Geneva, on the border of Switzerland with France, as shown in Fig.~\ref{fig:cern-aerial}.

\begin{figure}
    \centering
    \includegraphics[width=.5\linewidth]{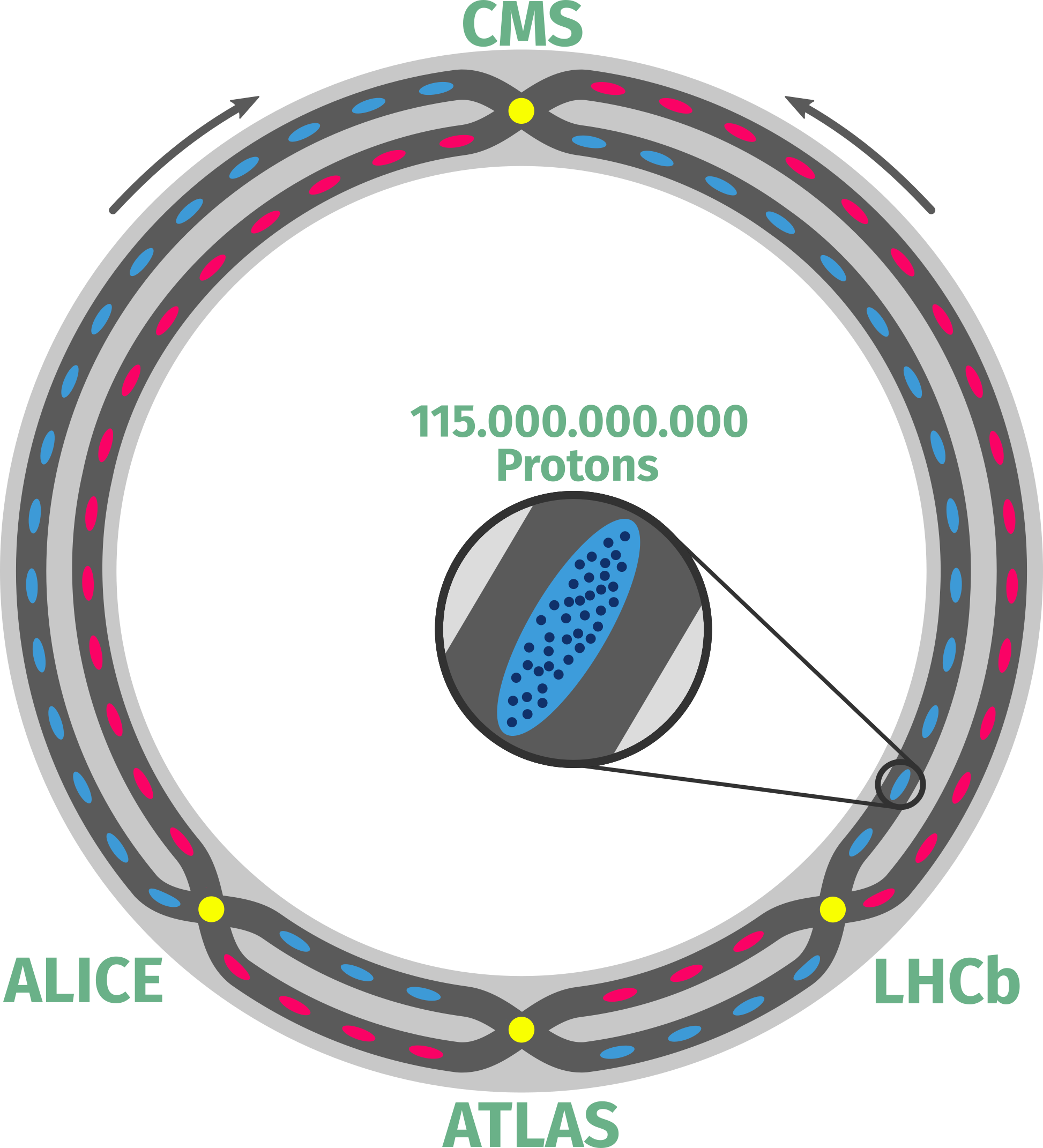}
    \caption{Sketch of the LHC showing its accelerator ring with two beam pipes and the four main CERN experiments. The beam pipes carry particle bunches that intersect at the interaction point of each experiment~\cite{hasse_simple_2023}.}
    \label{fig:lhc_sketch}
\end{figure}

\begin{figure}
    \centering
    \includegraphics[width=1\linewidth]{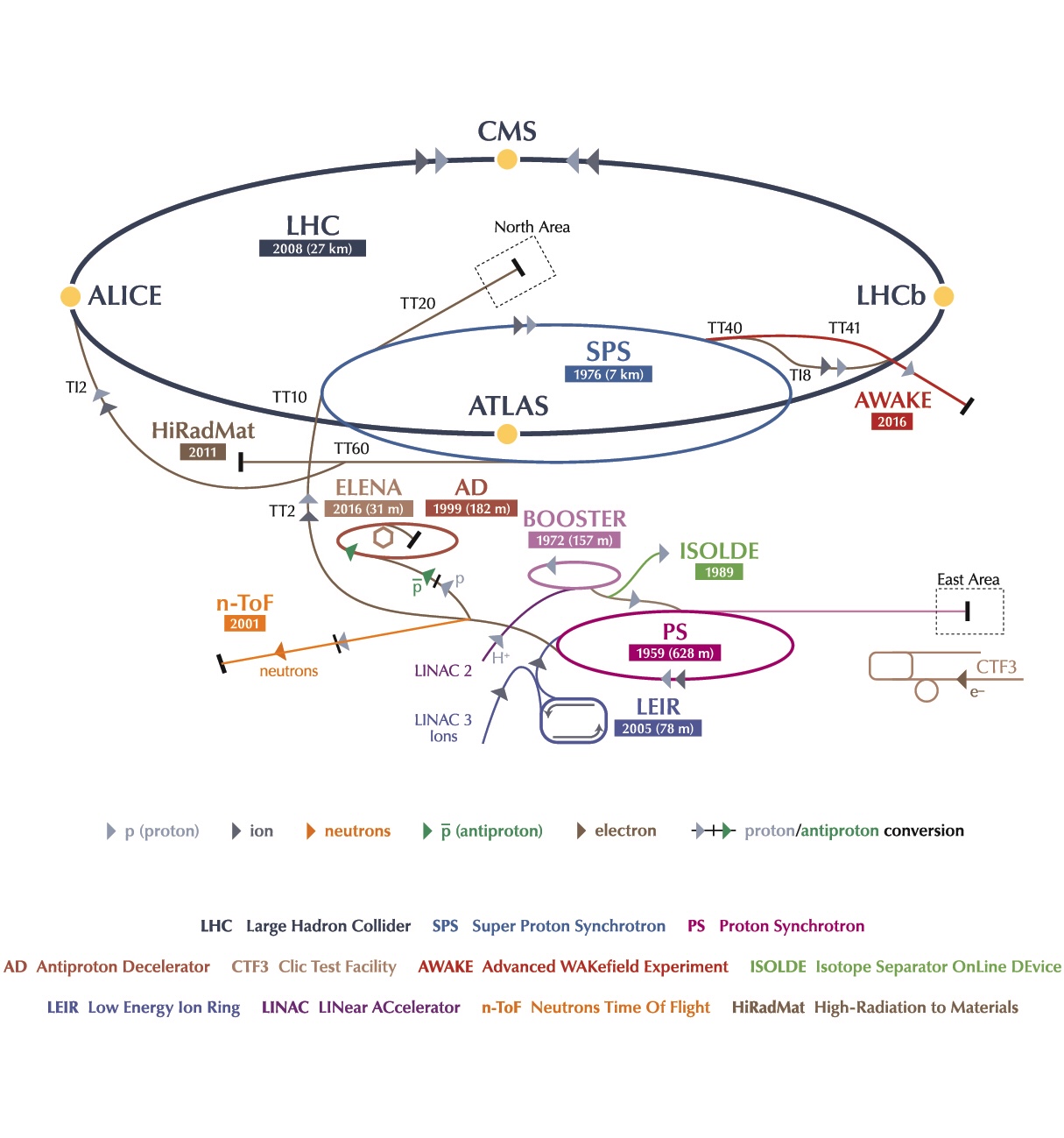}
    \caption{The CERN accelerator complex during Run~2. Figure from~\cite{de_melis_cern_2016}.}
    \label{fig:cern-complex}
\end{figure}

The LHC~\cite{evans_lhc_2008} at CERN is currently the world's biggest and most powerful particle accelerator located roughly 100 meters below ground. The accelerator has the form of a ring with a perimeter of 27~km and 90\% of its length is in molasse rock while the remaining 10\% is in limestone under the Jura mountain. This machine is mainly used to accelerate protons as well as heavy ions, such as lead, in order to collide them. The protons are accelerated in bunches, as illustrated in Fig.~\ref{fig:lhc_sketch}, in two superconducting magnet rings with opposite directions. After the particles have been accelerated, they are brought into collision at four interaction points hosting the detectors for the four main CERN experiments: ATLAS, CMS, ALICE and LHCb, featured on Fig.~\ref{fig:cern-complex}. These bunches of particles cross every 25~ns, or equivalently at a frequency of 40~MHz.

LHC operates in periods called ``Runs''. Run~1 took place between 2010--2012, Run~2 between 2015--2018 and Run~3 started in July 2022. Run~3 is planned to last until July of 2026 while Run~4 is scheduled for the summer of 2030. The LHC is designed to accelerate protons very close to the speed of light, reaching energies up to 7~TeV. During Run~3, the LHC collides protons with a center-of-mass energy of $\sqrt{s} =$ 13.6~TeV\footnote{Variable $s$ is the Mandelstam $s$ variable~\cite{mandelstam_determination_1958} defined as $s = (\mathbf{p}_1 + \mathbf{p}_2)^2 c^2$, where $\mathbf{p}_1$ and $\mathbf{p}_2$ are the four-momenta of the incoming particles. $\sqrt{s}$ is the center-of-mass energy, and is an observer-independent way to measure how hard the protons are being collided against each other.}. While ATLAS and CMS operate at a peak instantaneous luminosity of $\mathcal{L}=$ $2\times 10^{34}$~cm$^{-2}$s$^{-1}$ which decreases with time during an LHC fill, LHCb uses luminosity leveling~\cite{follin_implementation_2014} in order to keep the luminosity lower but constant. This is done in order to deliver steady conditions for physics analysis, but also because the processes that LHCb focuses on are difficult to record at high luminosities. For Run~3, LHCb aims for an instantaneous luminosity of $\mathcal{L} = 2\times 10^{33}$~cm$^{-2}$s$^{-1}$, which results in about five $p$--$p$ collisions per bunch crossing on average.

The CERN accelerator complex, serving as an injector to the LHC, is illustrated in Fig.~\ref{fig:cern-complex}. Protons are taken from a bottle of hydrogen gas, with the electrons stripped off the atoms using an electric field. The protons are then progressively accelerated through the accelerators complex, reaching the LHC at the end. Initially the protons are accelerated by the Linear Accelerator (LINAC) 2 followed by the Proton Synchrotron Booster (PSB), the Proton Synchrotron (PS) and the Super Proton Synchrotron (SPS). Finally protons leaving the SPS are injected into the LHC at an energy of 450~GeV. For heavy ions LINAC 3 is used. For Run~3, LINAC 2 has been replaced by LINAC 4.

\section{LHCb Detector Overview}
\label{sec:detector}

\begin{figure}
    \centering
    \includegraphics[width=1\linewidth]{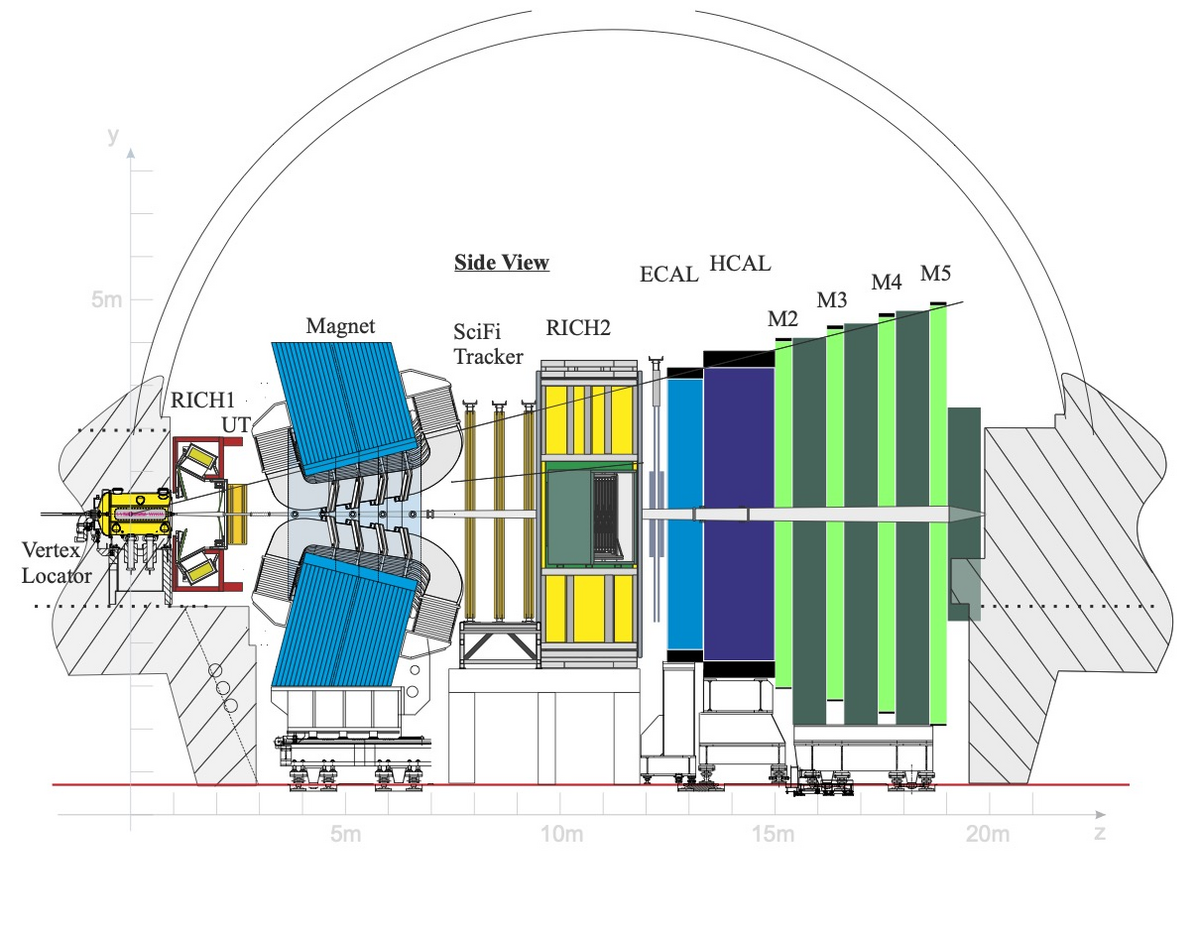}
    \caption{Layout of the upgraded LHCb detector. Figure from~\cite{lhcb_collaboration_lhcb_2024}.}
    \label{fig:lhcb-detector}
\end{figure}

The LHCb detector~\cite{lhcb_collaboration_lhcb_2008,lhcb_collaboration_lhcb_2015,lhcb_collaboration_lhcb_2024}, shown in Fig.~\ref{fig:lhcb-detector}, is a single-arm forward spectrometer covering the pseudorapidity range $2<\eta <5$, designed for the study of particles containing $b$ or $c$ quarks. The detector has been substantially upgraded prior to the Run~3 data-taking period, which started in 2022. The upgraded detector was designed to match the performance of the Run~1--2 detector, while allowing it to operate at approximately five times the luminosity. Simulation studies show the upgraded detector meeting these performance goals~\cite{lhcb_collaboration_lhcb_2024}.   

The high-precision tracking system has been fully replaced and consists of a silicon-pixel vertex detector, known as the Vertex Locator (VELO) and described in more detail in Section~\ref{sec:velo}, surrounding the $p$--$p$ interaction region~\cite{lhcb_collaboration_velo_2013}, a large-area silicon-strip detector~\cite{lhcb_collaboration_lhcb_2014}, known as the Upstream Tracker (UT), located upstream of a dipole magnet with a bending power of about 4~Tm, and three stations of scintillating fiber detectors~\cite{lhcb_collaboration_lhcb_2014}, collectively known as the Scintillating Fiber (SciFi) tracker. Different types of charged hadrons are distinguished using information from two Ring-Imaging Cherenkov (RICH) detectors~\cite{lhcb_rich_collaboration_performance_2013,lhcb_collaboration_particle_2013}, RICH 1 and 2. The whole photon detection system of the Cherenkov detectors has been renewed for the upgraded detector. With this configuration, the upgraded detector achieves a track momentum resolution of $\sigma_p /p \approx $ 0.5--1\%~\cite{lhcb_collaboration_track_2024}, over a broad range of momenta, where $\sigma_p$ is the measurement uncertainty of momentum $p$.

Photons, electrons and hadrons are identified by a calorimeter system consisting of electromagnetic (ECAL) and hadronic (HCAL) calorimeters. Muons are identified by a system of muon stations (M1--5) composed of alternating layers of iron and multiwire proportional chambers~\cite{alves_jr_performance_2013}.

Readout of all detectors into an all-software trigger~\cite{noauthor_lhcb_2014} is a central feature of the upgraded detector, facilitating the reconstruction of events at the maximum LHC interaction rate, and their selection in real time. The trigger system, described in further detail in Section~\ref{sec:trigger}, is implemented in two stages: a first inclusive stage based primarily on charged particle reconstruction which reduces the data volume by roughly a factor of 20, and a second stage, which performs the full offline-quality reconstruction and selection of physics signatures. A large disk buffer is placed between these stages to hold the data while the real-time alignment and calibration is being performed.

\section{Vertex Locator}
\label{sec:velo}

\begin{figure}
    \centering
    \includegraphics[width=1\linewidth]{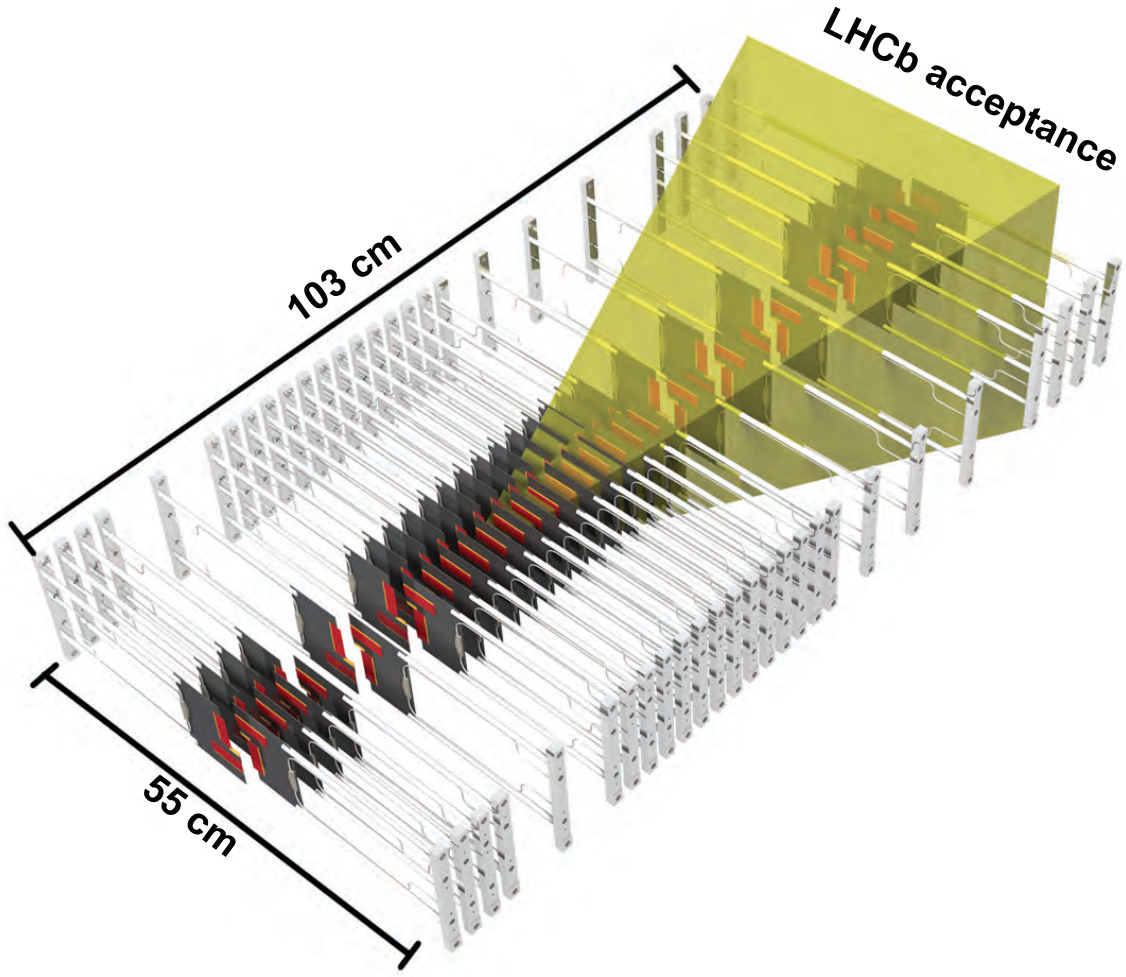}
    \caption{Upgrade VELO module layout, with the LHCb acceptance highlighted. This figure shows how different parts of the modules fall within the acceptance region for physics-quality tracks. Figure from~\cite{lhcb_collaboration_velo_2013}.}
    \label{fig:velo-3d}
\end{figure}

\begin{figure}
    \centering
    \includegraphics[width=1\linewidth]{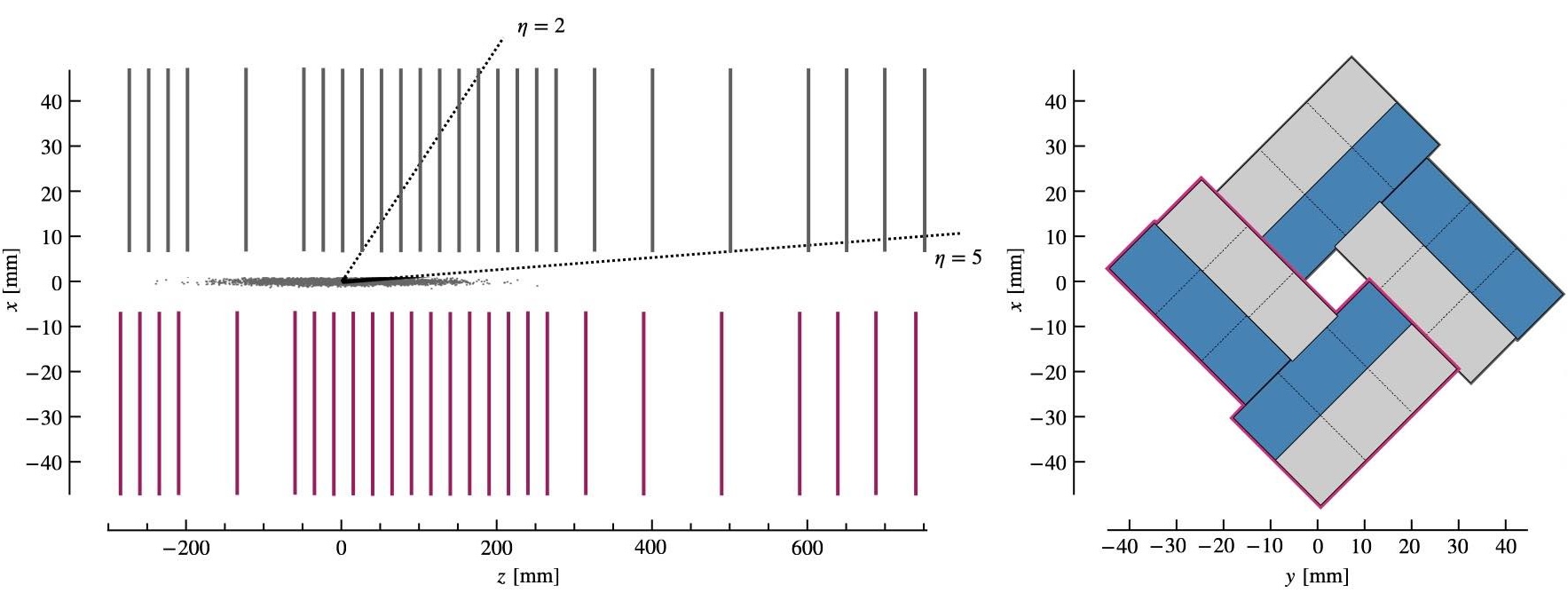}
    \caption{Left: Schematic top view of the $z$-$x$ plane at $y = 0$, illustrating the $z$-extent of the luminous region and the nominal LHCb pseudorapidity acceptance, $2 < \eta < 5 $. Right: Schematic of the nominal sensor layout around the $z$-axis in the closed VELO configuration. Half of the ASICs are positioned on the upstream module face (gray), while the other half are on the downstream face (blue). Figure from~\cite{ferrari_lhcb_2024}.}
    \label{fig:velo}
\end{figure}

The VELO, with dimensions as shown in Fig.~\ref{fig:velo-3d}, is the most important subdetector for the LHCb experiment. It detects particles created near the beam collision region and is used to locate primary and displaced collision vertices, the latter being characteristic of beauty and charm hadron decays. In addition, it helps in the reconstruction of tracks in the other subdetectors. The detector consists of 52 L-shaped modules, as shown in Fig.~\ref{fig:velo} on the left. The last station downstream is positioned at 751~mm while the first station upstream is at $-289$~mm.

The detector is divided into two movable halves, 26 modules on each side, which can be separated into what is known as the VELO open position during the injection and tuning of the LHC beams. When the beam conditions stabilize, the two halves can be closed and centered around the luminous region, as shown in Fig.~\ref{fig:velo}, on the right. The detector's vertex resolution improves as it gets closer to the interaction region. For this reason, the VELO modules are located at a record distance of 5.1~mm from the beam, when VELO is in its closed position.

\begin{figure}
    \centering
    \includegraphics[width=1\linewidth]{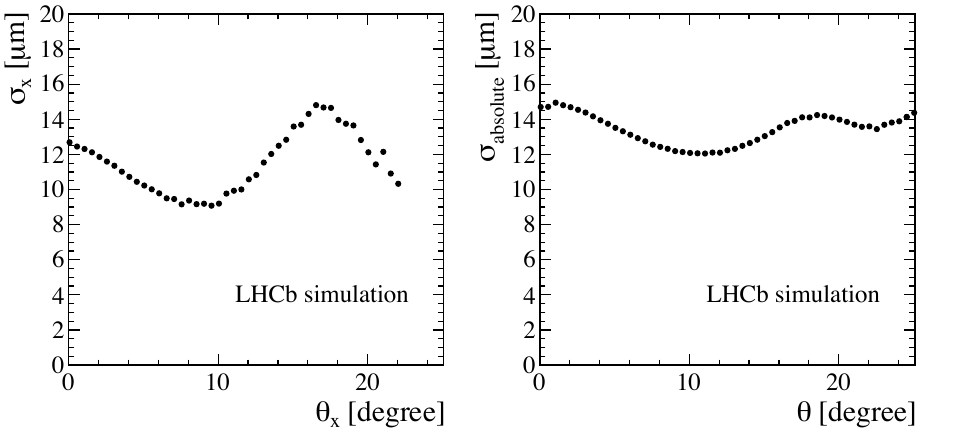}
    \caption{Dependence of hit resolution on the track polar angle. Left: Single hit resolution in $x$, defined as the RMS of the residual distribution, versus the projected angle $\theta_x$, for tracks with $| \theta_y|<2^\circ$. An optimal resolution is observed for tracks with angles close to $9^\circ$. Right: Absolute measurement error, defined as the average absolute distance between the true and reconstructed position, as a function of polar angle $\theta$, integrated over all azimuthal angles $\varphi$. Figure from~\cite{lhcb_collaboration_velo_2013}.}
    \label{fig:hit-resolution}
\end{figure}

Each VELO module consists of four sensors of 200~$\upmu$m thickness, each one containing three VeloPix~\cite{poikela_velopix_2017} chips. These chips have an active area of 256 $\times$ 256 pixels of size 55~$\upmu$m $\times$ 55~$\upmu$m~\cite{lhcb_collaboration_lhcb_2024}. The entire detector therefore totals almost 41 million channels. Pixels are clustered into hits on the readout cards during data acquisition. The achieved hit resolution is about 12.5~$\upmu$m in the $x$ and $y$ coordinates~\cite{lhcb_collaboration_velo_2013,buchanan_spatial_2022}. The dependence of this resolution on the track polar angle is shown in Fig.~\ref{fig:hit-resolution}.

The VELO modules are being cooled using evaporative bi-phase CO$_2$~\cite{jans_operational_2013} absorbing the heat generated by the VeloPix readout chips in order to ensure stable performance. The sensors operate in a secondary vacuum, which is separated from the LHC beam vacuum by an aluminum Radio Frequency (RF) box~\cite{jans_velo_2015}. The RF box further shields the detector electronics from RF pickup of the beams. The RF foil needs to be mechanically stable, in order to withstand pressure changes, and extremely radiation-hard due to the intense radiation environment very close to the LHC beam.

The RF foil, the cooling and the sensors all contribute to the material budget of the detector~\cite{lhcb_collaboration_velo_2013}. Particles traversing the detector material can undergo multiple scattering and lose energy, hence reducing the resolution of the detector. The extent of these effects depends on factors such as the track's angle of incidence and the number of VELO sensors crossed by the particle. The largest contribution, almost 53\%, comes in fact from the RF foil.

\begin{figure}
    \centering
    \includegraphics[width=1\linewidth]{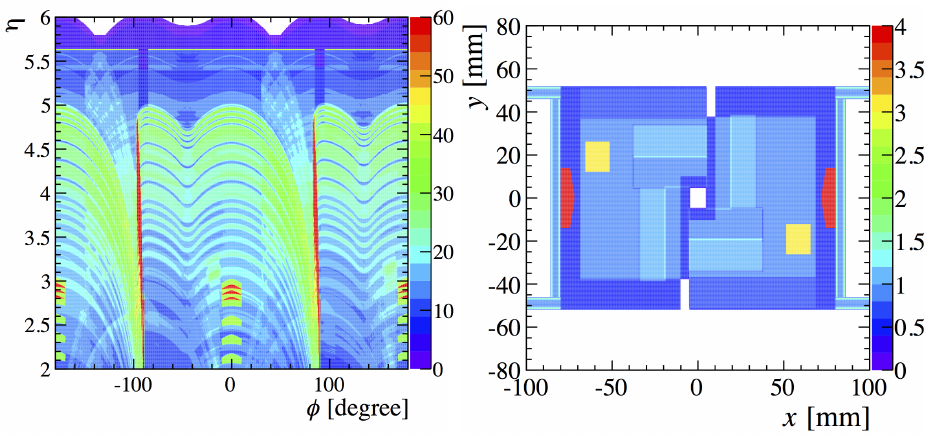}
    \caption{Left: Percentage of radiation length (between the origin $z=0$ and $z=835$~mm) seen by tracks traversing the VELO detector, as a function of pseudorapidity $\eta$ and azimuthal angle $\phi$. Right: Percentage of radiation length (between the origin $z=0$ and $z=835$~mm) seen by tracks crossing a VELO module (excluding the RF foil) at perpendicular incidence, as a function of the $x$ and $y$ coordinates. Figure from~\cite{lhcb_collaboration_velo_2013}.}
    \label{fig:material}
\end{figure}

Averaged over pseudorapidities $2 < \eta < 5$, the total material budget of the upgrade VELO is around 21.3\%~$X_0$, where $X_0$ is the radiation length of the material, i.e., the mean length into the material at which the energy of an electron is reduced by the factor $1/e$~\cite{gupta_calculation_2010}. In contrast, the material budget of a single module, averaged over its area, for perpendicularly incident tracks is roughly 1\%~$X_0$. The material budget for the detector as a whole as well as for a single module is shown in Fig.~\ref{fig:material}. The most prominent features in the material map are the ridges caused by the RF foil at $\phi = \pm \pi/2$, and the peaks associated with the cooling connectors at $\phi = 0$ and $\phi = \pi$.

In this configuration~\cite{lhcb_collaboration_velo_2013}, the VELO subdetector achieves: 

\begin{itemize}
    \item A primary vertex resolution of $(11.0 + 13.1/p_T[\text{GeV}/c])$~$\upmu$m, where $p_T$ is the transverse momentum expressed in GeV/c, and
    \item A B-meson decay-time resolution of 43~fs.
    
\end{itemize}

\section{Online System and Data Acquisition}

The upgraded online system builds upon and enhances the successful experiment control system from Run~1 and Run~2~\cite{lhcb_collaboration_lhcb_2001}. It introduces a new timing and fast signal control mechanism for distributing clock signals, as well as synchronous and asynchronous commands for the readout. Additionally, the data acquisition system has been significantly expanded. Other key hardware and software components include alignment and calibration frameworks, online monitoring, data storage, and the infrastructure required to operate these systems and the event-filter farm. The system's hardware consists of a powerful, custom-designed FPGA board, complemented by commercial off-the-shelf hardware. 

\begin{figure}
    \centering
    \includegraphics[width=1\linewidth]{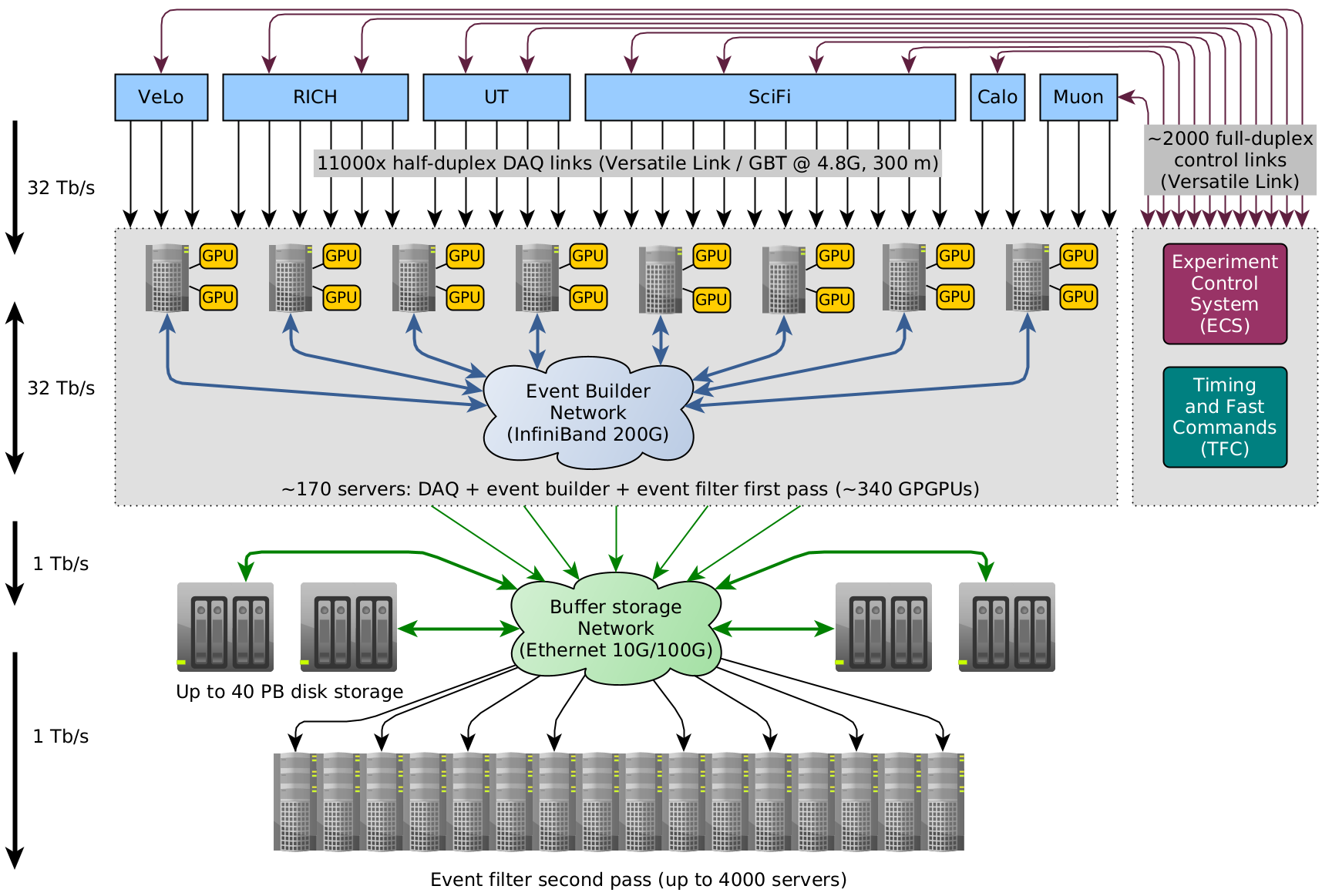}
    \caption{Upgraded LHCb online system. All system components are connected to the Experiment Control System (ECS) shown on the right, although these connections are not shown in the figure for clarity. Figure from~\cite{lhcb_collaboration_lhcb_2024}.}
    \label{fig:daq}
\end{figure}

\begin{figure}
    \centering
    \includegraphics[width=1\linewidth]{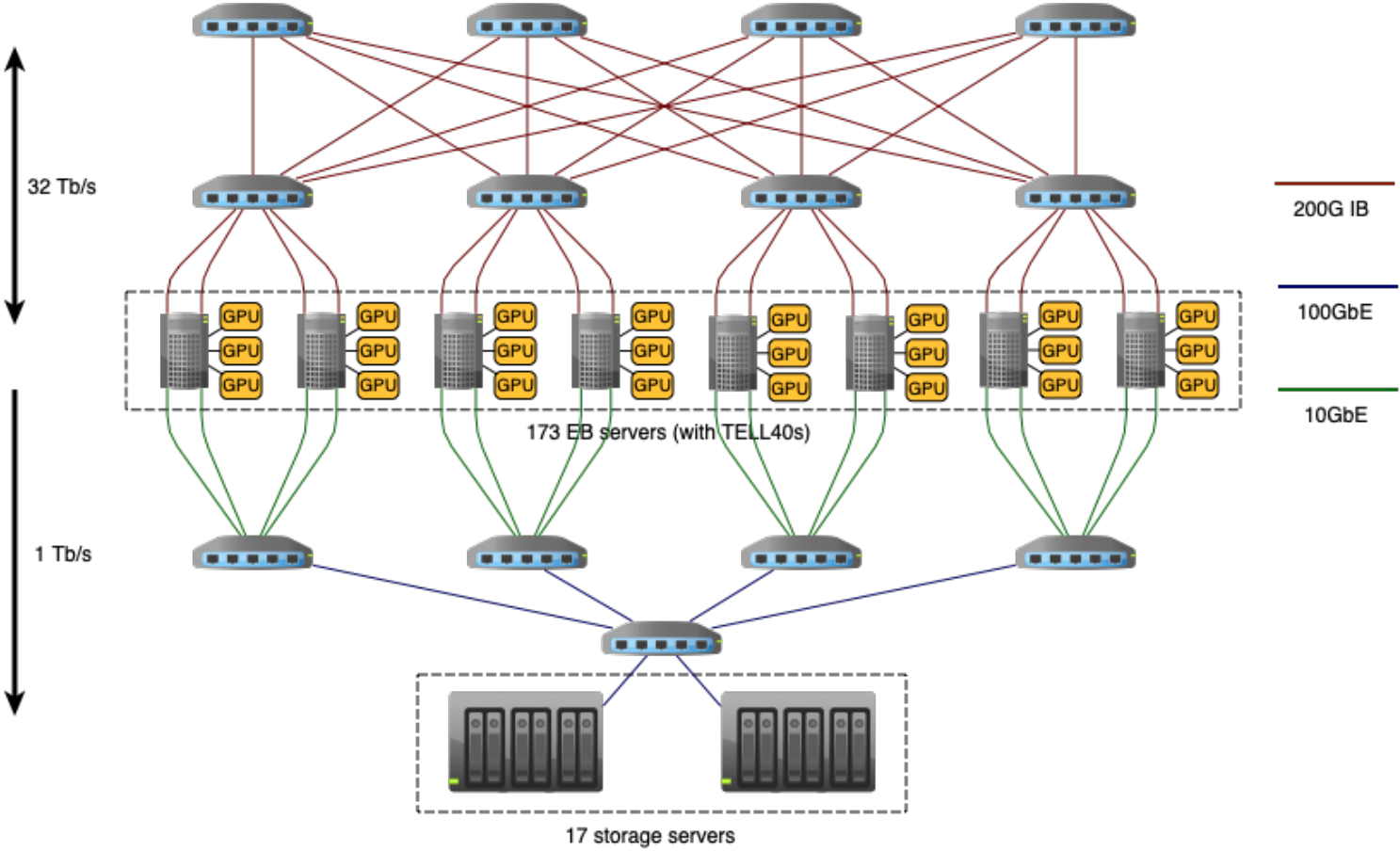}
    \caption{Upgraded LHCb online system with three GPUs running HLT1 on the EB servers. Figure from~\cite{lhcb_collaboration_lhcb_2020}.}
    \label{fig:daq2}
\end{figure}

The LHCb Data Acquisition (DAQ) system, illustrated in Fig.~\ref{fig:daq}, consists of 162 Event Builder (EB) servers with AMD EPYC 7502 32-core CPUs hosting custom-made FPGA detector readout boards (TELL40) and Nvidia RTX A5000 GPUs on the Peripheral Component Interconnect Express (PCIe) interface. The DAQ design is ``triggerless'', meaning that the detector is read out at the nominal LHC bunch-crossing frequency of 40~MHz. The more recently updated layout, with three GPU cards running HLT1 on the EB servers, and with the EB output data rate reduced by a factor between 30 and 60, is shown in Fig.~\ref{fig:daq2}.

A pool of DAQ readout supervisors centrally manages the readout of events, by generating synchronous and asynchronous commands, and by distributing the LHC clock. The system controls the FPGA cards used to receive the detector data~\cite{alessio_clock_2015}. It can also function as a very primitive trigger system and is used as such to e.g., downscale the rate of empty--empty, beam--empty, and empty--beam bunch crossings seen by the HLT.

Each TELL40 receives data from the different front-end electronics of the subdetectors, the so-called Multi-Fragment Packets (MFPs). The EB servers then collect the MFPs from all the subdetectors and group the information from the same events, creating the Multi-Event Packets (MEPs) containing 30\,000 events. By then transferring the MEPs directly from the EB server to the GPUs, a first event reconstruction and filtering is performed using Allen~\cite{lhcb_collaboration_allen_2004}. At the same time, in this way, the process minimizes the overhead associated with CPU-GPU data transfers. To maximize the efficiency of the TELL40s, certain aspects of event data reconstruction for the subdetectors can be integrated into the FPGA firmware. This approach is applied to the VELO, where the pixel data are clustered into hits directly on the TELL40s, as detailed in~\cite{bassi_fpga-based_2023}.

In summary, the EB nodes gather event fragments from all subdetectors, assemble them into complete events, and store these full-event packages in a shared memory buffer. From there, the Allen software trigger application (see Section~\ref{sec:trigger}) performs partial event reconstruction and selection using GPUs hosted in the EB servers. Events selected by Allen are then transferred to the buffer storage network, where they are temporarily stored before being accessed and processed by the alignment and trigger application, Moore~\cite{lhcb_collaboration_moore_2007}. Moore carries out full event reconstruction and selection, completing the real-time analysis of the event. The computing farm running Moore, known as the event-filter farm, consists of over 3000 general-purpose CPU servers of varying types and processing capacities. This farm operates as a computing cloud, with all nodes running the same operating system and lacking significant local storage, ensuring that any node can be easily replaced. Once processed, the data is sent to permanent storage, i.e., the Worldwide LHC Computing Grid.

\section{Software Framework}

The LHCb software framework underwent a major rewrite and update to support running offline-quality event reconstruction in real time within the trigger. This update unifies the codebase for the trigger and offline reconstruction, making offline processing simply a specific configuration of the same underlying algorithms. The backend code is primarily written in modern C++~\cite{stroustrup_c_1984}, while component configuration is handled in Python~\cite{python_developers_python_nodate}. CPU-targeted code is built on the Gaudi framework~\cite{barrand_gaudi_2001}, which is actively developed and used by both the LHCb and ATLAS collaborations. Meanwhile, the GPU codebase is implemented within the cross-architecture Allen framework~\cite{aaij_allen_2020,lhcb_collaboration_allen_2004}, which can be compiled for execution on both CPUs and GPUs, enabling its integration into Gaudi.

LHCb's real-time analysis strategy risks data loss if the software trigger implementation is erroneous. A code review system, maintained by a team of LHCb developers, includes automated unit and integration tests, and is in place to avoid breaking the functionality of the latest software stack~\cite{clemencic_lhcb_2015}. The software stack is released and deployed regularly to ensure the reproducibility of physics results.

The LHCb codebase consists of multiple independently versioned and managed projects, all using the Git version control system. The code is publicly available in~\cite{lhcb_collaboration_lhcb_nodate} and is distributed under the GNU General Public License v3, except for the Allen project, which is licensed under the Apache License v2.

\section{Simulation}

\begin{figure}
    \centering
    \includegraphics[width=1\linewidth,keepaspectratio]{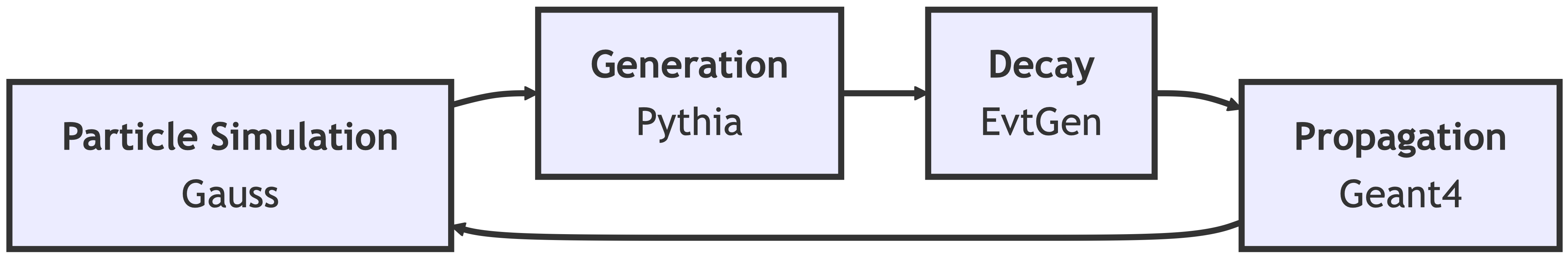}
    \caption{Illustration of the simulation process inside LHCb. Generated with~\cite{sveidqvist_mermaid_2014}.}
    \label{fig:simulation}
\end{figure}

Gauss~\cite{lhcb_collaboration_gauss_2007,clemencic_lhcb_2011} is the simulation framework used by LHCb to interface different event generators with decay engines and simulate the detector's response. It is used by LHCb to generate simulated events by coordinating several external tools. A typical event is produced through the following steps:

\begin{itemize}
    \item A ``production tool'' (such as Pythia~\cite{sjostrand_brief_2008} or GENXICC~\cite{chang_genxicc_2007}) generates an event containing the desired signal particle. The first way this can be done is by producing \textit{minimum bias} events until the signal particle appears. In a minimum bias sample, all events are generated by the production generator, with no requirement about their content. The second way a signal particle can be produced is by forcing its production in every event, so this sample is no longer minimum bias. The final generated event contains a mix of stable particles and unstable ones that can decay.

\item The signal particle is then decayed into the desired final state using a ``decay tool'' (usually EvtGen~\cite{lange_evtgen_2001}), while any other unstable particles are decayed separately.

\item The signal and its decay products may need to satisfy generator-level selection criteria, which are applied via a ``cut tool''.

\item Finally, the particles are propagated through the detector and the detector's response is modeled with Geant4~\cite{allison_geant4_2006}.

\end{itemize}
The process is summarized in Fig.~\ref{fig:simulation}.

\section{LHCb Trigger System}
\label{sec:trigger}

\subsection*{The Necessity for an RTA Trigger at LHCb}

As already discussed in Chapter~\ref{ch:motivation}, the amount of data produced by each of the main four LHC experiments is too massive to be stored on disk or tape. For example, a typical LHCb event under Run~3 conditions is around 100~kB of data, which at the LHC $p$--$p$ collision rate of 40~MHz results in a data rate of 4~TB/s. In 2017, for Run~2, the LHC delivered stable beams for 1634 hours~\cite{todd_lhc_2017}, so assuming the same availability for Run~3, the amount of data accumulates to 18~EB ($10^{18}$~B) per year. Together with the other three LHC experiments, the sum ends up at around $\mathcal{O}(100)$~EB per year, a number comparable to the monthly global internet traffic~\cite{cisco_public_cisco_nodate} and unfeasible to store permanently. This is the reason why all large HEP experiments have sophisticated triggers systems that filter the interesting collision events.

\begin{figure}
    \centering
    \includegraphics[width=1\linewidth]{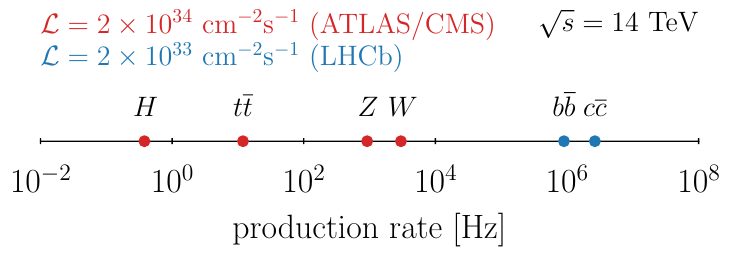}
    \caption{Production rates estimates for various Standard Model processes at the LHC in Run~3. Figure from~\cite{boettcher_allen_2022}.}
    \label{fig:trigger-rates}
\end{figure}

When the processes an experiment aims to investigate are sufficiently rare, hardware-based triggers designed to perform fast but relatively simple classifications of collision events are typically sufficient to reduce the data volume to a manageable level while maintaining high selection efficiency. For example, general purpose detectors, such as that of ATLAS and CMS, study processes such as Higgs or $W$ and $Z$ productions. With the production rates as shown in Fig.~\ref{fig:trigger-rates}, one can trigger efficiently from the full LHC collision rate of 40~MHz down to around 100~kHz with only a single detector systems, e.g., using high transverse energy $E_{\text{T}}$ calorimeter clusters. On the other hand, the primary focus of LHCb is on the production and decay of hadrons containing $b$ or $c$ quarks, such as $b\bar{b}$ or $c\bar{c}$ production, also shown on Fig.~\ref{fig:trigger-rates}. With these rates exceeding 1~MHz, it is obvious that the trigger rate cannot be lower than this threshold. Due to the physics of hadronic heavy flavor decays, and especially charm physics, these signals cannot be ``triggered'' in a classical way based on hardware because they are too abundantly produced in the first place. 

\begin{figure}
    \centering
    \includegraphics[width=1\linewidth]{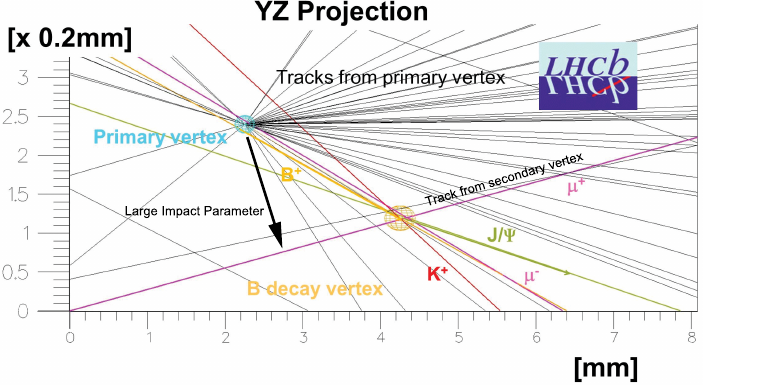}
    \caption{Illustration of a $B^+ \to J/\psi K^+$ candidate event in LHCb data, highlighting the tracks, primary and secondary vertices, and the impact parameter of the antimuon track. Figure from~\cite{lhcb_collaboration_velo_2013}.}
    \label{fig:pv-sv-impact}
\end{figure}

In addition, a characteristic signal for these processes is a displaced (secondary) vertex. As illustrated in Fig.~\ref{fig:pv-sv-impact}, in LHCb, tracks originating from secondary vertices with a large impact parameter are the principal signature of beauty and charm hadrons decaying. Because of the abundance of displaced vertices from lighter particle decays, in order to trigger on this signal efficiently, information is needed from the entire tracking system. Finally, the higher instantaneous luminosities of Run~3 result in a significant increase in combinatorial background, making its suppression one of the main challenges for reconstruction~\cite{calefice_effect_2022}.

In anticipation of this, LHCb had to redesign its trigger. The solution to the above problems was to remove the hardware trigger part and read out the full detector at 40~MHz in Run~3~\cite{gligorov_conceptualization_2018}. An offline-quality event reconstruction is performed on the incoming data directly, i.e., in real time. Some trigger lines select entire events while others only store the interesting reconstructed decay trees. In the latter case, only the relevant portion of the collision is extracted. By significantly reducing the average data size stored per event, bandwidth and storage challenges can be addressed~\cite{aaij_comprehensive_2019}. Performing offline-quality event reconstruction coupled with fine-grained selection is only achievable through a software-based trigger system operating on a detector that is both aligned and calibrated in real time. Processing the vast amounts of data described above within such a system presents significant challenges. The reconstruction software must deliver speed, precision, and efficiency all at once, which is only feasible when the detector is continuously aligned and calibrated during data acquisition. To address this, the High-Level Trigger (HLT) is divided into two stages, separated by a storage buffer. This architecture extends the effective real-time processing window to several days, enabling automated event analysis before selecting and writing the most relevant information to permanent storage.

\begin{figure}
    \centering
    \includegraphics[width=.6\linewidth]{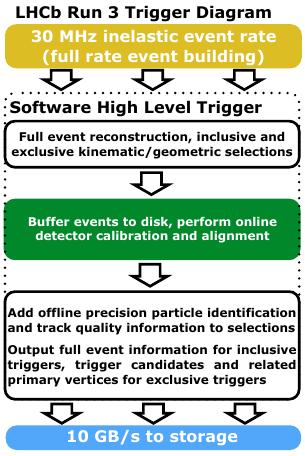}
    \caption{LHCb trigger diagram for Run~3. Figure from~\cite{lhcb_collaboration_rta_2020}.}
    \label{fig:lhcb-trigger}
\end{figure}

The trigger scheme of the LHCb experiment during Run~3 is depicted in Fig.~\ref{fig:lhcb-trigger}. The requirement of the trigger system is to reduce the data rate down to around 10~GB/s, in order for the data to be able to be saved to permanent storage. For comparison, for the signal rates shown in Fig.~\ref{fig:trigger-rates}, a 10~GB/s rate translates to a maximum event size of 10~kB and 10~MB, for a 1~MHz signal of $b \bar{b}$ and $c \bar{c}$ production, and a 1~kHz signal of $Z$ boson production, respectively. On the other hand, in our sample, there are on average 150 particles in the VELO acceptance, and around 2200 hits in each event. Since each hit has three coordinates of 32-bit floats, this would correspond to $3 \times 32 \times 2200 \approx $ 200~kB, only for the VELO part of the event.

\subsection*{LHCb Dataflow}

\begin{figure}
    \centering
    \includegraphics[width=1\linewidth]{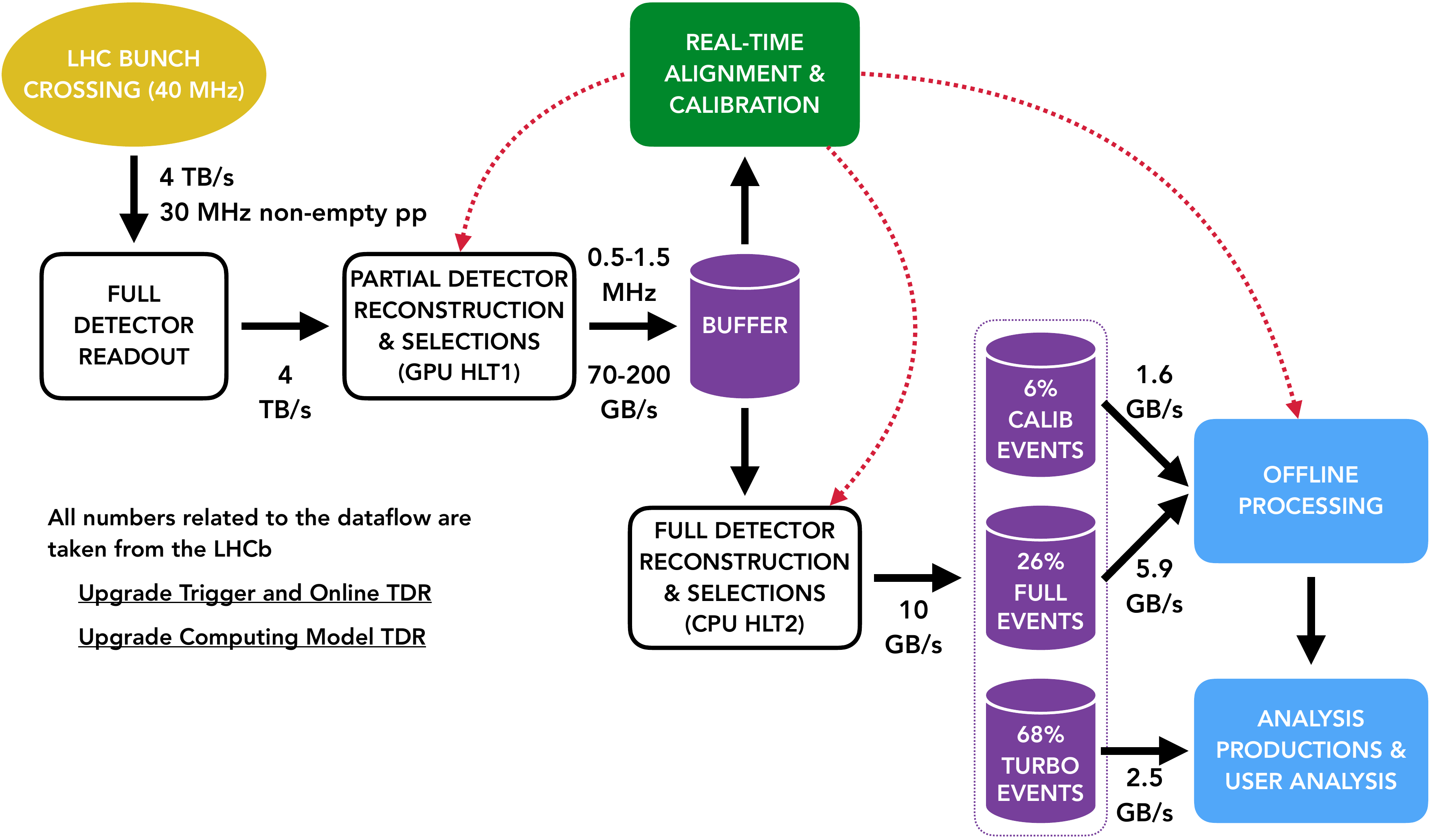}
    \caption{LHCb upgrade dataflow focusing on the real-time aspects. Figure from~\cite{lhcb_collaboration_rta_2020}.}
    \label{fig:dataflow}
\end{figure}

The software-based trigger developed and commissioned for LHCb in Run~3 is depicted in Fig.~\ref{fig:dataflow}. The detector data arriving from the subdetectors' frontend electronics are read out at the nominal LHC bunch-crossing frequency of 40~MHz and then partially reconstructed and filtered with Allen, the first level of the High-Level Trigger (HLT1), running on GPUs. Due to some bunches not being filled, and not all bunch crossing resulting in $p$--$p$ collisions, in practice, the incoming rate ends up at around 30~MHz. HLT1 reduces the event rate by approximately a factor of 30. The raw detector data, along with the trigger decision reports, are then stored in the buffer storage network of around 30~PB. This reduction factor varies depending on the buffer's capacity and the processing speed of the second trigger stage---that is, how quickly the buffer can be cleared. A subset of the HLT1 data is specifically chosen for detector alignment and calibration, which are carried out as soon as enough data has been gathered. The alignment and calibration includes a set of algorithms that measure with high precision the physical position and calibration parameters for the various subdetectors in order to provide the most accurate parameters for the reconstruction and selections. While this is performed, the buffer storage network holds the data. In subsequent runs, HLT1 benefits from an already aligned and calibrated detector, hence the term real-time alignment.

\begin{figure}
    \centering
    \includegraphics[width=1\linewidth]{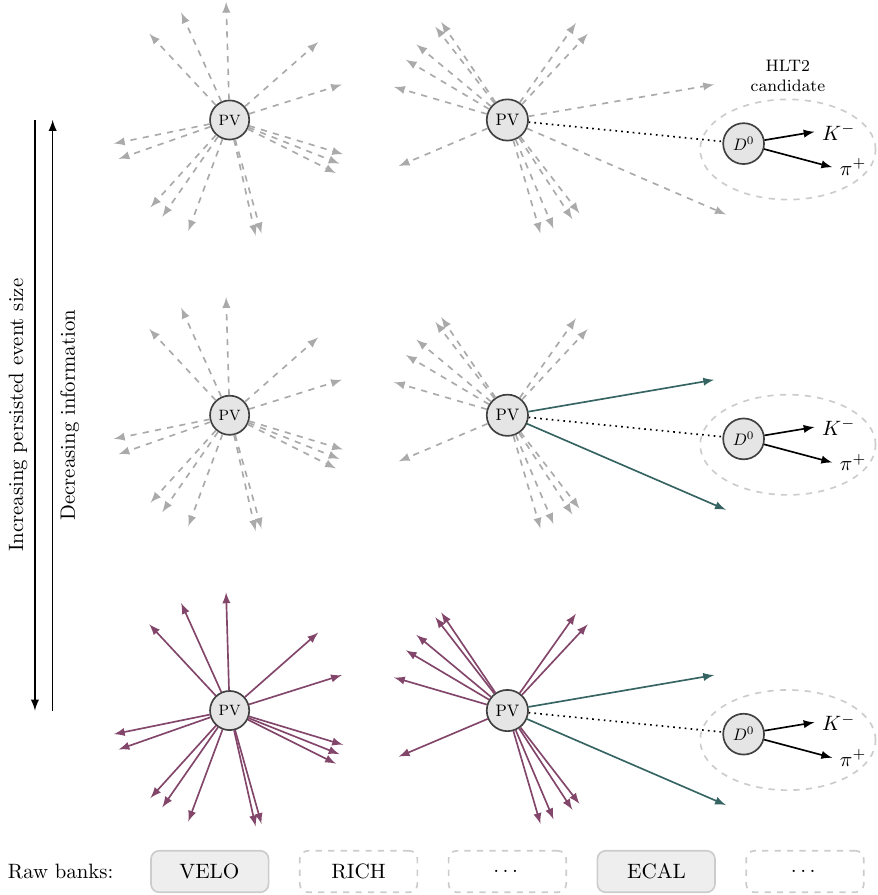}
    \caption{Illustration of the LHCb trigger computing model. The same reconstructed event is saved with varying levels of object persistence: turbo (top), selective persistence (middle), and complete reconstruction persistence (bottom). Top: A candidate $D^0 \to K^- \pi^+ $ is selected by HLT2; only the candidate and the Primary Vertex (PV) are persisted. Middle: Additional objects, e.g., pion tracks from candidate $D^{*+} \to D^0 \pi^+$, can also be persisted. Bottom: The full reconstruction event is persisted, including raw subdetector data banks. Solid lines and objects denote persisted information in each case. Raw banks are represented as rectangles.
 Figure from~\cite{lhcb_collaboration_computing_2018,aaij_comprehensive_2019}.}
    \label{fig:turbo}
\end{figure}

The second level of the High-Level Trigger (HLT2) uses the same detector alignment when processing the buffered output from HLT1. It carries out full event reconstruction on CPUs, including the tracking, calorimeter reconstruction, particle identification and the Kalman fit, and selects particle decay candidates based on numerous trigger lines (order of 1000) developed by data analysts targeting specific decays. By this stage, the data rate must be reduced to 10~GB/s, a bandwidth suitable for writing to modern permanent storage systems. This is achieved through three data streams. One stream collects data for data-driven calibrations, such as determining tracking and particle identification efficiencies \cite{lhcb_collaboration_measurement_2015,aaij_selection_2019}, which are essential for analysts. A significant portion of the bandwidth is allocated to storing full events---for example, those selected by inclusive topological triggers. However, both these events and the offline calibration data undergo post-processing to further reduce disk usage. The majority of events are available immediately after HLT2 for physics analysis (so-called turbo events), embodying the real-time nature of the data processing. This computing model is summarized in Fig.~\ref{fig:turbo}.

\subsection*{High-Level Trigger 1}

The first-level trigger is conceived as a first event rate filter in order to allow the events to be buffered to disk for real-time alignment and calibration and for further processing in the HLT2 stage. It operates under a strict requirement to process on average 30 million events per second. Before the start of Run~3, two solutions were developed. The first was targeting the x86 CPU architecture, while the second, Allen~\cite{lhcb_collaboration_allen_2004}, was running on GPU. A detailed comparison had to be performed in order for the experiment to decide which one it will be commissioning for Run~3~\cite{aaij_comparison_2021}. Both were viable options but the GPU solution ended up being commissioned and deployed for Run~3.

\begin{figure}
    \centering
    \includegraphics[width=1\linewidth]{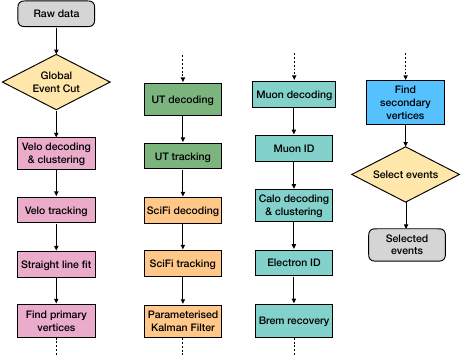}
    \caption{Schematic of the HLT1 reconstruction at LHCb. Figure from~\cite{scarabotto_tracking_2022}.}
    \label{fig:hlt1}
\end{figure}

Allen features various algorithms to perform the event reconstruction, from the raw data arriving from the LHCb subdetectors. The default sequence is presented in Fig.~\ref{fig:hlt1}. A global event cut rejects roughly 10\% of the busiest minimum bias events prior to any processing, based on UT and SciFi raw data information. The raw data is then decoded by the HLT1 framework, effectively transforming the raw information from the different tracking subdetectors to e.g., $(x, y, z)$ coordinates for each hit or energy cluster. The reconstruction of the VELO is performed using the ``Search by triplet'' algorithm~\cite{campora_perez_search_2021}, essentially an optimized straight line fit. The primary vertices are found and fit using the reconstructed VELO tracks to extrapolate back to their origin on the beamline. The VELO-UT tracking, with the ``CompassUT'' algorithm~\cite{fernandez_declara_parallel-computing_2019}, delivers the first measurements of charge and momentum, providing the earliest physics-relevant objects for selection.

Subsequently, ``Forward'' tracking~\cite{scarabotto_tracking_2022} is carried out by extrapolating the tracks from the VELO and UT detectors to the SciFi region, using a parametrization, for the sake of computational performance, of the magnetic field. Then a parametrized Kalman filter is used to improve the estimates of certain physics quantities, and the muon system is used to identify which of the extrapolated tracks included muons. The calorimeters are reconstructed in order to identify electrons and estimate their lost radiated energy with a Bremsstrahlung recovery algorithm, and secondary vertices from long-lived particle decays are found. Finally, events are selected with various trigger lines in order to accommodate the LHCb physics program.

\begin{figure}
    \centering
    \includegraphics[width=0.9\linewidth]{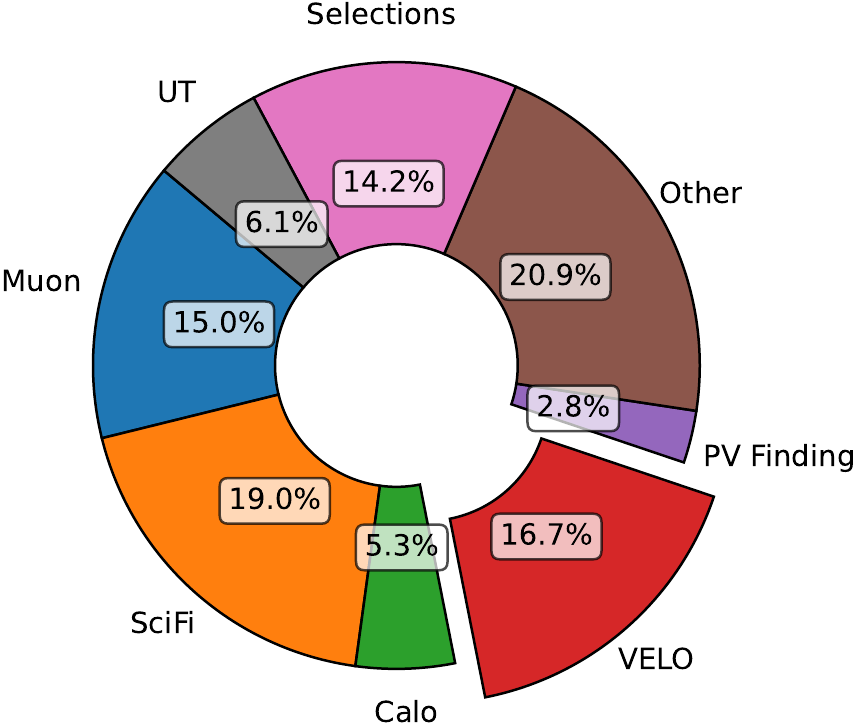}
    \caption{Breakdown of the default HLT1 reconstruction sequence published through Allen's continuous integration and performance regression system, as  measured on GPU on 2025. VELO tracking, roughly at 17\%, is highlighted. The different algorithms were separated and accumulated based on their objectives.}
    \label{fig:hlt1-pie}
\end{figure}

VELO tracking is an important part of the resource utilization related to HLT1. As shown in Fig.~\ref{fig:hlt1-pie}, as of 2025, it accounts for almost 17\% of the total throughput rate for Run~3. Tracking is discussed further in Chapter~\ref{ch:tracking}.

\section*{Conclusion}

In this chapter, I introduced the LHCb experiment at CERN, describing its detector, dataflow and online processing system. The LHCb trigger was also outlined. Finally, I motivated the need for a real-time trigger capable of operating at the full LHC collision rate of 40~MHz. 

We are finally ready to present the problem central to this thesis: charged particle track reconstruction. The next chapter introduces the general principles of tracking and examines how they are realized in the LHCb experiment. Special attention is given to the VELO subdetector, emphasizing its unique features and the particularities of performing tracking inside it.

\chapter{Track Reconstruction}
\label{ch:tracking}

\minitoc

\section*{Introduction}

This chapter explores the challenge of reconstructing the tracks of charged particles, also known as tracking, along with the specific requirements of the LHCb experiment in this context. I also describe how the reconstruction algorithms are evaluated and compared. This background establishes the framework within which our tracking algorithms were developed and assessed, and is essential for understanding the technical details of their evaluation.

\section{Track Reconstruction}

Track reconstruction, or tracking, in particle physics in general, is the process of reconstructing the trajectories of charged particles in a particle detector known as a tracker. The particles produced by the various processes involved in each particular experiment leave hits, precise records or deposits of their passage through the device, by interacting with the appropriately chosen components and materials. Often, the presence of a magnetic field curves the trajectories of these charged particles, and their momentum can be estimated using the local curvature of the particle track.

Tracking is usually split into two stages. The first stage, starting from the point cloud of all the hits left by all the particles in a single event, is where the hits, or clusters, suspected to originate from the same particle are identified and grouped together. The second is stage is where a curve is mathematically fit to these clusters in order to best approximate the particle's trajectory. This stage is known as track fitting. Based on this fit, important physics quantities can be inferred such as the charge and the momentum.

\begin{figure}
    \centering
    \includegraphics[width=1\linewidth]{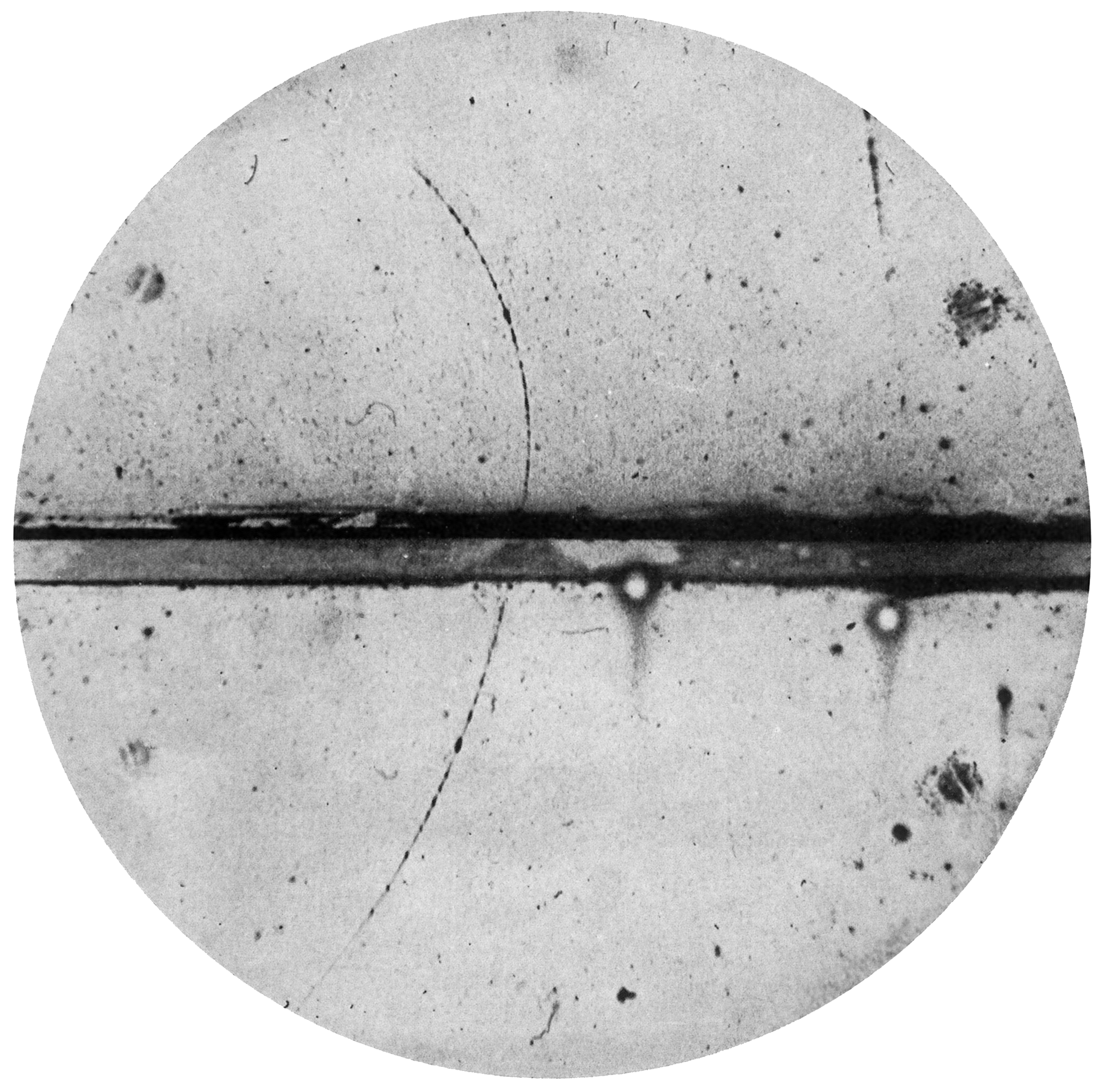}
    \caption{Cloud chamber photograph of the first positron ever observed. The thick horizontal line is a lead plate. The positron, the dark curved line, entered from the lower left, crossed the lead plate and was curved towards the upper left. The curvature is due to the applied magnetic field. The thickness of the track indicates that the particle has the mass of the electron, and the sign of the curvature that it is positively charged. Figure from~\cite{anderson_positive_1933}.}
    \label{fig:positron}
\end{figure}

Historically, there have been many devices used for tracking~\cite{martin_particle_2008,grupen_particle_2008}. The most important are summarized in Table~\ref{tab:detectors}. These include cloud chambers, for example, used by Carl Anderson in 1932~\cite{anderson_positive_1933}, as shown in Fig.~\ref{fig:positron}, to identify the first positron. Other technologies include bubble chambers, spark chambers, time projection chambers, and more recently silicon trackers. Indeed, solid state trackers have been used since the 1980s in experiments requiring compact, fast-readout and high precision, for colliders such as the LHC.

\begin{table}
\renewcommand{\arraystretch}{2}
\centering
\begin{tabular}{p{3.2cm}@{\hspace{2em}}p{7cm}@{\hspace{2em}}p{1cm}} \hline \hline
\textbf{Detector Type} & \textbf{Function} & \textbf{Use} \\ \hline
Cloud Chamber & Ionization causes condensation trails in supersaturated vapor & 1920--1950 \\
Bubble Chamber & Ionization leaves bubble tracks in superheated liquid hydrogen or similar & 1952-- \\
Spark Chamber & Sparks form along paths of ionization in a gas between charged plates & 1954-- \\
Wire Chamber & Detect charged particles and photons by tracking the trails of gaseous ionization & 1968-- \\
Time Projection Chamber & Use combination of electric and magnetic fields inside a sensitive volume of gas/liquid to perform 3D track reconstruction & 1974-- \\
Silicon Tracker & Semiconductor-based system (e.g., silicon strips or pixels) that collects charge from ionization & 1980-- \\ \hline \hline
\end{tabular}
\caption{Various tracking methods and a summary of their function.}
\label{tab:detectors}
\end{table}

\section{Track Reconstruction at LHCb}

As we saw in Chapter~\ref{ch:lhcb}, Section~\ref{sec:trigger}, tracking is a very important step of the LHCb pipeline. In LHCb, tracking is done using its tracking system, already described in Section~\ref{sec:detector}. It contains the Vertex Locator (VELO), the Upstream Tracker (UT) and the Scintillating Fiber (SciFi) tracker. Between the UT and the SciFi there is a dipole magnet which curves the trajectories of charged particles. From the curvature of these tracks, the momentum of the particles can be estimated. The magnetic field of the magnet does not reach into the VELO and hence the particles inside the VELO move in straight lines, unless they interact with the material of the detector or with other particles.

\begin{figure}
    \centering
    \includegraphics[width=0.9\linewidth]{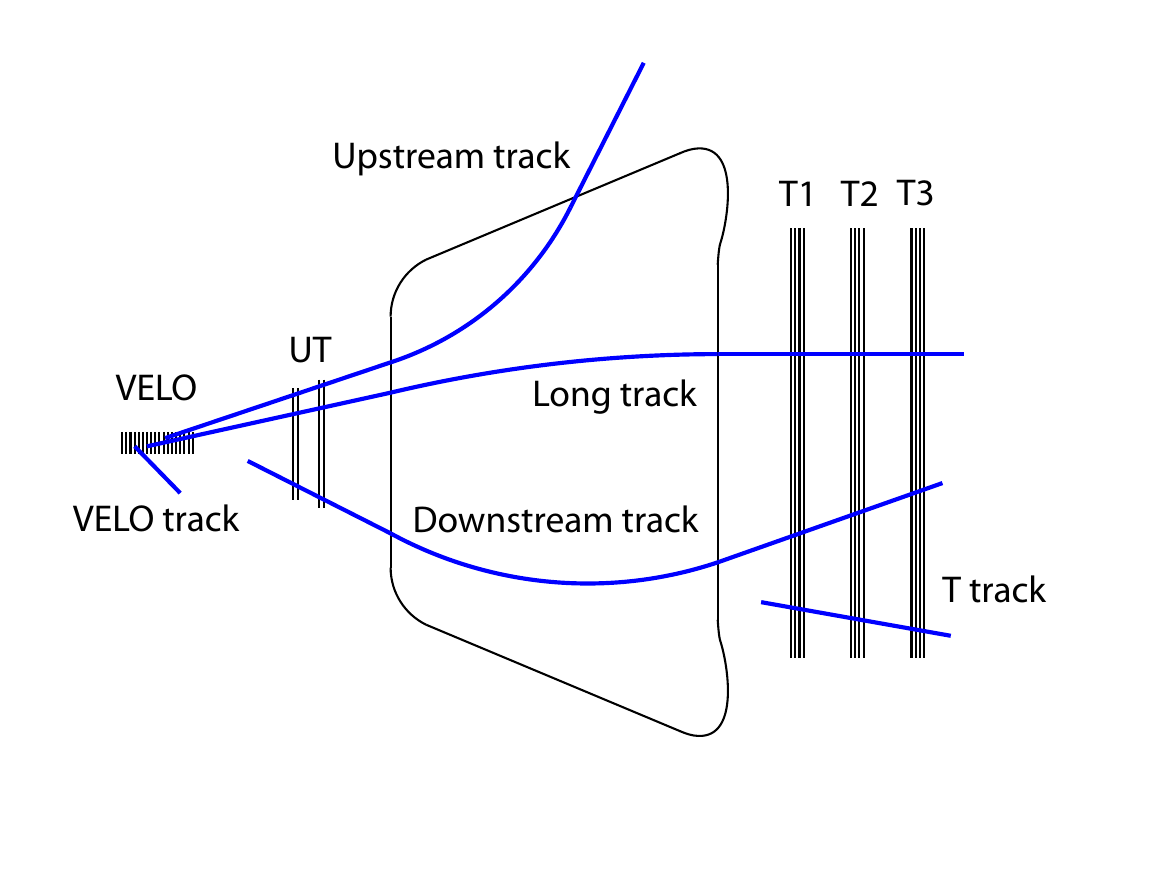}
    \caption{Depiction of the track types in the LHCb detector during Run~3. Figure from~\cite{li_tracking_2021,lhcb_collaboration_track_nodate}.}
    \label{fig:tracks}
\end{figure}

In the LHCb track event model, tracks are split according to the distribution of their hits throughout the detector into five categories, as shown in Fig.~\ref{fig:tracks}.

\begin{itemize}
    \item[] VELO Tracks: Tracks with hits only in the VELO subdetector.
    \item[] T Tracks: Tracks with hits solely on the SciFi.
    \item[] Long Tracks: Tracks with hits in both the VELO and the SciFi subdetectors. They may also contain additional hits in the UT.
    \item[] Upstream Tracks: Tracks with hits in the VELO and the UT only.
    \item[] Downstream Tracks: Tracks with hits in the UT and SciFi only.
\end{itemize}

Long tracks are of particular importance because they constitute the only category with hits before and after the magnet, and therefore their trajectories are maximally bent by the magnetic field. As a result, the estimate for the particle's momentum is the most precise, and momenta often have a large impact on the subsequent physics analyses.

When implementing a tracking algorithm, the developer needs to optimize the following three indicators, which are fully correlated with each other.

\begin{itemize}
    \item Efficiency: The ratio between the particles deemed as possible to be reconstructed and the particles actually correctly reconstructed.
    \item Ghost Rate: The fraction of incorrectly reconstructed tracks, i.e., tracks which are fake.
    \item Throughput: The number of events per second the algorithm is able to process, usually measured in Hz.
\end{itemize}

The conventions and definitions for track-finding performance in LHCb during Run~3 are outlined in~\cite{li_tracking_2021}. An important concept is \textit{reconstructibility}. In order to define the reconstructibility of a particle, different criteria are used for each tracking subdetector. For the VELO, a particle is considered as reconstructible if it has at least three hits on the VELO layers. Based on this, the efficiency and ghost rate are defined as:

\begin{equation}
    \text{efficiency} = \frac{N_{\text{reconstructed particles}}}{N_{\text{reconstructible particles}}} \,,
\end{equation}

\begin{equation}
    \text{ghost rate} = \frac{N_{\text{fake tracks}}}{N_{\text{reconstructed tracks}}} \,,
\end{equation}
where, in order for a track to be considered as properly reconstructed, 70\% of its hits have to come from the actual simulated particle. In this case, we say that the track is matched to the particle. Otherwise the track is deemed as fake.

Three additional indicators are commonly used to assess the performance of a reconstruction algorithm.

\begin{itemize}
    \item Clone Rate: This measures the proportion of reconstructed tracks that are duplicates, meaning they are associated with the same simulated particle as another track. It is defined as:
  \begin{equation}
      \text{clone rate} = \frac{N_{\text{clone tracks}}}{N_{\text{reconstructed tracks}}} \,.
  \end{equation}

  \item Hit Purity: This refers to the fraction of hits in a reconstructed track that originate from the true (simulated) particle. It is given by:
  \begin{equation}
      \text{hit purity} = \frac{N_{\text{track hits from true particle}}}{N_{\text{track hits}}} \,.
  \end{equation}
  
  \item Hit Efficiency: This indicates the fraction of the true particle's hits that are successfully included in a reconstructed track:
  \begin{equation}
      \text{hit efficiency} = \frac{N_{\text{track hits from true particle}}}{N_{\text{true particle hits}}} \,.
  \end{equation}
  
\end{itemize}

\subsection*{VELO Tracking}

The LHCb reconstruction starts by looking for tracks in the VELO, the subdetector closest to the $p$--$p$ collisions beamline. Since the magnetic field of the dipole magnet does not reach into the VELO volume, the tracks that particles trace are straight lines unless obstructed. Furthermore, VELO tracks are of crucial importance to the reconstruction of primary and secondary vertices as well as for evaluating variables for physics data analysis, such as the impact parameter. Their reconstruction performance impacts all later stages of the HLT1 sequence.

\begin{figure}
    \centering
    \includegraphics[width=0.95\linewidth]{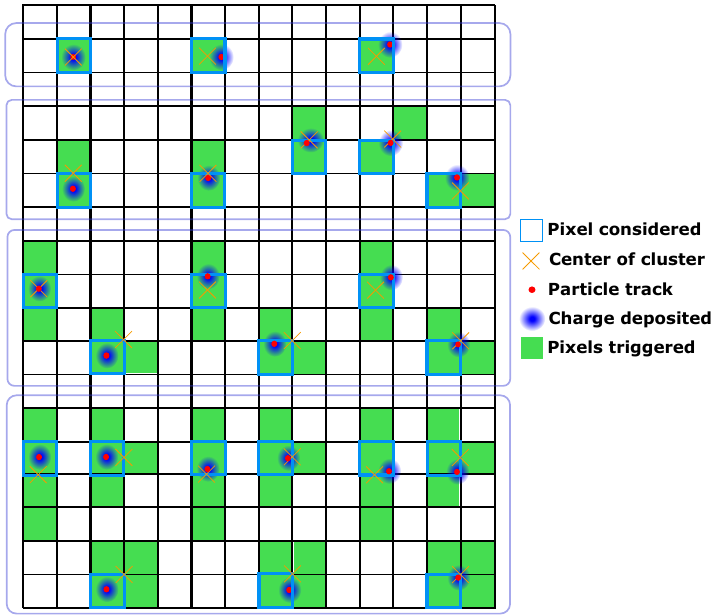}
    \caption{Examples of pixels being activated due to the passage of charged particles through the layers of a silicon detector. The ``deposited'' charge, due to ionization, is collected by the sensors and ``clusters'' of pixels are created. Examples are grouped by the number of pixels activated. Adapted, image courtesy of Paul Chabrillat, fellow PhD student at LPNHE.}
    \label{fig:clusters}
\end{figure}

The first stage is the clustering of the hits in the VELO~\cite{friday_lhcb_2025,bassi_real-time_2021}. Charged particles, passing through the VELO layers, interact with the silicon layers, ionizing the material and thus releasing charge. Depending on this interaction and the collection of the ionization charge at the electrodes, one or several pixels can be triggered, as shown in Fig.~\ref{fig:clusters}. The various examples are grouped by the number of pixels activated. The particle can pass through the center of the pixel, but it can also pass through one of its edges or even a corner, activating numerous adjacent pixels. This so-called ``charge sharing''~\cite{mathieson_charge_2002} also depends on the angle of incidence of the charged particle with respect to the pixels and the sensor thickness.

The resulting ``clusters'' of hits then need to be separated, a process known as Connected Component Labeling (CCL)~\cite{samet_efficient_1988,dillencourt_general_1992}. A connected component is defined as a set of pixels within which there is a relation of connectedness. For example, for each pair of pixels in the connected component there exists a path that connects the two pixels, completely inside the component. The clusters of pixels are finally transformed into hits, corresponding to a single set of coordinates $(x,y,z)$.

\begin{figure}
\begin{center}
  \subfloat[][]{\includegraphics[width=0.48\textwidth]{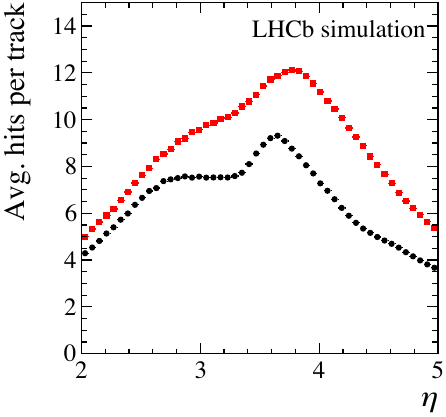}\label{fig:hits-per-track-1}}\quad
  \subfloat[][]{\includegraphics[width=0.48\textwidth]{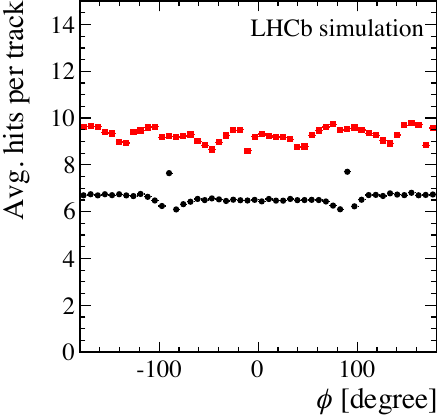}\label{fig:hits-per-track-2}}\\\vspace{4pt}
  \subfloat[][]{\includegraphics[width=0.48\textwidth]{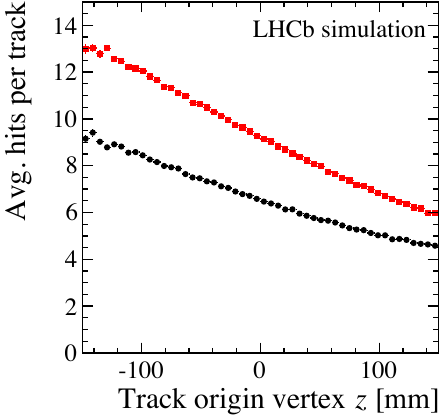}\label{fig:hits-per-track-3}}
\end{center}
\caption{Average number of VELO hits per track (at 7.6 interactions, and center-of-mass energy $\sqrt{s} =$ 14~TeV) as a function of \protect\subref{fig:hits-per-track-1} pseudorapidity $\eta$, \protect\subref{fig:hits-per-track-2} azimuthal angle $\phi$ and \protect\subref{fig:hits-per-track-3} track origin vertex
$z$-position. The current VELO is shown with black circles and the upgrade VELO with red squares. Figure from~\cite{lhcb_collaboration_velo_2013}.}
\label{fig:hits-per-track}
\end{figure}

Because of the structure of the VELO layers around the collision point, the VELO modules closer to it become more activated than the outer layers of the subdetector. Particles with high pseudorapidity $\eta > 3.5$, and hence high polar angle $\theta$, make it to the outer edge of the detector, traversing sometimes as many as 42 VELO stations. The average number of hits per track, for Run~3 conditions, as a function of various track parameters is shown in Fig.~\ref{fig:hits-per-track}. On the other hand, particles with low pseudorapidity move more perpendicularly to the beam direction and hence escape the detector only after crossing as few as 6 stations. For this reason, the hit density is higher for the layers closer to the luminous region, and progressively drops as we move away from it.

The main part of the track reconstruction is then performed. The Search by triplet~\cite{campora_perez_search_2021,campora_perez_fast_2019} algorithm achieves the required physics efficiency with the necessary computational performance. Optimized for GPUs, it uses parallelization on two levels, both event- and track-level. The algorithm utilizes a standard ``local track following'' method for reconstructing tracks. Tracks are first seeded from combinations of three hits in a region where the hit density is lowest---the outer stations of the subdetector---and the signal is easier to distinguish. Then these track seeds, or \textit{tracklets}, are extrapolated, or ``followed'', to regions with higher hit density---closer to the beamline---in order for the full reconstruction to be performed.

If we imagine a Cartesian system of coordinates, with the $z$-axis directed along the beamline, and the $x$-$y$ plane being perpendicular to the beam, as in Fig.~\ref{fig:velo} on the left, then the $p$--$p$ collisions tend to occur close to the $x$-$y$ plane origin. The particles coming out of these collisions end up traveling in straight lines, in the VELO, and with a constant phase angle $\varphi$, lying in the $x$-$y$ plane, in cylindrical coordinates. Therefore, the VELO hits are first sorted by $\varphi$ and $z$ before the start of the track reconstruction.

\begin{figure}
\begin{center}
  \subfloat[][]{\includegraphics[width=0.48\textwidth]{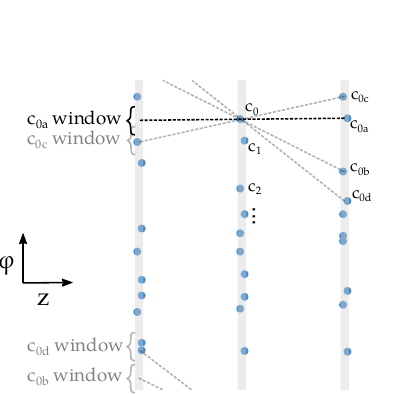}\label{fig:seeding}}\quad
  \subfloat[][]{\includegraphics[width=0.48\textwidth]{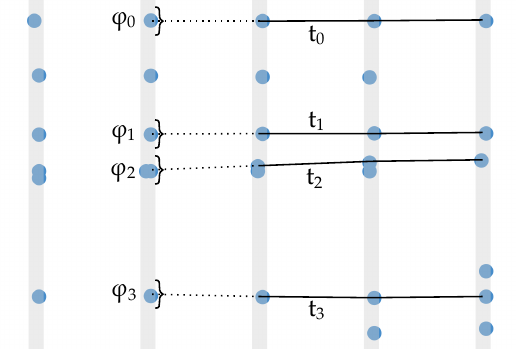}\label{fig:trace-a}}\\\vspace{4pt}
  \subfloat[][]{\includegraphics[width=0.48\textwidth]{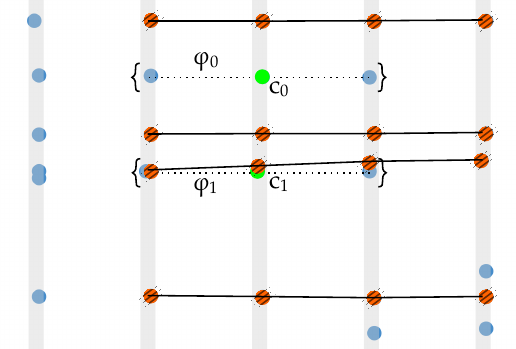}\label{fig:trace-b}}\quad
  \subfloat[][]{\includegraphics[width=0.48\textwidth]{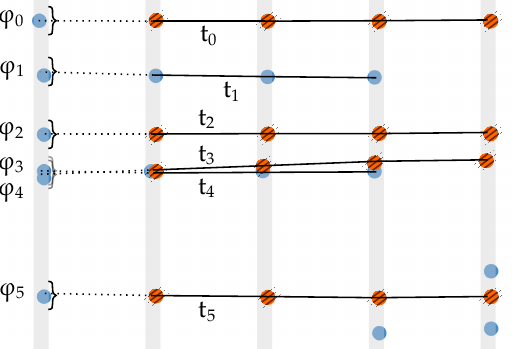}\label{fig:trace-c}}
\end{center}
\caption{Search by triplet~\cite{campora_perez_search_2021}, used for tracking in the VELO. It comprises iterative seeding and following stages, where modules are considered from right to left. \protect\subref{fig:seeding} Seeding stage: For the hit $c_0$, four candidate hits $c_{0a}$, $c_{0b}$, $c_{0c}$, and $c_{0d}$ are considered in the neighboring module to the right. Each resulting \textit{doublet} is then extrapolated to the neighboring module on the left, where hits within a specified $\varphi$ window are searched for. The $\varphi$ window allows for wrapping around. \protect\subref{fig:trace-a} Following stage: Developing tracks are extrapolated further, and candidate hits are searched for within a $\varphi$ window. \protect\subref{fig:trace-b}, \protect\subref{fig:trace-c} Subsequent seeding and following stages: Hits identified in the previous following stages are marked as flagged and are excluded from further consideration. Figure from~\cite{campora_perez_search_2021}.}
\label{fig:sbt}
\end{figure}

The steps involved in the algorithm are illustrated in Fig.~\ref{fig:sbt}, and are the following.

\begin{enumerate}
    \item \textbf{Seeding:} Triplets of hits are searched in consecutive VELO modules within search windows in angle $\varphi$. Used hits are flagged to avoid clone tracks.
    \item \textbf{Following:} The seeds found are extended with a straight line to the next modules in order to attach hits to them. Tracks are allowed to miss one VELO plane, allowing for possible inefficiencies of the sensors. If two consecutive layers are missed, the set of hits is stored as a tracklet candidate.
    \item \textbf{Iteration:} The process of seeding and following is iterated until all the hits have been flagged.
\end{enumerate}

\begin{figure}
    \centering
  \subfloat[][]{\includegraphics[width=1\textwidth]{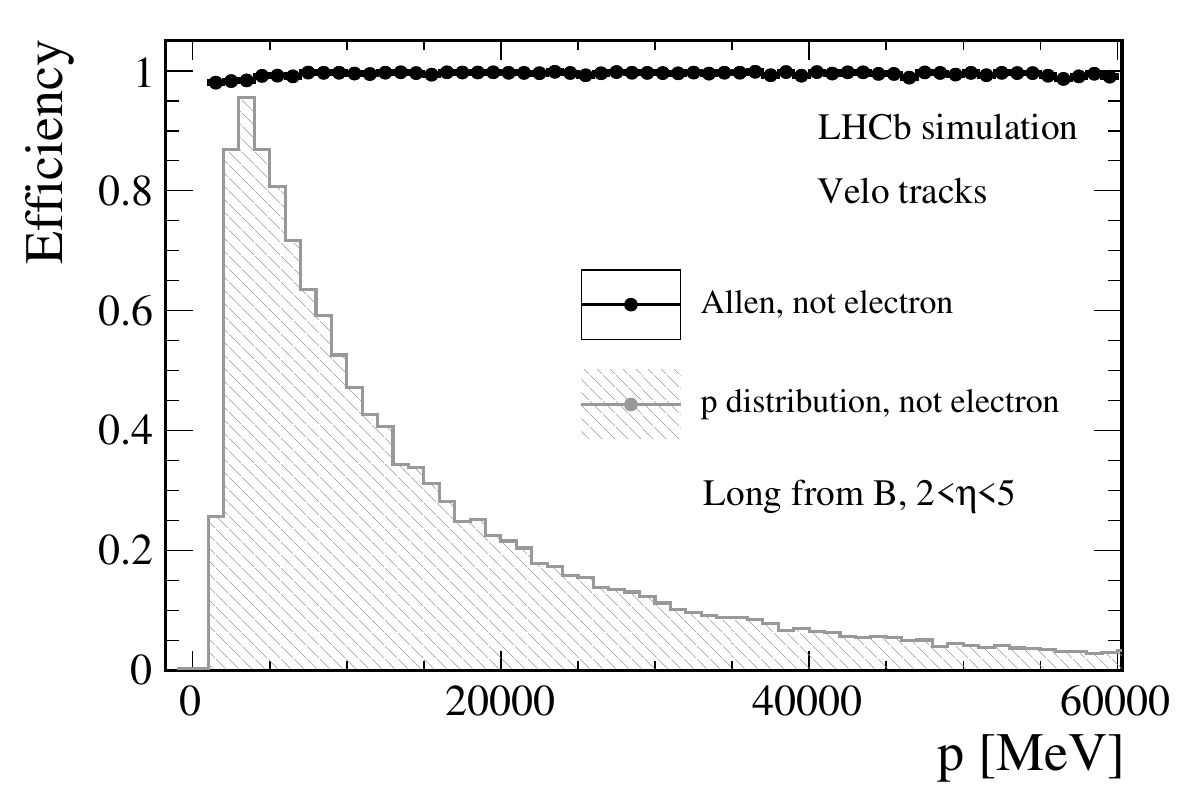}\label{fig:sbt-perf-a}}\\
  \subfloat[][]{\includegraphics[width=1\textwidth]{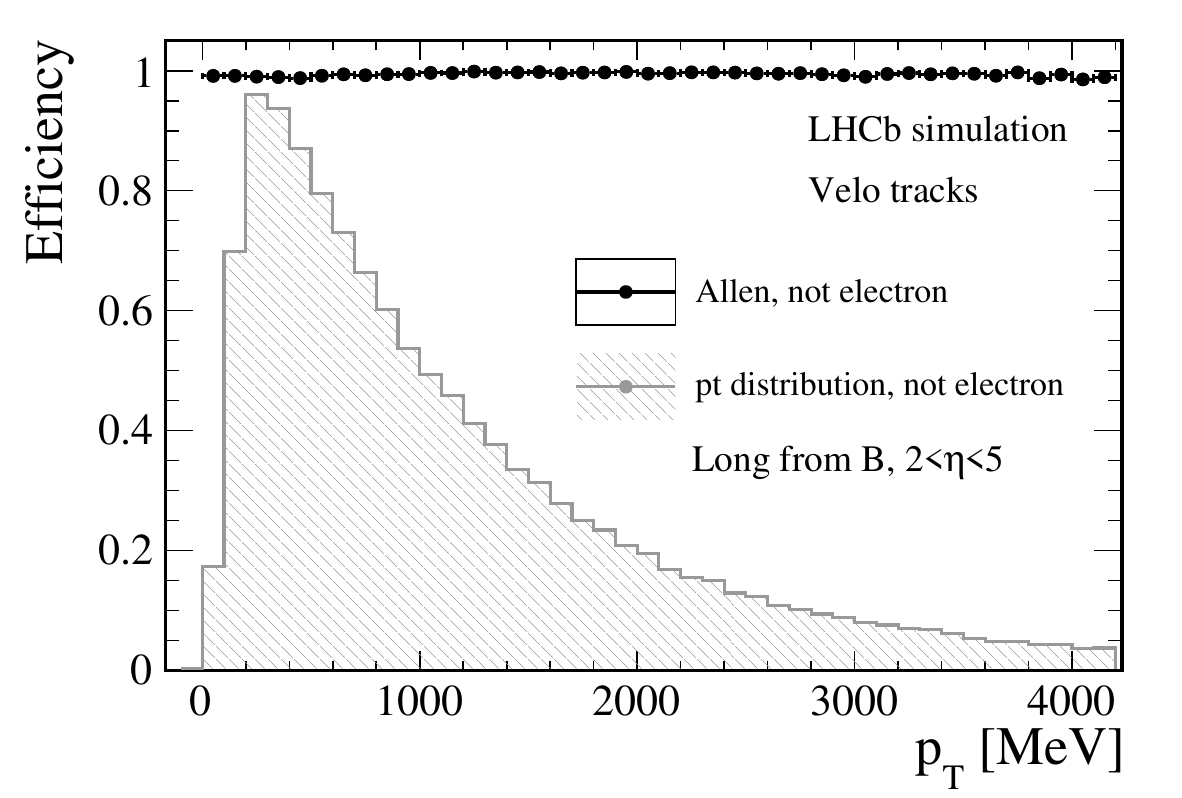}\label{fig:sbt-perf-b}}
    \caption{VELO tracking efficiency as a function of \protect\subref{fig:sbt-perf-a} momentum and \protect\subref{fig:sbt-perf-b} transverse momentum. Figure from~\cite{lhcb_collaboration_performance_2020}.}
    \label{fig:sbt-perf}
\end{figure}

A least mean squares straight line is fit on all track candidates built by the algorithm. Tracks with a $\chi^2$ above a certain threshold are accepted and considered as reconstructed tracks. Search by triplet achieves a tracking efficiency above 99\% across all ranges of momentum ($p$) and transverse momentum ($p_T$), as shown in Fig.~\ref{fig:sbt-perf}.

Electron and hadron trajectories have some key differences between them. Electrons, being significantly lighter, undergo significant multiple scattering. This causes them to zigzag slightly as they traverse the detector materials. The lower the energy, the larger these deviations are. Hadrons on the other hand, due to being much heavier, move in straighter and more predictable tracks.

In addition, electrons lose energy due to Bremsstrahlung (or braking radiation). Bremsstrahlung is emitted when the electrons are decelerated when deflected in the electric field of an atomic nucleus. The kinetic energy lost is converted into and emitted as photons. These radiative losses end up changing the curvature of the electrons and causing kinks in the tracks. Hadrons and muons, on the other hand, because of their mass, have negligible probability to emit Bremsstrahlung. Hadrons, mostly lose energy by ionization, which is smoother and more predictable. Because of this, low-energy electrons might not even make it to the full detector length, while hadrons usually do, unless they decay before reaching the outer parts of the detector. In fact, from simulations, we know that roughly 11\% of the kaons and about 14\% of the pions cannot be reconstructed due to hadronic interactions~\cite{lhcb_collaboration_measurement_2015}. For protons, the interaction losses caused by material interactions is found to be between 20\% and 30\% across the full kinematic range.

\section*{Conclusion}

In this chapter, I presented the problem of charged particle track reconstruction and explained its importance in accelerator-based experiments. The specifics of tracking inside LHCb's VELO subdetector were also outlined. This chapter concludes the background material necessary for understanding the work carried out in the course of this thesis. We now delve into the main results of this work, starting with presenting our graph neural network-based track reconstruction pipeline developed for the VELO subdetector.

\part{Main Results}
\label{part:main}

\chapter{ETX4VELO: Tracking with GNNs at LHCb}
\label{ch:etx4velo}

\minitoc

\noindent Parts of this chapter are adapted from~\cite{correia_graph_2024}. The repository of the project can be accessed at~\cite{gdl4hep_etx4velo_nodate}. I gratefully acknowledge the contributions of my co-author Anthony Correia, in general, for all the work on the ETX4VELO project that was done in common, and in particular, in the development of the triplet methodology for the pipeline, described in this chapter, the XDIGI2CSV~\cite{gdl4hep_xdigi2csv_nodate} and MonteTracko~\cite{gdl4hep_montetracko_nodate} projects, as well as in the writing of~\cite{correia_graph_2024}.

\section*{Introduction}

Tracking is a computationally crucial part of most HEP collider experiments. Performing the reconstruction of the tracks efficiently and precisely is important for various reasons. Firstly, the tracks may be used in triggering~\cite{aaij_allen_2020}. Secondly, tracks of charged particles can be used to estimate their momentum, which can be implied by their curvature. Momenta are frequently key parameters in the subsequent physics analyses, and hence the errors on their measurements have a direct impact on the final results.

Tracking, however, based on the algorithms currently in place, scales polynomially with the number of hits. For example, Search by triplet~\cite{campora_perez_search_2021}, currently used in LHCb in Run~3, scales roughly quadratically with respect to the number of hits. At the same time, with the numerous advancements in GPU hardware technology, a lot of research has been conducted considering the architecture, and where in the processing pipeline it would best fit~\cite{krasznahorkay_gpu_2020,fernandes_gpu_2023,mattmann_track_2012,ebrahim_optimising_2024,cms_collaboration_commissioning_2022,scarabotto_tracking_2022,aaij_allen_2020,rohr_usage_2021,rohr_usage_2024,rohr_gpu_2017,kiehn_trackml_2019}. However, although all major HEP experiments have effectively re-optimized their classical tracking algorithms to take advantage of modern parallel computing architectures, it is worth considering whether neural network-based tracking algorithms might offer a better long-term match for the hardware used in reconstruction.

This question has been heavily explored during the past decade~\cite{golling_trackml_2019,calafiura_trackml_2018,amrouche_tracking_2020,amrouche_tracking_2023,choma_track_2020,caillou_novel_2024,sharma_exploring_2019,stroud_transformers_2024,baranov_particle_2019,farrell_novel_2018,caron_trackformers_2025,marshall_developing_2022}, including feedforward neural networks and transformers. In particular, a specific direction towards Geometric Deep Learning (GDL)~\cite{bronstein_geometric_2017} and graphs has been highly favored. Graph Neural Networks (GNNs)~\cite{reuter_end--end_2025,biscarat_towards_2021,caillou_novel_2024,liu_hierarchical_2023,duarte_graph_2020,caillou_physics_2024,rougier_atlas_2022,dezoort_charged_2021,ju_graph_2020,lazar_accelerating_2023,shlomi_graph_2020,zhou_graph_2020} have gained attention largely due to the fact that the detector hits form a point cloud that can be naturally represented by a graph.

An application of special interest is that from the Exa.TrkX collaboration~\cite{exatrkx_exatrkx_nodate}. They developed a GNN-based pipeline for track finding~\cite{ju_performance_2021,hewes_graph_2021,ju_graph_2020}, designed with the ATLAS detector in mind. Interestingly, it exhibited a near-linear relationship between the throughput and the input event size~\cite{ju_performance_2021}, in contrast with the quadratic nature of conventional combinatorial algorithms. By adapting and modifying this pipeline for the VELO subdetector of the LHCb experiment, we developed the ETX4VELO pipeline. 

In this chapter, I present the pipeline, its development process, and its physics performance, including metrics such as tracking efficiency and fake rate. I also present the problem we had with the reconstruction of electrons, along with the solution we found based on the method of triplets.

\section{Early Version of ETX4VELO}
\label{sec:etx4velo-simplified}

This section outlines the early version of ETX4VELO, found in branch \texttt{simplified} of the ETX4VELO code on the GitLab repository~\cite{gdl4hep_etx4velo_nodate}. Early development, including the XDIGI2CSV~\cite{gdl4hep_xdigi2csv_nodate} and MonteTracko~\cite{gdl4hep_montetracko_nodate} libraries, is discussed in Appendix~\ref{app:early-dev}.

In our problem, the hits left by charged particles in the VELO detector form a point cloud in 3-dimensional space. However, they can also be represented as a graph, in which the successive hits (or nodes) of each particle traversing the detector are connected to each other. This graph can be thought of as the ``truth graph'' that perfectly describes the VELO event. At a high level, the end goal of the ETX4VELO pipeline is to produce a graph that closely approximates this truth graph.

More specifically, the basic idea of our GNN pipeline is to build an initial graph of possible connections between hits in the detector, accurately classify these connections as correct (genuine) or incorrect (fake), and then, after discarding the fake ones, transform them into a set of track objects---containers of the hits of each track---which can then be understood and used by the rest of the algorithms in LHCb's real-time pipeline. Since a graph with $N$ nodes consisting of all possible node connections would have $C(N,2) = N(N-1)/2$ edges, and thus would be prohibitively large, a key challenge is to construct this initial graph in such a way that nearly all initial connections made are part of the final graph, while as many fake connections as possible are not. This is why, for the construction of the first graph, a mapping to an embedding space is used.

The early version of our pipeline includes the following four steps. The steps are summarized here, and further details for each step are given in Section~\ref{sec:etx4velo}.

\begin{sidewaysfigure}
\begin{center}
\includegraphics[width=\linewidth,keepaspectratio]{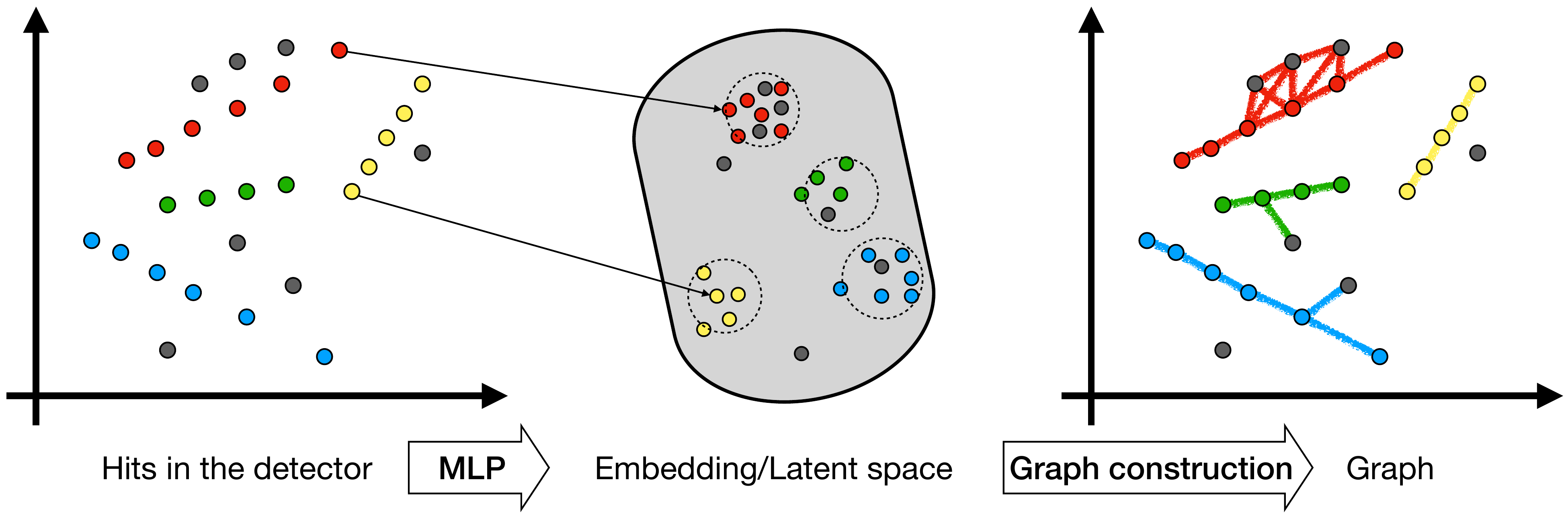}
\end{center}
\caption{Illustration of the process of moving from hits in the detector to the ``rough'' graph of the event. Colored hits correspond to the same particle, while gray hits represent noise. The hits in the detector are mapped to an embedding space with an MLP. The graph is then constructed using the mapping of the hits in the embedding space.}
\label{fig:hits-to-graph}
\end{sidewaysfigure}

\begin{figure}
    \centering
    \includegraphics[width=1\linewidth]{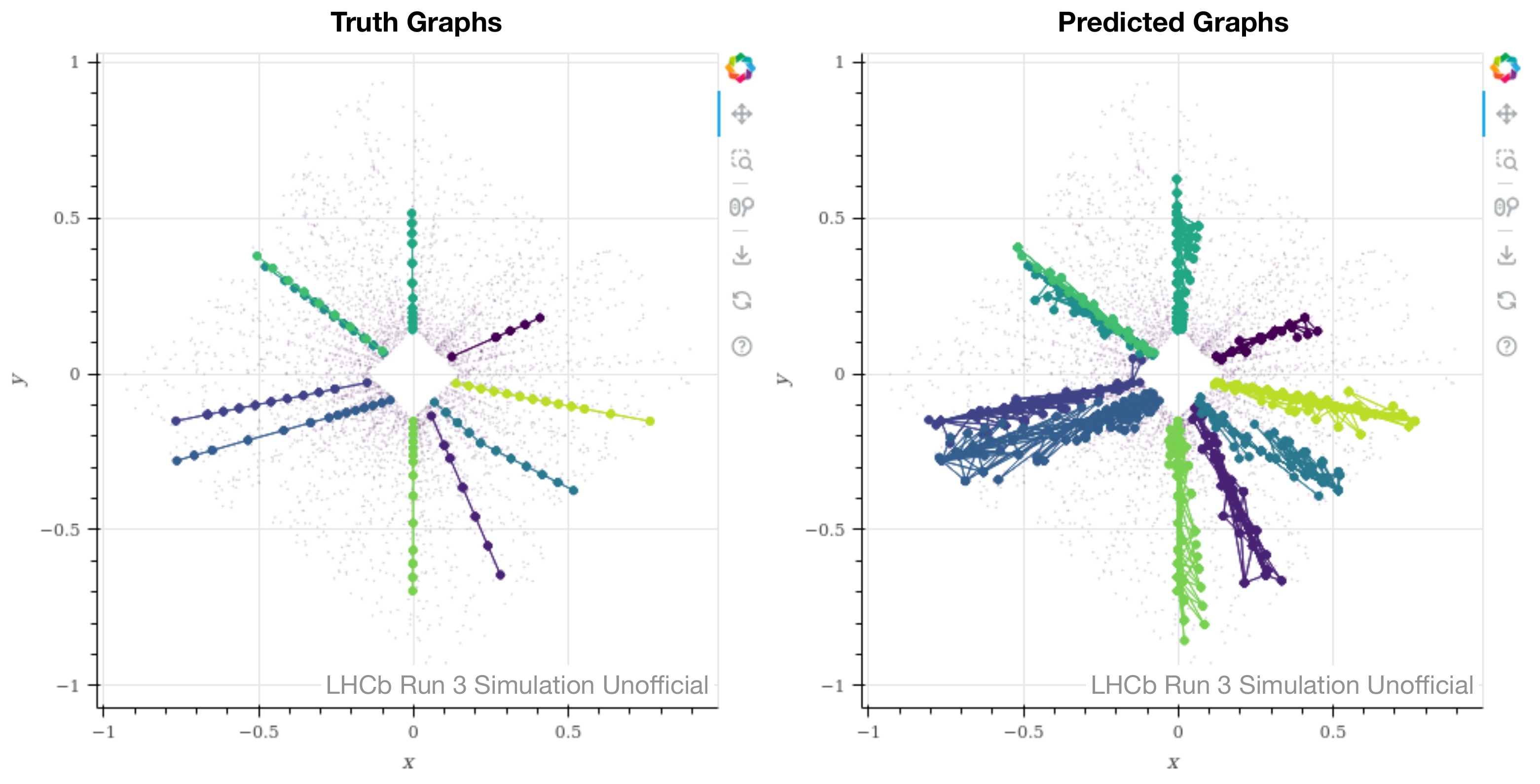}
    \caption{The process of graph construction for simulated LHCb data in the VELO subdetector from the early stages of the development of ETX4VELO. The $x$- and $y$-axis are the $x$- and $y$-directions perpendicular to the beamline and are shown in arbitrary units. A number of selected true particle tracks (left) are compared to their corresponding constructed graphs (right) using the graph construction process of the ETX4VELO pipeline. The gray dots are the activated VELO pixels over a number of treated events. This graph, generated using the Python Bokeh library, is based on the original quick start Exa.TrkX notebook.}
    \label{fig:embedding}
\end{figure}

\begin{sidewaysfigure}
\begin{center}
\includegraphics[width=\linewidth,keepaspectratio]{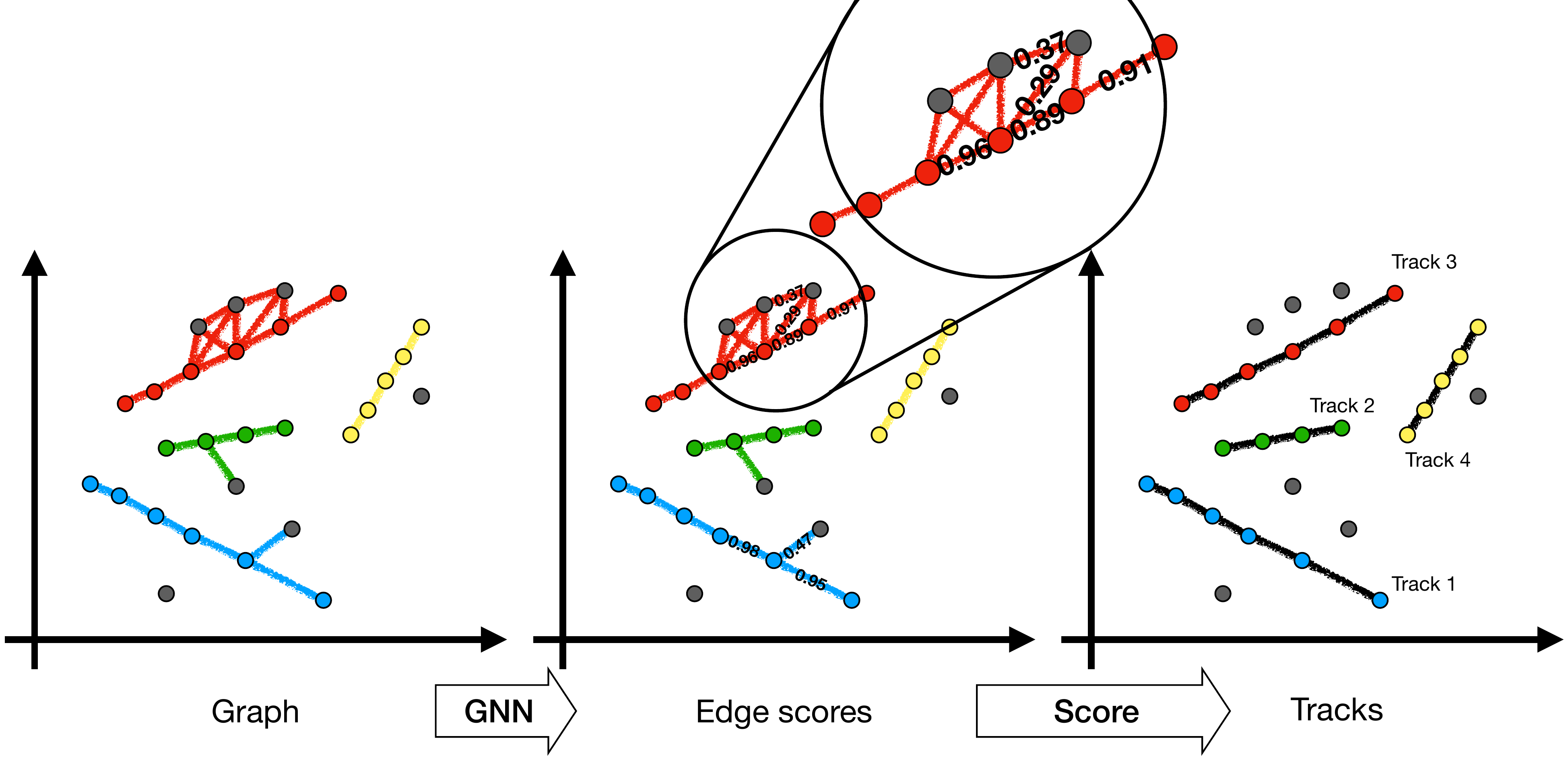}
\end{center}
\caption{Illustration of the process of moving from the event graph to the reconstructed tracks of the event. Colored hits correspond to the same particle, while gray hits represent noise. The graph of the event is then passed to the Graph Neural Network (GNN), which scores the edges between 0 (fake) and 1 (genuine). The edges having a score below a minimum edge score are removed, and finally, the tracks are reconstructed from the resulting graph, using a weakly connected components algorithm.}
\label{fig:graph-to-tracks}
\end{sidewaysfigure}

\begin{enumerate}
    \item \textbf{Embedding:} The hits in the detector are mapped to an embedding/latent space, the dimension of which is a hyperparameter, with an MLP. This neural network, using the truth information from the simulation about the hits, is trained to position hits that are likely to be connected by an edge close to each other in the embedding space. 
    
    \item \textbf{Graph construction:} The graph is constructed using the mapping of the hits in the embedding space. We create the graph in the following way. Around each target hit, we create a hypersphere of some fixed radius $r_\text{max}$, another hyperparameter of our pipeline, and after applying a k-NN algorithm, we identify the particles that are within this sphere. Finally, we connect the target hit---around which the hypersphere is created---with all the hits found to be inside this sphere. The result of this process for all the hits is our rough graph $G^{\text{hit}}_{\text{rough}}$. The event graph now contains most of the true edges, but it also contains fake edges. These edges will be hopefully removed by the next stages of the pipeline.
    
    \item[] This process of moving from hits in the detector to a graph of the event is illustrated in Fig.~\ref{fig:hits-to-graph}. An example of the process with simulated LHCb data is shown in Fig.~\ref{fig:embedding}.
    
    \item \textbf{GNN:} The graph of the event is then passed to the GNN. The GNN scores the edges between 0 (fake) and 1 (genuine). Again the training of the network is done using the truth information from the simulation.

    \item \textbf{Track Construction:} The edges having a score below a minimum edge score are removed, resulting in the purified hit graph $G^{\text{hit}}_{\text{purified}}$. Finally, the tracks are reconstructed from the resulting graph, after the score cut, by applying a Weakly Connected Components (WCC) algorithm~\cite{tarjan_depth-first_1972} to the purified hit graph $G^{\text{hit}}_{\text{purified}}$ in order to  interpret the different sets of connected hits as tracks.

    \item[] The process of moving from the event graph to the reconstructed tracks of the event is illustrated in Fig.~\ref{fig:graph-to-tracks}.
    
\end{enumerate}

At this point the architecture of the embedding MLP was four hidden layers of 256 neurons, ReLU~\cite{agarap_deep_2019} activations, and with about 200\,000 total number of parameters. On the other hand, the GNN had roughly two million parameters.

\begin{sidewaysfigure}
    \centering
    \includegraphics[width=1\linewidth]{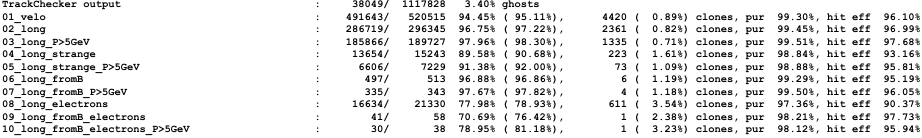}
    \caption{Evaluation of the early version of the ETX4VELO pipeline on 5000 events with the MonteTracko library. The evaluation is split across various particle categories (first column) and across various track-finding performance metrics: clone rate (\texttt{clones}), ghost rate (\texttt{ghosts}), hit purity (\texttt{pur}) and hit efficiency (\texttt{hit eff}).}
    \label{fig:etx4velo-simplified-performance}
\end{sidewaysfigure}

An example of the evaluation of the pipeline on 5000 events using MonteTracko is shown in Fig.~\ref{fig:etx4velo-simplified-performance}. MonteTracko is the custom evaluation suite developed specifically with the ETX4VELO pipeline in mind. It offers tools for matching simulated particle trajectories with reconstructed tracks and calculating performance metrics. TrackChecker, on the other hand, is the native set of algorithms used to check the tracks created by the Allen tracking system against the Monte Carlo (MC) truth information from the simulation. Similarly to the TrackChecker evaluation, the MonteTracko output is split across various particle categories---\texttt{01\_velo}, \texttt{02\_long}, etc.---and across various track-finding performance metrics: clone rate (\texttt{clones}), ghost rate (\texttt{ghosts}), hit purity (\texttt{pur}) and hit efficiency (\texttt{hit eff}). Therefore, the output is identical to Allen's VELO validation output.

As seen in Fig.~\ref{fig:etx4velo-simplified-performance}, the pipeline has an acceptable score for the majority of the categories as well as a slightly elevated ghost rate. However, the scores are not as good for the electron categories, categories 8, 9 and 10. The reasons for this discrepancy are studied in Section~\ref{sec:problem-with-electrons}.

\section{Reconstruction of Electrons}
\label{sec:problem-with-electrons}

As we saw in Fig.~\ref{fig:etx4velo-simplified-performance}, the pipeline is underperforming on the electron category. Electrons, and specifically long electrons, are important to the LHCb physics program because their tracks, extending from the VELO all the way to the SciFi and being curved by the magnet in between, offer a precise momentum measurement. However, the pipeline gives an efficiency of below 80\% for long electrons, while for the rest of the categories it gives an efficiency around 90--95\%.

The first naive attempt to solve this problem was to modify the training sample. For the rest of this section when we refer to electrons we mean both electrons and positrons, unless stated otherwise. One approach was to create events that contain only electrons, around 1500 each, by manually deleting all other particles, and training on them. The second approach was to select events that have a higher number of electrons, and then train on them. With these approaches the score for long electrons was improved to above 80\%, without any modifications to the architecture. However, the problem, as it turned out later, is more fundamental and is not really related to the balancing of the dataset.

\begin{sidewaysfigure}
    \centering
    \includegraphics[width=\linewidth]{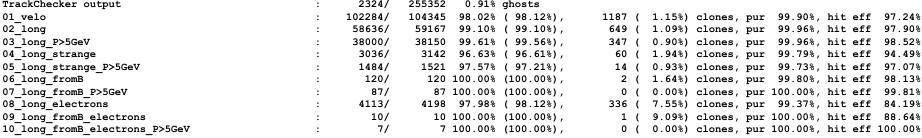}
    \caption{Evaluation of the early version of the ETX4VELO pipeline with the MonteTracko library, with a minimum track length of two. The evaluation is split across various particle categories (first column) and across various track-finding performance metrics: clone rate (\texttt{clones}), ghost rate (\texttt{ghosts}), hit purity (\texttt{pur}) and hit efficiency (\texttt{hit eff}).}
    \label{fig:2-hit-tracks}
\end{sidewaysfigure}

So far we have been building tracks with a minimum track length of three hits. However, something curious happens when the minimum track length is reduced to two. The result is shown in Fig.~\ref{fig:2-hit-tracks}. The efficiency on long electrons now jumps to 98\%, but the hit efficiency drops to about 84\%. At the same time, the clone rate for long electrons doubles. This result suggests that the hit efficiency issue for long electrons possibly comes from \textit{shared hits}: hits that belong to more than one particle.

\begin{figure}
    \centering
    \includegraphics[width=1\linewidth]{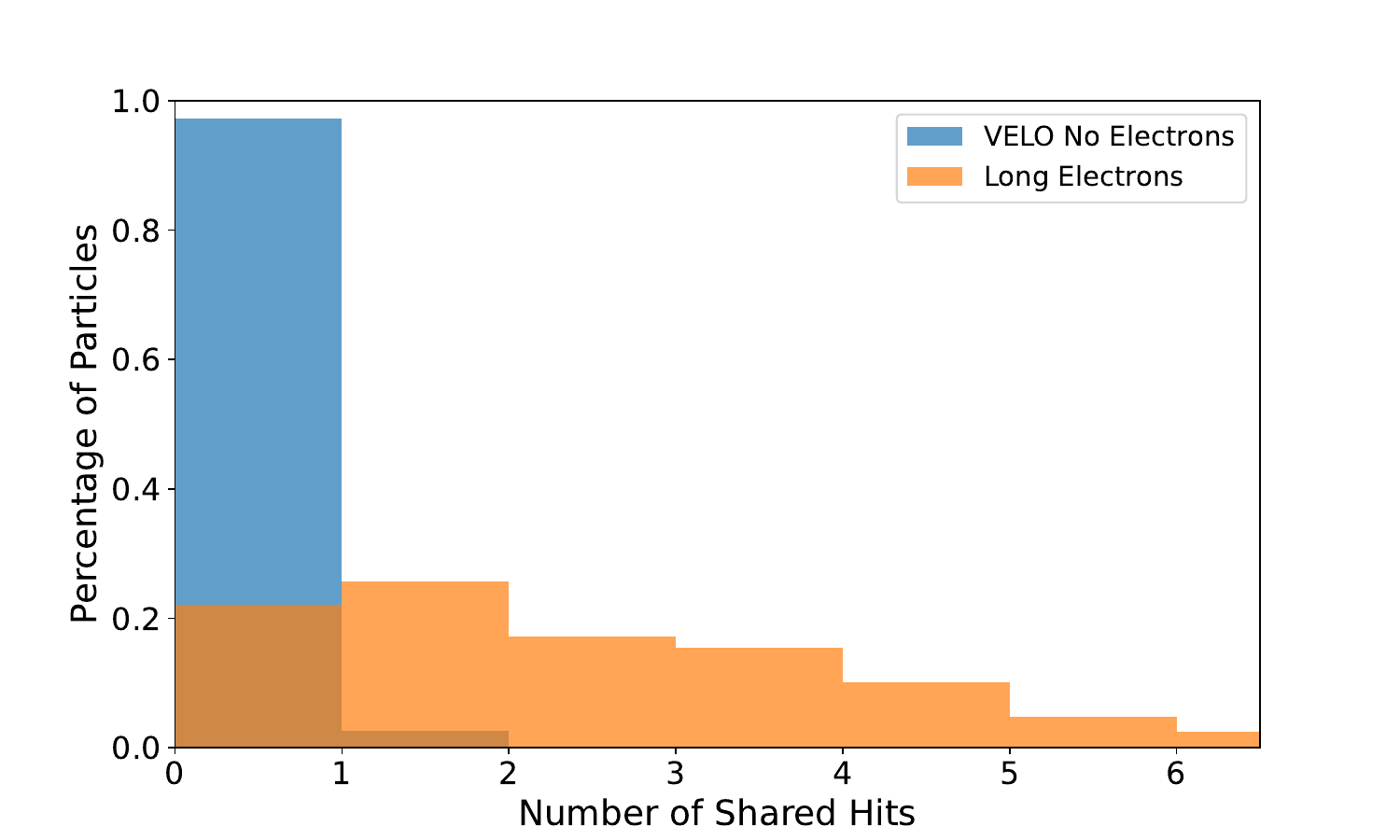}
    \caption{The percentage of particles versus the number of shared hits they have, in the simulated $p$--$p$ collision test sample. Particles reconstructible in the VELO, excluding electrons and antielectrons, are compared to long electrons---electrons reconstructible in the VELO and SciFi subdetectors.}
    \label{fig:shared-hits}
\end{figure}

\begin{figure}
    \centering
    \includegraphics[width=1\linewidth]{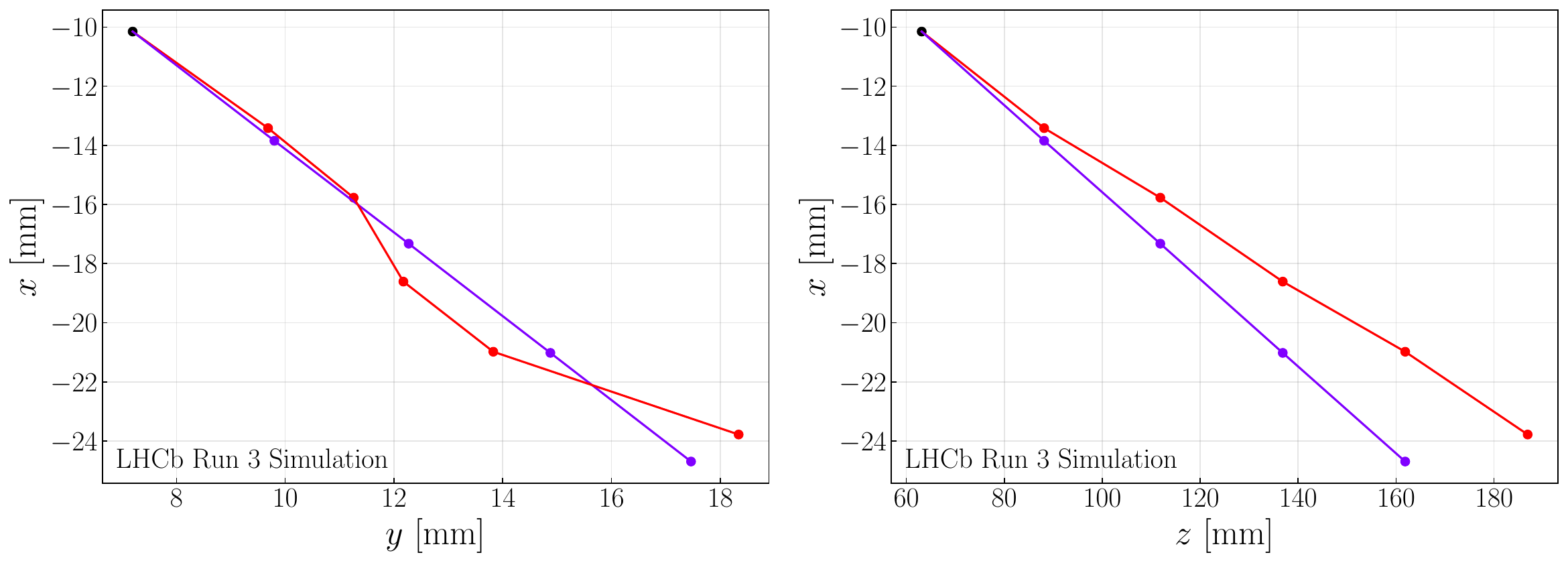}
    \caption{Example of 2 electrons (in red and purple) sharing their first hit (in black) within the simulated $p$--$p$ collision test sample, projected onto the $xy$- (left) and $xz$-planes (right). Figure from~\cite{lhcb_collaboration_performance_2023}.}
    \label{fig:minbias-sim10b-xdigi_v2_4_1498_shared_electron_017239190007272578_1082}
\end{figure}

\begin{figure}
    \centering
    \includegraphics[width=1\linewidth]{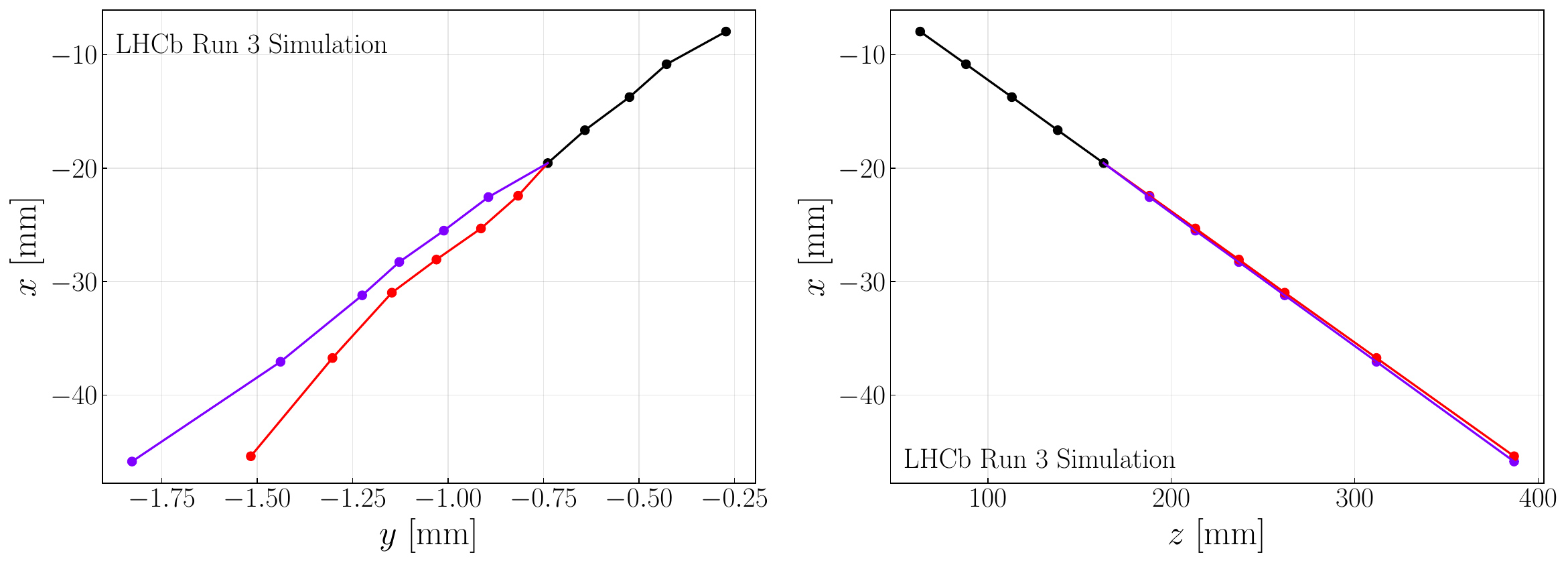}
    \caption{Example of 2 electrons (in red and purple) sharing their first five hits (in black) within the simulated $p$--$p$ collision test sample, projected onto the $xy$- (left) and $xz$-planes (right). Figure from~\cite{lhcb_collaboration_performance_2023}.}
    \label{fig:minbias-sim10b-xdigi_v2_4_1498_shared_electron_017239190007272578_1144}
\end{figure}

Indeed, long electrons are very special in the following sense: their tracks contain multiple shared hits, something illustrated by Fig~\ref{fig:shared-hits}. Interestingly, while 97\% of all VELO particles, excluding electrons and antielectrons, have no shared hits, only 22\% of long electrons (including positrons) do not have any shared hits. More specifically, due to material interactions resulting in electron--positron pairs, a large number of electrons and positrons, roughly 55\%, share hits with one another, and the two particles share at least 1 hit before splitting up. Examples of this phenomenon for long electrons are given in Figs.~\ref{fig:minbias-sim10b-xdigi_v2_4_1498_shared_electron_017239190007272578_1082} and \ref{fig:minbias-sim10b-xdigi_v2_4_1498_shared_electron_017239190007272578_1144}. 

Even though tracks that share hits in the beginning and then they diverge are the most common, various situations can lead to shared hits between tracks. A track might start from a hit that is part of another track, or end at a shared hit. Additionally, a particle may begin its trajectory from the same hit where another particle ends.

In Fig.~\ref{fig:minbias-sim10b-xdigi_v2_4_1498_shared_electron_017239190007272578_1082} for example, we have two different tracks/particles that share the first hit, the one shown in black. Since the GNN scores the edges, i.e., the hit--hit connections, it means that it is not capable to separate the two tracks from one another, because in the ideal scenario all the blue edges and all the red edges would receive a high score, close to 1. At the same time, the edges near the start of the tracks, that include the black hit, would also receive a high score. Therefore, in the end, the two tracks would come out as one merged track. These merged tracks are the reason behind the drop in hit efficiency in Fig.~\ref{fig:2-hit-tracks}.

\begin{figure}
    \centering
    \includegraphics[width=0.9\linewidth]{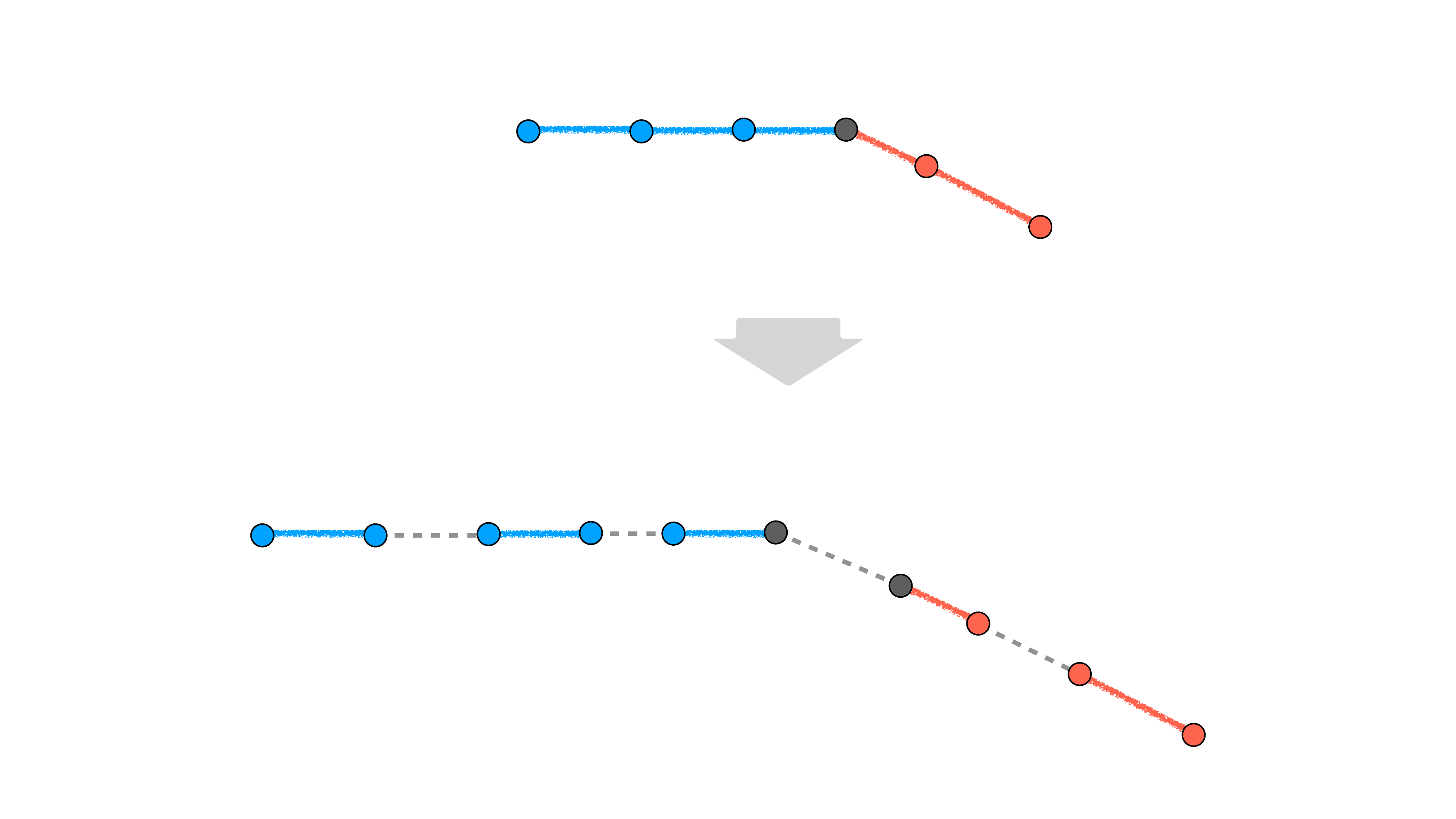}
    \caption{Illustration of the process of moving from hit--hit connections to edge--edge connections. Colored hits correspond to the same electron, while the gray hit in the middle represents the hit common to both particles. Figure from~\cite{giasemis_jrjc_2023}.}
    \label{fig:edge-edge-connections}
\end{figure}

However now, if instead of the hit--hit connections, the GNN scores the edge--edge connections, referred to as triplets because they are formed by three hits, in the ideal scenario, the GNN would score all the triplets with a high score, except for the edge--edge connection between the blue and the red edges, at the start of each track. In other words, the triplet with the shared hit in the middle, would be scored low, and hence the GNN would be able to separate the two tracks from each other. This process is illustrated in Fig.~\ref{fig:edge-edge-connections}.

This is the motivation behind the development of the triplet methodology, involving the transitioning from a graph of connected hits, $G^{\text{hit}}$, to a graph of edges, $G^{\text{edge}}$.

\section{The ETX4VELO Pipeline}
\label{sec:etx4velo}

The complete pipeline~\cite{gdl4hep_etx4velo_nodate} includes the steps outlined in Section~\ref{sec:etx4velo-simplified}, along with the triplet steps. I also describe the several optimizations applied to the pipeline. Initially, the pipeline begins with the hits, and the first two stages construct a preliminary graph, $G^{\text{hit}}_{\text{rough}}$. The hits are first embedded into a Euclidean space using an embedding MLP. The MLP is trained to group hits that are likely to be connected by an edge in close proximity within the embedding space. Subsequently, k-NN algorithms are applied in the embedding space to collect edges that are most likely to be genuine. When selecting edge candidates, nodes are considered for connection if they are no more than two planes apart.

The next steps involve the GNN, including the triplet methodology. The novelty here, as opposed to the early version of the pipeline, is the classification of edge connections in addition to individual edges. This allows for the separation of tracks with shared hits. This process involves moving from a graph of hit--hit connections, $G^{\text{hit}}$, to a graph of edge--edge connections, $G^{\text{edge}}$. These edge--edge connections, known as triplets, are formed by three hits. To reduce the number of edge--edge connections, which grows exponentially with the number of edges, it is crucial to first filter out fake edges.

First, a GNN encodes each edge in $G^{\text{hit}}_{\text{rough}}$ ($i \to j$) into a high-dimensional vector $\mathbf{e}_{i \to j}$, with dual training objectives: edge classification and triplet classification. Next, the \textit{edge classifier} network processes these encodings to generate edge scores ranging from 0 (fake) to 1 (genuine). Edges with scores below $s_{\text{edge, min}}$ are discarded, yielding the purified hit graph, $G^{\text{hit}}_{\text{purified}}$. Then, the graph of edges, $G^{\text{edge}}$, is constructed from the purified hit graph. Finally, the \textit{triplet classifier} network evaluates pairs of edge encodings (triplets), generating triplet scores. Triplets with scores below $s_{\text{edge, min}}$ are removed, resulting in the purified edge graph, $G^{\text{edge}}_{\text{purified}}$.

The final step involves constructing tracks from the purified edge graph $G^{\text{edge}}_{\text{purified}}$. The Exa.TrkX pipeline applies a WCC algorithm~\cite{tarjan_depth-first_1972} to the purified hit graph $G^{\text{hit}}_{\text{purified}}$ to break down the sets of connected hits into tracks. The classification of edge connections rather than only edges allows the WCC algorithm to be modified in such a way as to allow tracks to share multiple hits, which is particularly important for the efficiency of reconstructing electron--positron pairs.

\subsection{Datasets}
\label{sec:datasets}

To provide the necessary information for training the machine learning models, including the truth labels needed to compute the loss in a classification task, the pipeline is trained using simulated events. Likewise, evaluating the algorithm's accuracy also requires a simulation dataset to access detailed information about the particles produced in each event.

The results presented in this thesis are based on events generated using the full LHCb detector simulation. Proton--proton collisions are generated with Pythia~\cite{sjostrand_brief_2008} and particle decays are handled by EvtGen~\cite{lange_evtgen_2001}. The interactions of the generated particles with the detector, as well as the detector's response, are modeled using the Geant4 toolkit~\cite{allison_geant4_2006}, as detailed in~\cite{clemencic_lhcb_2011}. We utilize simulated minimum-bias samples that replicate typical LHCb data-taking conditions from 2022 to 2025, with an average of 5.3 inelastic proton--proton collisions per event. Each event features at least one particle with momentum above 2~GeV within the LHCb detector acceptance. In our sample, there are, on average, 150 particles in the VELO acceptance, and 2200 hits, in each event. Approximately 15\% of the hits in the simulation are spillover noise from previous events. The embedding network and GNN, described in Section~\ref{sec:etx4velo}, are trained on a dataset of 700\,000 events that meet the selection criteria outlined below.
\begin{enumerate}
    \item \textbf{Filtering non-linear particle tracks:} Tracks that exhibit significant non-linearity, often caused by multiple scattering (predominantly from low-energy electrons), are removed. This is determined by fitting a straight line to the particle's hits and applying a threshold on the average squared distance between the hits and the line. Even though we remove this special group of tracks from the training, the network is still able to later reconstruct these tracks. Furthermore, while this criterion enhances the training's physics performance, it excludes 2.5\% of tracks that are otherwise reconstructible in the VELO.
    \item \textbf{Minimum VELO hit count:} At least 500 genuine VELO hits are required to satisfy this criterion.
    \item \textbf{Exclusion of tracks with too few hits:} Tracks with fewer than three hits are omitted from consideration.
\end{enumerate}
These selection criteria are not applied to the test samples. For benchmarking both the physics and computational performance of ETX4VELO, Allen's existing tracking algorithms, used by LHCb during 2024 data-taking, serve as the reference. Detailed comparisons can be found in Sections~\ref{sec:etx4velo-performance} and \ref{sec:etx4velo-gpu-performance}.

\subsection{Hit Embedding and Rough Graph Construction}
\label{sec:graph-building}

To construct the graph \( G^{\text{hit}}_{\text{rough}} \), one could connect each hit to all hits on the next two planes, accounting for the possibility of a missing plane due to pixel inefficiencies. However, this results in an excessive number of edges, which increases the GNN's inference time and memory usage. To improve throughput, it is crucial to minimize the graph size at this stage. I begin by explaining the operation of our method, followed by the description of the training process and the loss functions in Eqs.~\eqref{eq:loss1}, \eqref{eq:loss2} and \eqref{eq:loss3}.

Most VELO tracks are produced directly from the initial proton--proton interactions, which occur in a relatively narrow interaction region with a spread of around 45~mm in $z$ and around 30~$\upmu$m in $x$ and $y$. This fact strongly constrains which edges have to be considered when constructing our graph. The embedding MLP captures this by accepting the hits as input and embedding them into an $n$-dimensional space. The hit embedding is denoted by $\mathbf{e}_h \in \mathbb{R}^{n}$. This embedding is done by passing the normalized cylindrical $\left(r, \varphi, z\right)$ coordinates to the MLP. In the embedding space, likely connected hits are positioned close together based on a reference squared distance $m = 1$, while unlikely connections are spaced apart. Here, the squared distance $d^2\left(a,b\right)$ between two hits $a$ and $b$ is defined as the usual Euclidean distance
\begin{equation}
    d^2\left(a,b\right) = |\mathbf{e}_a - \mathbf{e}_b|^2 = \sum_{i=1}^{n} \left(e_{a,i} - e_{b,i} \right)^2\,.
\end{equation}
Using this trained embedding MLP, a hit on plane $p$ is connected to hits on the next two planes, $p + 1$ and $p + 2$, if they are within a squared distance of $d^2_\text{max}$. To avoid an excessive number of edges, a maximum of $k_{\text{max}}$ edges per node is imposed. Consequently, the rough graph $G^{\text{hit}}_{\text{rough}}$ is constructed by applying a $k_{\text{max}}$-NN algorithm on plane $p$ between 0 and $n_\text{planes} - 2$ to the next two planes, $p + 1$ and $p + 2$, under a maximum squared distance of $d^2_\text{max}$. The k-NN implementation from Faiss~\cite{johnson_billion-scale_2021} is used for this purpose. The values of the hyperparameters $k_\text{max}$ and $d^2_\text{max}$ determine the size of the rough graph. There is a trade-off between the size of the graph, and the computational performance, since the number of edges is a critical parameter in the throughput of the pipeline. However, the larger the size of the graph, the higher the physics performance of the pipeline, as captured by metrics such as the efficiency and clone rate, described in Section~\ref{sec:etx4velo-performance}. Based on these considerations, the values of the hyperparameters are determined post-training to be $k_\text{max} = 50$ and $d^2_\text{max} = 0.9$.

In the training process, each step corresponds to one event, with noise hits removed as they are considered random and unrelated. The training set $T$ is composed of hit pairs from a query node $q$ to another node $a$ on the next two planes, representing $q \to a$ edge candidates. To focus on significant particles, a hit must belong to a reconstructible particle within acceptance and not be an electron to qualify as a query node. Almost all of the edges from electrons are identified by the network without training specifically on them and thus electrons are excluded. The training set $T = T_{\text{genuine}} \cup T_{\text{fake}}$ includes both connected pairs $T_{\text{genuine}}$ and disconnected pairs $T_{\text{fake}}$. It is constructed by merging three sets of pairs:
\begin{itemize}
    \item \textbf{Hard-negative mining:} Fake pairs are generated using the same $k_{\text{max}}^{\text{training}}$-NN procedure with $(d_{\text{max}}^{\text{training}})^2$ as during inference, representing fake pairs that would be classified as genuine during inference. The values $k_{\text{max}}^{\text{training}} = 50$ and $(d_{\text{max}}^{\text{training}})^2 = 1.5$ are used.
    \item \textbf{Random pairs:} For each query point, 1 pair is included.
    \item \textbf{Genuine edges:} All genuine edges from the query points are added to the training set.
\end{itemize}
To train the embedding MLP to reduce the distance of genuine pairs and increase that of fake pairs, the following loss function is minimized:
\begin{equation} \label{eq:loss1}
    \mathcal{L} = \mathcal{L}_{\text{fake}} + w_{\text{genuine}} \times \mathcal{L}_{\text{genuine}}\,,
\end{equation}
where $\mathcal{L}_{\text{genuine}}$ and $\mathcal{L}_{\text{fake}}$ are the normalized pairwise hinge embedding losses~\cite{cortes_support-vector_1995} for genuine and fake examples, respectively, defined as:
\begin{equation} \label{eq:loss2}
    \mathcal{L_{\text{genuine}}} = \frac{1}{\left|T_{\text{genuine}}\right|} \sum_{(q, a) \in T_{\text{genuine}}} d^2\left(q,a\right)\,,
\end{equation}
\begin{equation} \label{eq:loss3}
    \mathcal{L_{\text{fake}}} = \frac{1}{\left|T_{\text{fake}}\right|}  \sum_{(q, b) \in T_{\text{fake}}} \max{\left(0, m - d^2\left(q,b\right)\right)}\,.
\end{equation}
The margin $m$, representing a squared distance threshold, is fixed at $m = 1$. The parameter $w_{\text{genuine}} > 1$ reduces the likelihood of excluding true edges within this margin, thus favoring the inclusion of genuine edges over the exclusion of false ones.

\subsection{Graph Neural Network and Classifiers}
\label{sec:gnn}

\begin{figure}
    \centering
    \includegraphics[width=1\linewidth]{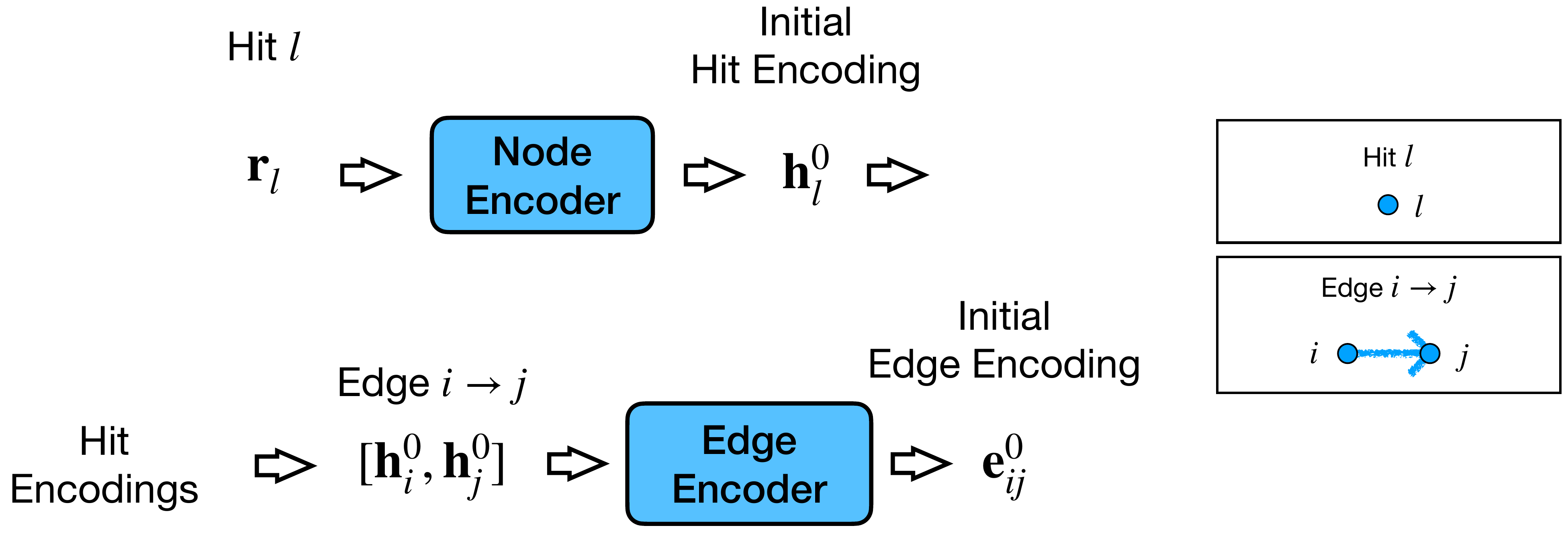}
    \caption{Schematic of the encoding step. The hit coordinates are processed through the node encoder network to produce hit encodings. These encodings are then used to generate the edge encodings. The notation $\left[\cdot, \cdot\right]$ represents concatenation of vectors.}
    \label{fig:gnn-1}
\end{figure}

\begin{figure}
    \centering
    \includegraphics[width=1\linewidth]{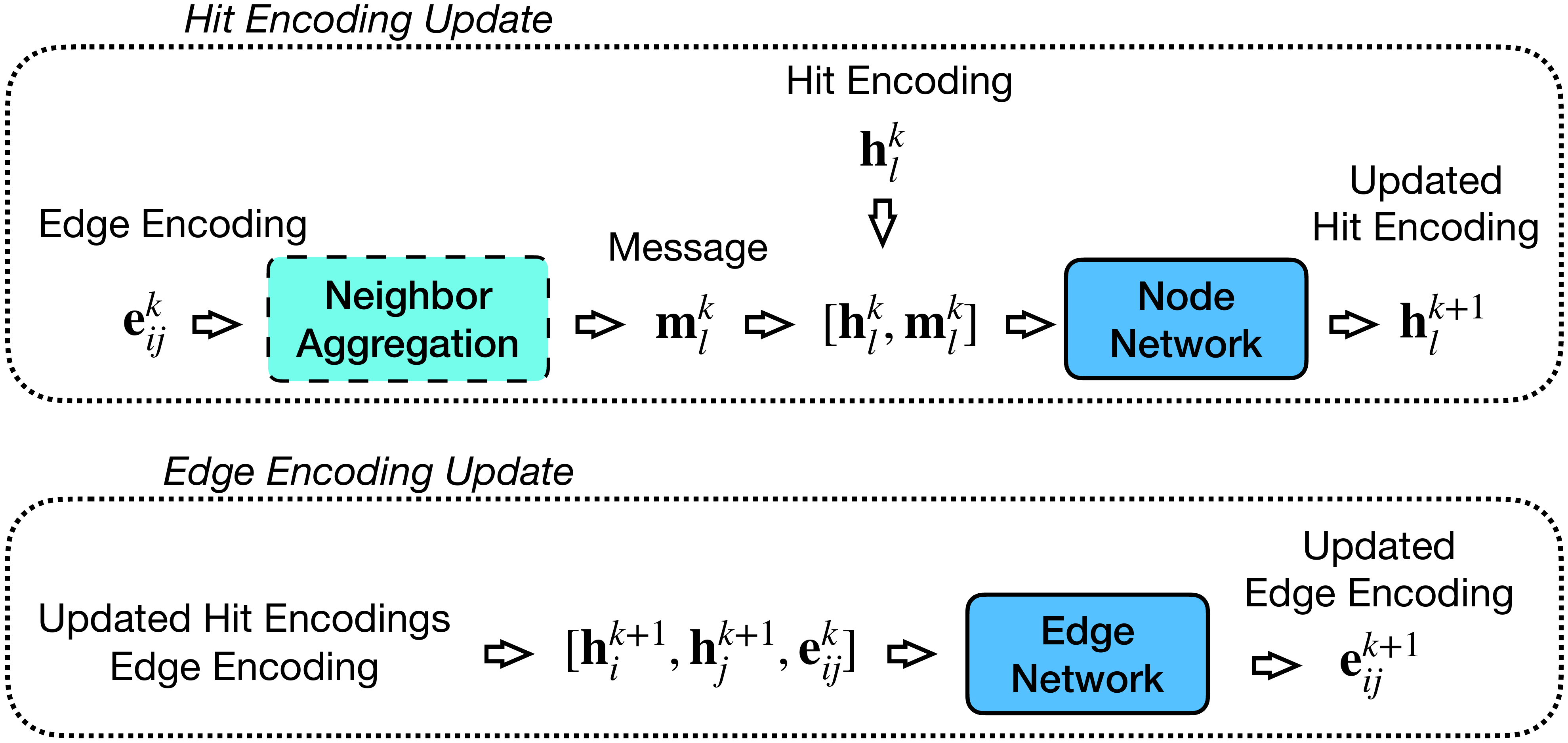}
    \caption{Schematic of the message passing step. The hit encodings are updated by incorporating the information from the graph structure through the message passing process and using the node network. The edge encodings are updated using the updated hit encodings and the edge network. The notations $\left[\cdot, \cdot\right]$ and $\left[\cdot, \cdot, \cdot \right]$ represent concatenation of vectors.}
    \label{fig:gnn-2}
\end{figure}

\begin{figure}
    \centering
    \includegraphics[width=1\linewidth]{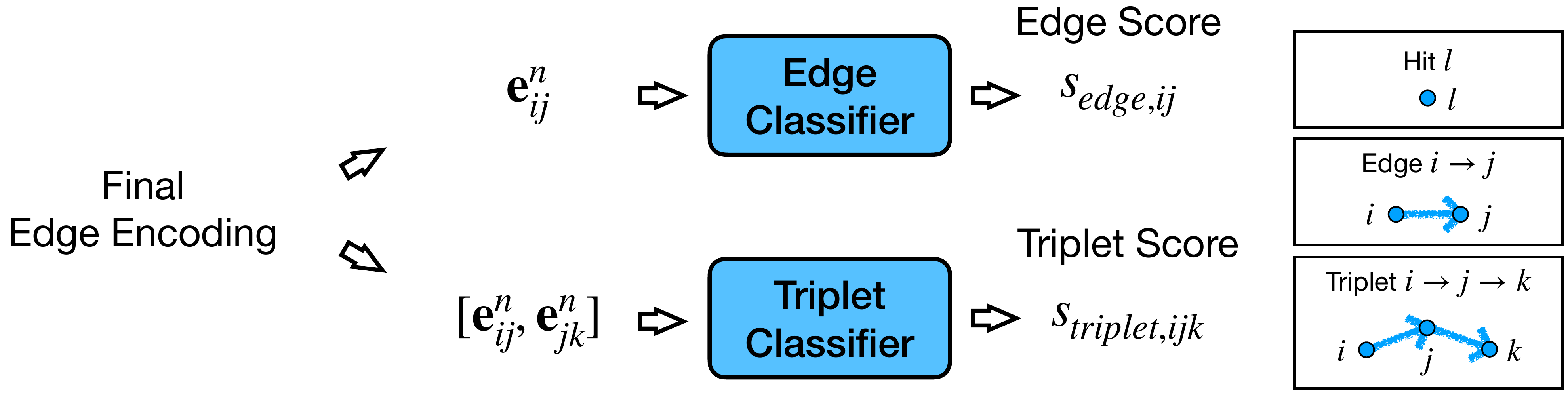}
    \caption{Schematic of the classification step. The final edge encodings are used to produce the edge scores and triplet scores, using the edge and triplet classifiers, respectively. The notation $\left[\cdot, \cdot\right]$ represents concatenation of vectors.}
    \label{fig:gnn-3}
\end{figure}

The GNN is employed to derive edge encodings used for both edge and triplet classification. The architecture of the GNN closely follows that of the Exa.TrkX collaboration, with minor deviations. The node and edge encoders, the node and edge networks and the edge and triplet classifiers are all MLPs, and should not be confused with the embedding network described in Section~\ref{sec:graph-building}, which also happens to be an MLP.

Initially, the hits (or nodes) $l$ are encoded from their normalized Cartesian coordinates $\mathbf{r}_l = (x_l, y_l, z_l)$ into an $n_h$-dimensional space $\mathbf{h}^0_l \in \mathbb{R}^{n_h}$ using the node encoder. The concatenated node features $\mathbf{h}^0_i$ and $\mathbf{h}^0_j$ of edges $i \to j$ are input to the edge encoder, producing the edge encodings $\mathbf{e}^{0}_{ij} \in \mathbb{R}^{n_e}$ in an $n_e$-dimensional space. This is illustrated in Fig.~\ref{fig:gnn-1}.

The hit and edge encodings are then iteratively updated over $n$ message passing steps. These steps allow the encodings to incorporate information from distant neighbors. During each message passing step $k \in \{0, \ldots, n - 1\}$, a message $\mathbf{m}^{k}_l$ is computed for each hit $l$ by aggregating the encodings of the edges connected to and from hit $l$. The message is computed as follows:
\begin{equation}
    \mathbf{m}^{k}_l = \left[\sum_{j\text{ s.t. }l \to j\text{ exists}} \mathbf{e}^{k}_{lj}, \sum_{i\text{ s.t. }i \to l\text{ exists}} \mathbf{e}^{k}_{il}\right]\,
\end{equation}
where $\left[\cdot,\cdot\right]$ denotes concatenation. This operation of aggregating edge encodings by summing them, using terminology from deep learning frameworks, will be referred to as \texttt{scatter\_add}. The node network updates the hit encodings $\mathbf{h}^{k+1}_{l}$ using the previous hit encodings $\mathbf{h}^{k}_{l}$ and the message $\mathbf{m}^{k}_{l}$, incorporating a residual connection. Similarly, the edge encodings are updated to $\mathbf{e}^{k+1}_{ij}$ using the previous edge encodings $\mathbf{e}^{k}_{ij}$ and the updated hit encodings $\mathbf{h}^{k+1}_{i}$ and $\mathbf{h}^{k+1}_{j}$, also with a residual connection. The message passing step is illustrated in Fig.~\ref{fig:gnn-2}.

Edge encodings are sufficient for both the edge classifier and triplet classifier. Therefore, the hit encodings are only utilized during the encoding and message passing steps to compute and update the edge encodings. The GNN is trained to classify both edges and triplets by minimizing the sum of the edge and triplet losses:
\begin{equation} \label{eq:gnn-loss}
    \mathcal{L} = \mathcal{L}_{\text{edges}} + \mathcal{L}_{\text{triplets}}\,.
\end{equation}
We use sigmoid focal losses~\cite{lin_focal_2020}, instead of Binary Cross-Entropy (BCE)~\cite{rumelhart_learning_1986}, for both the edge and triplet classification. Sigmoid focal loss is designed for class imbalanced datasets. It focuses more on hard examples, reduces the loss contribution from easy negatives and hence improves the learning capacity of the model. For the purposes of our pipeline, it outperforms the traditional BCE loss. 

To further ensure the GNN focuses on relevant triplets, edges with scores below 0.5 are discarded before triplet building and classification. The process of edge and triplet classification is illustrated in Fig.~\ref{fig:gnn-3}.

Several optimizations have been applied to the GNN to improve performance and efficiency. The size of the GNN was significantly reduced, with node and edge encodings now residing in a 32-dimensional space ($n_h = n_e = 32$), down from the initial 256 dimensions. The number of graph iterations was reduced from more than six to only five. Despite the reduction in network size, several changes and corrections were made to maintain reasonable physics performance. Notably, the node and edge networks used at each message passing step are distinct, making the GNN non-recurrent~\cite{correia_graph_2023}. This approach increases the number of trainable parameters but keeps the throughput unchanged while greatly improving the physics performance. 

\subsection{Triplet Building}
\label{sec:triplet-building}

\begin{figure}
  \begin{center}
  \subfloat[]{\includegraphics[height=0.25\textheight]{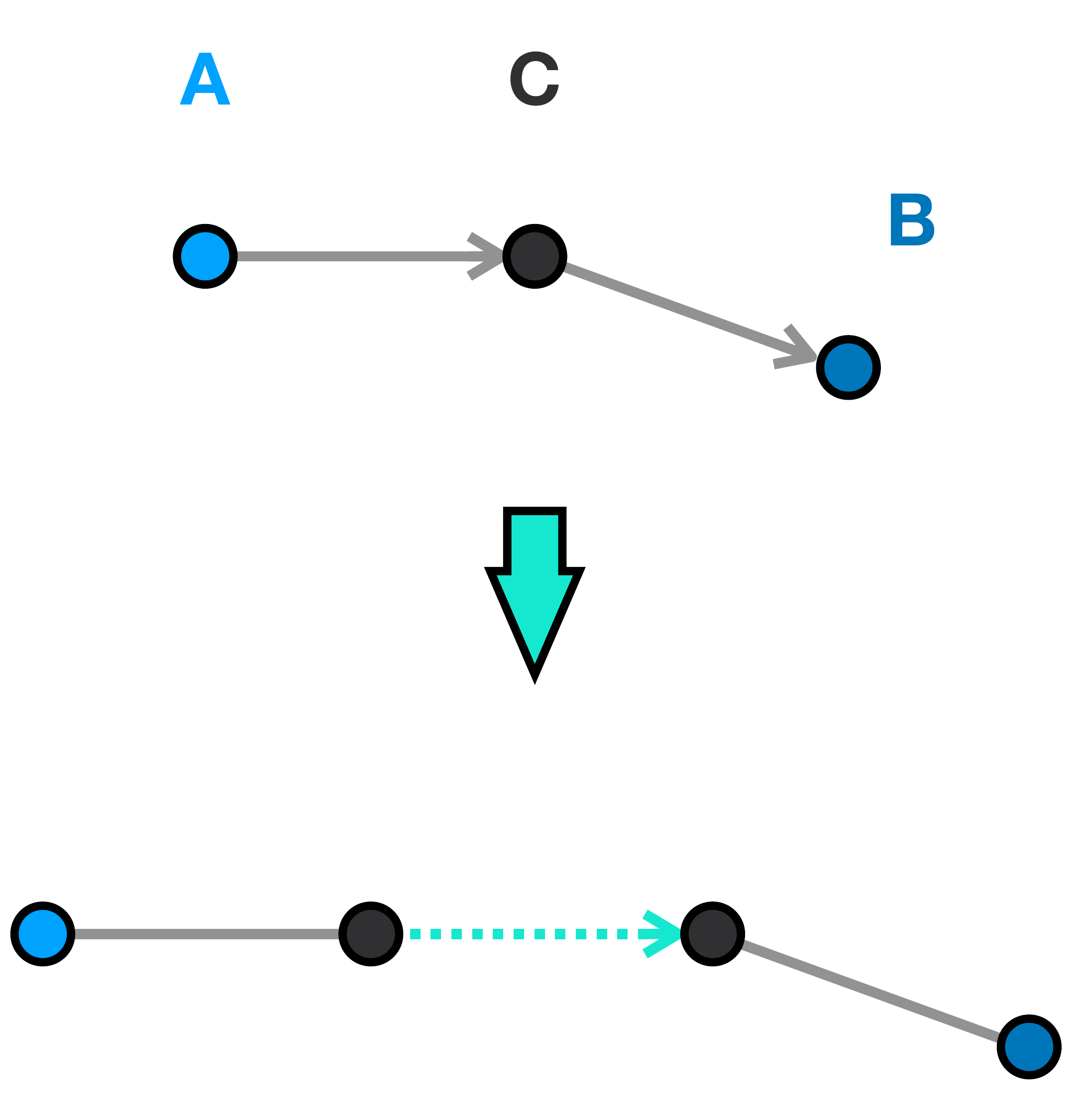}}
  \quad
  \subfloat[]{\includegraphics[height=0.25\textheight]{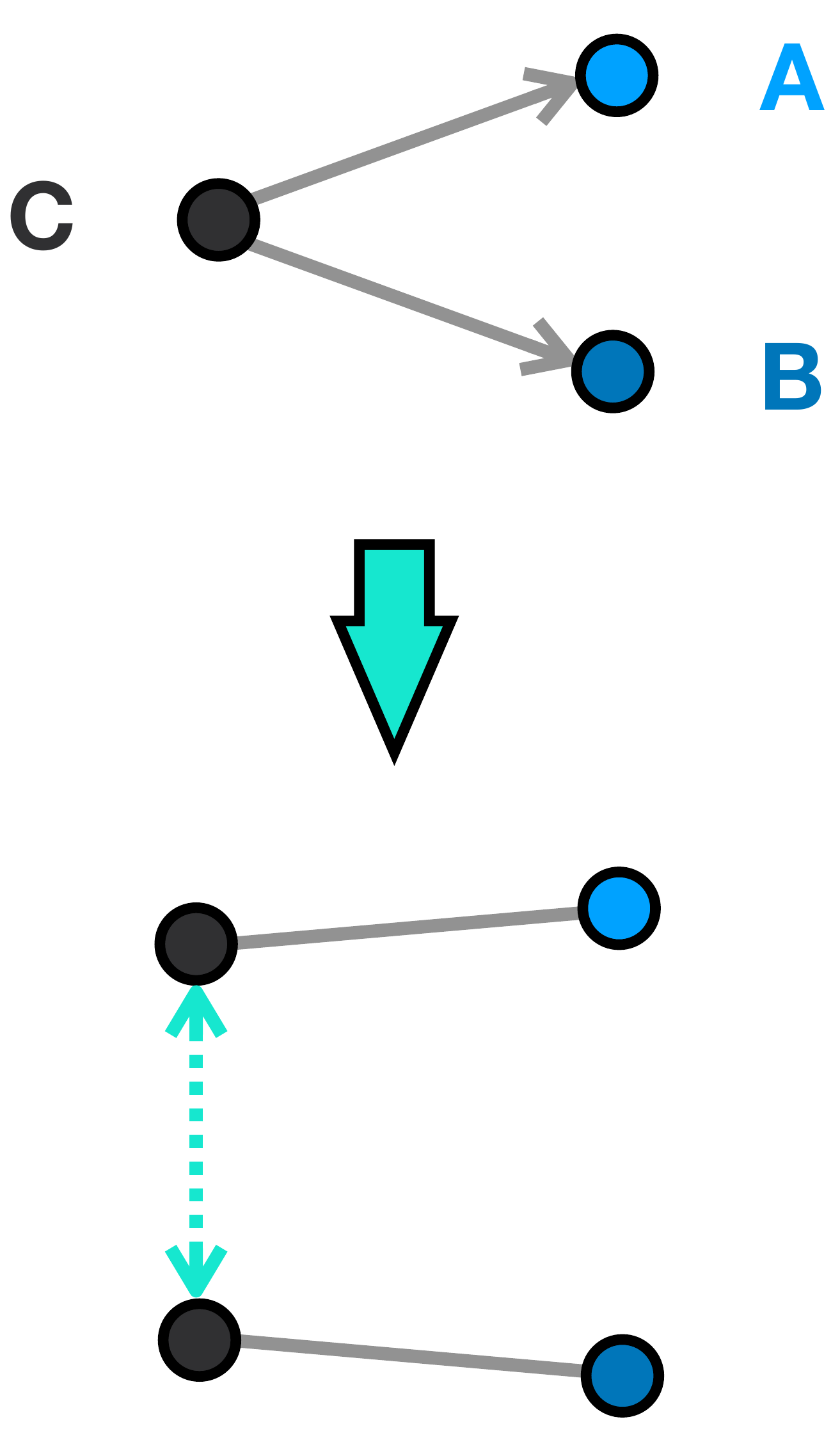}}
  \quad
  \subfloat[]{\includegraphics[height=0.25\textheight]{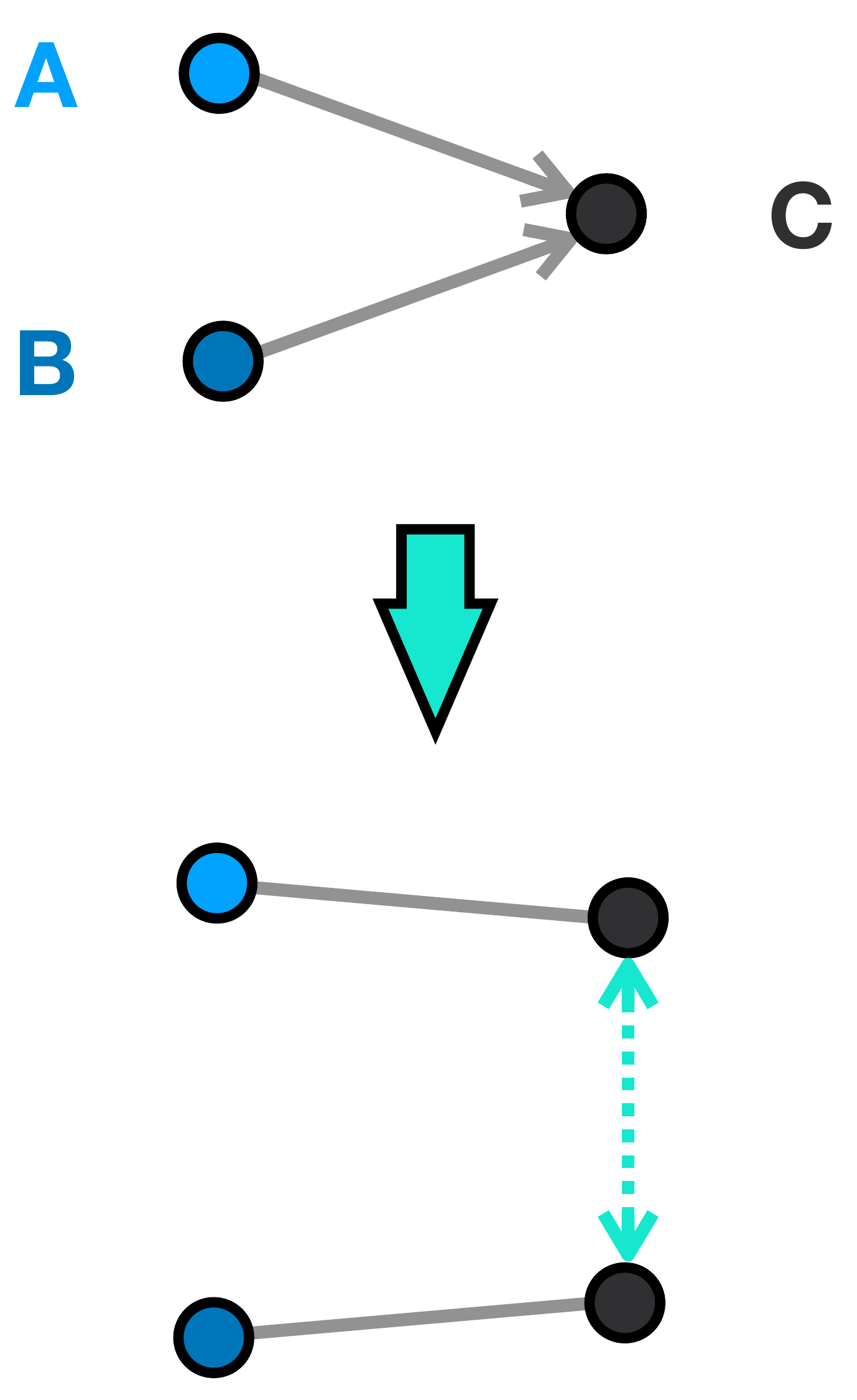}}
  \caption{Visual representation of the three triplet configurations in the edge graph: (a) the articulation, (b) the left elbow and (c) the right elbow. Adapted from~\cite{correia_graph_2024}.}
  \label{fig:triplets}
  \end{center}
\end{figure}

Edge connections, or triplets~\cite{choma_track_2020}, are derived from the purified hit graph $G^{\text{hit}}_{\text{purified}}$. Each triplet is composed of three hits: one shared hit, $C$, and two additional hits, $A$ and $B$. Only three distinct types of triplets can be formed, as shown in Fig.~\ref{fig:triplets}.
\begin{itemize}
    \item \textbf{Articulation:} Two consecutive edges, $A \to C$ and $C \to B$, with the common hit in the middle.
    \item \textbf{Left Elbow:} Edges $C \to A$ and $C \to B$, with the common hit on the left.
    \item \textbf{Right Elbow:} Edges $A \to C$ and $B \to C$, with the common hit on the right.
\end{itemize}
It was found that using separate triplet classifiers for articulations and elbows led to better performance. Each of these classifiers is an MLP ending with a layer with 1 unit, and a sigmoid activation function.

\subsection{Track Building}
\label{sec:track-building}

To reconstruct tracks from the purified edge graph $G^{\text{edge}}_{\text{purified}}$, applying the WCC algorithm alone would detect sets of connected edges, enabling the reconstruction of tracks that share a single hit. However, electron--positron pairs are produced with a very small opening angle between the two tracks, often resulting in multiple shared hits at the track origins. To address this, the process of building tracks from triplets is carried out in four steps. First, the left elbows and right elbows are connected, leaving only the articulations to connect. Duplicate edges resulting from elbow connections are removed, so that when two tracks share their initial hits, it is equivalent to two articulations sharing an edge. Second, a WCC algorithm is applied to the edge graph, excluding these shared articulations. The shared articulations now act as connections between two sets of connected edges, with one set being shared. Third, a new track is formed for each remaining articulation, effectively duplicating the shared set of connected edges. Finally, edges are replaced by their corresponding hits, converting the sets of connected edges into sets of connected hits, thereby representing the tracks.

\subsection{Training Process}

For the physics performance presented in Section~\ref{sec:etx4velo-performance}, the configuration files \texttt{velo-query-long\_h8} and \texttt{velo-h32\_e23444\_h23333-new-embeddings}, that can be found on the ETX4VELO codebase~\cite{gdl4hep_etx4velo_nodate}, were used for the embedding and the GNN, respectively.

The training, using the data described in Section~\ref{sec:datasets}, was conducted on the Convergence LIP6 server~\cite{lip6_convergence_nodate}, on Nvidia A100 80~GB GPUs. The embedding network was trained in roughly one day, while training of the GNN took approximately one week.

\begin{figure}
    \centering
    \includegraphics[width=1\linewidth]{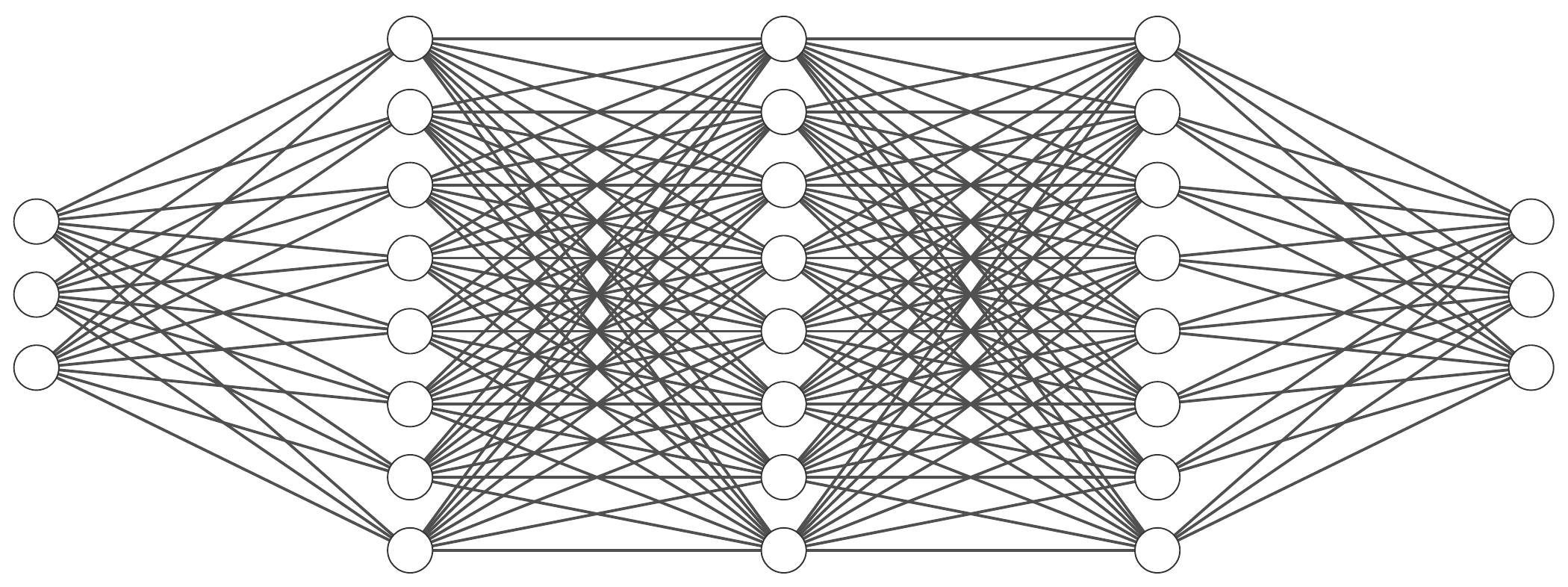}
    \caption{The architecture of the embedding network used for the physics performance presented in Section~\ref{sec:etx4velo-performance}. Generated using \cite{lenail_nn-svg_2019}.}
    \label{fig:mlp}
\end{figure}

The embedding network is visualized in Fig.~\ref{fig:mlp}. It is a fully-connected feed-forward neural network of 3-dimensional input, $(r, \varphi, z)$ as described in Section~\ref{sec:graph-building}, three hidden layers of 8 neurons each, ReLU activations, and a 3-dimensional output.

\begin{figure}
    \centering
    \includegraphics[width=0.75\linewidth]{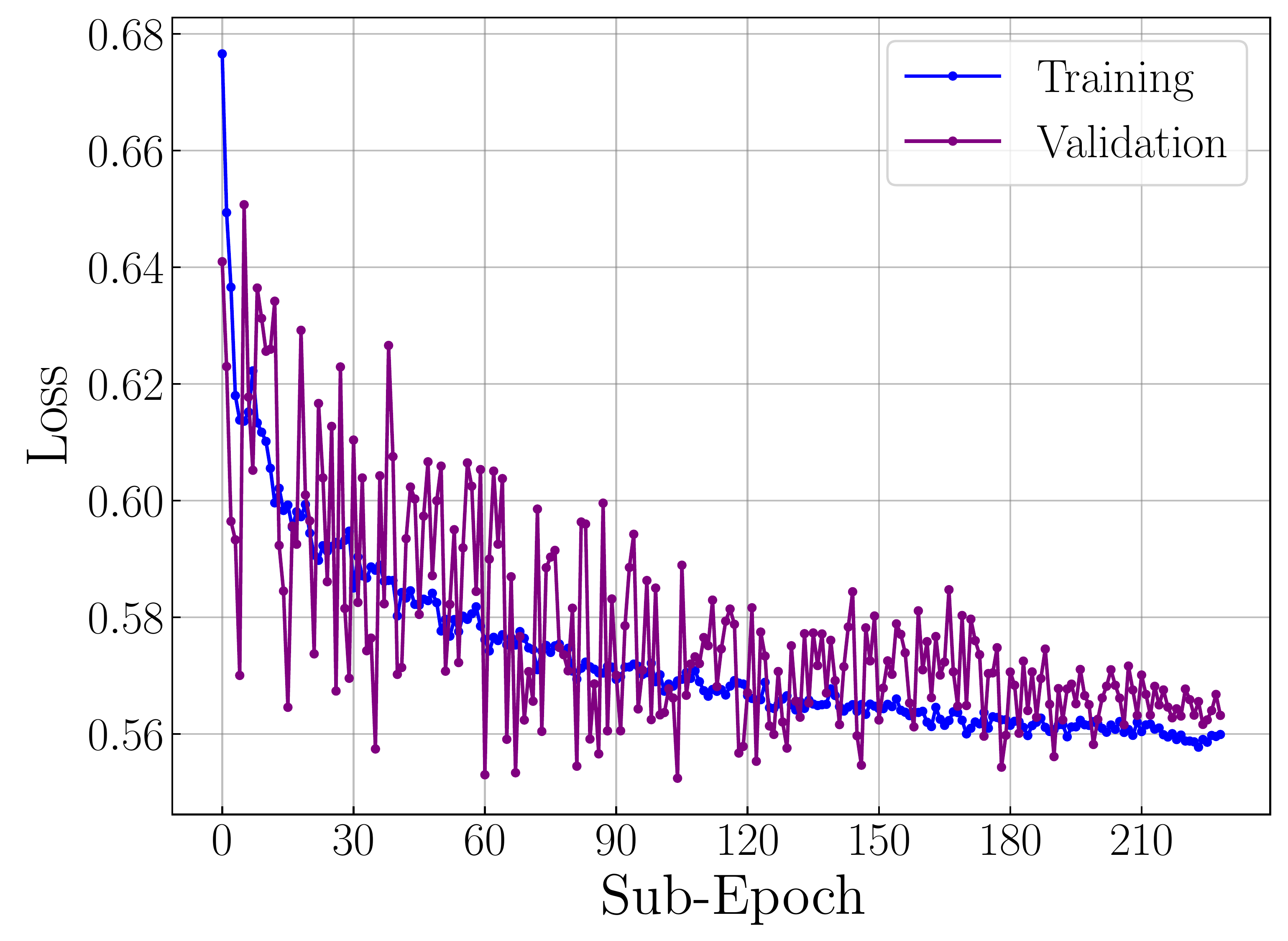}
    \caption{Training and validation losses for the Embedding MLP used for the physics performance presented in Section~\ref{sec:etx4velo-performance}.}
    \label{fig:mlp-loss}
\end{figure}

\begin{figure}
    \centering
    \includegraphics[width=0.75\linewidth]{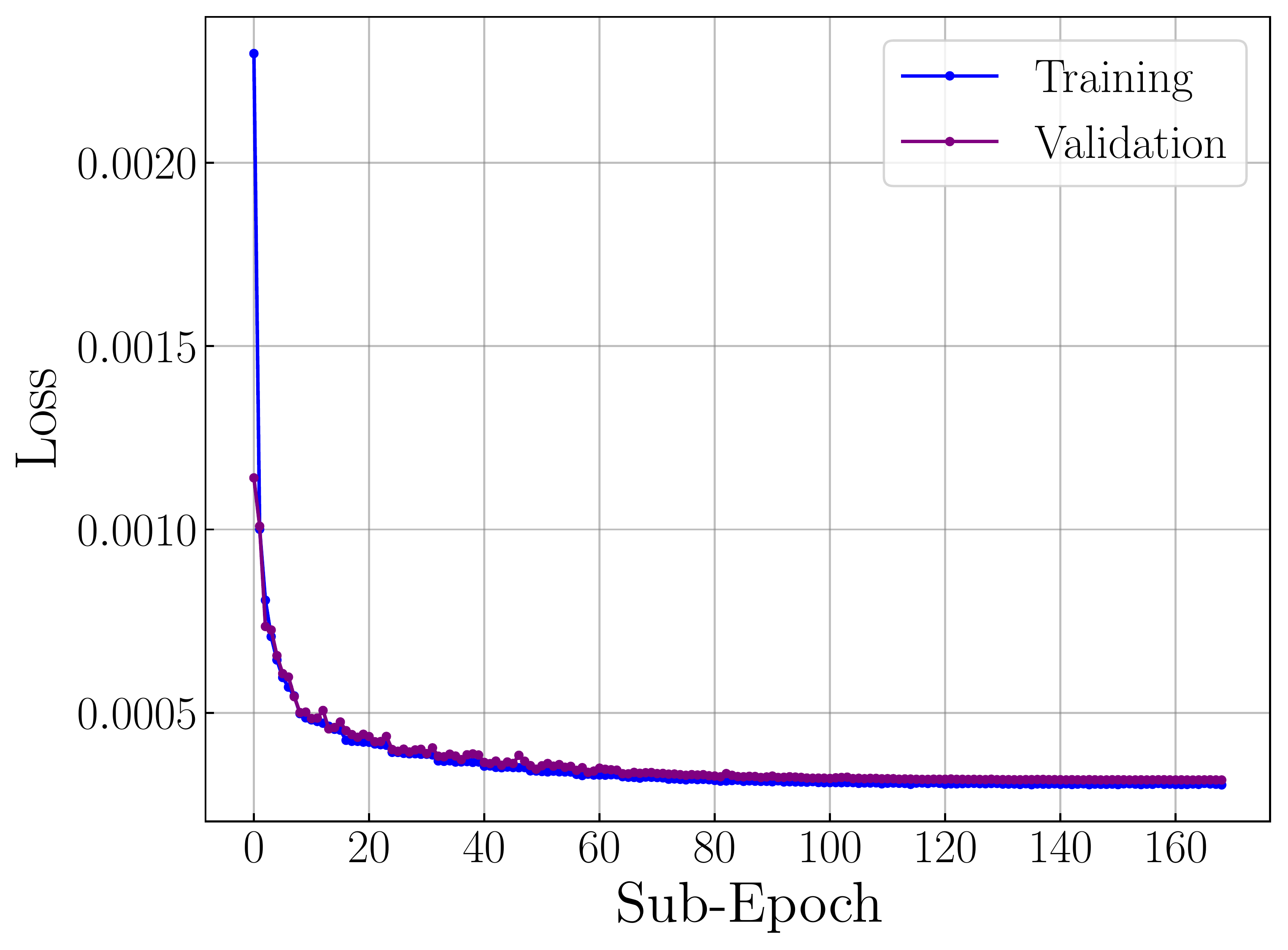}
    \caption{Training and validation losses for the GNN ending with the edge classifier---$\mathcal{L}_{\text{edges}}$ in Eq.~\eqref{eq:gnn-loss}.}
    \label{fig:gnn-edge-loss}
\end{figure}

\begin{figure}
    \centering
    \includegraphics[width=0.75\linewidth]{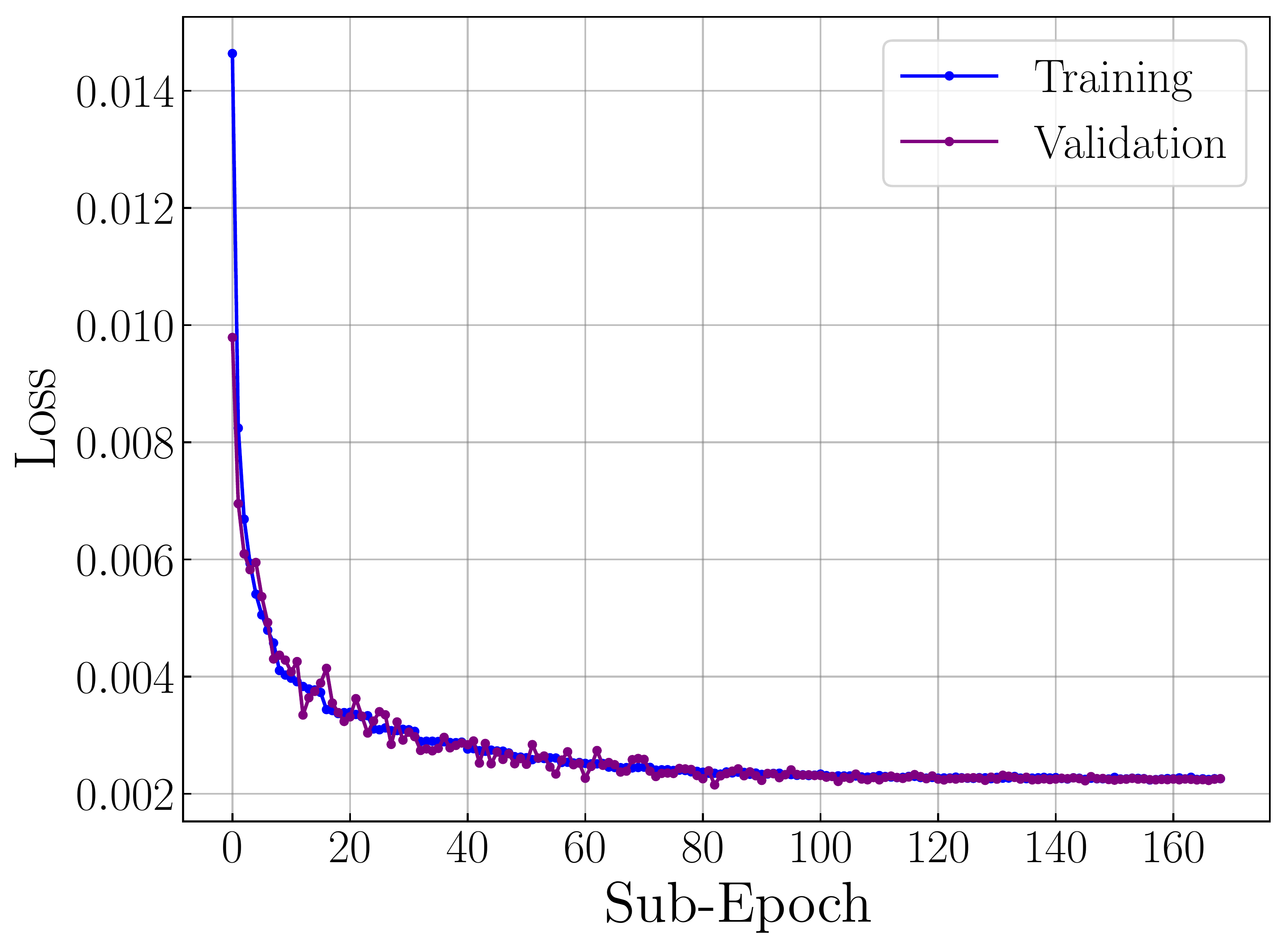}
    \caption{Training and validation losses for the GNN ending with the triplet classifier---$\mathcal{L}_{\text{triplets}}$ in Eq.~\eqref{eq:gnn-loss}.}    
    \label{fig:gnn-triplet-loss}
\end{figure}

The training and validation losses for the embedding are shown in Fig.~\ref{fig:mlp-loss}. The GNN's edge classification and triplet classification losses are shown in Figs~\ref{fig:gnn-edge-loss} and \ref{fig:gnn-triplet-loss}, respectively.

\section{Physics Performance}
\label{sec:etx4velo-performance}

The physics performance of the ETX4VELO pipeline is assessed using a sample of 1000 events and is compared to the default algorithm---Search by triplet---employed for track finding in the VELO within the Allen framework. The performance metrics for both ETX4VELO and Search by triplet are presented in Table~\ref{tab:long_performance} for long particles, Table~\ref{tab:velo_only_performance} for VELO-only particles, and Table~\ref{tab:fake_rate} for the fake track rate (the so-called ghost rate in LHCb). Additionally, these tables show the performance of the GPU implementation of the ETX4VELO pipeline, which does not include the triplet steps. In this implementation, tracks are determined by applying a WCC directly to the purified hit graph $G^{\text{hit}}_{\text{purified}}$. The implementation is described in detail in Chapter~\ref{ch:etx4velo-gpu}.

\begin{table}
\setlength{\tabcolsep}{3pt}
\begin{center}
\centerline{
\begin{tabular}{l|cc|cc|cc|cc}
\hline\hline
Long & \multicolumn{2}{c|}{Efficiency} & \multicolumn{2}{c|}{Clone Rate} & \multicolumn{2}{c|}{Hit Efficiency} & \multicolumn{2}{c}{Hit Purity} \\
\multicolumn{1}{l|}{} & Allen & ETX4VELO & Allen & ETX4VELO & Allen & ETX4VELO & Allen & ETX4VELO \\
\hline
\multicolumn{1}{l|}{No Electrons} & 99.35 & 99.35 (97.96) & 2.61 & 1.23 (0.88) & 96.34 & 98.58 (98.42) & 99.78 & 99.92 (99.95) \\
\multicolumn{1}{l|}{Electrons} & 95.21 & 98.10 (51.82) & 3.31 & 3.35 (0.93) & 95.69 & 97.33 (96.46) & 98.37 & 99.55 (95.05) \\
From Strange & 97.53 & 97.43 (92.23) & 2.70 & 1.62 (0.61) & 95.85 & 97.95 (96.39) & 99.44 & 99.59 (99.77) \\
\hline\hline
\end{tabular}
}
\caption{Track-finding performance (in percentages) of Search by triplet in Allen versus ETX4VELO for long particles. The values in parentheses correspond to the performance of the ETX4VELO pipeline without the triplet approach, as currently implemented in C++/CUDA and presented in Chapter~\ref{ch:etx4velo-gpu}. Reproduced from~\cite{correia_graph_2024}.}
\label{tab:long_performance}
\end{center}
\end{table}

\begin{table}
\setlength{\tabcolsep}{3pt}
\begin{center}
\centerline{
\begin{tabular}{l|cc|cc|cc|cc}
\hline\hline
VELO-only & \multicolumn{2}{c|}{Efficiency} & \multicolumn{2}{c|}{Clone Rate} & \multicolumn{2}{c|}{Hit Efficiency} & \multicolumn{2}{c}{Hit Purity} \\
\multicolumn{1}{l|}{} & Allen & ETX4VELO & Allen & ETX4VELO & Allen & ETX4VELO & Allen & ETX4VELO \\
\hline
\multicolumn{1}{l|}{No Electrons} & 97.03 & 97.05 (96.28) & 3.65 & 1.46 (0.87) & 94.07 & 97.68 (97.93) & 99.51 & 99.81 (99.92) \\
\multicolumn{1}{l|}{Electrons} & 67.84 & 83.60 (49.93) & 9.65 & 6.71 (3.51) & 79.57 & 90.83 (85.25) & 97.62 & 99.17 (98.25) \\
From Strange & 94.25 & 93.69 (84.33) & 5.16 & 4.09 (1.35) & 90.33 & 97.95 (90.79) & 99.43 & 99.49 (99.72) \\
\hline\hline
\end{tabular}
}
\caption{Track-finding performance (in percentages) of Search by triplet in Allen versus ETX4VELO for VELO-only particles. The values in parentheses correspond to the performance of the ETX4VELO pipeline without the triplet approach, as currently implemented in C++/CUDA and presented in Chapter~\ref{ch:etx4velo-gpu}. Reproduced from~\cite{correia_graph_2024}.}
\label{tab:velo_only_performance}
\end{center}
\end{table}

\begin{table}
\centering
\begin{tabular}{l|c|cc}
\hline\hline
 & \multirow{2}{*}{Allen} & \multicolumn{2}{c}{ETX4VELO} \\[1mm]
 &  & With Triplets & Without Triplets \\[1mm]
 \hline
Fake Rate & 2.18\% & 1.01\% & 2.07\%\\
\hline\hline
\end{tabular}
\caption{Fake rate of Search by triplet in Allen versus ETX4VELO, for the full pipeline with triplets and the pipeline excluding the triplet approach, as implemented in C++/CUDA and presented in Chapter~\ref{ch:etx4velo-gpu}. Reproduced from~\cite{correia_graph_2024}.}
\label{tab:fake_rate}
\end{table}

\begin{figure}
    \centering
    \includegraphics[width=1\linewidth]{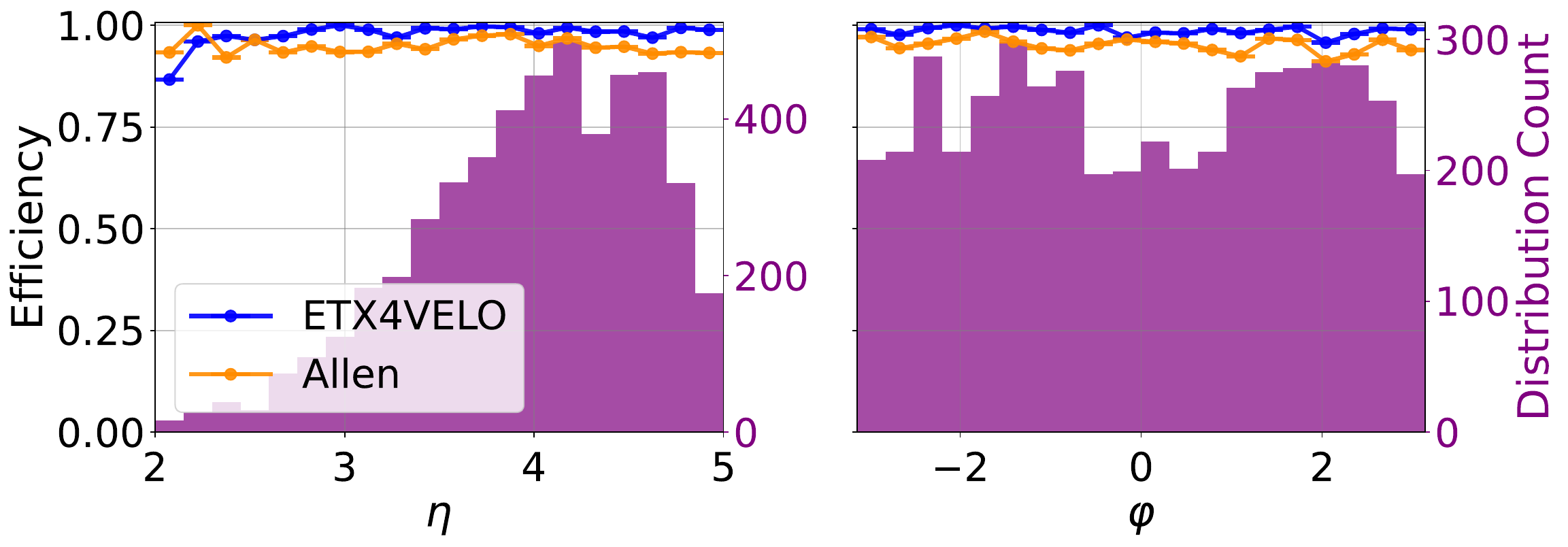}
    \caption{Comparison of ETX4VELO and Search by triplet in Allen, as a function of pseudorapidity $\eta$ (left) and track azimuthal angle $\varphi$ (right), and for long electrons, using Montetracko.}
    \label{fig:eta-phi}
\end{figure}

\begin{figure}
    \centering
    \includegraphics[width=1\linewidth]{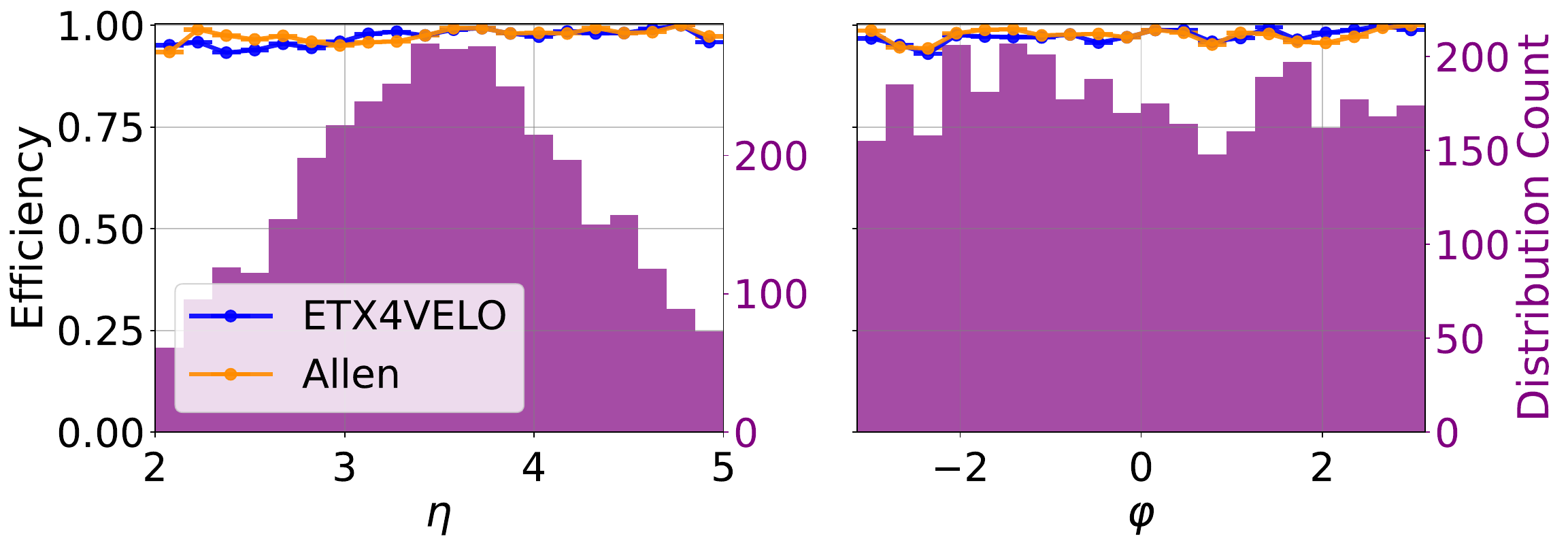}
    \caption{Comparison of ETX4VELO and Search by triplet in Allen, as a function of pseudorapidity $\eta$ (left) and track azimuthal angle $\varphi$ (right), and for long particles from strange decays, using Montetracko.}
    \label{fig:eta-phi-strange}
\end{figure}

\begin{figure}
    \centering
    \includegraphics[width=1\linewidth]{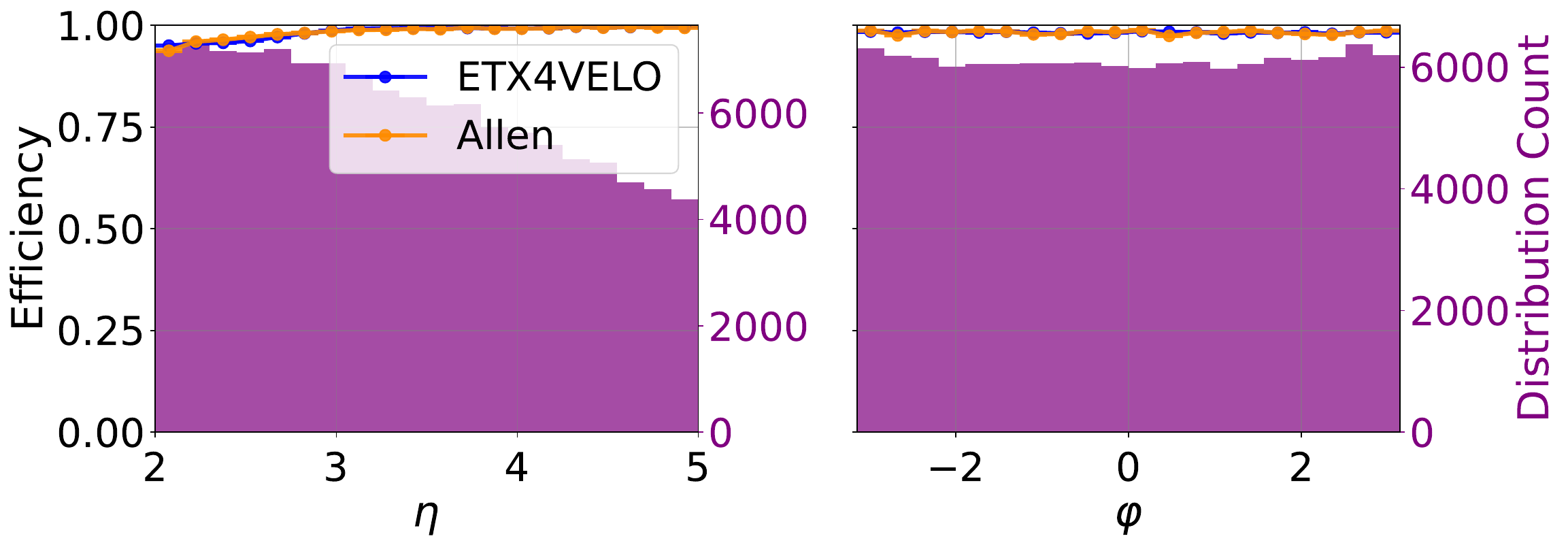}
    \caption{Comparison of ETX4VELO and Search by triplet in Allen, as a function of pseudorapidity $\eta$ (left) and track azimuthal angle $\varphi$ (right), and for particles in the VELO acceptance, excluding electrons, using Montetracko.}
    \label{fig:eta-phi-velo}
\end{figure}

\begin{figure}
    \centering
    \includegraphics[width=1\linewidth]{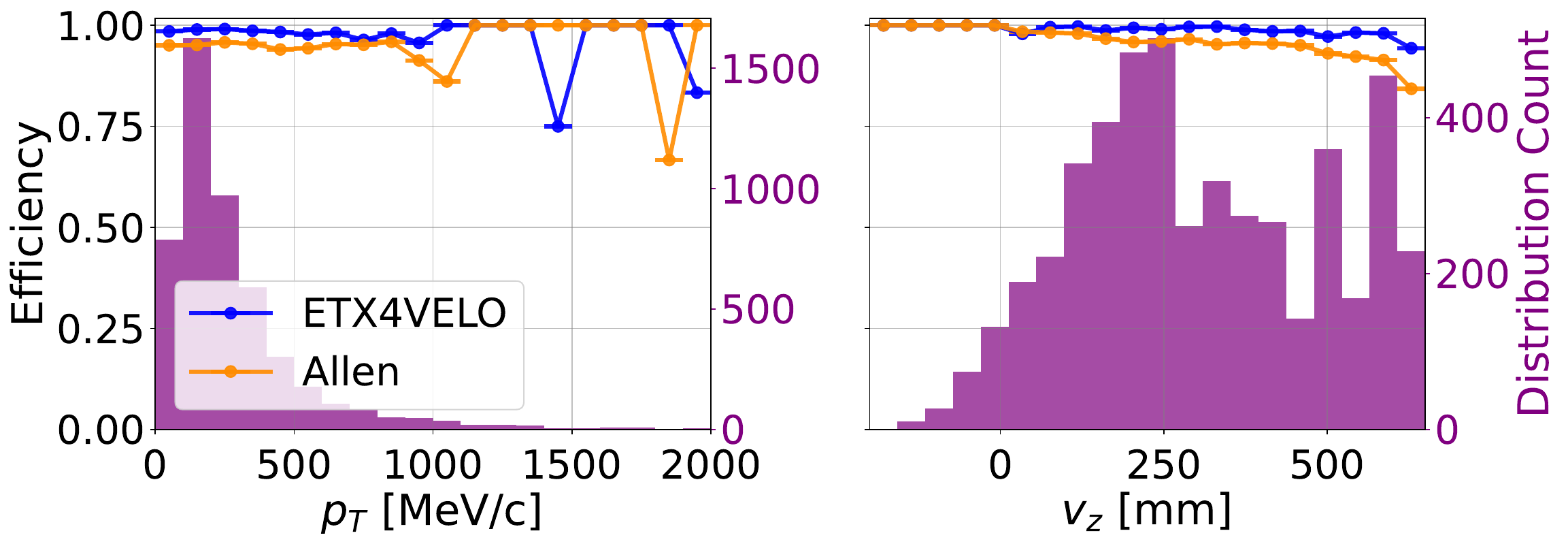}
    \caption{Comparison of ETX4VELO and Search by triplet in Allen, as a function of transverse momentum $p_T$ (left) and the $z$-coordinate of the origin vertex $v_z$ (right), and for long electrons, using Montetracko.}
    \label{fig:pt-vz}
\end{figure}

\begin{figure}
    \centering
    \includegraphics[width=1\linewidth]{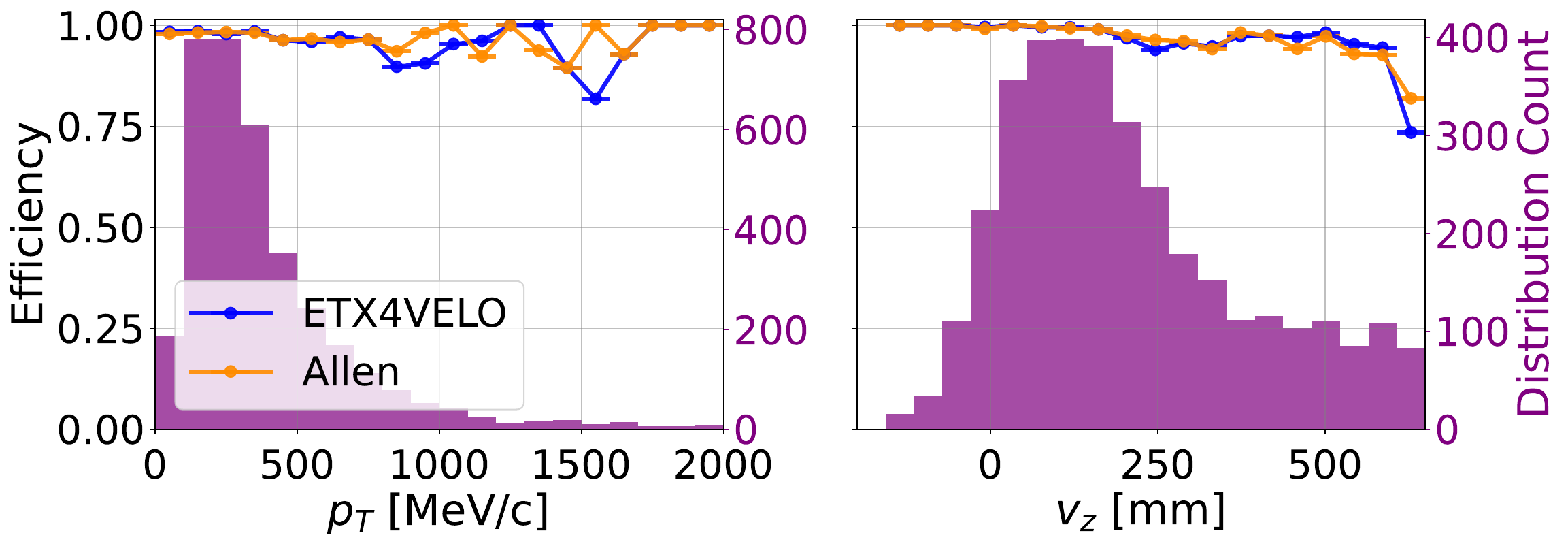}
    \caption{Comparison of ETX4VELO and Search by triplet in Allen, as a function of transverse momentum $p_T$ (left) and the $z$-coordinate of the origin vertex $v_z$ (right), and for long particles from strange decays, using Montetracko.}
    \label{fig:pt-vz-strange}
\end{figure}

\begin{figure}
    \centering
    \includegraphics[width=1\linewidth]{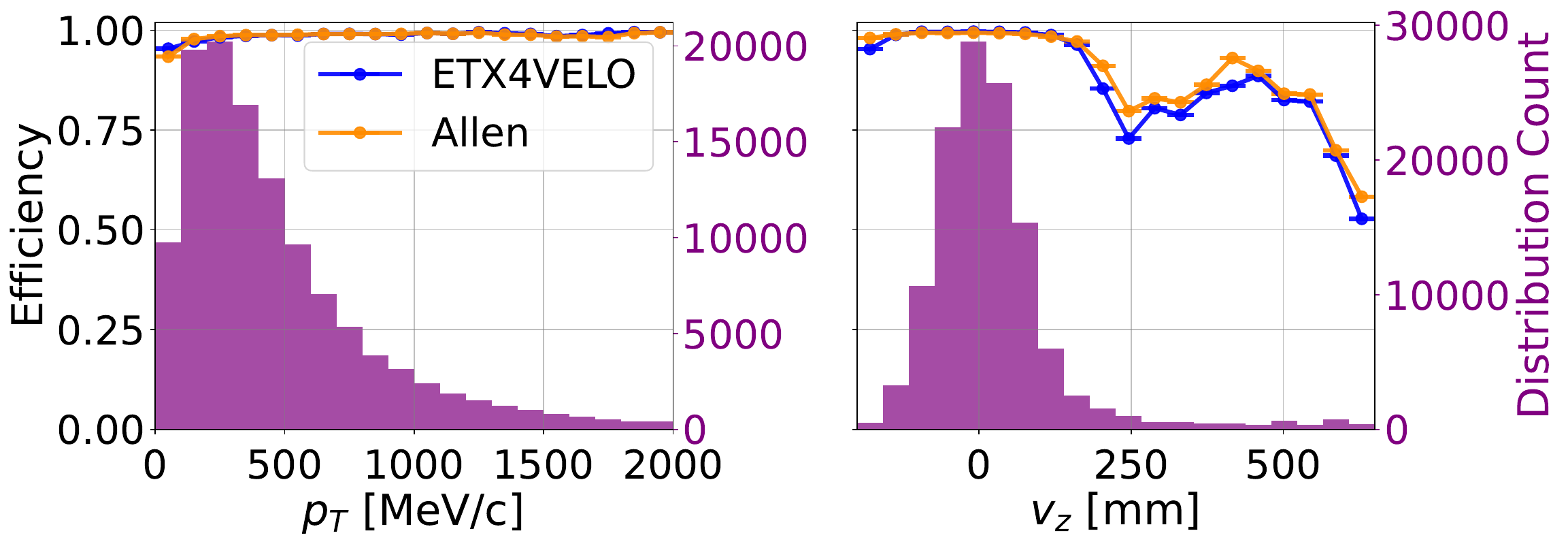}
    \caption{Comparison of ETX4VELO and Search by triplet in Allen, as a function of transverse momentum $p_T$ (left) and the $z$-coordinate of the origin vertex $v_z$ (right), and for particles in the VELO acceptance, excluding electrons, using Montetracko.}
    \label{fig:pt-vz-velo}
\end{figure}

The efficiency of the ETX4VELO pipeline and Allen is also compared with respect to pseudorapidity $\eta$ and track azimuthal angle $\varphi$ using the Montetracko library. This is done in Figs.~\ref{fig:eta-phi}, \ref{fig:eta-phi-strange} and \ref{fig:eta-phi-velo}, for long electrons, long particles from strange decays, and for particles in the VELO acceptance, excluding electrons, respectively. Similarly, the comparison is done as a function of transverse momentum $p_T$ and the $z$-coordinate of the origin vertex $v_z$ in Figs.~\ref{fig:pt-vz}, \ref{fig:pt-vz-strange} and \ref{fig:pt-vz-velo}.

\begin{figure}
    \centering
    \includegraphics[width=\linewidth]{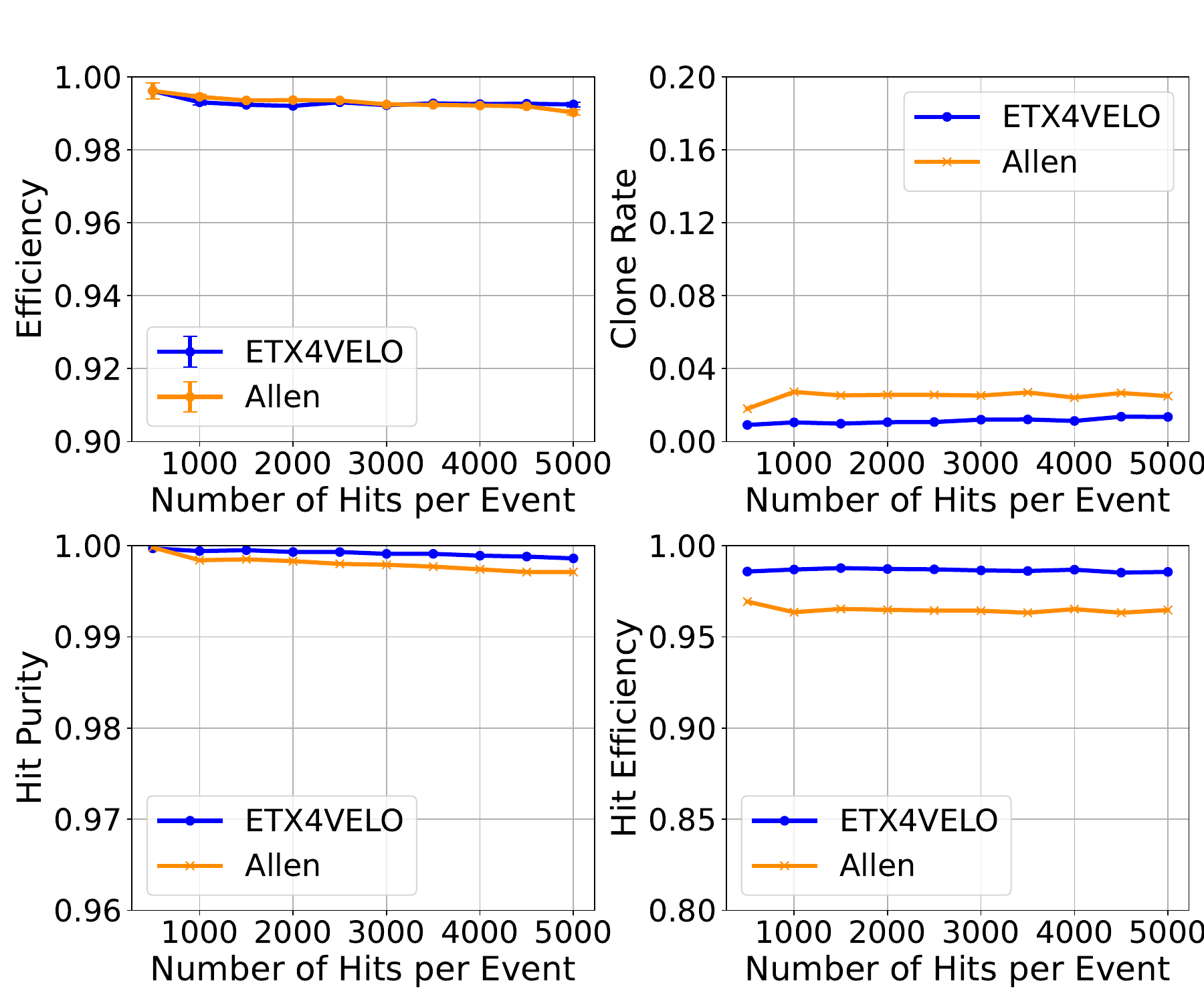}
    \caption{Track-finding performance comparison of Search by triplet in Allen versus ETX4VELO for long particles, excluding electrons, as a function of the occupancy of the detector. Reproduced from~\cite{correia_graph_2024}.}
    \label{fig:physics-vs-occupancy}
\end{figure}

\begin{figure}
    \centering
    \includegraphics[width=0.8\linewidth]{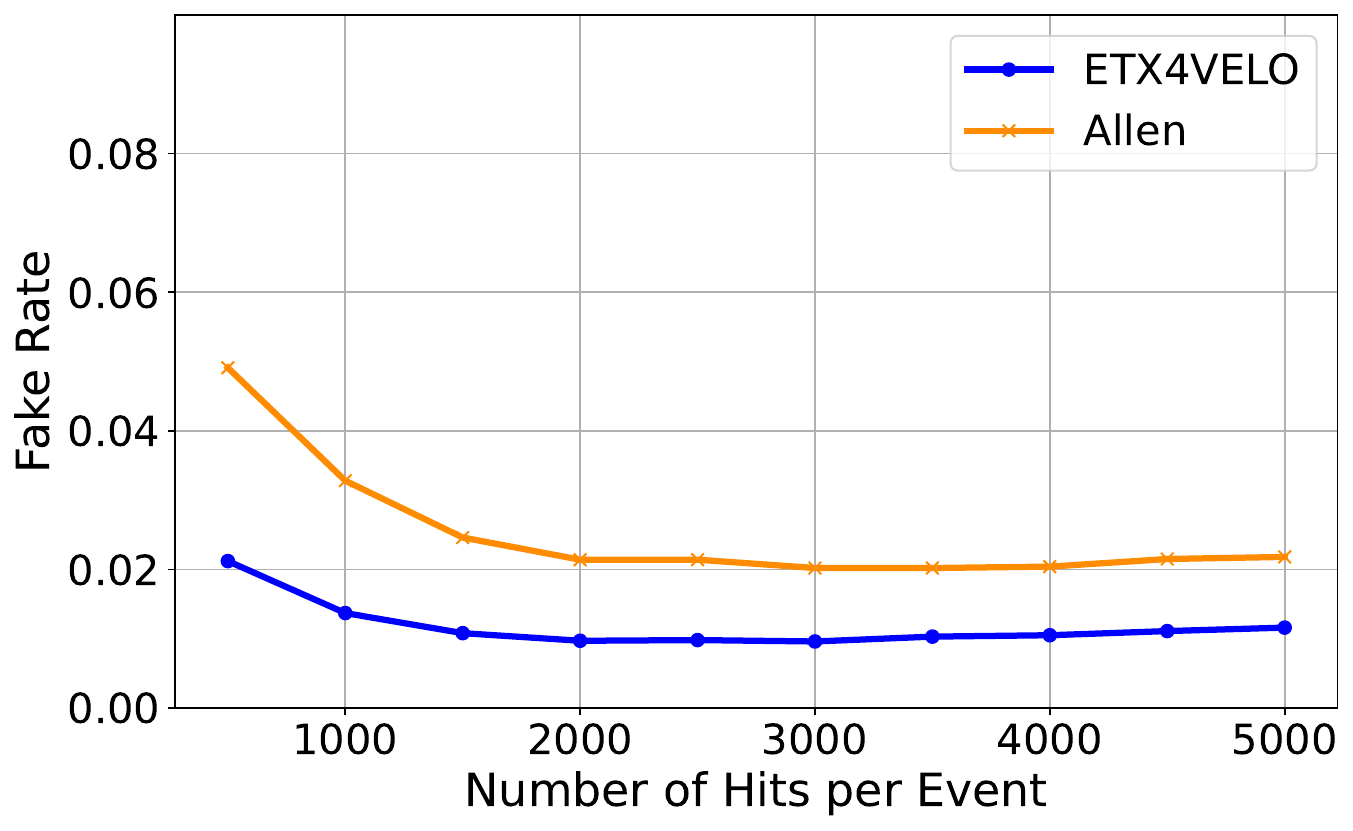}
    \caption{Fake rate comparison of Search by triplet in Allen versus ETX4VELO as a function of the occupancy of the detector.}
    \label{fig:ghost-vs-occupancy}
\end{figure}

Finally, the physics performance is also compared between the ETX4VELO pipeline and Allen for long particles, excluding electrons, as a function of the occupancy of the detector, i.e., the number of hits in each event, in Fig.~\ref{fig:physics-vs-occupancy}. Various track-finding performance metrics are plotted against the occupancy. Events are split into bins based on their occupancy, and the evaluation of the tracking algorithms is done on the events of each bin. The error bars for the efficiency are binomial errors~\cite{paterno_calculating_2004}. The fake rate is similarly compared in Fig.~\ref{fig:ghost-vs-occupancy}.

The ETX4VELO pipeline achieves performance on par with the Allen framework while more than halving the fake rate. It consistently delivers superior track quality across all categories, demonstrating higher hit efficiency and hit purity. Notably, it excels in reconstructing electron tracks. However, its performance is slightly weaker for particles originating from strange decays. This limitation is likely due to these particles being more tilted relative to the beamline, which may have led to their underrepresentation during the graph-building stage because of the selected squared maximum distance, $d^2_\text{max}$.

\section*{Conclusion}

In this chapter, I introduced ETX4VELO, and how it was gradually developed out of the foundational Exa.TrkX pipeline. I highlighted also its ability to separate tracks that share hits through a novel triplet-based approach. In addition, the physics performance of the pipeline was quantified and demonstrated to be on par with the current tracking algorithms in place inside LHCb's first-level trigger, Allen. Interestingly, the efficiency of electron reconstruction was significantly improved and the ghost rate was roughly halved to around 1\%.

The focus at this point shifted more towards the computational aspect of the pipeline. Its implementation on GPUs is presented in Chapter~\ref{ch:etx4velo-gpu}, while its partial implementation on FPGAs is presented in Chapter~\ref{ch:etx4velo-fpga}.

Future work could involve extending the pipeline to incorporate the remaining LHCb tracking detectors, namely the UT and SciFi. Preliminary efforts toward integrating the latter have already been initiated.

\chapter{Accelerating ETX4VELO on GPU}
\label{ch:etx4velo-gpu}

\minitoc

\noindent Parts of this chapter are based on work published in~\cite{correia_graph_2024}. The repository of the project can be accessed at~\cite{gdl4hep_etx4velo_cuda_nodate}.

\section*{Introduction}

In Chapter~\ref{ch:etx4velo}, the ETX4VELO pipeline was presented, along with its development process and physics performance. In this chapter, however, the focus shifts to the computational aspects of the pipeline.

ETX4VELO was aimed for tracking at the LHCb experiment. Since the beginning of Run~3 in July 2022, LHCb is triggering at the full collision rate of 30~MHz. The first-level trigger, Allen, among other things, performs an online partial detector reconstruction, including the reconstruction of charged particle tracks. This processing is done entirely on GPUs. Therefore, in order for ETX4VELO to be of interest to the collaboration, it has to be implemented in the same framework. Allen offers a pragmatic platform for benchmarking and comparing algorithms in a strict and fair way, aligned with the objectives of the experiment.

Furthermore, as we already saw, the GPU architecture is already being used for a wide range of tasks in HEP~\cite{bruch_real-time_2020,bauce_gap_2014,johnson_celeritas_2024,tognini_celeritas_2022,lund_accelerating_2025,kanzaki_application_2011,collazuol_fast_2012,bailly-reyre_looking_2024,pata_processing_2019,ammendola_real-time_2018,bruch_online_2017,chirkin_photon_2019,schreiner_goofit_2018}. Therefore, benchmarking of the architecture could be of even wider interest, not necessarily narrowed down to tracking.

Finally, performing machine learning tasks on GPUs is another area of potential interest. However, in the context of high-energy physics, it remains relatively underexplored at the time of writing~\cite{krupa_gpu_2021}.

Thus, I present the GPU implementation of ETX4VELO inside Allen and its computational performance benchmarked against the combinatorial tracking algorithms in Allen. The challenges encountered during the development process are also discussed.

\section{Development}

The main objective at this point was to start developing the tools needed for porting the PyTorch~\cite{paszke_pytorch_2019} models, and eventually the whole pipeline, to GPU.

Hard coding the models was not pursued because of two main reasons. Firstly, the GNN model, as opposed to the embedding, is extremely complex and thus the process of hard coding it would be very time-consuming. Secondly, the resulting implementation would not be easily reusable and customizable, in order to be adapted to other use cases.

Given that performance and reusability are central to our design goals, we based our implementation on two tools: TensorRT~\cite{tensorrt_developers_nvidia_nodate} and ONNX Runtime~\cite{onnx_runtime_developers_onnx_2021}, which prioritize these aspects, respectively.

\subsection*{GhostBuster Based Implementation on GPU}

The first attempt was using TensorRT. The model was exported with the ONNX~\cite{onnx_developers_onnx_nodate} open-source format. The code was adapted from an early version of the Ghost Probability neural network~\cite{zhuo_fastml_2024,de_cian_fast_2017} used for fake track rejection recently integrated in the first-level trigger of LHCb on GPUs. The exported MLP model was run on the GPU and the inference output was validated against the Python results. This process can be summarized as follows:
\begin{itemize}
    \item The input ONNX model is parsed and the network is created.
    \item The TensorRT builder is used to create a serialized CUDA inference engine.
    \item The TensorRT context is created.
    \item GPU memory buffers, matching the bindings defined by the engine, are allocated.
    \item Inference is performed by passing the input buffers to the \texttt{enqueue()} method.
    \item Validation of the output can be performed.
\end{itemize}
The throughput, as expected, was quite low. Only the embedding step, using the architecture at the time, comprising four hidden layers of 256 neurons, achieved a throughput of 8000 events per second. 

At this point, however, only the embedding MLP part of the pipeline was implemented, and the rest of the pipeline, including the k-NN, GNN and WCC steps, still had to be developed. At the time, the task of implementing all these steps from scratch seemed too ambitious to be pursued, so I decided to take a different approach and build upon an existing implementation, namely that from Exa.TrkX. I present this approach next.

\subsection*{Exa.TrkX Based Implementation on CPU}

The second approach was to use the Exa.TrkX implementation of their pipeline on C++~\cite{lazar_accelerating_2023}, which uses ONNX Runtime for the inference of the ML models, and integrate it into Allen. The focus now was to have something that works end-to-end, in order to be able to work on incrementally improving it. In this implementation ONNX Runtime is used to run the models on CPU, even though the initial design goal of the Exa.TrkX implementation was targeting GPU. By working on this implementation, I was able to further understand the intricacies of the C++ language, and this experience proved essential, later on, for implementing the various algorithms on GPU in CUDA.

\begin{figure}
    \centering
    \includegraphics[width=0.6\linewidth]{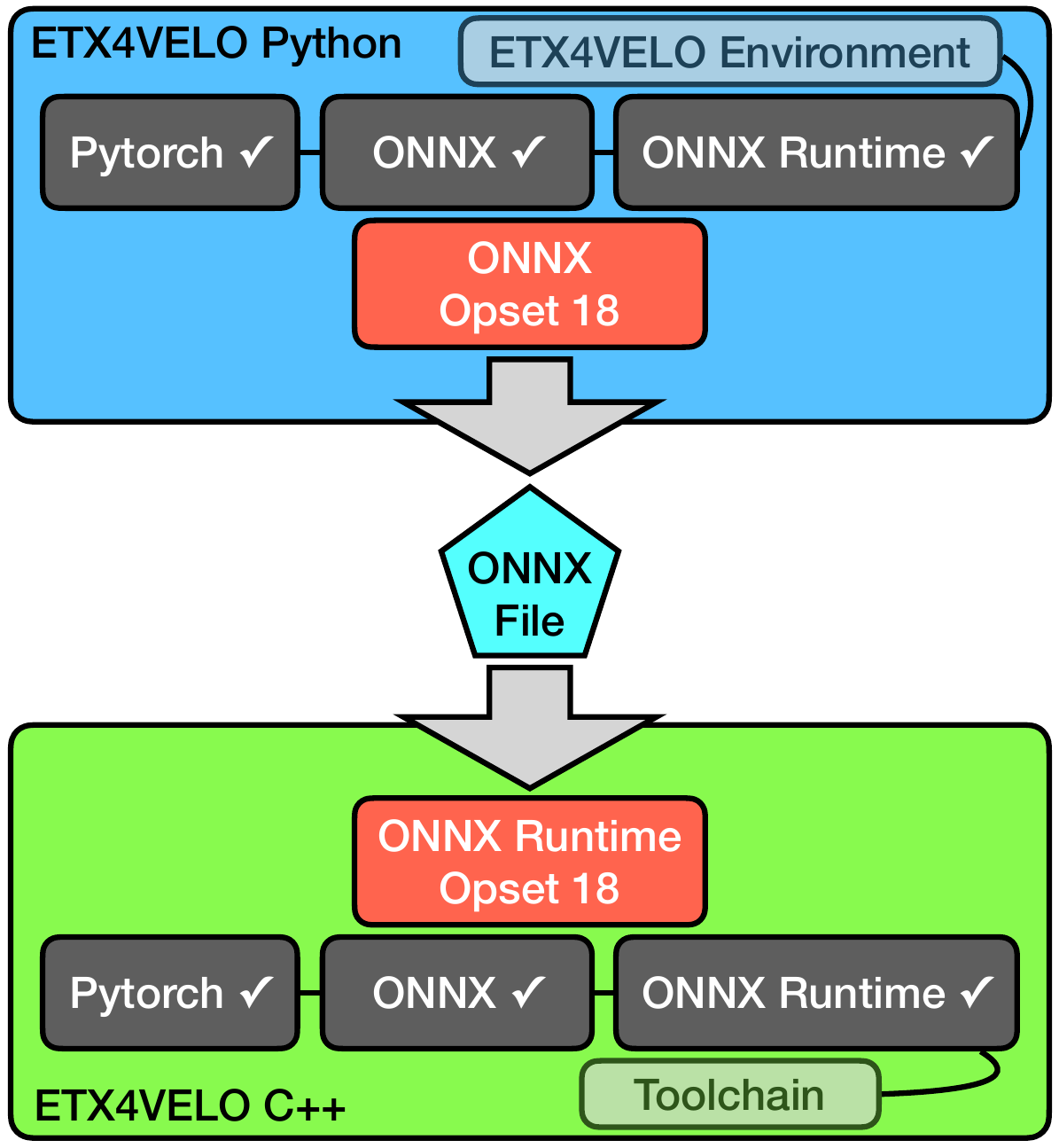}
    \caption{The process of passing the ETX4VELO models from the Python to the C++ side using ONNX and ONNX Runtime. On the C++ side, a cross-compilation toolchain was used~\cite{clemencic_cmake-based_2012}.}
    \label{fig:envs-python-vs-cpp}
\end{figure}

A major challenge was to resolve the environment for all the various dependencies of the implementation and Allen. More specifically the dependencies including ONNX and ONNX Runtime had to be compatible with each other between the Python and the C++ implementations. An illustration of the process of passing an ETX4VELO ONNX model from the Python to the C++ side is shown in Fig.~\ref{fig:envs-python-vs-cpp}. The versions for the dependencies as well as the necessary ``operator set (opset)'' number, used by ONNX to group together operator specifications, are in Table~\ref{tab:onnx-deps}. A minimum opset number of 18 is necessary because of the \texttt{scatter\_add} operation, used in the message passing step of the GNN, which is not supported in earlier opsets.

The pipeline was modified step by step to match the early version of the ETX4VELO pipeline and the models were exported to ONNX. The major differences are shown below.
\begin{enumerate}
    \item Our exported GNN was expecting an undirected graph, while the initial implementation was utilizing a directed graph.
    \item The self-loops, edges connecting a vertex to itself, had to be manually removed.
    \item Edges between hits that are on the same plane had to also be removed, since particles are expected to only leave hits on successive layers while traversing the detector.
    \item The cuGraph~\cite{rapids_developers_cugraph_nodate} dependency used in the track construction step was removed completely and the WCC algorithm was replaced by a custom implementation on CPU.
    \item The filtering step after the graph construction, used to reduce the number of edges in the rough graph, was completely removed.
\end{enumerate}
The k-NN step utilizes the FRNN library~\cite{xue_fixed_nodate} for fixed-radius nearest neighbor search. The Faiss method was also implemented and tested, and the results validated against the FRNN ones. There were no significant changes on the performance, however.

Finally, the Allen algorithms had to be modified in order to be able to ``consume'' the tracks created from the ETX4VELO pipeline. Effectively, the tracks have to be formatted in a specific way in order for the rest of the Allen algorithms to be able to use them. More specifically, the \texttt{Velo::TrackHits} struct is used, defined in the \texttt{VeloEventModel.cuh} header file. The hits of the tracks are organized in these data structures and then passed on to the evaluation algorithms of Allen.

The tracks constructed by the pipeline in Python and by the pipeline in C++ are evaluated on the same LHCb event. The results are shown in Figs.~\ref{tab:etx4velo-tracks} and \ref{tab:etx4velo-cpp-tracks} for the Python and C++ version, respectively. The two pipelines have identical physics performance, namely the number of tracks and the efficiencies are the same.

\begin{table}
    \centering
    \begin{tabular}{c|c|c}
    \hline \hline
        ONNX Runtime & ONNX & Opset\\ \hline
        $\geq 1.14$ & $\geq 1.13$ & 18 \\
    \hline \hline
    \end{tabular}
    \caption{Versions for the dependencies as well as the necessary opset number for passing the ETX4VELO models from the Python to the C++ side as illustrated in Fig.~\ref{fig:envs-python-vs-cpp}}
    \label{tab:onnx-deps}
\end{table}

\begin{table}
{\small
    \centering
    \setlength{\tabcolsep}{2pt}
    \centerline{
    \begin{tabular}{l|c|c|c}
    \hline \hline
        Category & Reconstructed/True & Efficiency (\%) & Clones (\%) \\ \hline
        VELO, No Electrons                   & 117 / 123  & 95.12 & 2.50 \\
        Long, No Electrons                   & 71 / 73    & 97.26 & 1.39 \\
        Long, No Electrons, $p>5$~GeV          & 50 / 52    & 96.15 & 0.00 \\
    \hline \hline
    \end{tabular}
    }
    \caption{Track reconstruction summary for the early version of the ETX4VELO pipeline in Python. For each particle category the number of reconstructed tracks is given along with the correct number of tracks, the clone rate, the purity and the hit efficiency.}
    \label{tab:etx4velo-tracks}
}
\end{table}

\begin{table}
{\small
    \centering
    \setlength{\tabcolsep}{2pt}
    \centerline{
    \begin{tabular}{l|c|c|c}
    \hline \hline
        Category & Reconstructed/True & Efficiency (\%) & Clones (\%) \\ \hline
        VELO, No Electrons                   & 117 / 123  & 95.12 & 2.50 \\
        Long, No Electrons                   & 71 / 73    & 97.26 & 1.39 \\
        Long, No Electrons, $p>5$~GeV           & 50 / 52    & 96.15 & 0.00 \\
    \hline \hline
    \end{tabular}
    }

    \caption{Track reconstruction summary for the early version of the ETX4VELO pipeline implemented in C++. For each particle category the number of reconstructed tracks is given along with the correct number of tracks, the clone rate, the purity and the hit efficiency.}
    \label{tab:etx4velo-cpp-tracks}
}
\end{table}

With these implementations at hand, we then moved onto creating a custom implementation, targeting GPU, for the ETX4VELO pipeline, writing the code from scratch.

\subsection*{Early Version of ETX4VELO on GPU}

The custom implementation of the simplified version of ETX4VELO on GPU was developed with the Allen architecture in mind. It followed the same design with the Exa.TrkX implementation, where the execution of the ML models is delegated to the ONNX Runtime inference engine. Furthermore, this version is implemented in a ``modular'' fashion, where the goal was to make the different steps of the pipeline as independent from each other as possible. The steps included are the following:

\begin{itemize}
    \item Calculation of the features of the data, recentering and rescaling, filling of the buffers.
    \item Inference of the embedding with ONNX Runtime.
    \item Construction of the edges of the graph based on the embedding.
    \item Preprocessing of the edges in order to put them in the desired format.
    \item Batching of the edges in order to be used by the batched version of the GNN.
    \item Inference of the GNN in batches in order to get the edge scores.
    \item Construction of the tracks.
    \item Formatting and consolidation of the tracks. 
    \item Validation with Allen.
\end{itemize}
For the batching of the GNN, the number of events we can batch together is around a maximum of 10 events, assuming the memory of a Nvidia GeForce RTX 2080 Ti.

\begin{figure}
    \centering
    \includegraphics[width=0.75\linewidth]{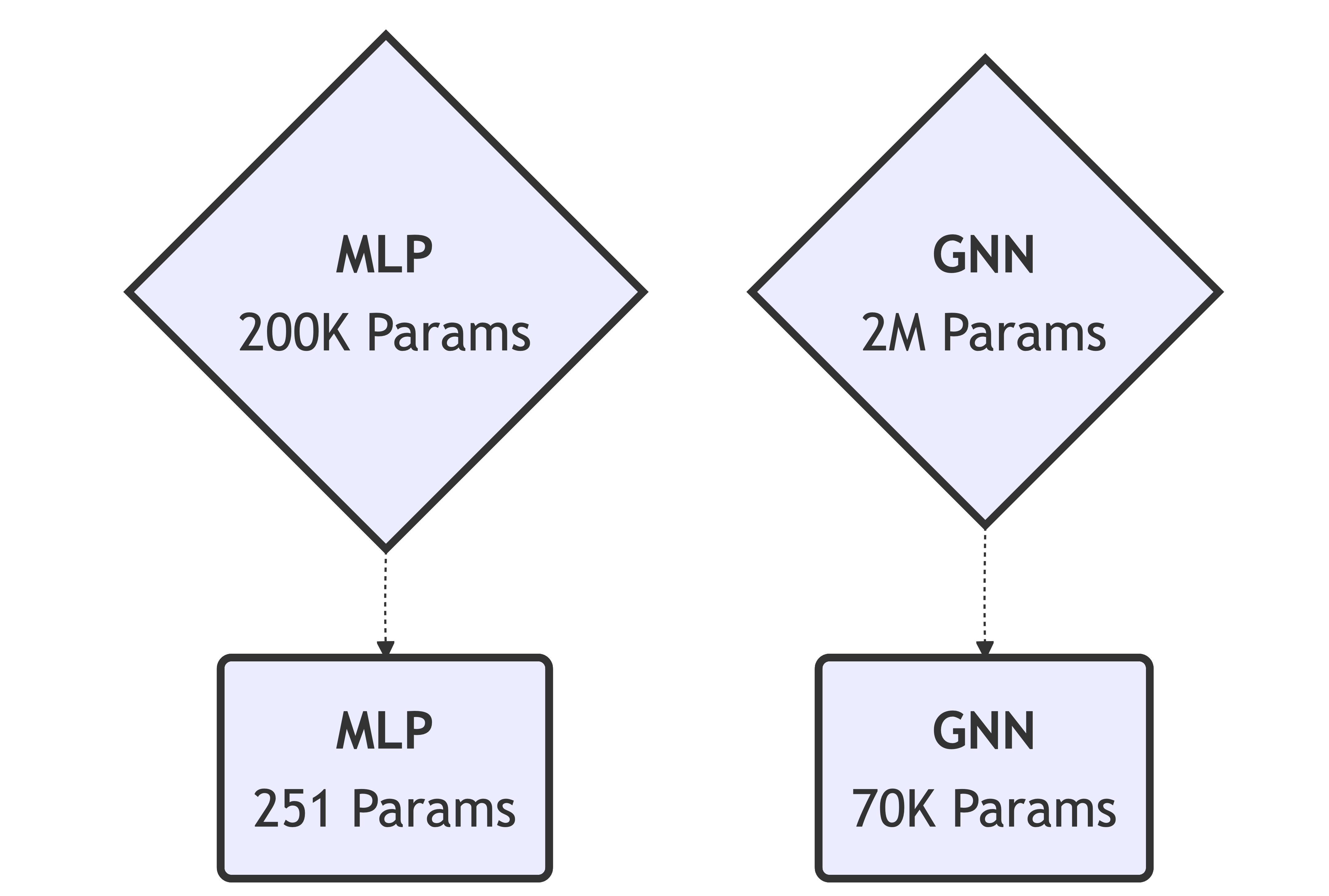}
    \caption{Starting from the original Exa.TrkX model architectures, the ETX4VELO models, the embedding MLP and the GNN, used in Chapter~\ref{ch:etx4velo}, Section~\ref{sec:etx4velo-simplified}, were reduced down to the minimum size possible, while keeping the physics performance within acceptable levels. Generated with~\cite{sveidqvist_mermaid_2014}.}
    \label{fig:model-size}
\end{figure}

This end-to-end implementation, with the exported models from the early version of the pipeline, results in a throughput of around 20 events per second. Having started from the original Exa.TrkX pipeline, the models were minimally modified, and hence were extremely heavy at this point, resulting in low throughput. Particles within the ATLAS and CMS detectors exhibit curved and helical trajectories, which means that the reconstruction models need to be significantly deeper in order to learn the corresponding patterns. In the VELO detector, however, due to the lack of any magnetic field, the particles move in straight lines and hence the pattern recognition task becomes simpler. Because of this, we were able to reduce the size of the models. The embedding MLP was reduced from around 200\,000 to 251 parameters, while the GNN was reduced from two million parameters down to only 70\,000, as shown in Fig.~\ref{fig:model-size}. With these smaller models, and after a series of optimizations of the algorithms for throughput, we arrive at the final ETX4VELO pipeline, which is presented next.

\section{The ETX4VELO Pipeline on GPU}
\label{sec:etx4velo-gpu}

I present the most performant version of the ETX4VELO pipeline implemented on GPUs~\cite{gdl4hep_etx4velo_cuda_nodate} and discuss the various optimizations and algorithmic reimplementations within the pipeline that contributed to its improvement. It is implemented in C++/CUDA within the Allen framework, the first-level trigger of LHCb on GPUs. This implementation utilizes the optimized models described in Section~\ref{sec:etx4velo} as well as Allen's components for tasks such as memory management, event loading, dispatching, and VELO hit decoding. The classical Allen reconstruction pipeline operates under specific constraints, processing 500 events across 16 CUDA streams while allocating 500 MB of GPU memory per stream. To ensure fair comparison, the ETX4VELO pipeline adheres to these same parameters.

Computational throughput is evaluated using Nvidia GeForce RTX 2080 Ti and GeForce RTX 3090 GPUs, with 50 repetitions of the pipeline to minimize the influence of I/O overhead on the measurements. The pipeline currently includes the following steps: (1) embedding network inference, (2) k-NN algorithm, (3) GNN inference up to the edge classifier, and (4) WCC algorithm. The track-building step from edge triplets, however, remains a future task.

\subsection{Structure of Data in Allen}

\begin{figure}
    \centering
    \includegraphics[width=0.7\linewidth]{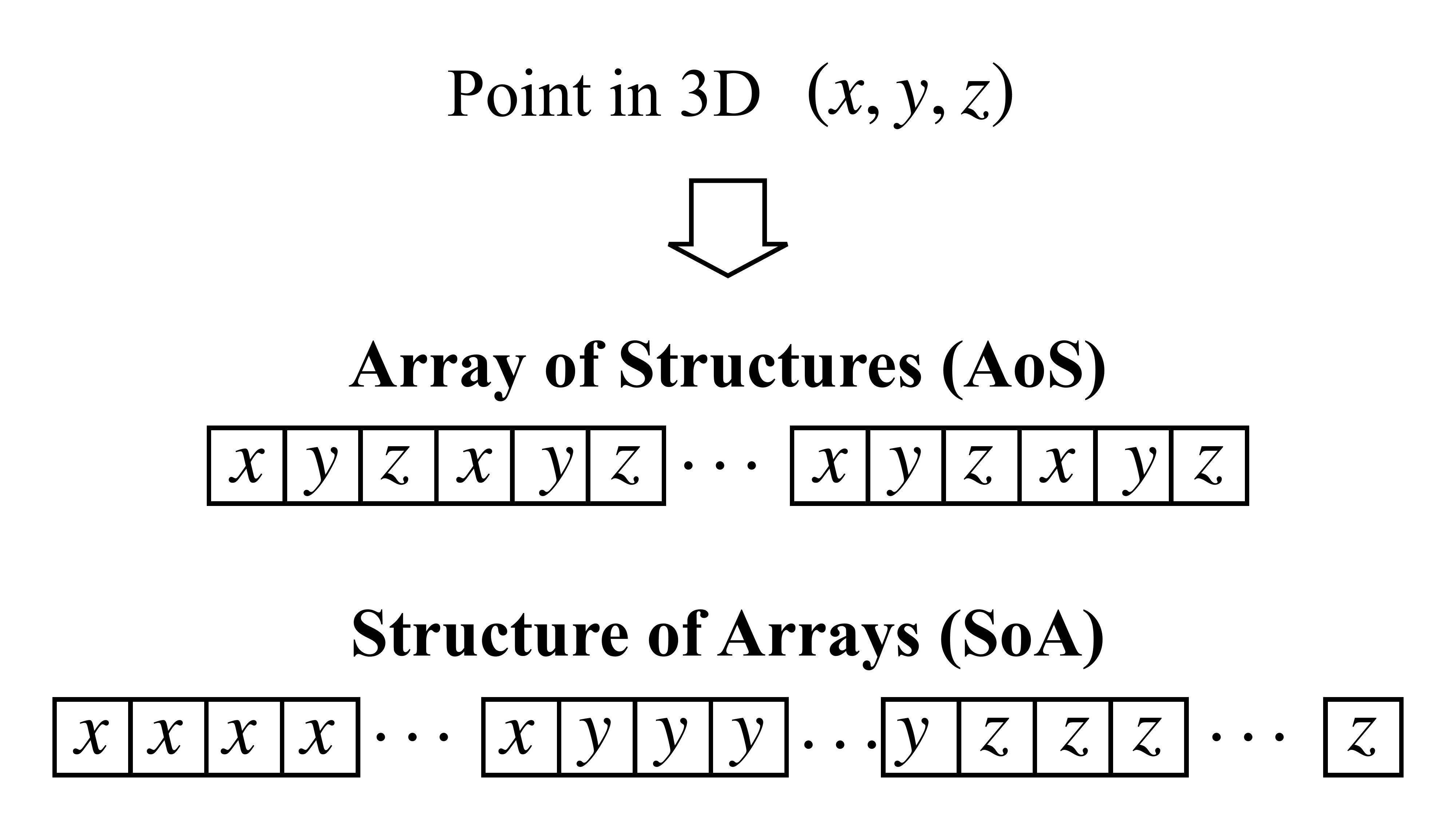}
    \caption{Illustration of combining and storing various points in 3-dimensional space in physical memory under two different memory layouts. Inspired by~\cite{mniszewski_enabling_2021}.}
    \label{fig:aos-soa}
\end{figure}

In order for Allen to leverage the parallelization capabilities of the GPU, the data need to be in a certain format. The various CUDA threads perform the same operations on, for example, various hits at the same time. This happens under the Single Instruction, Multiple Threads (SIMT) execution model, which is similar to SIMD instructions with extra per-thread autonomy. For this to be efficient, the hit data need to be contiguous in memory, so that with a single instruction they can be loaded onto the registers. This layout is known as Structure of Arrays (SoA) and most data in Allen are stored in this format~\cite{vom_bruch_workshop_2021,esen_fast_2023}. SoA is often contrasted with the Array of Structures (AoS) layout, which is more intuitive and frequently used in object-oriented programming. AoS and SoA are shown in Fig.~\ref{fig:aos-soa}.

\begin{figure}
    \centering
    \includegraphics[width=1\linewidth]{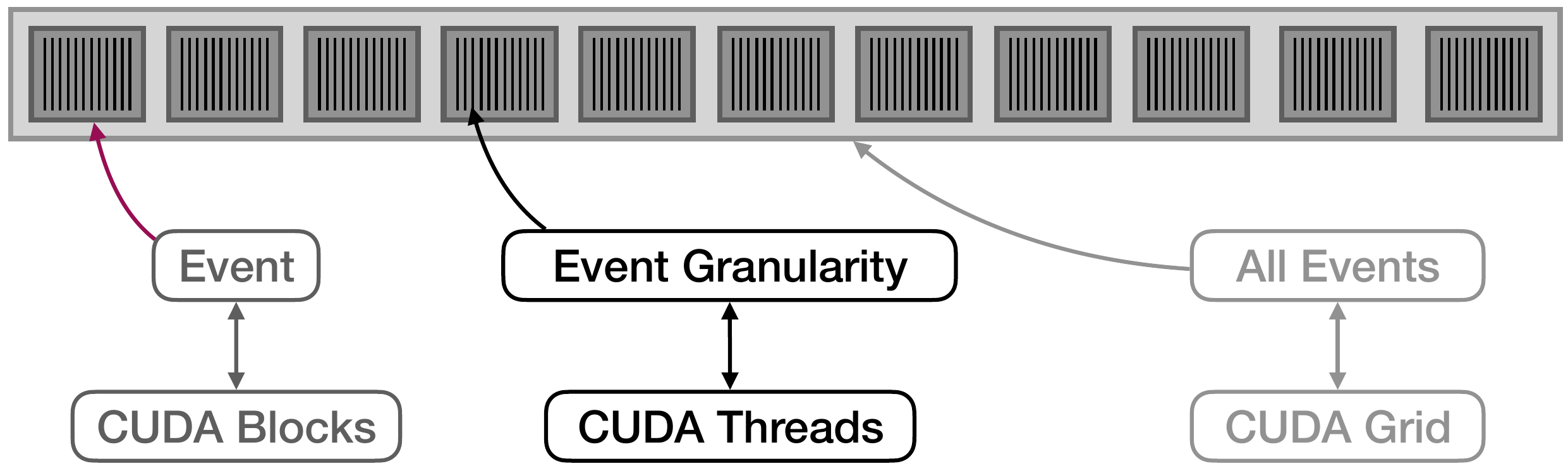}
    \caption{The conventional two-level parallelization scheme used in Allen. Events are mapped to CUDA blocks and executed in parallel. The processing is accelerated further by utilizing parallelism within each event to perform operations at finer granularities.}
    \label{fig:allen-parallel-1}
\end{figure}

In this way, most Allen algorithms benefit from a two-level parallelization scheme. The convention usually is as follows: LHCb events, being entirely independent physics events, are each mapped to CUDA blocks that are executed concurrently. Within each block, intra-event parallelism is exploited to accelerate operations at finer granularities---such as performing operations involving clusters or tracks within the event~\cite{campora_perez_ep_2019}. This scheme is illustrated in Fig.~\ref{fig:allen-parallel-1}. In fact, the number of events processed by a single GPU turns out to be a key factor in the LHCb workload.

In addition, because of this one-dimensional layout of data in Allen, ``offsets'' are needed to know where specifically the data of interest can be found within the array. This can be efficiently implemented using prefix sums, and that is why prefix sums repeatedly appear throughout the Allen codebase. Typically, prefix sums are performed on the host, but since recently, some of these calculations are also deployed on the GPU side~\cite{lhcballen_merge_2024}.

Furthermore, to avoid race conditions, locks need to be used in order to provide mutual exclusion between the threads. In other words, while a thread is executing a critical set of operations, the other threads should be locked out of accessing and modifying the corresponding data. This, in Allen, is implemented with \textit{atomic operations:} operations that need to be executed without interruption and as a single, indivisible unit~\cite{barry_modern_2012}. In this way, consistency and integrity of shared data structures is ensured.

Finally, Allen, as most components of the event reconstruction, use single precision data types, namely 32-bit floats. This improves performance because using single precision allows twice as many numbers to fit in both the cache line and registers compared to 64-bit double precision.

\subsection{Network Inference}

\begin{figure}
    \centering
    \includegraphics[width=0.6\linewidth]{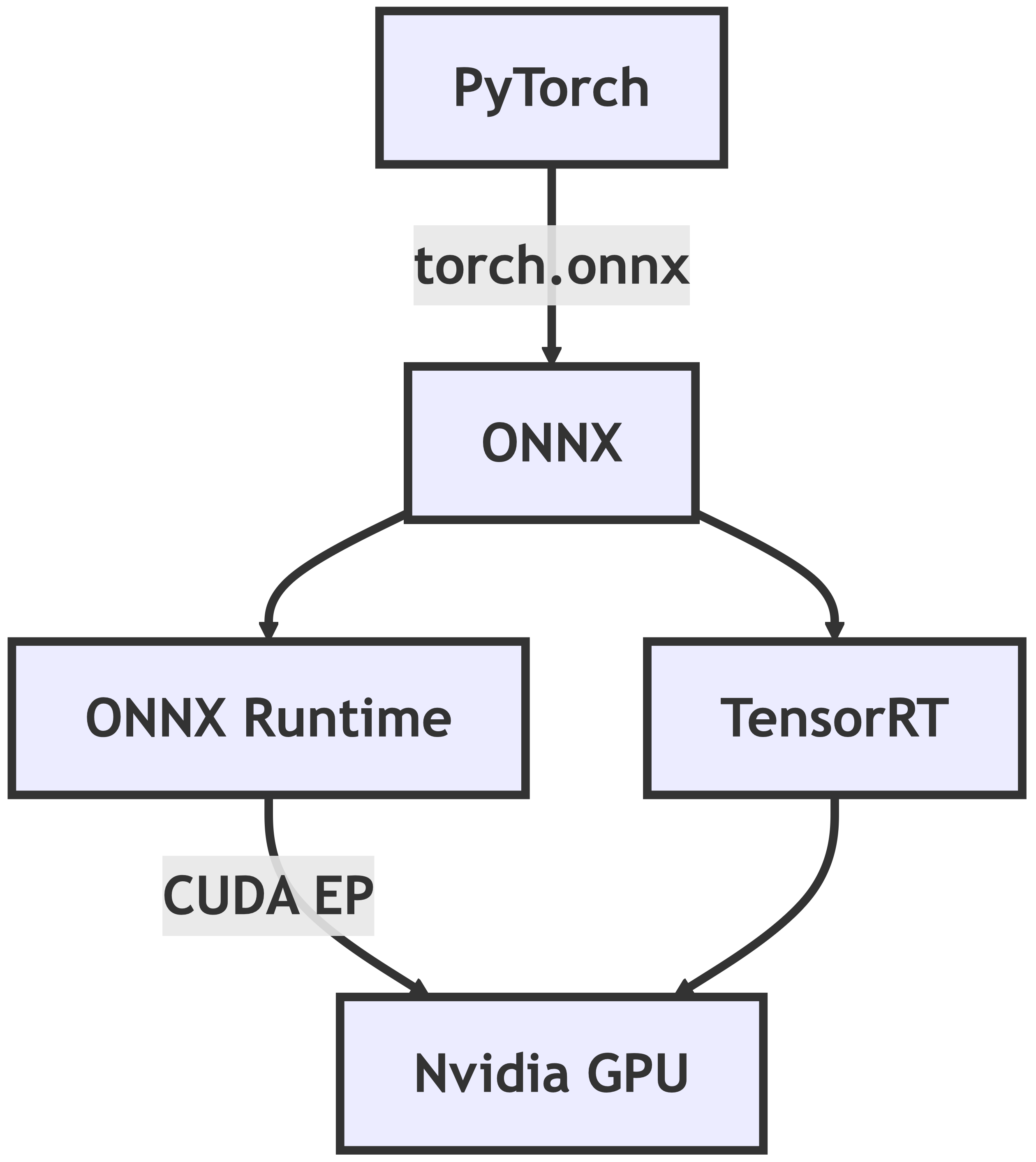}
    \caption{The process of deploying ML models trained in PyTorch on an Nvidia GPU using the ONNX format, and the ONNX Runtime and TensorRT inference engines. ONNX Runtime's CUDA Execution Provider (EP) is used. Generated with~\cite{sveidqvist_mermaid_2014}.}
    \label{fig:inference}
\end{figure}

Inference for the embedding network and the GNN, both trained in PyTorch, is performed using inference engines. We experimented with ONNX Runtime (ORT) leveraging its CUDA backend and TensorRT (TRT), as shown in Fig.~\ref{fig:inference}. Both engines require the PyTorch models to be exported with ONNX, using the \texttt{torch.onnx} method. TensorRT integrates seamlessly with the Allen framework for memory management, making it particularly well-suited for real-time inference and production workflows. ONNX Runtime, on the other hand, does not natively support this feature but provides the flexibility of a CPU backend, enabling easy transitions between GPU and CPU-based pipelines. To set up the environment dependencies for model inference in Allen, a CUDA-enabled cross-compilation toolchain was used~\cite{clemencic_cmake-based_2012}.

\begin{figure}
    \centering
    \includegraphics[width=1\linewidth]{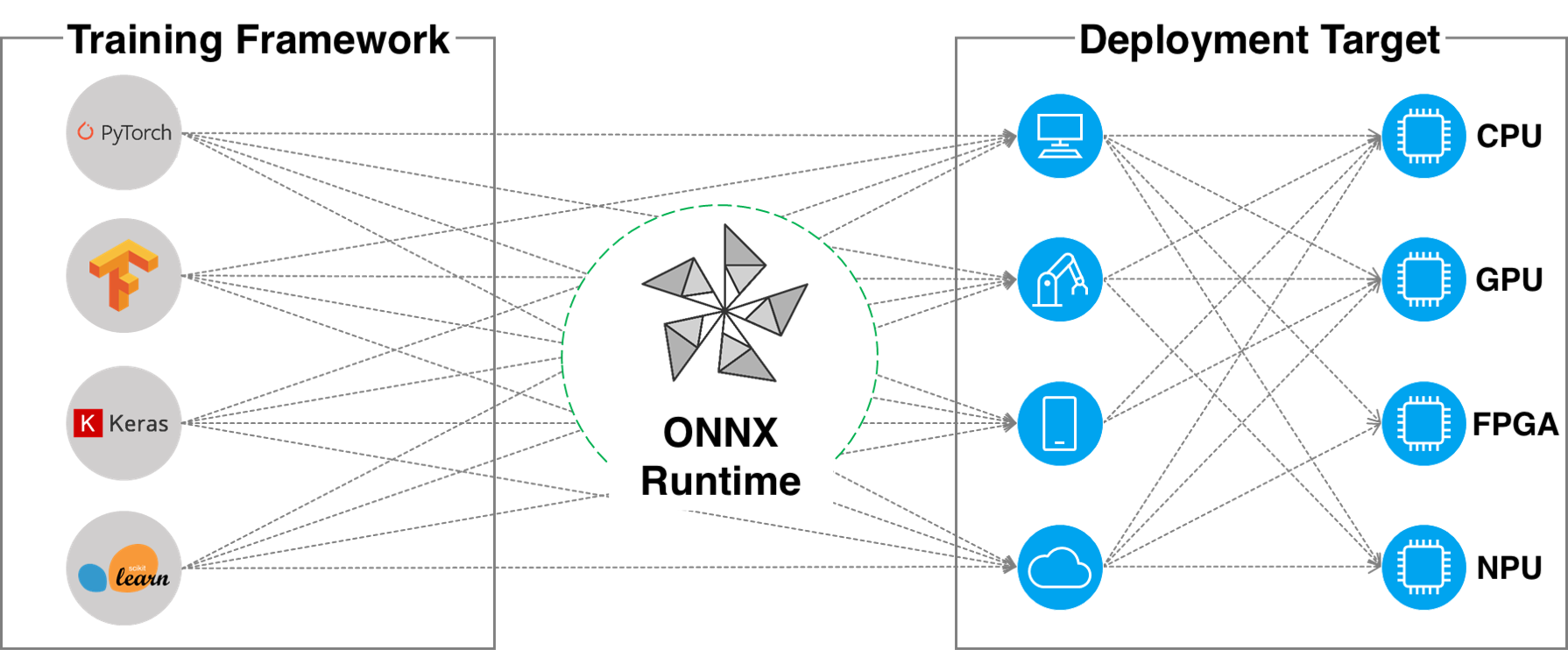}
    \caption{Illustration of the function of ONNX Runtime for different training frameworks and different deployment targets, including CPUs, GPUs, FPGAs and Neural Processing Units (NPUs)~\cite{tan_deep_2021}. Adapted from~\cite{onnx_runtime_execution_nodate}.}
    \label{fig:ep}
\end{figure}

\begin{figure}
    \centering
    \includegraphics[width=1\linewidth]{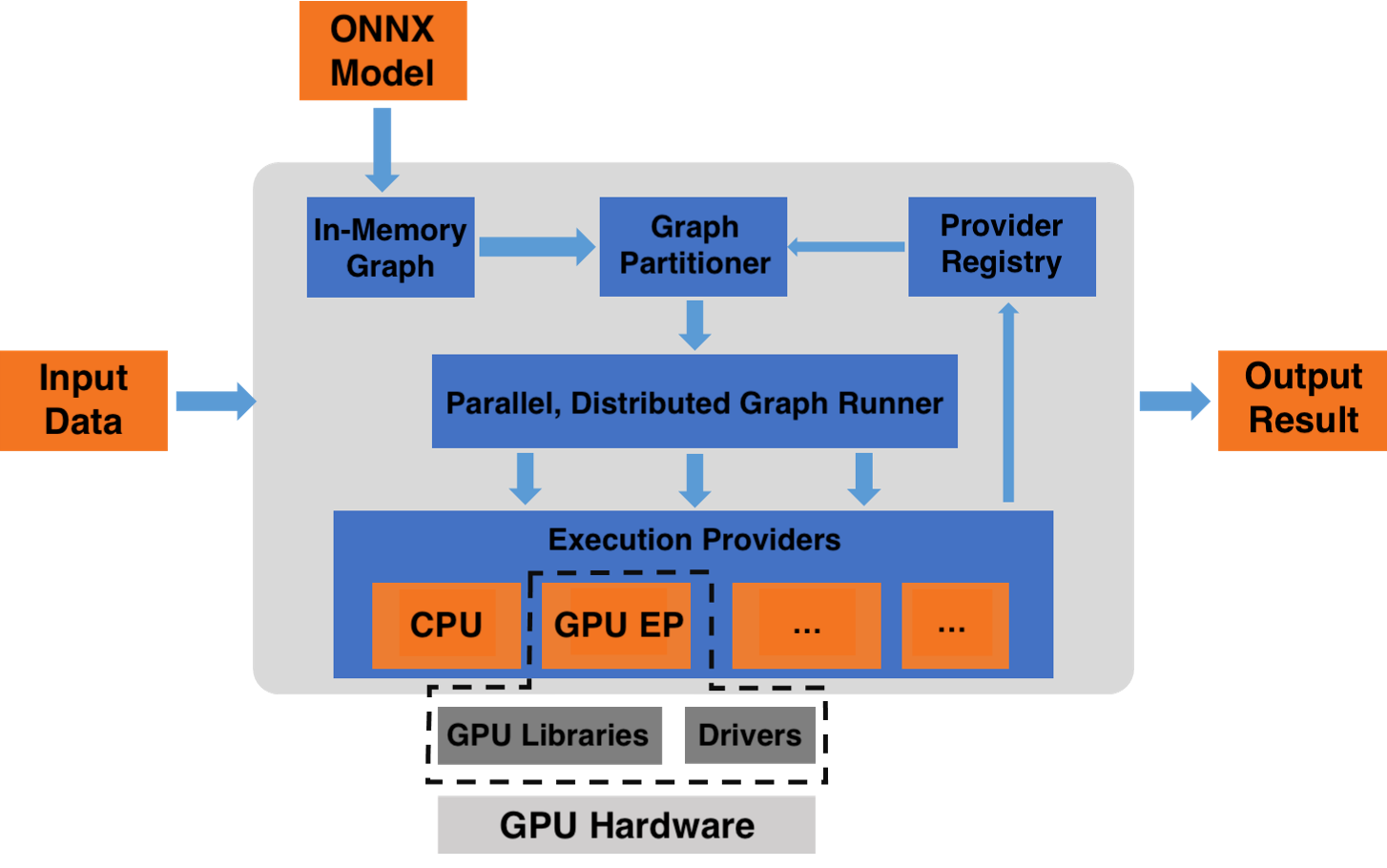}
    \caption{The process of executing an ONNX exported model using ONNX Runtime on a GPU using the GPU Execution Provider (EP). Adapted from~\cite{onnx_runtime_execution_nodate}.}
    \label{fig:ep-2}
\end{figure}

ONNX Runtime uses its extensible Execution Providers (EPs) framework to integrate with a variety of hardware acceleration libraries, ensuring ONNX models run on any platform, as shown in Figs.~\ref{fig:ep} and \ref{fig:ep-2}. This flexible interface lets application developers deploy their models across both cloud and edge environments while fully exploiting each platform's compute capabilities.

\begin{figure}
    \centering
    \includegraphics[width=1\linewidth]{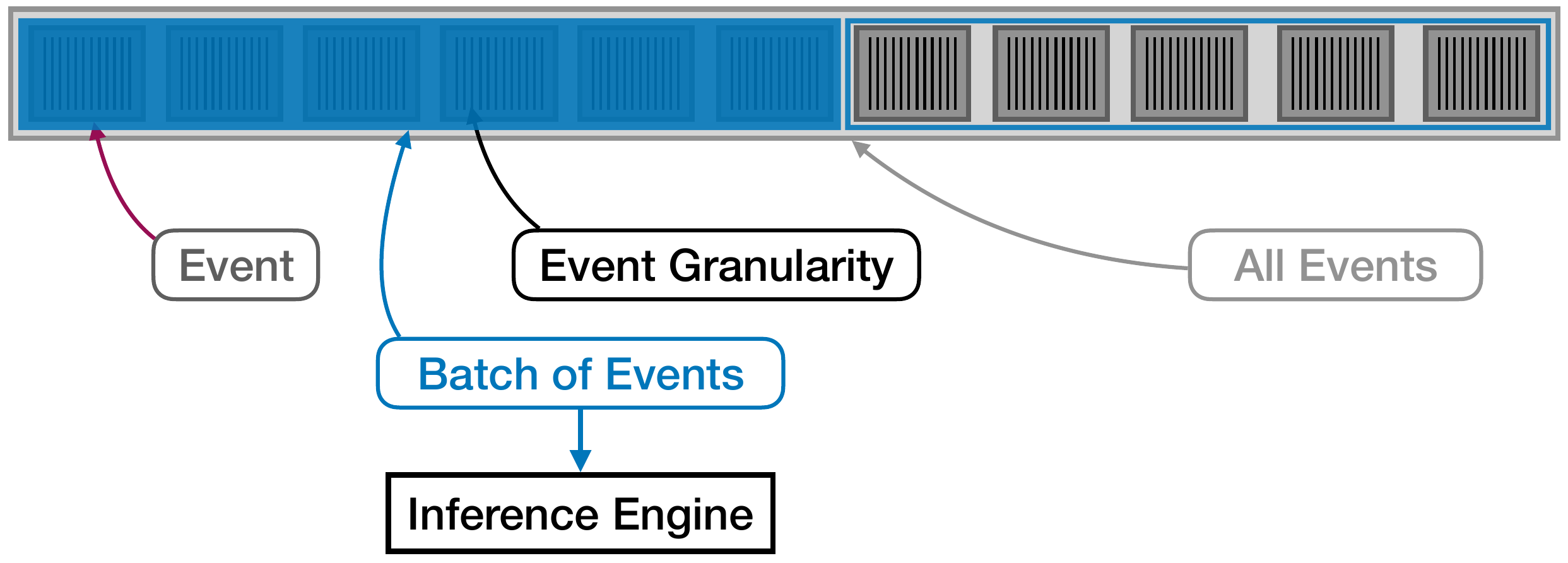}
    \caption{The different LHCb events are batched together and passed on to the corresponding inference engine, such as ONNX Runtime or TensorRT. While staying within memory constraints, this ensures maximum parallelization and acceleration.}
    \label{fig:allen-parallel-2}
\end{figure}

For embedding inference, Allen processes 500 events per CUDA stream, passing all the hits from these events directly to ONNX Runtime or TensorRT. This approach ensures maximum parallelization while staying within memory constraints. For GNN inference, events are grouped into batches containing up to $2^{20}$ hits and $2^{22}$ edges, the largest feasible size for GPU memory, allowing the pipeline to handle variations in event dimensions. This process of batching events together in order to accelerate the model inference is illustrated in Fig.~\ref{fig:allen-parallel-2}.

Support for the \texttt{scatter\_add} operation during message-passing proved challenging for both ONNX Runtime and TensorRT. The latest ONNX Runtime release (v18) now includes native support for this operation. In contrast, TensorRT required the implementation of a custom plugin to handle it. Although TensorRT 10.0 and newer versions provide partial support for \texttt{scatter\_add}, this support is limited to specific data type combinations, excluding certain configurations like INT8.

\subsection{k-NN Implementation}

The algorithm is based on the standard k-nearest neighbor search algorithm, where the top $k$ nearest neighbors need to be identified, as shown in Algorithm~\ref{alg:knn}.

\begin{algorithm}
\caption{$k$ Nearest Neighbors Search}
\label{alg:knn}
\begin{algorithmic}[1]
\REQUIRE Dataset $X = \{x_i\}_{i=1}^n$, $x_i \in \mathbb{R}^d$, query point $x_q$, number of neighbors $k$
\FOR{$i = 1$ to $n$}
    \STATE $d_i \gets \sqrt{ \sum_{j=1}^{d} (x_i^{(j)} - x_q^{(j)})^2 }$
\ENDFOR

\STATE Create list $L = \{(i, d_i)\}_{i=1}^{n}$

\STATE Sort $L$ by $d_i$ in ascending order

\STATE Select first $k$ elements from $L$ to get $k$ nearest neighbors

\RETURN Indices of $k$ nearest neighbors
\end{algorithmic}
\end{algorithm}

A sequential loop is used to iterate through the hits in planes $p + 1$ and $p + 2$ for each node in plane $p$, calculating the squared distance in the embedding space. These iterations are performed in parallel both across different hits on a single plane and across multiple planes. When the squared distance is smaller than the maximum squared distance $d^2_\text{max}$, the hit is added to the list of the $k_\text{max}$ nearest neighbors. If the list is already full, the farthest neighbor in the list is replaced by the new hit if it is closer. However, this replacement occurs rarely, with less than 0.1\% of hits having more than 50 neighbors.

\subsection{WCC Implementation}

After all the previous steps of the pipeline, we end up with a big graph that contains various components, the different tracks, that are disconnected with each other. In order to get the tracks, we have to efficiently break this graph apart into its constituents. In other words, we need to identify the ``weakly connected components'' of this graph, and for this we can use various graph traversal algorithms. Our implementation is based on the Depth-First Search (DFS) algorithm~\cite{cormen_introduction_2009}, shown in Algorithms~\ref{alg:dfs-1} and \ref{alg:dfs-2}. For a graph $G = (V, E)$, $\text{Adj}[u]$ is the set of vertices adjacent to vertex $u \in V$. The graph is traversed by starting at some root node and exploring as far as possible along each branch before backtracking.

\begin{algorithm}
\caption{Depth-First Search}
\label{alg:dfs-1}
\begin{algorithmic}[1]
\REQUIRE Graph $G = (V, E)$
\FOR{each vertex $u \in V$}
    \STATE $u.\text{color} \gets \text{WHITE}$
    \STATE $u.\text{parent} \gets \text{NULL}$
\ENDFOR
\FOR{each vertex $u \in V$}
    \IF{$u.\text{color} = \text{WHITE}$}
        \STATE \textbf{DFS-Visit}$(u)$
    \ENDIF
\ENDFOR
\end{algorithmic}
\end{algorithm}

\begin{algorithm}
\caption{DFS-Visit$(u)$}
\label{alg:dfs-2}
\begin{algorithmic}[1]
\STATE $u.\text{color} \gets \text{GRAY}$
\FOR{each $v \in \text{Adj}[u]$}
    \IF{$v.\text{color} = \text{WHITE}$}
        \STATE $v.\text{parent} \gets u$
        \STATE \textbf{DFS-Visit}$(v)$
    \ENDIF
\ENDFOR
\STATE $u.\text{color} \gets \text{BLACK}$
\end{algorithmic}
\end{algorithm}

Our custom implementation takes advantage of the planar structure of the VELO detector. The goal is to assign a unique connected component label to each node, indicating which track the node is part of. Initially, each node is assigned a distinct label, usually its own index. Then, in parallel, the label of each node in plane $p$ is updated to the smallest label among its connected nodes on the left, with the process progressing sequentially from plane 1 to plane 25 (where the 26 VELO planes are numbered from 0 to 25). If a node on the right is connected to more than two nodes on the left, it will only update its label to match one of these nodes, leaving one left-side node without the correct label. To address this, the procedure is repeated in reverse order, from plane 24 to plane 0, taking into account the connections on the right.

\subsection{Quantization}
\label{sec:mlp-quantization}

The embedding MLP model was quantized to INT8 precision. I performed Post-Training Quantization (PTQ) using Nvidia's PyTorch-Quantization library~\cite{nvidia_developers_pytorch-quantization_nodate}, targeting the TensorRT backend. This approach utilizes the 8-bit tensor cores on Nvidia GPUs, instead of the standard CUDA cores for matrix-multiplication tasks, resulting in higher computational throughput.

Quantization in TensorRT is handled in the following way~\cite{nvidiatensorrt_working_nodate}. When a model is quantized, the operators ``QuantizeLinear (Q)'' and ``DequantizeLinear (DQ)'' are added inside the ONNX computation graph, in order to simulate quantization. Later, when this graph is processed by the TensorRT builder, all possible optimizations and fusions are done. For example, when the builder sees a series (DQ, DQ) $\rightarrow$ Node $\rightarrow$ Q, e.g., for a node that takes in two input tensors and outputs a single tensor, it fuses it into a ``Quantized Node (QNode)''. That is, while Node operates on full float values, QNode operates on quantized values. Therefore, during inference, the calculation happens on quantized values and hence is sped up.

\begin{figure}
    \centering
    \includegraphics[width=1\linewidth]{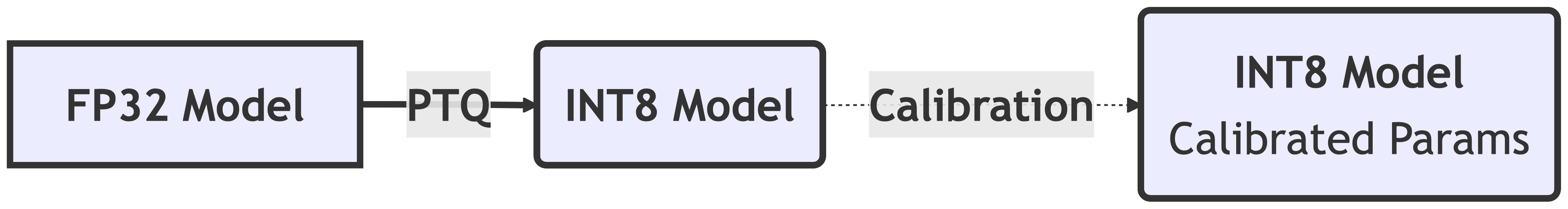}
    \caption{Illustration of the process of performing Post-Training Quantization (PTQ) and calibration of a model with 32-bit floating-point precision, down to a model with 8-bit integer precision. After PTQ, the quantization parameters of the quantized model are calibrated using a representative data sample that reflects the intended deployment scenario. Generated with~\cite{sveidqvist_mermaid_2014}.}
    \label{fig:calib}
\end{figure}

Without calibration, the model's drop in precision resulted in the creation of roughly 80\% more edges in the rough graph---for example, increasing from 29\,000 to around 52\,000 edges in a single event. This proved detrimental to the throughput of the pipeline, since the large number of edges is a major aspect of the throughput limitations of the pipeline. To address this, I calibrated the model's quantization parameters using 5000 events. The performance of the pipeline with the INT8 version of the embedding is presented in Table~\ref{tab:quantization_performance}, with a fake rate of 1.72\%. In this setup, the rest of the pipeline remains in FP32 precision. After calibration, the embedding increases the number of edges by only roughly 5--10\%. The overall pipeline performance is therefore minimally reduced. The quantization and calibration process is illustrated in Fig.~\ref{fig:calib}.

\begin{table}
\begin{center}
\centerline{
\begin{tabular}{l|cc|cc|cc|cc}
\hline\hline
Long & \multicolumn{2}{c|}{Efficiency} & \multicolumn{2}{c|}{Clone Rate} & \multicolumn{2}{c|}{Hit Efficiency} & \multicolumn{2}{c}{Hit Purity} \\
\multicolumn{1}{l|}{} & INT8 & FP32 & INT8 & FP32 & INT8 & FP32 & INT8 & FP32 \\
\hline
\multicolumn{1}{l|}{No Electrons} & 97.66 & 97.96 & 0.77 & 0.88 & 99.95 & 98.42 & 98.23 & 99.95 \\
\multicolumn{1}{l|}{Electrons} & 58.50 & 51.82 & 2.41 & 0.93 & 96.39 & 96.46 & 92.39 & 95.05 \\
From Strange & 89.03 & 92.23 & 1.27 & 0.61 & 99.73 & 96.39 & 94.07 & 99.77 \\
\hline\hline
\end{tabular}
}
\caption{Track-finding performance (in percentages) of the ETX4VELO pipeline for long particles using the FP32 embedding MLP versus the INT8 version. For the INT8 case, the rest of the pipeline remains in FP32 precision. Reproduced from~\cite{correia_graph_2024}.}
\label{tab:quantization_performance}
\end{center}
\end{table}

Quantization of the GNN has not yet been accomplished. This is because \texttt{scatter\_add}, an operation needed for message passing and hence integral to the functioning of the GNN, at the time of development, was not natively supported by the TensorRT 10.0 release. A custom plugin for this operation has been developed for single precision, but for INT8 quantization to be applied to the GNN, the plugin needs to support this precision as well. Since, INT8 was also not supported~\cite{nvidiatensorrt_scatterelements_nodate}, and enabling this support would be non-trivial~\cite{nvidiatensorrt_issue_nodate}, I left it for future work. Nevertheless, the quantization of the GNN holds the most potential for significant throughput improvements.

\subsection{Physics Performance}

The physics performance comparison between the GPU implementation, without the inclusion of the triplet methodology, and the PyTorch trained version in Chapter~\ref{ch:etx4velo}, is summarized in the same chapter in Figs.~\ref{tab:long_performance}, \ref{tab:velo_only_performance} and \ref{tab:fake_rate} in Section~\ref{sec:etx4velo-performance}.

\section{Computational Performance}
\label{sec:etx4velo-gpu-performance}

\begin{figure}
    \centering
    \includegraphics[width=\linewidth]{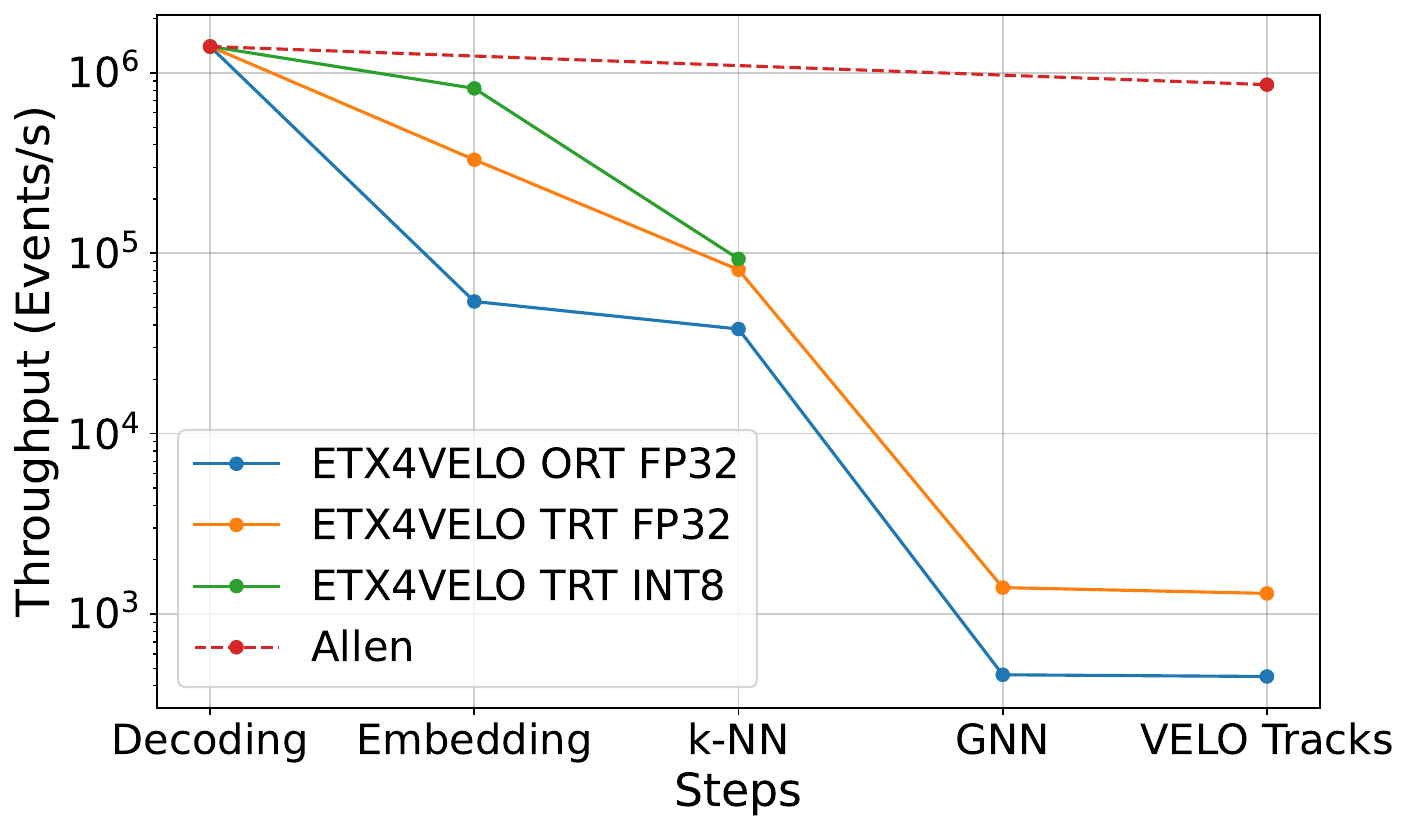}
    \caption{Throughput comparison of track reconstruction in the VELO on an Nvidia GeForce RTX 3090. Adapted from~\cite{correia_graph_2024}.}
    \label{fig:throughput-comparison-3090}
\end{figure}

The current throughput of the pipeline is detailed in Table~\ref{tab:gpu-results-1} and Table~\ref{tab:gpu-results-2}, which present results for two different GPU cards. Fig.~\ref{fig:throughput-comparison-3090} provides a visual representation of the throughput progression across the pipeline steps for one of these cards, alongside a comparison with Allen. The architectures of the GPU cards, which are based on the TU102~\cite{techpowerup_nvidia_nodate-1} and GA102~\cite{techpowerup_nvidia_nodate-2} dies, are compared in Table~\ref{tab:gpu-architecture}. 

The reported throughput values are obtained using Allen's built-in throughput timer. Throughput measurements correspond to specific stages of the pipeline, captured under the ``up to step'' column. The comparison starts from the decoding of the VELO, which involves the unpacking and translation of the binary information from the subdetector readout into meaningful hits or clusters corresponding to particle interactions with the detector. It is then followed by the usual embedding, k-NN, GNN and track building steps. Additional data includes the number of streams and memory usage per stream. For ETX4VELO, throughput is presented in three categories: ``ORT FP32'', ``TRT FP32'', and ``TRT INT8'', indicating the inference engine and precision used for the machine learning models.

The TensorRT implementation in FP32 demonstrates a substantial performance advantage over the ONNX Runtime equivalent. This is particularly evident in the embedding step, where TensorRT achieves a throughput of 260\,000 events per second compared to ONNX Runtime's 46\,000. Moreover, TensorRT requires less memory, facilitating simultaneous execution of the GNN across multiple streams. However, both implementations see a sharp decline in throughput after the k-NN and GNN stages, dropping below 100\,000 and 1000 events per second, respectively. Notably, the INT8-quantized TensorRT implementation of the embedding MLP achieves an impressive 540\,000 throughput after the embedding step. Despite this, the throughput falls significantly to 67\,000 after the k-NN stage. The current k-NN implementation lacks parallelization over neighbors, suggesting potential for future optimization to address this bottleneck.

\begin{table}
\setlength{\tabcolsep}{3pt}
\begin{center}
\centerline{
\begin{tabular}{l|c|c|ccc}
\hline\hline
Up to Step & Streams & Memory per Stream (MB) & \multicolumn{3}{c}{Throughput (Events/s $\times 10^{\rule{0pt}{1.5ex}3}$)} \\
\multicolumn{1}{l|}{} & & & ORT FP32 & TRT FP32 & TRT INT8 \\
\hline
VELO Decoding & 16 & 500 & \multicolumn{3}{c}{770} \\
Embedding & 16 & 500 & 46 & 260 & 540 \\
k-NN & 16 & 500 & 28 & 53 & 67 \\
GNN & 4 (1) & 2000 (9600) & 0.32 & 0.86 & - \\
VELO Tracks (WCC) & 4 (1) & 2000 (9600) & 0.32 & 0.85 & - \\
\hline\hline
\end{tabular}
}
\caption{Throughput of the GPU implementation of ETX4VELO on Nvidia GeForce RTX 2080 Ti. The number of streams and memory used for the GNN and WCC step by the ORT pipeline is shown in parentheses. These throughputs should be compared to 530\,000 for the full Allen pipeline ending in VELO tracks. Adapted from~\cite{correia_graph_2024}.}
\label{tab:gpu-results-1}
\end{center}
\end{table}

\begin{table}
\setlength{\tabcolsep}{3pt}
\begin{center}
\centerline{
\begin{tabular}{l|c|c|ccc}
\hline\hline
Up to Step & Streams & Memory per Stream (MB) & \multicolumn{3}{c}{Throughput (Events/s $\times 10^{\rule{0pt}{1.5ex}3}$)} \\
\multicolumn{1}{l|}{} & & & ORT FP32 & TRT FP32 & TRT INT8 \\
\hline
VELO Decoding & 16 & 500 & \multicolumn{3}{c}{1400} \\
Embedding & 16 & 500 & 54 & 330 & 820 \\
k-NN & 16 & 500 & 38 & 81 & 93 \\
GNN & 8 (1) & 2500 (9600) & 0.46 & 1.4 & - \\
VELO Tracks (WCC) & 8 (1) & 2500 (9600) & 0.45 & 1.3 & - \\
\hline\hline
\end{tabular}
}
\caption{Throughput of the GPU implementation of ETX4VELO on Nvidia GeForce RTX 3090. The number of streams and memory used for the GNN and WCC step by the ORT pipeline is shown in parentheses. These throughputs should be compared to 860\,000 for the full Allen pipeline ending in VELO tracks. Adapted from~\cite{correia_graph_2024}.}
\label{tab:gpu-results-2}
\end{center}
\end{table}

\begin{table}
\centering
\begin{tabular}{lcc}
\hline \hline
Specification & GeForce RTX 2080 & RTX 3090 \\
\hline
Microarchitecture       & Turing    & Ampere   \\
Die       & TU102    & GA102   \\
Die size                & 754 mm\textsuperscript{2} & 628 mm\textsuperscript{2} \\
Transistors             & 18.6 billion & 28.3 billion \\
Streaming Multiprocessors (SMs) & 72 & 84 \\
CUDA cores              & 4608     & 10\,752   \\
Tensor cores            & 576       & 336      \\
L1 cache (total)        & 6.75~MB & 10.5~MB \\
L1 cache per SM        & 96~KB & 128~KB \\
L2 cache                & 6~MB      & 6~MB     \\
Thermal design power    & 280~W     & 350~W    \\
Memory    & 11~GB     &   24~GB  \\
\hline \hline
\end{tabular}
\caption{Comparison of the architecture of the GeForce RTX 2080 Ti and RTX 3090 Nvidia GPU cards.}
\label{tab:gpu-architecture}
\end{table}

\subsection*{Hardware Specifications}

All benchmarks were performed on the LHCb online network with the configuration below.

\begin{itemize}
    \item CPU: 2$\times$ AMD EPYC 7502, 32 cores each (64 cores total, 128 threads), 2.5~GHz base clock
    \item RAM: 503~GiB
    \item NUMA Configuration: 2 nodes 
    \begin{itemize}
        \item Node 0: CPUs 0--31,64--95
        \item Node 1: CPUs 32--63,96--127
    \end{itemize}
    \item L2 Cache: 32 MiB (64 instances)
    \item L3 Cache: 256~MiB (16 instances)
    \item GPU: CUDA 12.1, driver version 530.30.02
    \item Storage: 894~GB SSD
    \item Operating System: RHEL 9.1, Linux kernel 5.14.0
\end{itemize}

\subsection*{ONNX Runtime vs. TensorRT}

The two inference engines have both pros and cons. Firstly, on one hand, ONNX Runtime demonstrated better support for most operations, without any need to manually implement or customize them. Specifically the \texttt{scatter\_add} operation was supported, while for the TensorRT implementation a plugin had to be implemented. Secondly, ONNX Runtime enables changing the backend architecture, from a GPU to a CPU for example, making it easier to compare between different implementations, while TensorRT is only targeting Nvidia GPUs. 

On the other hand, TensorRT is better documented so the implementations were easier to do. Also, since TensorRT is targeting only GPU, the produced implementations are more optimized, having lower memory footprint and higher throughput. Finally, for the TensorRT memory allocation, Allen's memory manager was used, while we were unable to do this with ONNX Runtime. The comparison is summarized in Table~\ref{tab:ort-trt-comparison}

\begin{table}
\centering
\begin{tabular}{l|l}
\hline \hline
\textbf{ONNX Runtime}                & \textbf{TensorRT}                        \\ \hline
Better out-of-the-box support        & Better documentation                     \\ \hline
CPU backend                          & Lower memory footprint                   \\ \hline
                                     & Higher throughput                        \\ \hline
                                     & Memory managers reconciled more easily   \\ \hline \hline
\end{tabular}
\caption{Comparison between the ONNX Runtime and TensorRT inference engines.}
\label{tab:ort-trt-comparison}
\end{table}

\section{Throughput Scaling Comparison}

\begin{figure}
    \centering
    \includegraphics[width=1\linewidth]{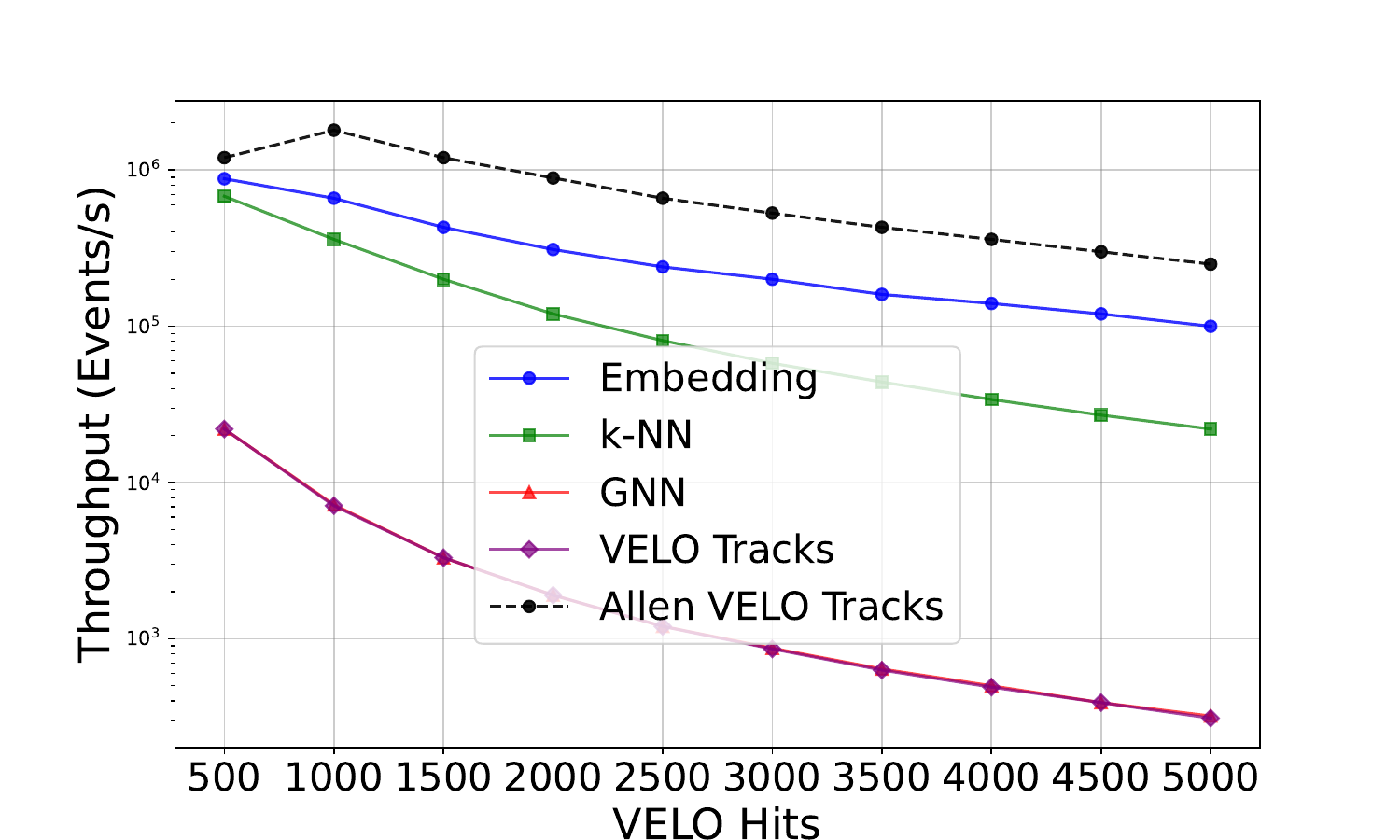}
    \caption{Comparison of the scaling of the throughput as a function of occupancy between the ETX4VELO pipeline and Allen.}
    \label{fig:scaling-comparison-a5000}
\end{figure}

We now study the scaling of the ETX4VELO pipeline with the occupancy of the detector, the number of hits in each event. Events are split into bins based on their occupancy, and the throughput of the tracking algorithms is measured on the events within each bin. The measurement is done for both the ETX4VELO pipeline and Allen on the Nvidia RTX A5000 GPU. For the ETX4VELO pipeline, the throughput is measured for the intermediary and final steps. The comparison is shown in Fig.~\ref{fig:scaling-comparison-a5000}.

\begin{figure}
    \centering
    \includegraphics[width=1\linewidth]{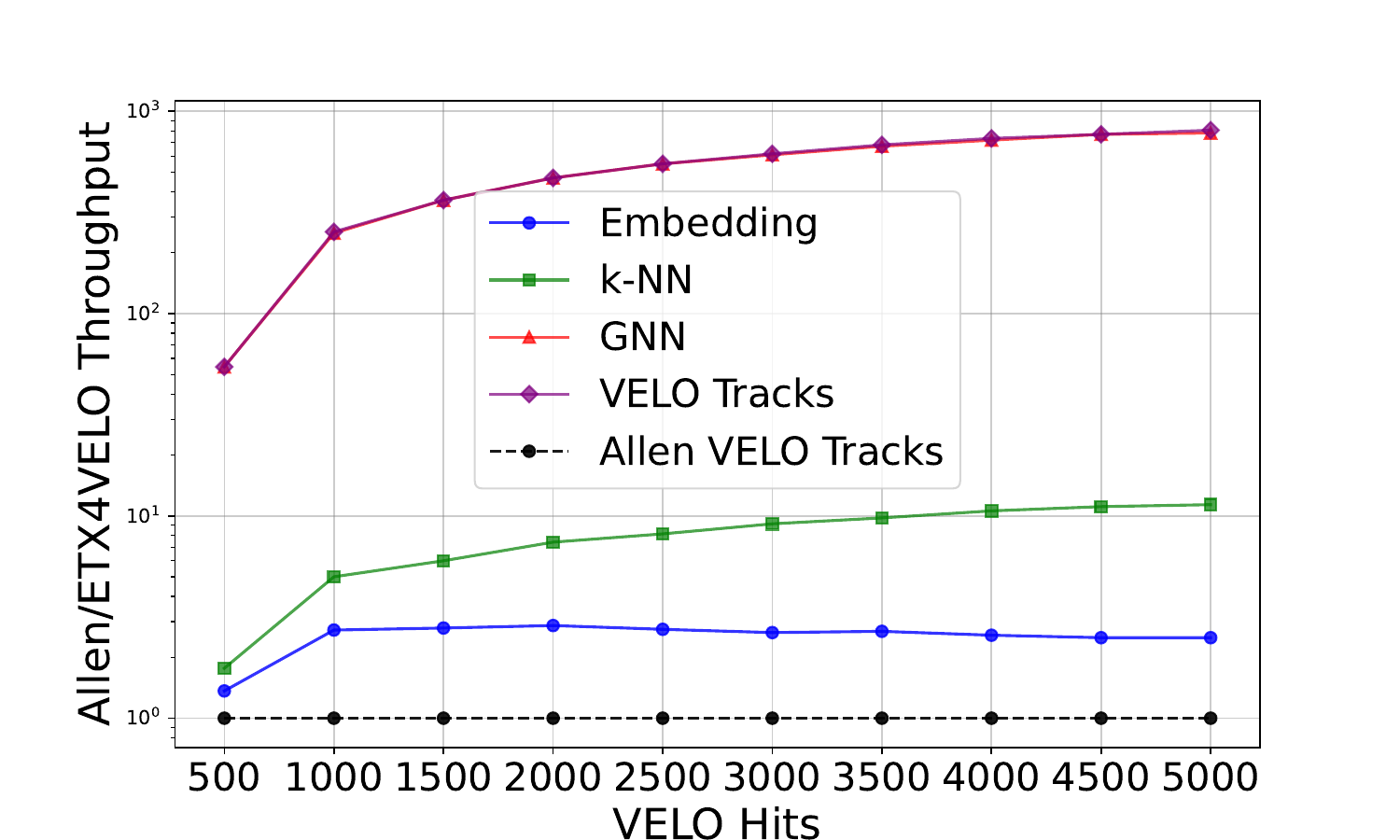}
    \caption{Comparison of the ETX4VELO throughput as a function of occupancy with the Allen one. We plot the ratio of the Allen Throughput divided by the ETX4VELO Throughput.}
    \label{fig:scaling-comparison-relative-a5000}
\end{figure}

\begin{figure}
    \centering
    \includegraphics[width=1\linewidth]{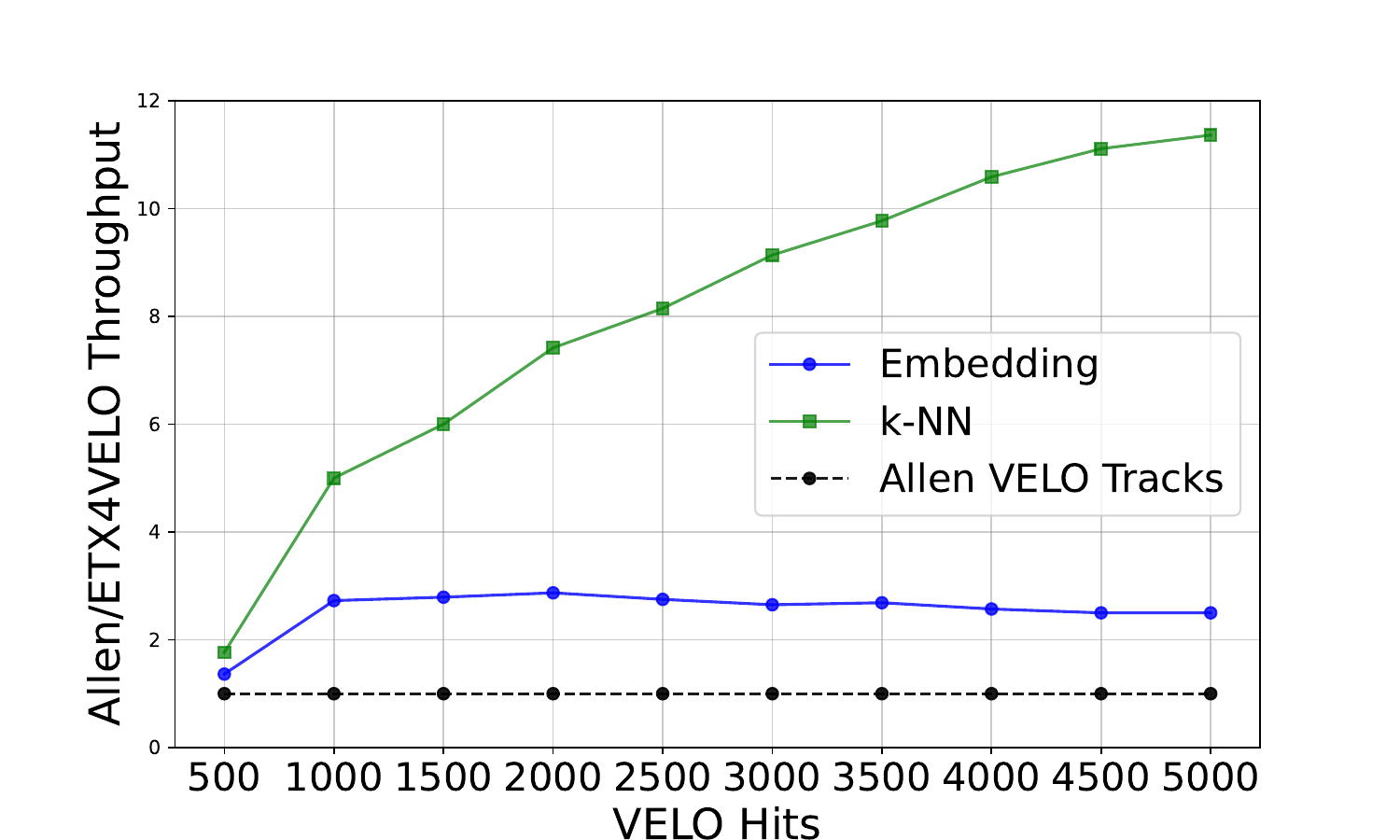}
    \caption{Comparison of the ETX4VELO throughput as a function of occupancy with the Allen one. We plot the ratio of the Allen Throughput divided by the ETX4VELO Throughput.}
    \label{fig:scaling-comparison-relative-linear-a5000}
\end{figure}

Interestingly, the gap between the embedding step and the k-NN is widening with increasing occupancy. In order to study this, in Fig.~\ref{fig:scaling-comparison-relative-a5000} we plot the ratio
\begin{equation}
    \frac{\text{Allen throughput}}{\text{ETX4VELO throughput}}\,.
\end{equation}
The same comparison, on linear axes is shown in Fig~\ref{fig:scaling-comparison-relative-linear-a5000}. It is obvious that the k-NN is one major problem in the scaling of the ETX4VELO throughput. As seen in Fig.~\ref{fig:scaling-comparison-relative-linear-a5000}, the embedding is scaling as well, and possibly better, than the combinatorial Allen algorithms. However, with the increasing number of hits in each event, the k-NN is doing increasingly worse. This is to be expected, since with a larger number of hits in each event, the k-NN has a larger number of distance calculations and comparisons to do.

\section*{Conclusion}

In this chapter, the ETX4VELO GPU implementation inside Allen was presented. The computational performance of the pipeline, excluding the triplet-based methodology presented in Chapter~\ref{ch:etx4velo}, was compared against the classical tracking algorithms currently in place inside the first-level trigger of LHCb. The pipeline was found to significantly underperform in terms of throughput.

In addition, the parallel algorithms---including the k-NN and WCC steps---implemented as CUDA kernels were described. The partial quantization of the pipeline was described, along with the reasons why the quantization of the GNN was ultimately not pursued.

Further work can include the finalization of the GNN quantization, which bears significant promise in increasing the throughput of the pipeline, and potentially narrowing the performance gap between the two algorithms. Finally, the implementation of the triplet-based methodology on GPU would also be a worthwhile pursuit.

\chapter{Accelerating ETX4VELO on FPGA}
\label{ch:etx4velo-fpga}

\minitoc

\noindent Parts of this chapter are adapted from~\cite{giasemis_comparative_2025}. The repository of the project can be found at~\cite{gdl4hep_hls4ml-etx4velo_nodate}. I would like to express my sincere gratitude to my co-author, Vladimir Lončar, for the multiple useful discussions throughout the course of this work.

\section*{Introduction}

In Chapters~\ref{ch:etx4velo} and \ref{ch:etx4velo-gpu}, we saw the ETX4VELO pipeline and its implementation on GPUs inside LHCb's first-level trigger. In this context, Allen offers a platform for deploying and benchmarking ML-based algorithms on GPU, enabling high-throughput inference. However, beyond GPUs, FPGAs also represent a compelling hardware option for such applications. Given the stringent real-time processing requirements of HEP experiments, FPGAs are commonly employed for tasks such as data compression, data acquisition, and high-speed data transmission~\cite{musa_fpgas_2008,summers_application_2018,ahmad_khan_fpga_2023,campora_improvements_2014,cachemiche_pcie-based_2016,mitra_common_2016}. They offer the potential for improved computational and energy efficiency, as they are specifically configured and optimized for a specific task. Moreover, with the growing interest in machine learning within the HEP community, it becomes crucial to assess the suitability of FPGAs for deploying neural network-based algorithms.

Indeed, ML inference on FPGAs has attracted significant attention~\cite{nurvitadhi_can_2017,suda_throughput-optimized_2016,wojcicki_accelerating_2022,bassi_fpga-based_2023,lazzari_fpga-based_2022,fantechi_real-time_2023,haefeli_fpga-based_2006,bassi_real-time_2021,bartz_fpga-based_2020,boser_cnns_2018,denby_fast_2003,blaiech_survey_2019,guo_survey_2019,shawahna_fpga-based_2019}, with several efforts emerging within the high-energy physics community. Hybrid GPU-FPGA designs have also been explored~\cite{liu_hybrid_2020}. Several ML techniques have already been implemented in the hardware components of the LHC trigger system. Notable examples include the use of Boosted Decision Trees (BDTs) for muon momentum inference in the Level-1 CMS trigger~\cite{acosta_boosted_2017}, and a convolutional neural network that replaces a traditional pattern-finding algorithm for hit processing in the Level-0 ATLAS trigger~\cite{giagu_fast_2020}.

More specifically, in LHCb, FPGAs are used for the detector readout. With Upgrade~II of the LHCb detector planned in the 2030s~\cite{lhcb_collaboration_physics_2016}, it is therefore interesting to explore to what extent parts of the pattern recognition algorithms involved in the trigger, currently mostly classical but potentially incorporating more and more machine learning methods, can be moved ``closer'' to the detectors by performing them on the data acquisition FPGA boards, potentially improving the cost-effectiveness and energy efficiency of the experiment.

As machine learning models used in high-level triggers for various experiments grow increasingly complex, potentially incorporating architectures like GNNs, it becomes essential to investigate their foundational components---the Multilayer Perceptron (MLP). A comparative analysis of FPGAs and other processing architectures across different contexts can give valuable insights into their suitability and performance for high-energy physics applications. In this chapter, I focus on the initial stage of the ETX4VELO track reconstruction pipeline for the VELO detector, that includes an MLP. Prior to attempting the acceleration of the full GNN, it is important to test and optimize the associated workflow and tools using a simpler model. The MLP serves as an ideal candidate for this purpose.

The acceleration of the GNN on FPGAs was not attempted for various reasons. Firstly, also considering time constraints, implementing the GNN is significantly harder than the MLP, given the size and complexity of the model. Secondly, even though there have been various attempts to deploy GNNs on FPGAs~\cite{elabd_graph_2022,zhang_review_2024,abi-karam_gengnn_2022,que_ll-gnn_2023,neu_real-time_2024,iiyama_distance-weighted_2021,heintz_accelerated_2020,gimenes_ample_2025}, the architecture might not be the most suitable one. This is largely due to the irregular memory accesses of the algorithm, resulting from the sparse structure of graphs~\cite{que_ll-gnn_2023,gimenes_ample_2025}. 

In greater detail, GNNs face two computational challenges, hindering their applicability in real-time scenarios~\cite{tsoi_symbolic_2023}. Firstly, creating the graph needed as input to the GNN is time-consuming: GNNs often use k-NN or other similar methods to construct these graphs~\cite{stark_equibind_2022,lieret_high_2023}. With brute-force methods, creating these graphs from $n$ points scales like $\mathcal{O}(n^2)$, which can significantly limit the scalability and applicability of the approach. Although more efficient k-d tree-based algorithms reduce the complexity to $\mathcal{O}(n\log n)$, their limited potential for parallelization makes them impractical for real-time use cases~\cite{wieschollek_efficient_2016}.

Secondly, the irregular topology of graphs and the neighborhood aggregation process in GNNs result in non-uniform computations and unpredictable memory access patterns. These factors pose significant challenges to conventional hardware accelerators~\cite{jang_exploiting_2011,hashemi_learning_2018,abadal_computing_2021}, rendering GNNs less appropriate for real-time point cloud processing.

To address these challenges, novel approaches are being explored, including the machine learning technique known as symbolic regression~\cite{soybelman_accelerating_2024,tsoi_symbolic_2023}. This approach replaces the graph-based neural network by substituting each network block with a symbolic function, preserving the graph structure of the data and enabling message passing. Additionally, more modern, transformer-based architectures are being investigated as potential alternatives to GNNs~\cite{miao_locality-sensitive_2024}. Whether a full FPGA implementation of the GNN is ultimately worthwhile still remains to be decided.

In addition, due to the architecture of FPGAs, operations at low precision are considerably faster and more efficient than floating-point operations, and hence quantization is quite standard in the field~\cite{baskin_uniq_2019,coelho_automatic_2021,coelho_ultra_2020,pappalardo_qonnx_2022,di_guglielmo_compressing_2020}. Therefore, 8-bit integer precision was used for the implementation of the embedding. Moreover, pruning is also proven to give remarkable results, reducing the size of the model while keeping the accuracy almost, if not exactly, the same~\cite{ramhorst_fpga_2023}, but this was left for future work.

Furthermore, since various attempts have already been made to abstract out the low-level nuances of FPGA design~\cite{venieris_fpgaconvnet_2016,venieris_toolflows_2018,guan_fp-dnn_2017,noronha_leflow_2018,sharma_high-level_2016}, and since many researchers in the HEP community are not experts in FPGA programming, I used HLS4ML~\cite{duarte_fast_2018,fastml_team_hls4ml_nodate} for the deployment of the models. The Python library is designed to be user-friendly, even for individuals without extensive experience in FPGA design. By lowering the barrier to entry, HSL4ML enables a broader range of scientists to leverage the benefits of FPGA acceleration in their research. At the same time, this abstraction of the hardware enables more complex systems to be integrated onto FPGAs.

The framework consumes the model representation from various frameworks like Keras/TensorFlow~\cite{keras_developers_keras_nodate,abadi_tensorflow_2016} or PyTorch~\cite{paszke_pytorch_2019}, and generates the code used by the high-level synthesis tool in order to generate the Verilog~\cite{gateway_design_automation_verilog_1984} or VHDL~\cite{us_department_of_defense_vhdl_1987} code, effectively hiding all the difficulties of writing low-level RTL code. It is designed for applications where low-latency and high-throughput implementations are critical. 

To explore these considerations concretely, I implemented the ETX4VELO embedding MLP on the PYNQ-Z2 board. Moreover, the GPU implementation presented in Chapter~\ref{ch:etx4velo-gpu} was compared against the Alveo U50 and U250 data center accelerator cards.

\section{Implementation of the Embedding}
\label{sec:fpga-mlp}

\begin{figure}
    \centering
    \includegraphics[width=0.4\linewidth]{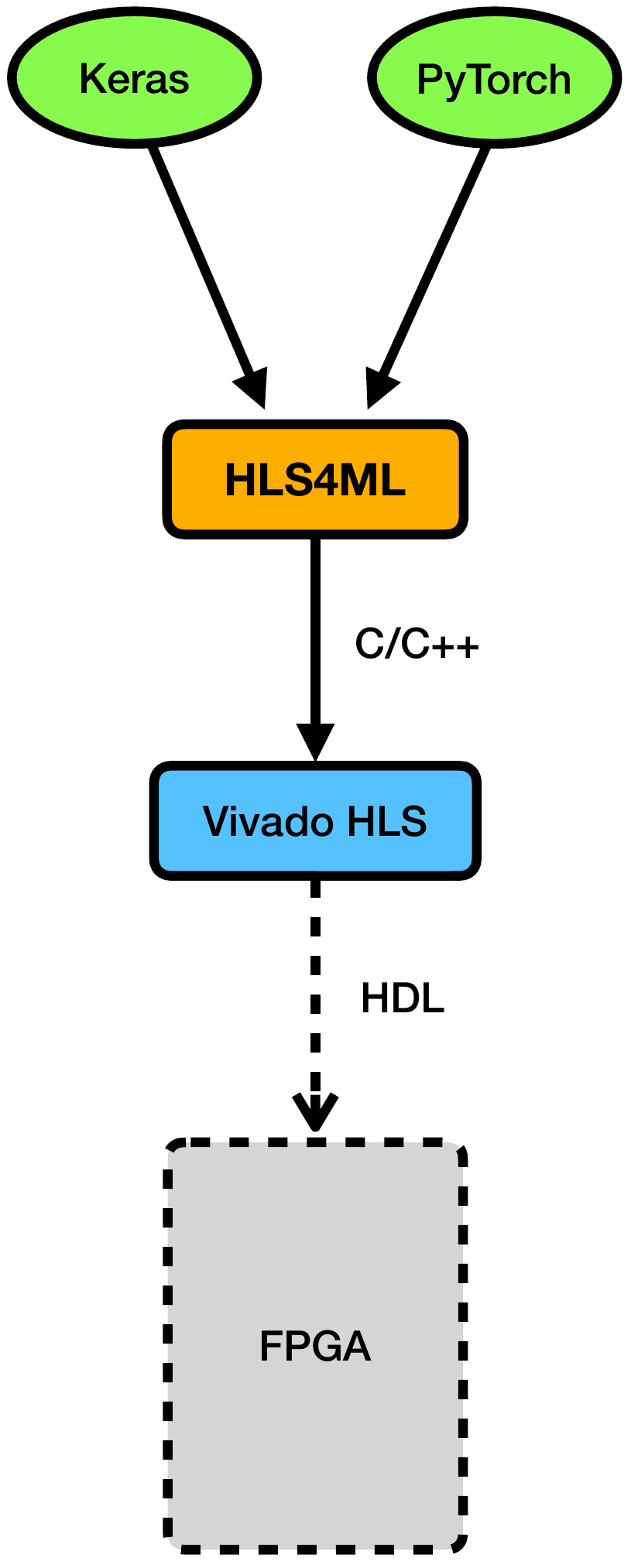}
    \caption{Illustration of the process of converting an ML model trained in PyTorch or Keras to a firmware implementation for FPGAs using the package HLS4ML. An important step of the process is the High-Level Synthesis (HLS) conversion using Vivado or some other HLS tool towards the Hardware Description Language (HDL) implementation on the FPGA.}
    \label{fig:hls4ml}
\end{figure}

The implementation of the ETX4VELO embedding MLP on FPGA hardware is investigated by benchmarking its throughput against the GPU counterpart using the HLS4ML library~\cite{duarte_fast_2018,fastml_team_hls4ml_nodate,fahim_hls4ml_2021}. HLS4ML is a Python package designed for machine learning inference on FPGAs. The models, designed and trained using common ML platforms such as Keras and PyTorch, are converted into firmware implementations for supported FPGA boards through High-Level Synthesis (HLS) tools, such as Vivado~\cite{amd_developers_vivado_nodate} or Vitis~\cite{amd_developers_vitis_nodate}. The code is first transformed from Python to the high-level description of the neural network in HLS C/C++, creating the HLS project to be synthesized. After the HLS synthesis, the model is described in Register-Transfer Level (RTL) description using a Hardware Description Language, such as Verilog or VHDL. Finally, the bitstream can be loaded onto the FGPA in order for it to be configured. This process is summarized in Fig.~\ref{fig:hls4ml}.

The models initially targeted by the HLS4ML project were feedforward neural networks, but has since extended to BDTs~\cite{summers_fast_2020} and CNNs~\cite{aarrestad_fast_2021}. Its viability has also been demonstrated in resource constrained and safety-critical applications, in the context of autonomous vehicles~\cite{ghielmetti_real-time_2022}.

Using this library, I converted the trained model into FPGA firmware and deployed it on the PYNQ-Z2 board through AMD's open-source PYNQ framework~\cite{amd_developers_pynq_nodate}.

\subsection{PYNQ Framework}

PYNQ provides a Jupyter-based environment with Python APIs, facilitating the use of AMD Xilinx Adaptive Computing platforms.

\begin{figure}
    \centering
    \includegraphics[width=1\linewidth]{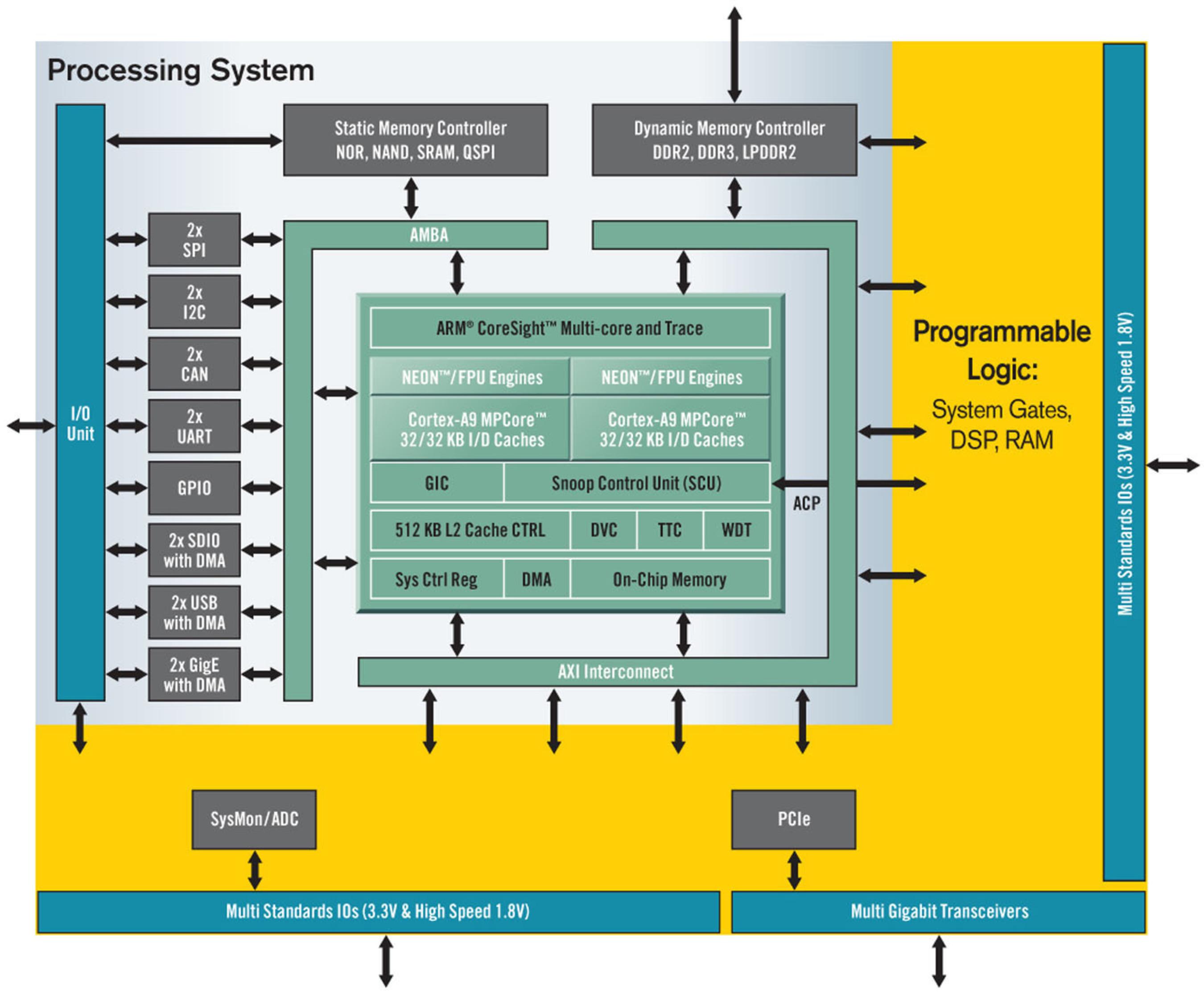}
    \caption{Block diagram of the Zynq-7000 family, highlighting the processing system and the programmable logic of the SoC. Figure from~\cite{xilinx_xilinx_2011}.}
    \label{fig:zynq}
\end{figure}

The AMD/Xilinx Zynq, found in PYNQ-Z1 and PYNQ-Z2 boards, is an SoC based on a dual-core ARM Cortex-A9 processor, referred to as the Processing System (PS), integrated with the traditional reconfigurable FPGA fabric, referred to as the Programmable Logic (PL). The block diagram is shown in Fig.~\ref{fig:zynq}, in line with what we saw in Chapter~\ref{ch:hpc}, Section \ref{sec:intro-fpga}. The PS subsystem features a range of dedicated peripherals---such as memory controllers, USB, UART, IIC, and SPI---and can be expanded with additional hardware IPs using a PL overlay.

\textit{Overlays}~\cite{amd_pynq_2022}, also known as hardware libraries, are configurable FPGA designs that expand the functionality of a user application by extending the processing system of the Zynq into the programmable logic region. They can be used to accelerate a software application, or to customize the hardware configuration for a particular application.

Similarly to software libraries, which can be called by the programmer to perform certain tasks while avoiding the intricate details of the implementation, overlays can be loaded to the FPGA dynamically. For example, for an image processing application, the various image processing functions, such as edge detection, compression, etc., could be implemented in different overlays and loaded from Python, as required.

The PYNQ framework provides a Python interface for this: allowing PL overlays to be controlled from the Python module running on the PS. FPGA design requires hardware engineering knowledge and expertise, but PYNQ overlays abstract out the low-level details of the implementation. In this way, overlays can be used by software developers working at the application level, without necessarily hardware design knowledge.

\subsection{PYNQ-Z2 Development Board}

\begin{figure}
    \centering
    \includegraphics[width=1\linewidth]{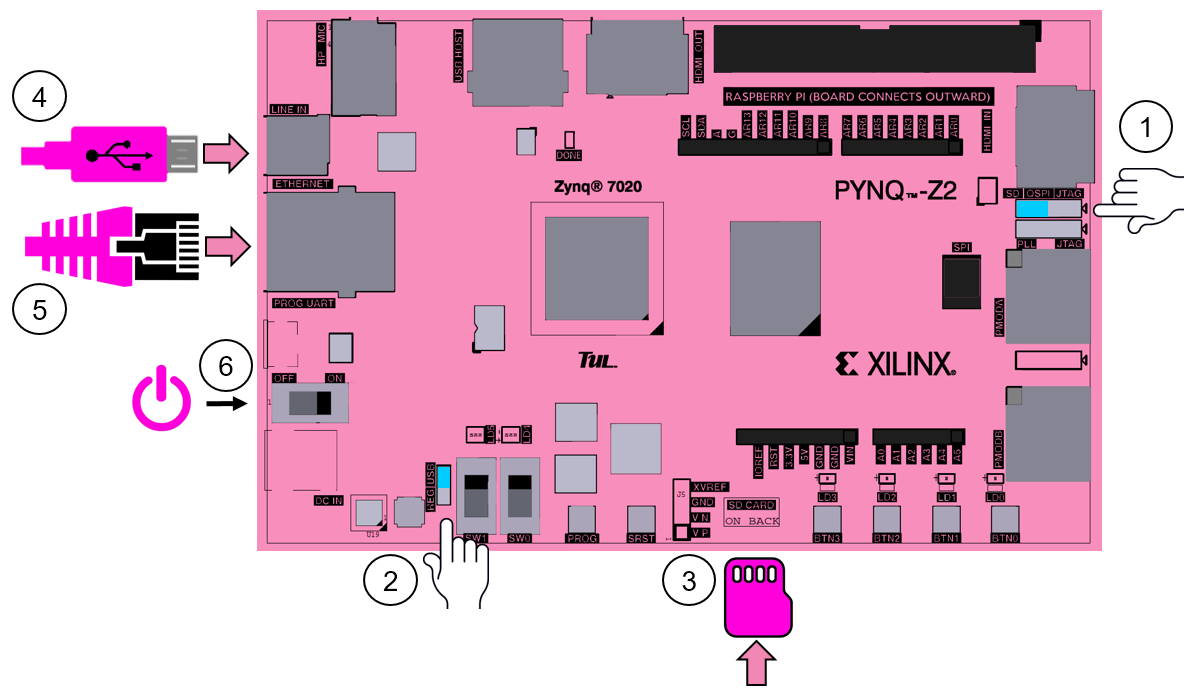}
    \caption{Setup of the PYNQ-Z2 board. 1: The board is set to be booted from the micro SD storage by setting the boot jumper to SD. 2: The board is set to be powered from the micro USB by setting the power jumper to USB. 3: The micro SD card, loaded with the PYNQ-Z2 image, is inserted. 4: The USB cable is connected to the computer, and the PROG-UART micro USB port on the board. 5: The board is connected to the network via the Ethernet port. 6: The board is turned on. Figure from~\cite{amd_pynq_2022}.}
    \label{fig:pynq-z2}
\end{figure}

The PYNQ-Z2 is a low-cost Zynq 7000 development board from TUL~\cite{tul_pynq-z2_nodate} suitable for exploring the capabilities of the PYNQ framework. It features a Zynq Z7020, a Double Data Rate 3 Synchronous Dynamic RAM (DDR3 SDRAM) of 512~MB, micro SD storage, HDMI I/O ports, Ethernet and USB ports, and various LEDs and pushbuttons. The setup of the board, from~\cite{amd_pynq_2022}, is illustrated in Fig.~\ref{fig:pynq-z2}. First, the board is set to boot from the micro SD storage by setting the boot jumper to the SD position. The board is set to be powered from the micro USB by setting the power jumper to the USB position. Then, the micro SD card, loaded with the PYNQ-Z2 image, is inserted into the micro SD card slot. Next, the USB cable is connected to the computer, and the PROG-UART micro USB port on the board. Finally, the board is connected to the network via Ethernet and turned on.

\subsection{Workflow}

With HLS4ML, the trained embedding MLP can be converted into HLS code, which can then subsequently synthesized into Verilog or VHDL using Vivado or Vitis HLS. For boards supported by the PYNQ project, the workflow consists of the following steps.

\subsubsection{Model Import} 
The trained PyTorch model is saved in PyTorch's native checkpoint format and is then imported. Since HLS4ML, at the time of development, supported mainly Keras/TensorFlow~\cite{keras_developers_keras_nodate, abadi_tensorflow_2016}, and had limited support for PyTorch, I had to replace the tanh activations with ReLU, and remove the layer normalization~\cite{ba_layer_2016}, used for stabilizing and speeding up the training process, in order for the model to be processed by the HLS4ML library without errors.

\subsubsection{Model Configuration}  
The model parameters and FPGA target settings are specified in HLS4ML. For instance, the optimization strategy is determined by setting the \texttt{strategy} keyword to either ``latency'' or ``resource'', depending on whether the design prioritizes latency or resource utilization. In my case, I opted for the former. Additionally, the precision of inputs, outputs, weights, and biases is defined. Here, I used \texttt{ap\_fixed<16,6>}, where 16 represents the total number of bits, and 6 specifies the number of bits allocated to the integer part (i.e., the signed number above the binary point).

\subsubsection{HLS Conversion}  
The model is converted to HLS code using HLS4ML. This process involves providing the model, the input data shape, and the target FPGA to the HLS4ML PyTorch converter. The converter parses the layers of the MLP, interprets them, and generates the corresponding HLS project. In this case, the model, shown in Fig.~\ref{fig:mlp}, is a fully connected feed-forward neural network with a 3-dimensional input, three hidden layers consisting of 8 neurons each, ReLU activations, and a 3-dimensional output. The resulting HLS project is then compiled using Vivado HLS. In my case, I used version 2020.1 of Vivado.

In HLS4ML, there are the concepts of the frontend and the backend. The frontend is responsible for parsing the input neural network into an internal model graph, while the backend determines the type of output generated from this graph. These frontends and backends can be selected independently. For example, frontends include parsers for Keras or ONNX, while backends include Vivado HLS, Intel HLS, and Vitis HLS. Here I chose the VivadoAccelerator backend. The VivadoAccelerator backend of HLS4ML leverages the PYNQ software stack to easily deploy models on supported devices. For this backend, the I/O type, the hardware part that is being targeted, the clock period, etc. has to be specified. The target FPGAs for my implementations are the PYNQ-Z2 board, which contains a Xilinx Zynq-7020 FPGA, and the Alveo U50 and U250 featuring the UltraScale+ and XCU250 FPGAs, respectively. It is important to note that the PYNQ-Z2 board is designed for educational purposes, whereas the Alveo cards are significantly larger and intended for use in data centers. All three cards are supported by the PYNQ project.

\subsubsection{Synthesis and Implementation} 
The HLS code is synthesized to Verilog/VHDL. The model is finally ready to be synthesized with Vivado HLS. At this point, we can optionally perform the C simulation of the code, a process where the code is validated for errors and segmentation faults. The IP core is exported and the bitstream is saved.

\subsubsection{Deployment}  
The RTL implementation is deployed on the FPGA. The process involves transferring the bitstream generated by Vivado, along with the hardware handoff file (used for building a platform for the target device), the driver, and some data, to the FPGA. The model is then executed using PYNQ overlays, accessed through a Python interface running on the PS of the FPGA, allowing the user to reconfigure the PL part of the FPGA.  

With HLS4ML, a custom neural network overlay is created to facilitate data transfer via the Advanced eXtensible Interface (AXI)-Stream communication bus protocol---part of the AMBA specification. The target board is configured using the bitstream file generated by the VivadoAccelerator backend. Finally, in Python, a \texttt{NeuralNetworkOverlay} object is instantiated to load the bitstream onto the FPGA's PL. Additionally, the input and output data shapes must be defined to allocate the necessary buffers for data transfer. The \texttt{predict} method is then used to send input data to the PL and retrieve the corresponding output data.

\subsubsection{Computational Performance}
I benchmarked the throughput of the model on the FPGA.  The inference can be timed using the built-in profiling tool of PYNQ Overlays. The current 16-bit implementation on the PYNQ-Z2 board achieves a throughput of approximately 1.2 million inferences per second. For an average LHCb event of 2200 VELO hits for our sample, the effective throughput comes out to 550 events per second.

\subsection{Evaluation of Precision Loss}

To assess the quality of the model's inference on the FPGA, each step of the workflow must be evaluated. The model, originally implemented in PyTorch, uses single-precision floating-point format (FP32), whereas the chosen FPGA implementation relies on 16-bit precision. This inevitably introduces some loss of precision, as techniques such as Quantization-Aware Training (QAT) or detailed profiling have not yet been applied. These techniques are reserved for future work, and the current evaluation is based on untuned quantization parameters. Furthermore, since the model is not a classifier, conventional evaluation metrics, like the Receiver Operating Characteristic (ROC) curve, are not applicable.

Using HLS4ML's method \texttt{predict} on the compiled HLS model of the MLP, we can get the predictions for the input array of approximately 200\,000 hits from the sample used in the GPU implementation. This is similar to doing the C simulation of the code, but the prediction results are more easily accessed. This can be particularly useful when prototyping different configurations for a model. 

\begin{figure}
    \centering
    \includegraphics[width=0.9\linewidth]{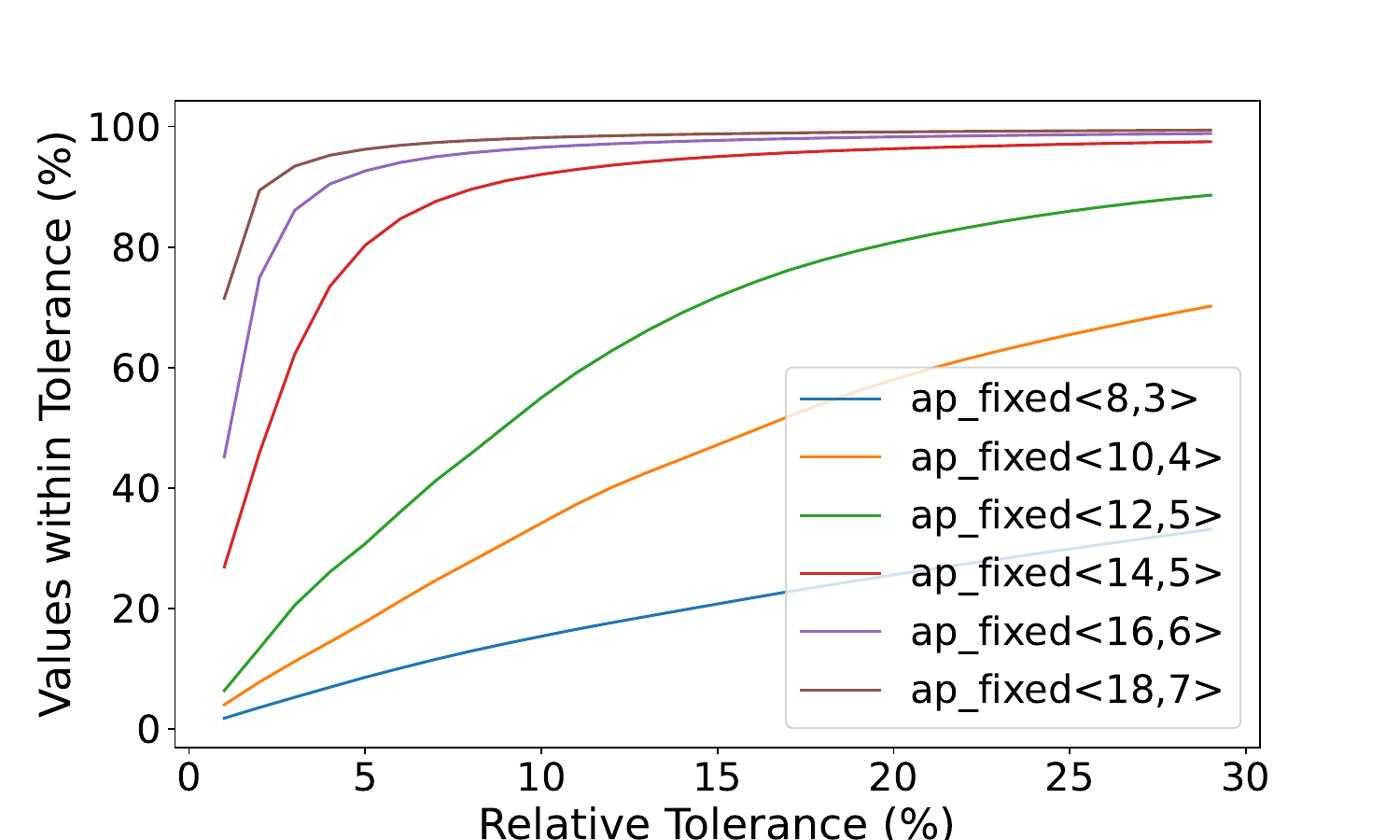}
    \caption{Percentage of values predicted, using the untuned, compiled HLS model, within a specific tolerance window away from the PyTorch inference (in 32-bit precision) values. The model is compiled for various precisions between 8 and 18 bits.}
    \label{fig:pct_vs_tol}
\end{figure}

In Fig.~\ref{fig:pct_vs_tol}, I compare these predictions with the expected predictions of the model in PyTorch. I plot the percentage of the coordinates predicted that lie within a specified window around the expected values. The model is compiled for various precisions between 8 and 18 bits. The integer part bit widths were chosen in such a way as to preserve the ratio of integer to total bit width of the \texttt{<16,6>} implementation.

Here, the \texttt{<16,6>} implementation is chosen, where 97\% of values are predicted within 10\% of correct values. Similarly to what we saw in Chapter~\ref{ch:etx4velo-gpu}, Section~\ref{sec:mlp-quantization}, when the precision of the embedding slightly decreases, the GNN is still able to maintain the physics performance of the pipeline almost unchanged.

Finally, the HLS4ML \texttt{predict} output was validated against the inference of the model on the hardware. The predictions on the PYNQ-Z2 card match perfectly the HLS4ML predictions on CPU.

\section{Latency Comparison of ML Model Inference}
\label{sec:fpga-gpu-latency}

One way to compare the inference of the ML model between GPU and FPGA is the latency. The estimates of the FPGA latency are provided by the Xilinx suite. However for the GPU, we have the direct measurements.

By profiling the ETX4VELO pipeline, using the implementation described in Chapter~\ref{ch:etx4velo-gpu}, up to the embedding step with NSight Systems~\cite{nvidia_developers_nsight-systems_nodate}, the individual latency of every kernel launched can be seen, as in Table~\ref{tab:cuda_gpu_kernels}. Using also Nsight Compute~\cite{nvidia_developers_nsight-compute_nodate}, the kernels related to the inference of the MLP can be identified. In our case, we have four linear layers, and four activations. Therefore, that is eight different kernels that can be matched from the Nsight Systems output. By adding the average time for the \texttt{sm70\_xmma\_gemm\_f32f32...} kernel four times, three times for the kernel \texttt{\_\_myl\_bb0\_4\_AddMeaSub...}, and one time for kernel \texttt{\_\_myl\_bb0\_1\_SliAdd...}, we get a total inference latency of 1\,574\,146.9~ns, according to Table~\ref{tab:cuda_gpu_kernels}. 

\begin{figure}
    \centering
    \includegraphics[width=0.9\linewidth]{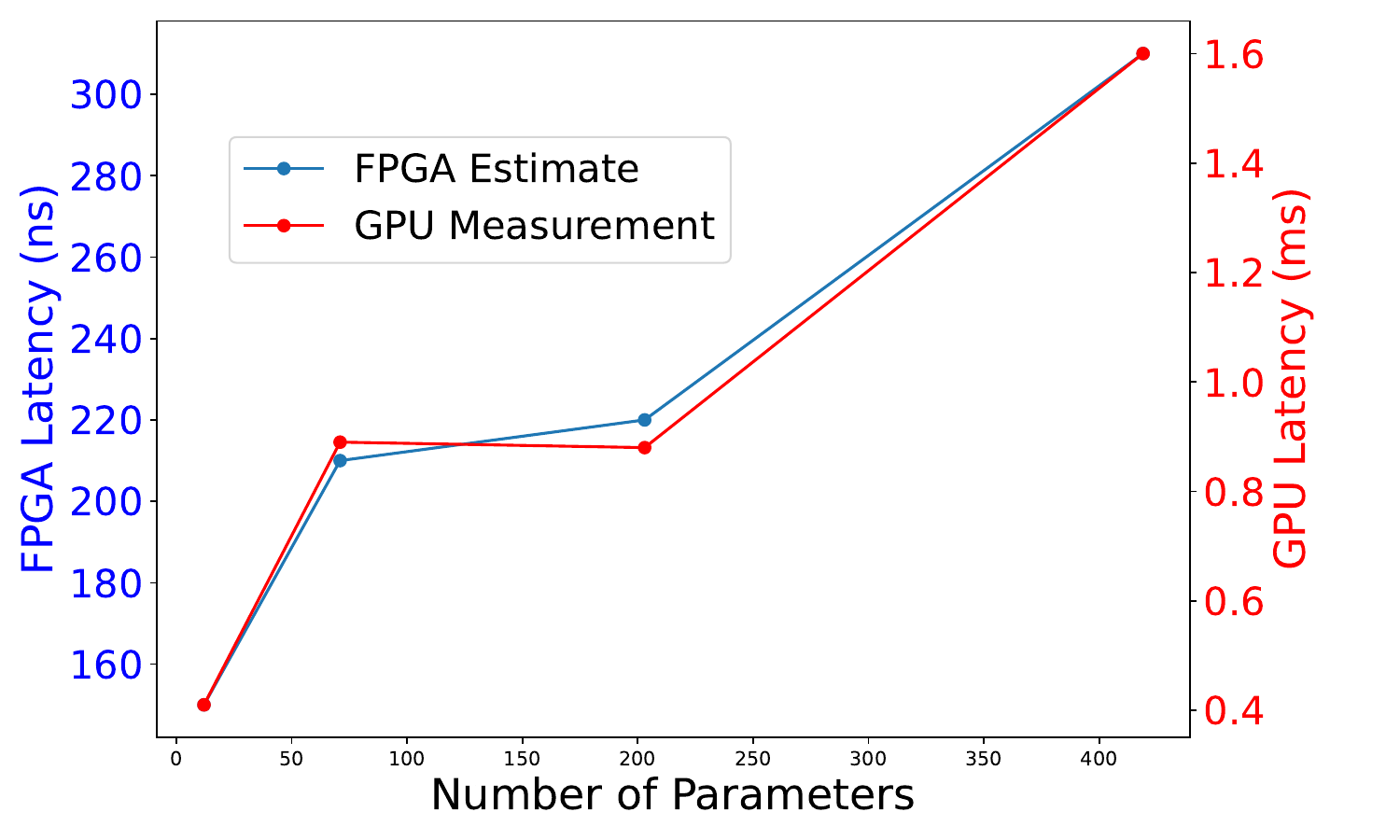}
    \caption{Comparison between the FPGA and GPU ML model inference latency for various model sizes.}
    \label{fig:model_size_vs_latency}
\end{figure}

By repeating this procedure for various model sizes, I ended up with the comparison in Fig.~\ref{fig:model_size_vs_latency}. As expected, the GPU is much slower in terms of latency, but, as we will see in Section~\ref{sec:fpga-gpu-throughput}, a decent competitor to the FPGA in terms of throughput. The other interesting remark is that despite the huge difference in latency, the profiles of the two curves almost overlap.

\begin{table}
\centering
\begin{tabular}{llll}
\hline \hline
Time (\%) & Total Time (ns) & Avg (ns) & Name \\
\hline
28.9 & 365\,609\,912 & 190\,421.8 & \texttt{sm70\_xmma\_gemm\_f32f32...} \\
28.2 & 356\,800\,240 & 247\,777.9 & \texttt{\_\_myl\_bb0\_4\_AddMeaSub...} \\
27.1 & 343\,299\,213 & 715\,206.7 & \texttt{velo\_calculate...} \\
4.3 & 54\,892\,741 & 114\,359.9 & \texttt{decode\_retinaclusters...} \\
4.2 & 53\,294\,246 & 111\,029.7 & \texttt{velo\_calculate...} \\
3.2 & 40\,903\,940 & 85\,216.5 & \texttt{etx4velo\_fill\_input...} \\
2.6 & 33\,180\,491 & 69\,126.0 & \texttt{\_\_myl\_bb0\_1\_SliAdd...} \\
0.9 & 11\,398\,889 & 23\,747.7 & \texttt{calculate\_number\_of...} \\
0.4 & 5\,086\,183 & 10\,596.2 & \texttt{populate\_module...} \\
\hline \hline
\end{tabular}
\caption{CUDA GPU kernel summary from profiling the ETX4VELO pipeline with Nsight Systems.}
\label{tab:cuda_gpu_kernels}
\end{table}

\section{Throughput Comparison of ML Model Inference}
\label{sec:fpga-gpu-throughput}

Since a throughput comparison would be more fair and informative, we proceed by comparing the Alveo boards with the GPU, again using the implementation presented in Chapter~\ref{ch:etx4velo-gpu}. In Table~\ref{tab:clk-u250-16b} we can see the clock performance for the implementation of the MLP on the Alveo U250 card. In Table~\ref{tab:lat-u250-16b}, the latency is estimated to be at 90~ns. From this we can compute the effective theoretical throughput achievable to 11.1 million inferences per second. And given there are on average 2200 hits in an LHCb event from our sample, we get an effective throughput estimate of 5100 events per second. 

The utilization estimates are summarized in Table~\ref{tab:util-u250-16b}. We can extrapolate the theoretically maximum achievable throughput by estimating the number of available IPs that can be deployed on this high-end card, based on the resource usage estimates from Vivado.

\begin{table}
    \centering
    \begin{tabular}{l|c|c|c}
        \hline \hline
        Clock   & Target & Estimated & Uncertainty \\ \hline
        \texttt{ap\_clk} & 5.00 ns   & 4.365 ns    & 0.62 ns       \\ \hline \hline
    \end{tabular}
    \caption{Clock performance estimates from Vivado for the 16-bit implementation of the ETX4VELO embedding MLP on the Alveo U250 card.}
    \label{tab:clk-u250-16b}
\end{table}

\begin{table}
    \centering
    \begin{tabular}{cc|cc|cc|c}
        \hline \hline
        \multicolumn{2}{c|}{Latency (Cycles)} & \multicolumn{2}{c|}{Latency (Absolute)} & \multicolumn{2}{c|}{Interval} & Pipeline \\
        Min & Max & Min & Max & Min & Max & Type \\ \hline
        18 & 18 & 90.000 ns & 90.000 ns & 18 & 18 & none \\
        \hline \hline
    \end{tabular}
    \caption{Vivado synthesis report for the latency from Vivado for the 16-bit implementation of the ETX4VELO embedding MLP on the Alveo U250 card.}
    \label{tab:lat-u250-16b}
\end{table}

\begin{table}
\centering
\begin{tabular}{c|c|c|c|c|c}
\hline \hline
Name & BRAM & DSP & FF & LUT & URAM \\ \hline
DSP & - & - & - & - & - \\
Expression & - & - & 40 & 2301 & - \\
FIFO & - & - & - & - & - \\ 
Instance & - & 132 & 990 & 5457 & - \\
Memory & - & - & - & - & - \\
Multiplexer & - & - & - & 152 & - \\
Register & - & - & 338 & - & - \\
Total & 0 & 132 & 1368 & 7910 & 0 \\ \hline
Available & 5376 & 12\,288 & 3\,456\,000 & 1\,728\,000 & 1280 \\
Utilization (\%) & 0 & 1 & $\sim$0 & $\sim$0 & 0 \\ \hline \hline
\end{tabular}
\caption{Vivado synthesis report for resource utilization for the 16-bit implementation of the ETX4VELO embedding on the Alveo U250 card.}
\label{tab:util-u250-16b}
\end{table}

This card has 12\,288 DSP slices, and the current 16-bit implementation of the ETX4VELO model on this card is using 132 of them. Assuming that the number of DSP blocks needed is not going to change dramatically we can assume that we should be able to launch at most $12\,288/132 \approx 93 $ IPs on this card. Consequently, the maximum throughput we could achieve would be $93 \times 5100 = 470\,000$ events per second. 

Although the 8-bit implementation is less accurate than the 16-bit version, a similar calculation is performed for it, as the appropriate HLS4ML tools can potentially minimize the precision loss. The utilization estimates for the 8-bit implementation on the Alveo U250 are summarized in Table~\ref{tab:util-u250-8bit}. Since the implementation uses 8-bit precision, the DSP blocks were not allocated, and the LUTs, being the most utilized resource, determine the maximum number of IPs that can be deployed on this platform. Therefore, with approximately 205 IPs and a latency of 85~ns, the maximum achievable throughput is calculated to be $1.1 \times 10^6$ events per second.

\begin{table}
\centering
\begin{tabular}{c|c|c|c|c|c}
\hline \hline
Name & BRAM & DSP & FF & LUT & URAM \\ \hline
DSP & - & - & - & - & - \\
Expression & - & - & 40 & 1785 & - \\
FIFO & - & - & - & - & - \\
Instance & - & 0 & 410 & 6462 & - \\
Memory & - & - & - & - & - \\
Multiplexer & - & - & - & 149 & - \\
Register & - & - & 272 & - & - \\
Total & 0 & 0 & 722 & 8396 & 0 \\ \hline
Available & 5376 & 12\,288 & 3\,456\,000 & 1\,728\,000 & 1280 \\
Utilization (\%) & 0 & 0 & $\sim$0 & $\sim$0 & 0 \\ \hline \hline
\end{tabular}
\caption{Vivado synthesis report for resource utilization for the 8-bit implementation of the ETX4VELO embedding on the Alveo U250 card. Adapted from~\cite{giasemis_comparative_2025}.}
\label{tab:util-u250-8bit}
\end{table}

Finally, the same calculation is performed for the smaller card Alveo U50. In this case, with a latency of 85~ns, and resources for approximately a maximum of 103 IPs, the throughput comes out to $550 \times 10^3$ events per second. The utilization estimates for the 8-bit implementation on this board are summarized in Table~\ref{tab:util-u50-8bit}. 

\begin{table}
\centering
\begin{tabular}{c|c|c|c|c|c}
\hline \hline
Name & BRAM & DSP & FF & LUT & URAM \\ \hline
DSP & - & - & - & - & - \\
Expression & - & - & 40 & 1785 & - \\
FIFO & - & - & - & - & - \\
Instance & - & 0 & 410 & 6462 & - \\
Memory & - & - & - & - & - \\
Multiplexer & - & - & - & 149 & - \\
Register & - & - & 272 & - & - \\
Total & 0 & 0 & 722 & 8396 & 0 \\ \hline
Available & 2688 & 5952 & 1\,743\,360 & 871\,680 & 640 \\
Utilization (\%) & 0 & 0 & $\sim$0 & $\sim$0 & 0 \\ \hline \hline
\end{tabular}
\caption{Vivado synthesis report for resource utilization for the 8-bit implementation of the ETX4VELO embedding on the Alveo U50 card.}
\label{tab:util-u50-8bit}
\end{table}

When running with the specified flags, the GPU reaches its maximum power consumption of 350~W. The measured idle power, with no processes running and the GPU fan at 0\% utilization, is 50~W. For the Alveo cards, we refer to the official specifications, which quote a maximum total power consumption of 75~W and 225~W for the U50 and U250, respectively, assuming the implementation will utilize the hardware close to its maximum capacity. In comparison, the PCIe40~\cite{cachemiche_aces_2018} readout board for LHCb, which contains an Intel Arria 10 FPGA, is estimated to consume 150~W during normal operation. For the idle power consumption of both Alveo cards, I used the value of 24~W provided in the official AMD documentation for benchmark results on the Alveo U50 FPGA, assuming that the idle power will not differ significantly between the two models.

We can also compare the energy per event using the throughput and the power. Dividing the Thermal Design Power (TDP) of each device, in joules per second, by the throughput expressed in events per second results in the energy cost of a single event. With a power of 350~W and throughput $0.82 \times 10^6$ events per second, the GPU results in 430~$\mu$J per event. On the other hand, the Alveo U50 results in 140~$\mu$J per event. Finally, the same calculation for the Alveo U250 results in 210~$\mu$J per event.

The prices of the accelerators are also considered. At the time of writing, the Alveo U50 is listed at 2965~USD on the official AMD website~\cite{amd_alveo_nodate-1}, while for the GPU, the launch price of 1499~USD~\cite{techpowerup_nvidia_nodate} is used. The price of the Alveo U250 is not listed on the official website~\cite{amd_alveo_nodate}, so I estimated its current market price at approximately 10\,000~USD based on publicly available sources.

We now compare with the GPU throughput, which does not include memory transfers between the host and the device, since the data always reside on the device due to the Allen architecture. On the one hand, the 16-bit implementation on the Alveo U250 is, unsurprisingly, twice as slow as the GPU's 8-bit implementation. On the other hand, the 8-bit implementation on the Alveo U250 is on par with the 8-bit implementation on the GeForce RTX 3090, with the potential to slightly outperform it, while consuming just over 60\% of the power used by the GPU counterpart. However, it should be noted that the price of the U250 is roughly ten times the price of the GPU. Furthermore, the implementation on the Alveo U50 is slightly slower than the GPU counterpart, while the power usage is almost $5\times$ lower. Interestingly, the choice between FPGAs and GPUs involves a trade-off between their upfront cost and the long-term expense of power consumption over their operational lifetime.

The results are summarized in Table~\ref{tab:fpga-gpu-comparison}. For the Alveo implementations, \texttt{<a,b>} refers to the precision being \texttt{ap\_fixed<a,b>}, where \texttt{b} is the integer bit width, and \texttt{a} is the total bit width. The comparison of the cards is in Table~\ref{tab:accelerators}.

\begin{table}
\centering
    \begin{tabular}{c|c|cc|c}
        \hline \hline
        Accelerator & Alveo U50 & \multicolumn{2}{c|}{Alveo U250} & RTX 3090 \\
        Implementation & \texttt{<8,3>} & \texttt{<8,3>} & \texttt{<16,6>} & TRT INT8 \\
        \hline
        Throughput (Events/s $\times 10^{\rule{0pt}{1.5ex}6}$) & 0.55 & 1.10 & 0.47 & 0.82 \\
        Active Power Draw (W) & 75 & \multicolumn{2}{c|}{230} & 350 \\
        Idle Power Draw (W) & 24 & \multicolumn{2}{c|}{24} & 50 \\
        Energy per Event ($\mu$J) & 140 & 210 & 490 & 430 \\
        Energy Gain & 3.1x & 2.0x & - & 1.0x \\ \hline
        Price (USD) & 3000 & \multicolumn{2}{c|}{$\sim$ 10\,000} & 1500 \\
        \hline \hline
    \end{tabular}
\caption{Comparison of the embedding MLP throughput between the theoretical performance of the Alveo FPGA implementations and the GeForce RTX 3090 GPU implementation. For the Alveo implementations, \texttt{<a,b>} refers to the precision being \texttt{ap\_fixed<a,b>}. The power usage, while running the inference and while idle, the energy cost of the inference of a single event, and the price are also compared. The gain is given with respect to the GPU implementation. Adapted from~\cite{giasemis_comparative_2025}.}
\label{tab:fpga-gpu-comparison}
\end{table}

\begin{table}
    \centering
    \setlength{\tabcolsep}{2pt}    
    \begin{tabular}{p{5cm}c|p{5cm}} \hline \hline
        \multicolumn{2}{c|}{Device} & Specifications \\ \hline
         \vspace{-1.5cm} AMD\newline Alveo U50 & \includegraphics[height=2cm]{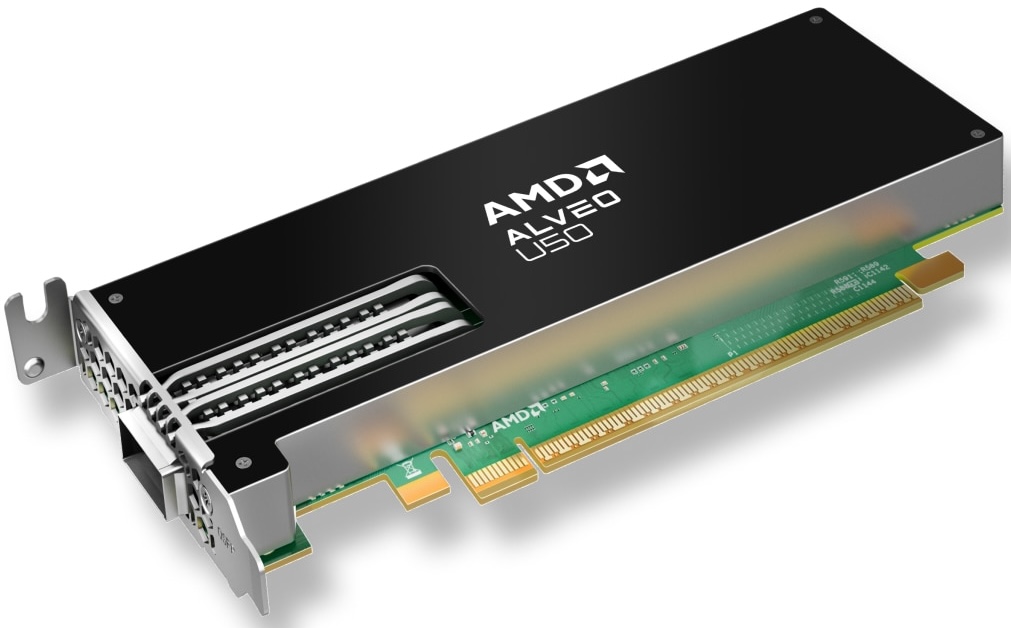} & \vspace{-1.75cm} Memory: 8~GB \newline Maximum Power: 75~W\newline Price: 3000~USD \\ \hline
         \vspace{-1.5cm} AMD\newline Alveo U250 & \includegraphics[height=2cm]{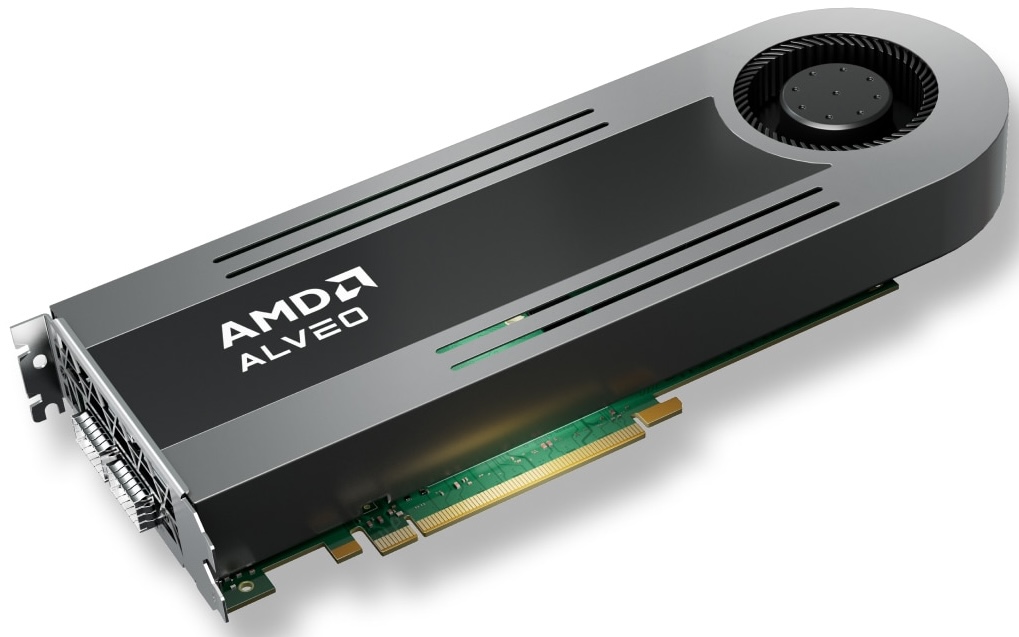} & \vspace{-1.75cm} Memory: 64~GB\newline Maximum Power: 230~W \newline Price: $\sim$10\,000~USD \\ \hline
         \vspace{-1.5cm} Nvidia\newline GeForce RTX 3090 & \includegraphics[height=2cm]{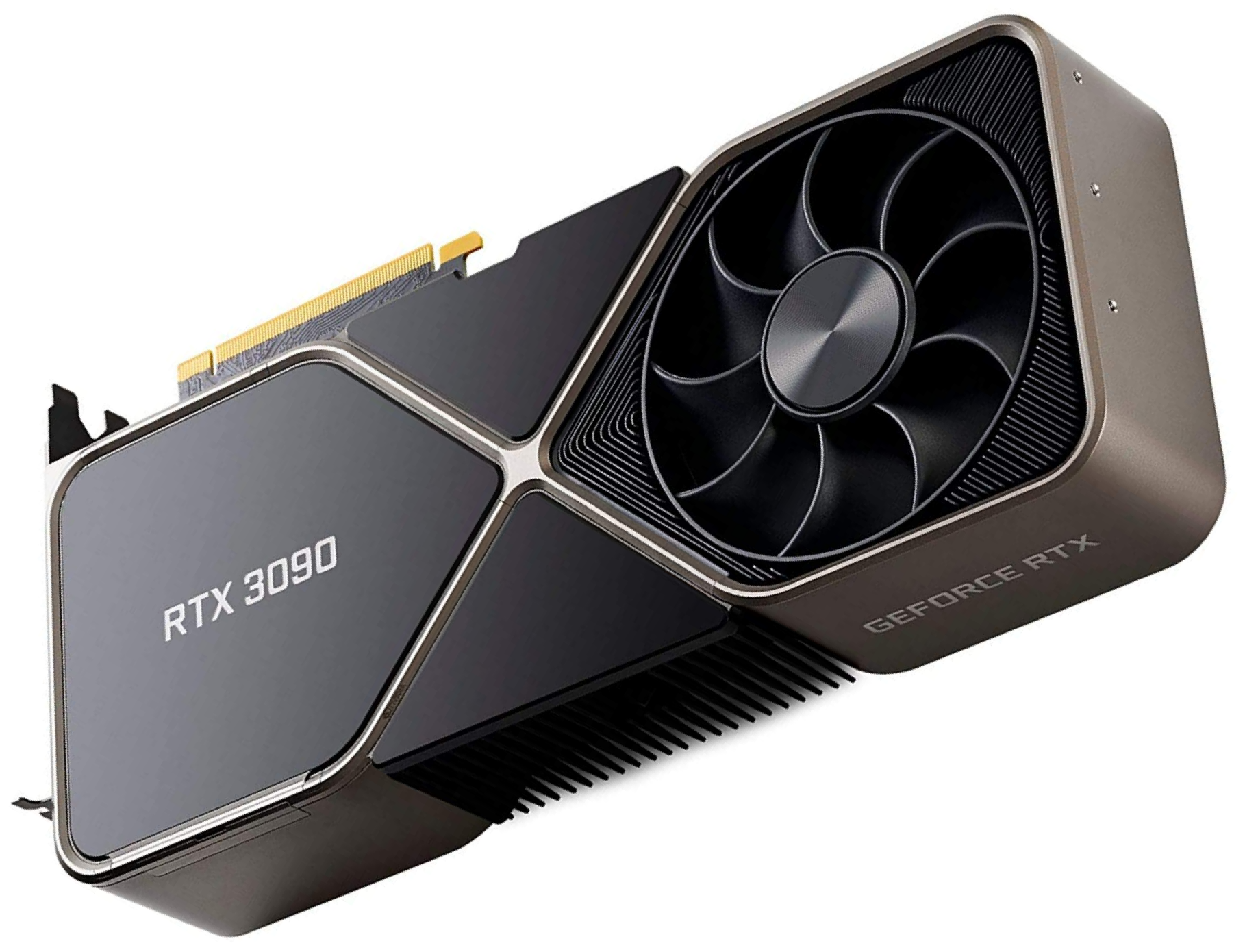} & \vspace{-1.75cm} Memory: 24~GB\newline Maximum Power: 350~W \newline Price: 1500~USD \\ \hline \hline
    \end{tabular}
    \caption{Comparison of the FPGA and GPU cards used for the different implementations of the ETX4VELO embedding.}
    \label{tab:accelerators}
\end{table}

\section{Purchase vs. Operating Cost}

We now proceed to compare the initial purchase cost of the accelerators listed in Fig.~\ref{tab:accelerators} with their operating costs, specifically focusing on electricity expenses. I used the cost of 125.2~EUR/MWh at the time of writing from the Swiss Federal Office of Energy website on electricity prices~\cite{swiss_federal_office_of_energy_energy_nodate}. In order to simplify the calculation, I assumed that the devices will be used at their maximum capacity throughout the day, 7 days a week and throughout the entire year. The total hours will thus be 24~h $\times$ 365 = 8760~h. Therefore, running the GPU for one year at 350 W, would consume approximately 3.1 MWh, costing around 380 EUR. In contrast, the Alveo U50 would be about 3.1 times more cost-efficient, as shown in Table~\ref{tab:fpga-gpu-comparison}. Specifically, the Alveo U50 would cost 120 EUR to operate for a year, saving roughly 260 EUR annually compared to the GPU.

Continuing this comparison, recovering the 1500~USD price difference between the Alveo U50 and the GeForce RTX 3090---assuming, for the sake of simplicity, a USD/EUR exchange rate of 1:1---would take approximately 6 years. Therefore, it would seem sensible to invest on the more expensive but less power-demanding accelerator when its operation is planned over a period of more than 5--10 years, and when the device is planned to be utilized near its full capacity consistently throughout the year.

Therefore, at the end of a 5 year period, the total monetary cost would be the same but the energy consumed by the FPGA would be significantly lower. Indeed, the GPU will have consumed 15.33~MWh while the FPGA only 3.29~MWh, resulting in the saving of approximately 12~MWh of electricity, or equivalently $4.32 \times 10^{10}$~J.

We can further convert this saving of electricity into greenhouse gas emissions equivalencies. Based on the calculator from the United States Environmental Protection Agency~\cite{us_environmental_protection_agency_greenhouse_nodate}, 12~MWh are equivalent to 4.7 metric tons of Carbon Dioxide (CO$_2$). This is in turn equivalent to the CO$_2$ emissions from roughly one gasoline-powered passenger vehicle driven for one year. This conversion is based on delivered electricity and already incorporates average generation inefficiencies.

For a more complete comparison, maintenance, upgrade, development, and optimization costs should also be considered; however, this is left for future work.

\subsection*{Server Operation}

We can now imagine creating a server, similar to LHCb's DAQ server~\cite{aaij_evolution_2021}, containing multiple of these devices, starting with a server containing eight GeForce RTX 3090s. With a TDP of 350~W, the eight GPUs would consume 2800~W of power. Assuming other components such as the CPU, RAM, storage, motherboard and fans require around 1000~W, the total system power consumption could be around 3800~W.

The electrical power consumed by the GPUs is almost entirely converted into heat. Therefore, the cooling system would need to be able to extract this amount of energy per unit time in order to keep the temperature stable in the server room. The Coefficient of Performance (CoP) of a heat pump is the ratio of useful heating or cooling provided to the energy required, and most air conditioners have a CoP between 3.5 and 5~\cite{team_air_conditioning_gc_air_2018}. Here, assuming that our cooling system has a CoP of 3.5, we can calculate the needed power for its operation roughly at 1100~W. Therefore, the cost of the operation of the server would be around 4900~W.

On the other hand, for a similar server of eight Alveo U50 FPGAs, assuming that the power required for all the other components apart from the FPGA boards is the same, the total power would sum up to around 2100~W, including a cooling system of similar properties. Finally, if the FPGAs were Alveo U250s, the total power would be around 3700~W. The comparison is summarized in Table~\ref{tab:energy-costs}. Similarly to the calculation before, I extrapolated the cost over a period of one and five years.

\begin{table}
    \centering
    \begin{tabular}{c|c|c|c} \hline \hline
        Server & Alveo U50 & Alveo U250 & RTX 3090 \\ \hline
        Power (W) & 2100 & 3700 & 4900 \\
        Energy per Year (MWh) & 18 & 32 & 43\\
        1-Year Energy Cost (EUR) & 2300 & 4100 & 5400 \\
        5-Year Energy Cost (EUR) & 12\,000 & 20\,000 & 27\,000 \\ \hline \hline
    \end{tabular}
    \caption{Comparison of the costs of a server containing eight GPUs or FPGAs.}
    \label{tab:energy-costs}
\end{table}

\section*{Conclusion}

In this chapter, I presented a detailed comparison of machine learning inference on FPGAs and GPUs in the context of future HEP experiments in the HL-LHC era, where scalability, power efficiency, and computational performance will be increasingly critical. These findings underscore the potential of FPGAs as viable alternatives for high-throughput applications in particle physics, especially when energy is a crucial consideration. The combination of HLS4ML's ease of use and the inherent advantages of FPGAs makes this approach a compelling choice for researchers aiming to deploy ML models in hardware without deep expertise in FPGA design.

This work not only emphasizes the FPGA's strengths in energy efficiency and throughput but also identifies avenues for future improvement, such as incorporating quantization-aware training to preserve essential physics performance and enhancing computational efficiency through the use of HLS directives. These optimizations could further unlock the full potential of FPGAs, making them even more appropriate for the high-performance environment of real-time data processing at the LHC.

For future work, firstly, the theoretical extrapolation done on the Alveo cards can be practically implemented, and a comparison including memory and I/O overheads can be performed. Secondly, other interesting hardware can be explored, such as the Intel-based LHCb DAQ board, PCIe40~\cite{cachemiche_aces_2018}. Thirdly, the implementation of the GNN on the FPGA might be an interesting avenue. Whether its pursuit is worthwhile is still to be decided, depending on considerations of its size, the time needed for its implementation, and the potential for throughput increases. Finally, since the beginning of the experiments in this chapter, the support of PyTorch by HLS4ML has been greatly improved. The new functionalities could be included and the results re-evaluated.

\chapter{Conclusion and Outlook}
\label{ch:conclusion}

With the HL-LHC in the near future, the high-energy physics community is preparing for a new era of real-time processing at unprecedented data rates. Aggressive R\&D is needed in order to redevelop the computational infrastructure of the collaborations, since triggering more efficiently, and in real time, will be increasingly in demand.

Machine learning, and especially deep learning, is increasingly drawing the attention of the HEP community, as is the case in many other fields. Its ability to efficiently learn representations and adapt to specific problems offers hope for making better use of the available computational resources.

In this thesis, I presented our graph neural network-based pipeline, ETX4VELO, and demonstrated that the required physics performance is reachable. In particular, the pipeline is on par with the classical tracking algorithms currently in place in the first-level trigger of LHCb for Run~3. This includes stringent requirements for the reconstruction efficiency based on various metrics.

Moreover, the end-to-end implementation of the pipeline inside Allen on GPUs was presented. The trained embedding and GNN neural networks were exported using the ONNX format, and deployed on Nvidia GPUs using the TensorRT and ONNX Runtime inference engines. The classical algorithms, including the k-NN and the WCC, were implemented as compute kernels in the C++ extension of CUDA, utilizing the parallelization capabilities of the hardware. The computational performance was compared against Allen, highlighting the weakness of the pipeline in terms of throughput with respect to Allen.

Finally, the pipeline was partially implemented on FPGAs. Since the GNN comprises multiple MLPs, the implementation focused on one of its fundamental components: the MLP. Furthermore, due to time constraints and the scope of the challenge, the implementation of the GNN was deferred for future work. The embedding MLP, on 8 bits, was benchmarked against the GPU implementation, and the implementations were compared based on energy efficiency considerations.

The ETX4VELO work would be interesting to be extended to the other tracking detectors of the LHCb experiment: the SciFi or the UT. However, compared to the VELO, these detectors, and hence their tracking algorithms~\cite{aiola_hybridseeding_2021,bailly-reyre_looking_2024,jashal_standalone_2022}, have different configurations and operating principles, making such an extension non-trivial. Secondly, the triplet-based methodology, currently left out of the GPU implementation, could be implemented, and the impact on the computational and physics performance re-evaluated. Thirdly, once the compatibility and support issues have been resolved, the quantization of the GNN could be pursued. This avenue provides the most hope for throughput gains for the ETX4VELO pipeline. Finally, the k-NN and WCC algorithms can potentially be redeveloped to leverage more efficiently the GPU resources.

Regarding the FPGA side of the work, whether the implementation of the GNN on FPGAs is worthwhile remains an open question. The size and complexity of the network, along with the algorithm's irregular memory access patterns, may render FPGAs unsuitable architectures for this task. Alternative methods may also need to be explored. Symbolic regression, for example, is an interesting avenue. It replaces the graph-based neural network by substituting each network block with a symbolic function, preserving the graph structure of the data and enabling message passing. This approach makes the implementation on FPGAs significantly easier. Other architectures, avoiding the computational challenges facing GNNs, like transformer-based models, may be worth exploring as alternatives. 

Furthermore, a complete implementation, instead of an extrapolation of packing multiple embedding neural networks on Alveo cards, can be done. In this way, the large-scale, high-throughput environment needed in the context of an LHC trigger can be simulated, and the comparison between the two hardware architectures can be conducted more fairly, taking into account data transfers and I/O overheads.

The work related to the ETX4VELO pipeline can be further extended to other applications, such as primary vertex finding, building upon the work in~\cite{akar_progress_2021,akar_advances_2023}, or graph clustering for electromagnetic calorimeters~\cite{canudas_graph_2022}. The end of the pipeline can be minimally modified to match the task at hand, leaving the graph architecture unchanged, and trained accordingly to leverage the new datasets.

All in all, this work contributes to the understanding of how machine learning models can be deployed in high-frequency data environments for the purpose of real-time decision-making at the LHC, on heterogeneous architectures, including GPUs and FPGAs. Nevertheless, the field of high-energy physics still has progress to make before the community's computational know-how is sufficiently mature to meet the challenges posed by the increasing event complexity and the data rates of the HL-LHC and post-HL-LHC era.

\appendix

\part{Appendices}

\chapter{Notations, Units and Physical Constants}
\label{app:notations}

\section{Notations}

\begin{table}[H]
\centering
\renewcommand{\arraystretch}{1.5}
\begin{tabular}{lp{11cm}}
\hline \hline
Example & Description \\
\hline
\textbf{x}, \textbf{y}, \textbf{z} & With bold lowercase letters we denote \textit{vectors}. \\
\textbf{X}, \textbf{Y}, \textbf{Z} & With bold uppercase letters we denote \textit{matrices}. \\
$x$, $a$, $t$ & \textit{Scalars} are denoted with characters in non-bold, italics font. \\
$f$, $g$, $h$ & Non-bold, italics font  is used for \textit{functions with scalar output}. \\
$\mathbf{f}$, $\mathbf{g}$, $\mathbf{h}$ & Boldface font  is used for \textit{functions with vector output}.\\
$\mathbb{R}$ & The set of \textit{real numbers}. \\
\hline \hline
\end{tabular}
\label{tab:notations}
\caption{Notations.}
\end{table}
\raggedbottom

\pagebreak

\section{Units and Abbreviations}

\begin{table}[H]
\centering
\renewcommand{\arraystretch}{1.5}
\begin{tabular}{lp{11cm}}
\hline \hline
Symbol & Meaning \\
\hline
m & \textit{Meter}, unit of length. \\
s & \textit{Second}, unit of time. \\
kg & \textit{Kilogram}, unit of mass. \\
C & \textit{Coulomb}, unit of charge. \\
fb & \textit{Femtobarn}, unit of area, $1~\text{fb} = 10^{-43}~\text{m}^2$. \\
fb$^{-1}$ & \textit{Inverse femtobarn}, unit of integrated luminosity.\\
J & \textit{Joule}, unit of energy.\\
W & \textit{Watt}, unit of power. \\
Wh & \textit{Watt-hour}, unit of energy. \\
eV & \textit{Electron volt}, unit of energy, $1~\text{eV} = 1.602\,18 \times 10^{-19}~\text{J}$. \\
T & \textit{Tesla}, unit of magnetic flux density. \\
EUR & \textit{Euro}, currency unit. \\
USD & \textit{US dollar}, currency unit. \\
\hline \hline
\end{tabular}
\label{tab:units}
\caption{Units and abbreviations.}
\end{table}
\raggedbottom

\section{Physical Constants}

\begin{table}[H]
\centering
\renewcommand{\arraystretch}{1.5}
\begin{tabular}{p{5cm}p{1.5cm}p{5cm}}
\hline \hline
\textbf{Constant} & \textbf{Symbol} & \textbf{Value} \\
\hline
Speed of light in vacuum & $c$ & $2.997\,924\,58 \times 10^8 \ \text{m/s}$ \\
Elementary charge & $e$ & $1.602\,176\,634 \times 10^{-19} \ \text{C}$ \\
\hline \hline
\end{tabular}
\caption{Physical constants.}
\label{tab:constants}
\end{table}
\raggedbottom

\chapter{Early ETX4VELO Development}
\label{app:early-dev}

The ETX4VELO project started from the Exa.TrkX pipeline~\cite{gdl4hep_exatrkx_nodate} in Python, containing an embedding MLP and a GNN. The pipeline was gradually adapted and modified specifically for data from the VELO subdetector.

In order for the VELO data to be imported to the pipeline, they have to be converted from the data formats used by LHCb to CSV. For this purpose, the XDIGI2CSV library~\cite{gdl4hep_xdigi2csv_nodate} was developed. The XDIGI2CSV repository is designed for reproducible execution of the Allen and Moore algorithms. Its primary purpose is to convert DIGI or XDIGI files from the grid into CSV or Parquet formats. However, it offers additional functionality, such as converting DIGI/XDIGI files into MDF or ROOT~\cite{brun_root_1997} formats and transforming MDF files into CSV.

Using XDIGI2CSV, the VELO data were split into two files. The \texttt{hits\_velo.csv} file contains the VELO cluster coordinates as well as the Monte Carlo (MC) identifier of the origin particle of this cluster. The file \texttt{mc\_particles.csv} associates, for each event, the MC identifier with the properties of the corresponding particle, e.g., momentum, pseudorapidity,, etc. 

Next, the data were split per event, organized as in the TrackML Particle Tracking Challenge~\cite{cern_trackml_nodate}, and the Cartesian coordinates transformed to cylindrical. The true edges of the graph were calculated as follows.
\begin{itemize}
    \item The clusters with the same MC identifier are found.
    \item The clusters are then ordered with respect to the distance from the origin vertex of the particle $(v_x,v_y, v_z)$.
    \item The true edges are the edges between these ordered, successive hits.
\end{itemize}
The data were then fed into the GDL4HEP Exa.TrkX fork~\cite{gdl4hep_exatrkx_nodate}, where the code was modified in order to make it compatible with the VELO data. 

\begin{figure}
    \centering
    \includegraphics[width=1\linewidth]{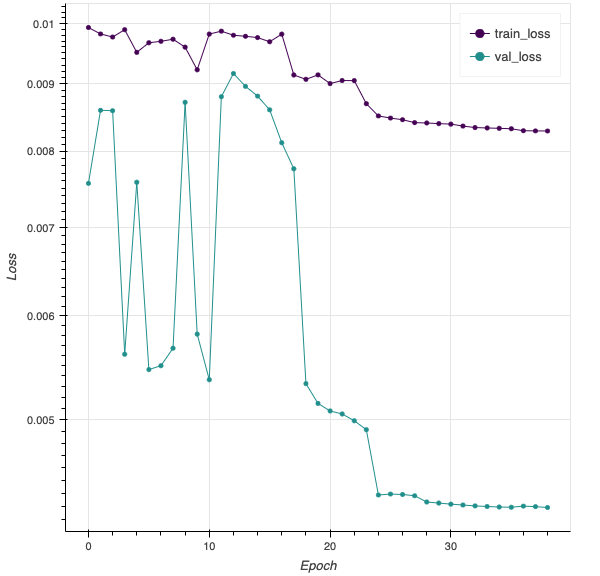}
    \caption{Train and validation losses for the training of the Embedding MLP, described in Section~\ref{sec:etx4velo-simplified}, for a reconstruction efficiency of 67\%.}
    \label{fig:mlp-training}
\end{figure}

\begin{figure}
    \centering
    \includegraphics[width=1\linewidth]{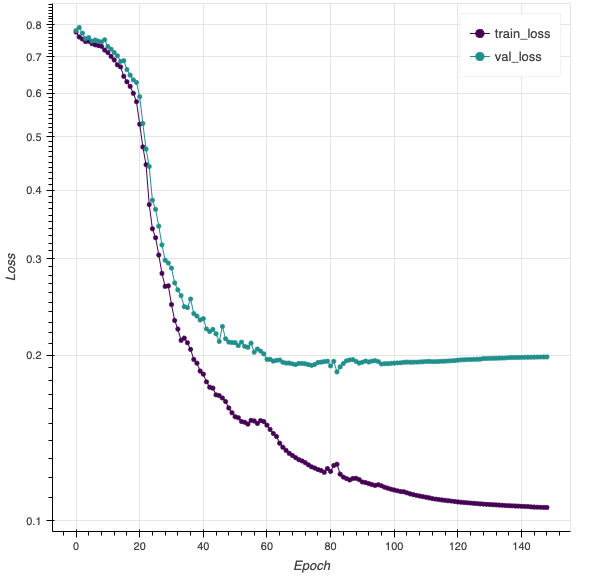}
    \caption{Train and validation losses for the training of the GNN, described in Section~\ref{sec:etx4velo-simplified}, for a reconstruction efficiency of 67\%.}
    \label{fig:gnn-training}
\end{figure}

The training was done on 100 LHCb events. Plots of the trainings are shown in Figs.~\ref{fig:mlp-training} and \ref{fig:gnn-training}. The train and validation losses are plotted as a function of the epochs of the training. This comparison is a frequent diagnostic tool for overfitting and underfitting. The fact that the two losses initially drop simultaneously demonstrates that the model is learning. The apparent plateauing, especially in Fig.~\ref{fig:gnn-training}, suggests that the model has reached a local minimum in the loss function without significant overfitting.

The initial reconstruction efficiency was as low as 67\%. The efficiency was improved to around 76\% by increasing the number of iterations of the GNN. In addition, training only on particles that are reconstructible resulted in the efficiency jumping to 84\%. Finally, excluding electrons during training and evaluation resulted in an efficiency of roughly 90\%.

However, in order to have a better insight into how the algorithms can be improved, a more comprehensive evaluation is necessary. For this reason, the MonteTracko library~\cite{gdl4hep_montetracko_nodate} was created. MonteTracko offers tools for matching simulated particle trajectories with reconstructed tracks, calculating performance metrics, and visualizing the results in multiple formats. The library has been tested and validated to yield identical results as the Allen VELO validation sequence. An example of the printout from the MonteTracko evaluation is shown in Chapter~\ref{ch:etx4velo}, Section~\ref{sec:etx4velo-simplified}.

\chapter{Further Resources}
\label{app:further}

\begin{center}
    \textbf{Scan the QR code to visit my personal website:} \href{https://www.fotisgiasemis.com}{\textbf{fotisgiasemis.com}}
\end{center}

\vspace{1cm}

\setlength{\fboxrule}{1pt}
\begin{center}
    \fbox{\qrcode[height=4.2cm]{https://www.fotisgiasemis.com}} \\
    \vspace{0.3cm}
    \texttt{\small https://fotisgiasemis.com/}
\end{center}

\backmatter

\cleardoublepage
\printbibliography[
    heading=bibintoc,
    title={Bibliography}
]

\end{document}

%% file: glossary_acronyms.tex
\newglossaryentry{hyperparameter}
{
        name=Hyperparameter,
        description={A parameter that defines a configurable aspect of a model's learning process and that has to be externally defined.}
}

\newglossaryentry{trigger}
{
        name=Trigger,
        description={The system used in high-energy physics experiments to filter the vast volume of raw data produced by particle collisions.}
}

\newglossaryentry{positron}
{
        name=Positron,
        description={The antiparticle of the electron.}
}

\newglossaryentry{montetracko}
{
        name=MonteTracko,
        description={A library developed for evaluating the performance of track reconstruction algorithms according to the LHCb definitions and conventions.}
}

\newglossaryentry{cuda}
{
        name=CUDA,
        description={A parallel computing platform and application programming interface by Nvidia that allows software to use certain types of GPUs for accelerated general-purpose processing.}
}

\newglossaryentry{onnx}
{
        name=Open Neural Network Exchange (ONNX),
        description={Open-source AI ecosystem defining open standards for representing machine learning algorithms and tools.}
}

\newglossaryentry{bitstream}
{
        name=Bitstream,
        description={Term frequently used to describe the sequence of bits loaded on an FPGA for its configuration.}
}

\newglossaryentry{communication-protocol}
{
        name=Communication protocol,
        description={A system of rules defined to enable two or more entities of a communication system to transmit information between them, such as UART, IIC and SPI.}
}

\newglossaryentry{databus}
{
        name=Databus,
        description={A communication system that transfers data between internal components of a computer or between computers.}
}

\newglossaryentry{physics-performance}
{
        name=Physics performance,
        description={The effectiveness of a reconstruction algorithm in enabling accurate physics analyses, often evaluated based on metrics such as the tracking efficiency, fake rate, etc.}
}

\newglossaryentry{computational-performance}
{
        name=Computational performance,
        description={A measure of a reconstruction algorithm's efficiency in terms of throughput, latency, resource usage, etc.}
}

\newglossaryentry{latency}
{
        name=Latency (FPGA),
        description={The time delay between the input and the corresponding output for an FPGA design, typically measured in clock cycles or nanoseconds.}
}

\newglossaryentry{throughput}
{
        name=Throughput,
        description={The amount of data passing through, or processed by, a system for a given period of time.}
}

\newglossaryentry{kernel}
{
        name=Compute kernel,
        description={An optimized routine designed to run on parallel computing hardware such as GPUs and FPGAs.}
}

\newglossaryentry{pynq}
{
        name=PYNQ,
        description={An open-source platform from AMD that enables FPGA development on Zynq SoCs in the Python ecosystem.}
}

\newglossaryentry{cern}
{
        name=CERN,
        description={\textit{Conseil Européen pour la Recherche Nucléaire}, the European Organization for Nuclear Research.}
}

\newglossaryentry{sm-term}
{
        name=Standard Model,
        description={The theory describing the three out of the four know fundamental forces in the universe and classifying all the known elementary particles in existence.}
}

\newglossaryentry{fermion}
{
        name=Fermion,
        description={Subatomic particles, such as the electron, with half-integer spin and which follow the Fermi--Dirac statistics.}
}

\newglossaryentry{boson}
{
        name=Boson,
        description={Subatomic particles, such as the W boson, with integer spin and which follow Bose--Einstein statistics.}
}

\newglossaryentry{spin}
{
        name=Spin,
        description={An intrinsic form of angular momentum carried by elementary particles, and consequently also by composite ones.}
}

\newglossaryentry{lepton}
{
        name=Lepton,
        description={A fermion that does not experience the strong nuclear force.}
}

\newglossaryentry{hadron}
{
        name=Hadron,
        description={A composite subatomic particle, such as the proton and the neutron, made out of various quarks bound together by the strong nuclear force.}
}

\newglossaryentry{quark}
{
        name=Quark,
        description={A type of elementary particle that is one of the fundamental constituents of composite particles called hadrons---such as the proton and the neutron.}
}

\newglossaryentry{meson}
{
        name=Meson,
        description={A hadronic subatomic particle usually consisting of a quark and an antiquark bound together by the strong nuclear interaction.}
}

\newglossaryentry{baryon}
{
        name=Baryon,
        description={A hadronic subatomic particle, such as the proton and the neutron, usually consisting of three valence quarks.}
}

\newglossaryentry{pdf}
{
        name=Probability density function,
        description={A function that assigns a probability density to each possible value of a given continuous random variable.}
}
\newglossaryentry{iid}
{
        name=Independent and Identically Distributed (IID),
        description={Random variables that have the same probability distribution and are all mutually independent.}
}
\newglossaryentry{tdp}
{
        name=Thermal Design Power (TDP),
        description={The maximum amount of heat a computing device is expected to generate under operation at full capacity.}
}
\newglossaryentry{lip6}
{
        name=LIP6,
        description={\textit{Laboratoire d'Informatique de Sorbonne Université}, computer science lab associated with Sorbonne University and CNRS.}
}
\newglossaryentry{lpnhe}
{
        name=LPNHE,
        description={\textit{Laboratoire de Physique Nucléaire et des Hautes Énergies}, nuclear and high energy physics lab associated with CNRS and IN2P3.}
}
\newglossaryentry{cnrs}
{
        name=CNRS,
        description={\textit{Centre National de la Recherche Scientifique}, the French National Centre for Scientific Research, France’s main public research organization.}
}

\newglossaryentry{in2p3}
{
        name=IN2P3,
        description={\textit{Institut National de Physique Nucléaire et de Physique des Particules}, the coordinating body for nuclear and particle physics in France, a CNRS division.}
}

\newglossaryentry{luminosity}
{
        name=Luminosity,
        description={The number of collisions detected, in a certain period of time, and across a certain cross-sectional area.}
}

\newglossaryentry{reconstructible}
{
        name=Reconstructible,
        description={In LHCb, a particle is considered as reconstructible in the VELO subdetector, if it has at least three hits on the VELO layers.}
}

\newglossaryentry{long}
{
        name=Long tracks,
        description={In LHCb, tracks that have hits in both the VELO and the SciFi subdetectors.}
}

\newglossaryentry{quantization}
{
        name=Quantization,
        description={A method used to lower the computational and memory demands of machine learning model inference by encoding weights and activations using data types with precisions lower than the standard 32- or 64-bit floats.}
}

\newglossaryentry{acceptance}
{
        name=Detector acceptance,
        description={The region of the detector where particles can be detected.}
}

\newglossaryentry{occupancy}
{
        name=Detector occupancy,
        description={The number of activated sensors in a detector during one event.}
}

\newglossaryentry{event}
{
        name=Event,
        description={In accelerator physics, the result of fundamental interactions, typically collisions, from a bunch crossing.}
}

\newglossaryentry{bunch-crossing}
{
        name=Bunch crossing,
        description={The crossing of two particle accelerator beams arranged in bunches and circulating in opposite directions.}
}

\newglossaryentry{pile-up}
{
        name=Pile-up,
        description={The number of proton--proton interactions during a bunch crossing.}
}

\newglossaryentry{block-diagram}
{
        name=Block diagram,
        description={For an FPGA, a high-level schematic representation of the FPGA, showing the main functional components and their interconnections.}
}

\newglossaryentry{memory}
{
        name=Computer memory,
        description={Often synonymous with the term RAM, computer memory stores information temporarily between the processor and the primary storage, and is essential to the functioning of the computer.}
}

\newglossaryentry{ram-gls}
{
        name=Random Access Memory (RAM),
        description={A form of electronic computer memory that can be read and modified in any order.}
}

\newglossaryentry{storage}
{
        name=Computer storage,
        description={Technology comprising computer components and recording media used for long-term retention of digital data.}
}

\newglossaryentry{lut-gls}
{
        name=Lookup Table (LUT),
        description={A lookup table is an array that associates input values with corresponding output values, effectively approximating a mathematical function.}
}

\newglossaryentry{ff-gls}
{
        name=Flip-Flop (FF),
        description={One of the most important components in FPGAs and capable of storing one bit of information, FFs are used to keep track of the state inside the chip.}
}

\newglossaryentry{primary-vertex}
{
        name=Primary vertex,
        description={A point in space where a particle collision occurred, reconstructed from the tracks of particles emerging directly from the collision.}
}

\newglossaryentry{secondary-vertex}
{
    name=Secondary (or displaced) vertex,
    description={A point displaced from the primary vertex, where the decay of a long-lived particle occurred, reconstructed from the tracks of decay products that do not originate from the primary interaction.}
}

\newacronym{lip6-acro}{LIP6}{\textit{Laboratoire d'Informatique de Sorbonne Université}}
\newacronym{lpnhe-acro}{LPNHE}{\textit{Laboratoire de Physique Nucléaire et des Hautes Énergies}}
\newacronym{cern-acro}{CERN}{\textit{Conseil Européen pour la Recherche Nucléaire}}
\newacronym{cuda-acro}{CUDA}{Compute Unified Device Architecture}
\newacronym{sm}{SM}{Standard Model}
\newacronym{bsm}{BSM}{Beyond the Standard Model}
\newacronym{rta}{RTA}{Real-Time Analysis}
\newacronym{ieee}{IEEE}{Institute of Electrical and Electronics Engineers}
\newacronym{ml}{ML}{Machine Learning}
\newacronym{ai}{AI}{Artificial Intelligence}
\newacronym{hep}{HEP}{High-Energy Physics}
\newacronym{gdl4hep}{GDL4HEP}{Geometric Deep Learning for High-Energy Physics}
\newacronym{etx4velo}{ETX4VELO}{Exa.TrkX for VELO}
\newacronym{lhc}{LHC}{Large Hadron Collider}
\newacronym{hl-lhc}{HL-LHC}{High Luminosity Large Hadron Collider}
\newacronym{lhcb}{LHCb}{Large Hadron Collider beauty}
\newacronym{atlas}{ATLAS}{A Toroidal LHC Apparatus}
\newacronym{cms}{CMS}{Compact Muon Solenoid}
\newacronym{alice}{ALICE}{A Large Ion Collider Experiment}
\newacronym{cdf}{CDF}{Collider Detector at Fermilab}
\newacronym{mlp}{MLP}{Multi Layer Perceptron}
\newacronym{gnn}{GNN}{Graph Neural Network}
\newacronym{cpu}{CPU}{Central Processing Unit}
\newacronym{gpu}{GPU}{Graphics Processing Unit}
\newacronym{fpga}{FPGA}{Field-Programmable Gate Array}
\newacronym{asic}{ASIC}{Application-Specific Integrated Circuit}
\newacronym{knn}{k-NN}{k-Nearest Neighbors}
\newacronym{wcc}{WCC}{Weakly Connected Components}
\newacronym{ort}{ORT}{ONNX Runtime}
\newacronym{onnx-acro}{ONNX}{Open Neural Network Exchange}
\newacronym{trt}{TRT}{TensorRT}
\newacronym{velo}{VELO}{Vertex Locator}
\newacronym{ut}{UT}{Upstream Tracker}
\newacronym{scifi}{SciFi}{Scintillating Fiber}
\newacronym{hls4ml}{HLS4ML}{High Level Synthesis for Machine Learning}
\newacronym{hls}{HLS}{High Level Synthesis}
\newacronym{mc}{MC}{Monte Carlo}
\newacronym{relu}{ReLU}{Rectified Linear Unit}
\newacronym{api}{API}{Application Programming Interface}
\newacronym{ip}{IP}{Intellectual Property}
\newacronym{hdl}{HDL}{Hardware Description Language}
\newacronym{rtl}{RTL}{Register Transfer Level}
\newacronym{ps}{PS}{Processing System}
\newacronym{pl}{PL}{Programmable Logic}
\newacronym{ptq}{PTQ}{Post-Training Quantization}
\newacronym{qat}{QAT}{Quantization-Aware Training}
\newacronym{dsp}{DSP}{Digital Signal Processing}
\newacronym{lut}{LUT}{Lookup Table}
\newacronym{ff}{FF}{Flip-Flop}
\newacronym{linac}{LINAC}{Linear Accelerator}
\newacronym{psb}{PSB}{Proton Synchrotron Booster}
\newacronym{ps-cern}{PS}{Proton Synchrotron}
\newacronym{sps}{SPS}{Super Proton Synchrotron}
\newacronym{daq}{DAQ}{Data Acquisition}
\newacronym{dl}{DL}{Deep Learning}
\newacronym{gdl}{GDL}{Geometric Deep Learning}
\newacronym{fnn}{FNN}{Feedforward Neural Network}
\newacronym{cnn}{CNN}{Convolutional Neural Network}
\newacronym{bdt}{BDT}{Boosted Decision Tree}
\newacronym{mle}{MLE}{Maximum Likelihood Estimation}
\newacronym{mse}{MSE}{Mean Square Error}
\newacronym{sgd}{SGD}{Stochastic Gradient Descent}
\newacronym{tdp-acro}{TDP}{Thermal Design Power}
\newacronym{soc}{SoC}{System on a Chip}
\newacronym{ram}{RAM}{Random Access Memory}
\newacronym{esr}{ESR}{Early-Stage Researcher}
\newacronym{msca}{MSCA}{Marie Sk\l{}odowska-Curie Actions}
\newacronym{cp}{CP}{Charge-Conjugation Parity}
\newacronym{pv}{PV}{Primary Vertex}
\newacronym{sv}{SV}{Secondary Vertex}
\newacronym{sisd}{SISD}{Single Instruction Stream, Single Data Stream}
\newacronym{misd}{MISD}{Multiple Instruction Stream, Single Data Stream}
\newacronym{simd}{SIMD}{Single Instruction Stream, Multiple Data Stream}
\newacronym{mimd}{MIMD}{Multiple Instruction Stream, Multiple Data Stream}